\documentclass[oneside,letterpaper,oldfontcommands]{memoir}
\usepackage{./sty/citthesis}

\usepackage{mVersion}
\increaseBuild

\usepackage{textcomp}

\usepackage[T1]{fontenc}
\usepackage[]{lmodern}
\usepackage[sc]{mathpazo}

\usepackage[boxed,linesnumbered]{algorithm2e}
\usepackage{graphicx}
\usepackage{pdflscape}
\usepackage{float}
\usepackage{units}
\usepackage{url}
\usepackage{multirow}
\usepackage[group-separator={,},text-rm]{siunitx}
\usepackage[normalem]{ulem}
\usepackage{color}
\usepackage{setspace}
\usepackage{gensymb} 

\makeatletter
\def\bstctlcite{\@ifnextchar[{\@bstctlcite}{\@bstctlcite[@auxout]}}
\def\@bstctlcite[#1]#2{\@bsphack
  \@for\@citeb:=#2\do{%
    \edef\@citeb{\expandafter\@firstofone\@citeb}%
    \if@filesw\immediate\write\csname #1\endcsname{\string\citation{\@citeb}}\fi}%
  \@esphack}
\makeatother

\DeclareSIUnit{\bit}{b}
\DeclareSIUnit{\month}{month}
\DeclareSIUnit{\months}{months}
\DeclareSIUnit{\day}{day}

\usepackage{pifont} 
\usepackage{tikz} 
\newcommand*\circled[1]{\tikz[baseline=-0.75ex]{
  \node[shape=circle,draw,inner sep=0.25pt] (char) {\textsf{\tiny #1}};}}

\usepackage{tipa}     

\usepackage{stmaryrd}

\usepackage[colorlinks,linkcolor=blue,filecolor=blue,citecolor=blue,urlcolor=blue,backref=page]{hyperref}

\newsubfloat{figure}

\setauthor{Justin J. Meza}
\settitle{Large Scale Studies of Memory, Storage, and Network Failures \\ in a Modern Data Center}
\doctors
\setchair{Dr.\ Onur Mutlu}
\setdept{Electrical and Computer Engineering}
\setdegrees{B.S., Computer Science, University of California at Los Angeles\\
M.S., Electrical and Computer Engineering, Carnegie Mellon University}
\setdefdate{May 2018}
\setgraddate{December 2018}
\setcopyyear{2010--2018}

\newcommand{\onur}[2]{#2}
\newcommand{\more}[2]{#2}
\newcommand{\also}[2]{#2}
\newcommand{\change}[2]{#2}

\newcommand{\squishlist}{
 \begin{list}{$\bullet$}
  { \setlength{\itemsep}{0pt}
     \setlength{\parsep}{1pt}
     \setlength{\topsep}{0pt}
     \setlength{\partopsep}{0pt}
     \setlength{\leftmargin}{1.0em}
     \setlength{\labelwidth}{1em}
     \setlength{\labelsep}{0.5em} } }

\newcounter{Lcount}
\newcommand{\squishlisttwo}{
\begin{list}{\arabic{Lcount}. }
{ \usecounter{Lcount}
\setlength{\itemsep}{0pt}
\setlength{\parsep}{0pt}
\setlength{\topsep}{0pt}
\setlength{\partopsep}{0pt}
\setlength{\leftmargin}{2em}
\setlength{\labelwidth}{1.5em}
\setlength{\labelsep}{0.5em} } }

\newcommand{\squishend}{
  \end{list}  }

\begin{document}

\bstctlcite{IEEEexample:BSTcontrol}

\frontmatter

\thetitlepage
\copyrightpage

\section*{Acknowledgements}

\vspace{1em}

\noindent
I gratefully acknowledge:

\vspace{1em}

\sl

\noindent
My advisor, Onur Mutlu (who also chaired my doctoral committee),

\noindent
\hspace{2em}who had confidence in me even when I did not.

\vspace{1em}

\noindent
My doctoral committee---Greg Ganger, James Hoe, and Kaushik Veeraraghavan---

\noindent
\hspace{2em}who were always there to listen to me and guide me.

\vspace{1em}

\noindent
The SAFARI group at CMU,

\noindent
\hspace{2em}for lifelong friendships.

\vspace{1em}

\noindent
My family, friends, and colleagues (of which there are too many to list!)

\noindent
\hspace{2em}who kept me going.

\vspace{2em}

\rm
\noindent
I also acknowledge generous partial support throughout my PhD from:

\begin{itemize}

\item National Science Foundation grants 0953246, 1065112, 1147397, 1212962, 1320531, 1409723, and 1423172; Gigascale Systems Research Center; Intel Corporation URO Memory Hierarchy Program; Intel Science and Technology Center for Cloud Computing; Semiconductor Research Corporation; Carnegie Mellon CyLab.

\item Equipment and gift support from AMD, Facebook, Google, Hewlett-Packard Labs, IBM, Intel, Qualcomm, Samsung, Oracle, and VMware.

\end{itemize}

\newpage
\section*{Abstract}

The workloads running in the modern data centers of large scale Internet service providers (such as Alibaba, Amazon, Baidu, Facebook, Google, and Microsoft) support billions of users and span globally distributed infrastructure. Yet, the devices used in modern data centers fail due to a variety of causes, from faulty components to bugs to misconfiguration. Faulty devices make operating large scale data centers challenging because the workloads running in modern data centers consist of interdependent programs distributed across many servers, so failures that are isolated to a \emph{single} device can still have a \emph{widespread} effect on a workload.

In this dissertation, we measure and model the device failures in a large scale Internet service company, Facebook. We focus on three device types that form the foundation of Internet service data center infrastructure: DRAM for main memory, SSDs for persistent storage, and switches and backbone links for network connectivity. For each of these device types, we analyze long term device failure data broken down by important device attributes and operating conditions, such as age, vendor, and workload. We also build and release statistical models of the failure trends for the devices we analyze.

\textbf{\emph{For DRAM devices,}} we analyze the memory errors in the \emph{entire fleet} of servers at Facebook over the course of fourteen months, representing billions of device days of operation. The systems we examine cover a wide range of devices commonly used in modern servers, with DIMMs that use the modern DDR3 communication protocol, manufactured by 4 vendors in capacities ranging from 2GB to 24GB.  
We observe several new reliability trends for memory systems that have not been discussed before in literature, develop a model for memory reliability, show how system design choices such as using lower density DIMMs and fewer cores per chip can reduce failure rates of a baseline server by up to 57.7\%. We perform the first implementation and real-system analysis of page offlining at scale, on a cluster of thousands of servers, identify several real-world impediments to the technique, and show that it can reduce memory error rate by 67\%. We also examine the efficacy of a new technique to reduce DRAM faults, \emph{physical page randomization}, and examine its potential for improving reliability and its overheads.

\textbf{\emph{For SSD devices,}} we perform a large scale study of flash-based SSD reliability at Facebook. We analyze data collected across a majority of flash-based solid state drives over nearly four years and many millions of operational hours in order to understand failure properties and trends of flash-based SSDs. Our study considers a variety of SSD characteristics, including: the amount of data written to and read from flash chips; how data is mapped within the SSD address space; the amount of data copied, erased, and discarded by the flash controller; and flash board temperature and bus power.  
Based on our field analysis of how flash memory errors manifest when running modern workloads on modern SSDs, we make several major observations and find that SSD failure rates \emph{do not} increase monotonically with flash chip wear, but instead they go through several distinct periods corresponding to how failures emerge and are subsequently detected.

\textbf{\emph{For network devices,}} we perform a large scale, longitudinal study of data center network reliability based on operational data collected from the production network infrastructure at Facebook. Our study covers reliability characteristics of both intra and inter data center networks. For intra data center networks, we study seven years of operation data comprising thousands of network incidents across two different data center network designs, a cluster network design and a state-of-the-art fabric network design. For inter data center networks, we study eighteen months of recent repair tickets from the field to understand the reliability of Wide Area Network (WAN) backbones. In contrast to prior work, we study the effects of network reliability on software systems, and how these reliability characteristics evolve over time. We discuss the implications of network reliability on the design, implementation, and operation of large scale data center systems and how the network affects highly-available web services.

Our key conclusion in this dissertation is that we can gain a deep understanding of why devices fail---and how to predict their failure---using measurement and modeling. We hope that the analysis, techniques, and models we present in this dissertation will enable the community to better measure, understand, and prepare for the hardware reliability challenges we face in the future.

\newpage

\tableofcontents

\newpage

\listoftables

\newpage

\listoffigures

\mainmatter

\chapter{Introduction}
\label{sec:introduction}

We introduce why device failures in modern data centers are a problem and propose our solution for tolerating device failures better. We also state the thesis of this dissertation, our contributions and how we organize this dissertation.

\section{The Problem: Device Failures Affect the Workloads Running in Data Centers}

When we write a program, we do not think about how a hardware failure in our computer may cause our program to behave incorrectly. This allows us to write simpler programs, because we do not need to account for the many ways that hardware devices may fail. We can write programs this way because hardware manufacturers work hard to ensure devices fail infrequently.

Similarly, the engineers who write the programs running in modern data centers can also write simpler programs by not worrying about device failures. However, the programs running in modern data centers for large scale Internet service providers (such as Amazon, Baidu, Facebook, Google, and Microsoft) have three properties that make device failures problematic:

\begin{description}

\item[Property 1 (Interdependence)] \textbf{\emph{The programs running in modern data centers make up larger workloads.\footnote{We call the collection of programs that communicate to perform some work a \emph{workload}.}}} The programs in a workload share state (e.g., data stored in caches or databases), and thus \emph{depend} on one another. Because the programs in a workload are interdependent, a failure that affects one program in a workload can propagate (e.g., through shared memory, message passing, or network communication) and affect other programs in the workload.

\item[Property 2 (Distribution)] \textbf{\emph{The workloads running in modern data centers are distributed across many servers.\footnote{We call a specially-designed computer running in a data center a \emph{server}. We discuss the architecture of the servers we examine in this dissertation in \S\ref{subsec:Data Center Server Design}.}}} Distribution is an important property of a data center workload that allows the workload to satisfy its need for high availability and high scalability. Because distribution involves spreading a workload across many (potentially tens of thousands of) servers, distribution also increases the likelihood of one program in a workload to experience a device failure.

\item[Property 3 (Commodity hardware)] \textbf{\emph{The servers running in modern data centers consist of commodity hardware.}} Unlike some other computing environments, such as transaction processing mainframes and high performance computing, modern data center designers do \emph{not} try to protect all devices from failures. This allows modern data centers to use simpler, commodity server hardware---while trading off some amount of reliability.

\end{description}

Together, these three properties lead to a fundamental problem: \textbf{\emph{Because the workloads running in modern data centers consist of interdependent programs distributed across many servers with unknown reliability, failures that are isolated to a single device can have a widespread effect on a workload.}}

\section{The Solution: Measure and Model Device Failures to Better Tolerate Them}

In this dissertation, we seek to \textbf{\emph{understand how the devices in modern data centers fail, so that we can design solutions to prevent device failures from becoming workload failures.}} We approach our solution in three steps. First, we build tools, design techniques, and perform field studies to \emph{measure} device failures in large scale data centers, specifically focusing on Facebook's data centers which we have access to. Second, we use the data that we collect on device failures in such data centers to \emph{model} how and when devices might fail again. Third, we use our measurements and models to recommend best practices for data center operators to tolerate device failures, and examine some solution ideas.


\section{Thesis Statement}

Our thesis in this dissertation is, \textbf{\emph{if we measure the device failures in modern data centers, then we can learn the reasons why devices fail, develop models to predict device failures, and learn from failure trends to make recommendations to enable workloads to tolerate device failures.}}

\newpage

\section{Contributions}

We make the following major contributions in this dissertation:

\begin{enumerate}

  \item We measure the memory (DRAM), storage (SSD), and network failures across the entire fleet of servers in the modern data centers at Facebook, and examine how the failures affect the systems running on them. Our study covers many millions of operational device-hours across fourteen months of DRAM failures, four years of SSD failures, seven years of switch failures. While we \emph{do not} measure how other devices in data centers fail (such as CPUs, GPUs, or emerging non-volatile memory), we \emph{do} describe methodologies for future work to perform studies similar to ours, but for other devices.

\item Using our measurements, we develop and provide models for DRAM, SSD, and network failures. For DRAM devices, we provide a tutorial on how data center hardware designers can use our models to build more failure-tolerant servers by comparing the relative failure rate of servers using different types of DRAM devices.

\item We use the reliability trends that we observe across the devices in Facebook's fleet of servers to provide insights into how data center operators can better tolerate such failures. We examine two fully-implemented solutions to tolerate DRAM failures at Facebook, \emph{page-offlining} and \emph{physical page randomization}, propose best practices for SSD and network device failures, and identify new areas for future research in better tolerating data center device failures.

\end{enumerate}

We also make additional contributions in our field studies of modern data center device failures, which we comprehensively summarize next.

\subsection{DRAM Device Failure Contributions}

In Chapter~\ref{chp:dramfailures}, which is an expanded version of our study~\cite{meza15-2} entitled, ``Revisiting memory errors in large-scale production data centers: analysis and modeling of new trends from the field,'' we examine all of the DRAM failures in Facebook's fleet of servers over fourteen months. We make the following major contributions:

\begin{itemize}

\item Memory errors follow a power-law distribution, specifically, a Pareto distribution with decreasing hazard rate, with average error rate exceeding median error rate by around $55\times$. [\S\ref{sec:incidence}]

\item Non-DRAM memory failures from the memory controller and memory channel contribute the majority of errors and create a kind of \emph{denial of service attack} in servers. [\S\ref{sec:component}]

\item More recent DRAM cell fabrication technologies (as indicated by chip density) show higher failure rates (prior work that measured DRAM \emph{capacity}, which is not closely related to fabrication technology, observed inconclusive trends). [\S\ref{subsec:DIMM Capacity and DRAM Density}]

\item DIMM architecture decisions affect memory reliability: DIMMs with fewer chips and lower transfer widths have the lowest error rates, likely due to their lower electrical noise. [\S\ref{subsec:DIMM Architecture}]

\item While CPU and memory utilization do not show clear trends with respect to failure rates, workload type can influence server failure rate by up to $6.5\times$. [\S\ref{sec:workload}]

\item We develop a model for memory failures and show how system design choices such as using lower density DIMMs and fewer processors can reduce failure rates of baseline servers by up to 57.7\%.
	
\item We perform the first analysis of page offlining in a real-world environment, showing that error rate can be reduced by around 67\%. We also identify and fix several challenges that others might face if they implement page offlining at scale.

\item We evaluate the efficacy of a new technique to reduce DRAM faults, \emph{physical page randomization}, and examine its potential for improving reliability and its overhead.

\end{itemize}

\subsection{SSD Device Failure Contributions}

In Chapter~\ref{chp:ssdfailures}, which is an expanded version of our study~\cite{meza15} entitled, ``A large-scale study of flash memory errors in the field,'' we examine all of the SSD failures in Facebook's fleet of servers across a variety of internal and external characteristics and examine how these characteristics affect the trends for uncorrectable errors. We make the following major contributions:

\begin{itemize}

\item We observe that SSDs go through several distinct periods---early detection, early failure, usable life, and wearout---with respect to the factors related to the amount of data written to flash chips. Due to pools of flash blocks with different reliability characteristics, failure rate in a population does not monotonically increase with respect to amount of data written.  This is unlike the failure rate trends seen in raw flash chips. [\S\ref{sec:corr}]
  
\item We must design techniques to help reduce or tolerate errors \emph{throughout} SSD operation, not only toward the end of life of the SSD. For example, additional error correction at the beginning of an SSD's life could help reduce the failure rates we see during the \emph{early detection} period. [\S\ref{sec:corr}]

\item  We find that the effect of read disturbance errors is not a predominant source of errors in the SSDs that we examine. While prior work has shown that such errors can occur under certain access patterns in controlled environments~\cite{brand-irps93, mielke-irps08, cai-date12, cai-dsn15}, we do {\em not} observe this effect across the SSDs we examine.  This corroborates prior work which showed that the effect of write errors in flash cells dominate error rate compared to read disturbance~\cite{mielke-irps08, cai-date12}.  It may be beneficial to perform a more detailed study of the effect of disturbance errors in flash-based SSDs used in servers. [\S\ref{sec:read}]

\item Sparse logical data layout across an SSD's physical address space (e.g., non-contiguous data) greatly affects SSD failure rates; dense logical data layout (e.g., contiguous data) can also negatively impact reliability under certain conditions, likely due to adversarial access patterns. [\S\ref{subsec:DRAM Buffer Usage}]

\item Further research into flash write coalescing policies with information from the system level may help improve SSD reliability.  For example, information about write access patterns from the operating system could potentially inform SSD controllers of non-contiguous data that is accessed most frequently, which may be one type of data that adversely affects SSD reliability and is a candidate for storing in a separate write buffer. [\S\ref{subsec:DRAM Buffer Usage}]

\item Higher temperatures lead to increased failure rates, but do so most noticeably for SSDs that do not employ throttling techniques.  We find techniques like \emph{throttling}, which help reduce SSD temperature, to be effective at reducing the failure rate of SSDs.  We also find that SSD temperature is correlated with the power used to transmit data across the PCIe bus that connects the SSD to the server's central processors. Power can thus potentially be used as a proxy for temperature in the absence of SSD temperature sensors. [\S\ref{sec:external}]

\item We find that the amount of data written by the system software overstates the amount of data written to flash cells. This is because the operating system and SSD controller buffer certain data, so not every write in the system software translates to a write to SSD cells. Simply reducing the rate of software-level writes without considering the qualities of the write access pattern to system software is not sufficient for improving SSD reliability.  Studies seeking to model the effects of reducing software-level writes on flash reliability should also consider how other aspects of SSD operation, such as system-level buffering and SSD controller wear leveling, affect the actual data written to SSDs. [\S\ref{subsec:Data Written by the System Software}]

\end{itemize}

\subsection{Network Device Failure Contributions}

In Chapter~\ref{chp:networkfailures}, which is an extended version of our study~\cite{Meza17} entitled, ``A large scale study of data center network reliability,'' we examine seven years of network failures \emph{within data centers} and eighteen months of network failures \emph{between data centers}, focusing on how network failures lead to software-level \emph{incidents} that affect the workloads in Facebook data centers. We make the following major contributions:

\begin{itemize}

  \item We observe that most failures that software cannot repair involve maintenance, faulty hardware, and misconfiguration. We find that humans cause $2\times$ more errors than hardware. We attribute this trend to the increasing complexity required for humans to operate and maintain the network devices in modern data centers. [\S\ref{sec:root-causes}]

  \item When network devices with higher bandwidth fail, the network devices have a higher likelihood of affecting software systems. Network devices built from commodity chips have much lower software incident rates compared to devices from third-party vendors due to the devices' integration with automated failover and remediation software. Software incidents due to rack switches are increasing over time and are currently around 28\% of all incidents. [\S\ref{sec:incident_rate}]

  \item Although high bandwidth \emph{core} network devices have the most software incidents, the incidents they have are low severity. Newer fabric network devices that run in modern data centers cause software incidents of lower severity than older cluster network devices. [\S\ref{sec:incident_impact}]

  \item Incidents from the older cluster network design increased steadily over time until the adoption of newer fabric networks, with cluster networks currently having $2.8\times$ the software incidents compared to fabric networks. [\S\ref{sec:incidents-by-topology}]

  \item While high reliability is essential for widely-deployed devices, such as rack switches, software incident rates vary by three orders of magnitude across device types. Larger networks tend to have longer incident remediation times. [\S\ref{sec:mtbi}]

  \item We develop models for the reliability of Facebook's Wide Area Network (WAN), which consists of a diverse set of network edge nodes and links that form a backbone. We find that time to failure and time to repair closely follow \emph{exponential} functions. We provide models for these phenomena so that future studies can build on our models and use them to understand the nature of backbone failures. [\S\ref{sec:mean-time-between-failures}--\S\ref{sec:reliability_fiber_vendor}]

  \item Backbone edge nodes that convey traffic between data centers fail on the order of months and recover on the order of hours. However, there is high variance in edge node failure rate and recovery rate.  Path diversity in the backbone network topology ensures that large scale networks can tolerate failures with long repair times. [\S\ref{sec:mean-time-between-failures}]

  \item Links supplied by backbone vendors typically fail on the order of months, with links in big cities failing less frequently. Both failure rate and recovery rate for links span multiple orders of magnitude among vendors. [\S\ref{sec:reliability_fiber_vendor}]

  \item Backbone edge node failure rate varies by months across continents in the world. Edges nodes recover within \SI{1}{\day} on average on all continents. [\S\ref{sec:reliability_geo_location}]

\end{itemize}

\section{Dissertation Organization}

This dissertation is organized in seven chapters. In Chapter~\ref{chp:backgroundrelated}, we provide a background on modern large scale data center design; describe the fundamental reasons why DRAM, SSD, and network hardware fails; and related work in these fields. In Chapters~\ref{chp:dramfailures} to \ref{chp:networkfailures}, we analyze years of DRAM, SSD, and network device failures across Facebook's entire data center infrastructure. In each of these chapters, we describe how to measure device failures. We also discuss what we learned about hardware failures in Facebook's data centers. We discuss the lessons learned from our studies in Chapter~\ref{sec:Lessons Learned Across Many Billions of Device-Hours}. We conclude the dissertation in Chapter~\ref{chp:conclusions} and discuss future research directions.

\chapter{Background and Related Research}
\label{chp:backgroundrelated}

Our study focuses on the devices and systems running in a modern data center, so, we begin by discussing modern data center architecture and server design, using Facebook's data centers as an example (\S\ref{sec:Modern Data Center Design}). We then discuss the systems that we examine and the fundamental reasons why their DRAM, SSD, and network devices fail (\S\ref{sec:DRAM Devices}, \S\ref{sec:SSD Devices}, \S\ref{sec:Network Devices}). Our study sheds light on new causes and effects of data center device failure (\S\ref{subsec:How DRAM Devices Fail}, \S\ref{subsec:How SSD Devices Fail}, \S\ref{subsec:How Network Devices Fail}), and to put it into perspective, we also discuss related studies of DRAM, SSD, and network device failures in data centers (\S\ref{subsec:Related Research in DRAM Failures in Modern Data Centers}, \S\ref{subsec:Related Research in SSD Failures in Modern Data Centers}, \S\ref{subsec:Related Research in Network Failures in Modern Data Centers}).

We distinguish between \emph{failures} (also called \emph{faults}) and \emph{errors}: Failures and faults are the underlying reason why a device malfunctions and errors are how failures and faults manifest in software that uses a malfunctioning device. A hard, or permanent, fault is present all the time: when the faulty component is used, it causes an error. A soft, or transient/intermittent, fault appears intermittently (i.e., is not present all the time): it causes an error when it is present \emph{and} when the faulty component is used. Due to error correction it is possible for a failure/fault to occur but not cause an error in the software. Our studies typically examine failures and errors within the devices we examine.

\section{Modern Data Center Design}
\label{sec:Modern Data Center Design}

Modern web services comprise software systems running in multiple data centers that cater to a global user base. At this scale, it is important to use all available data center resources as efficiently as possible. Effective resource utilization is challenging due to three reasons:

\begin{enumerate}

\item \textbf{\emph{Evolving workloads:}} The workload of a web service constantly changes as its user base grows and new products are launched. Further, individual software systems might be updated several times a day~\cite{FacebookRelease2012} or even continually~\cite{QuoraRelease2013}. While modeling tools~\cite{Gmach:2009:RPM, Liu2005, Urgaonkar:2008, XiaoyunZhu:Islands} can estimate the initial capacity needs of a system, an evolving workload can quickly render models obsolete.

\item \textbf{\emph{Infrastructure heterogeneity:}} Constructing data centers at different points in time leads to a variety of networking topologies, different generations of hardware, and other physical constraints in each location that each affect how systems scale.

\item \textbf{\emph{Changing bottlenecks:}} Each data center runs hundreds of software systems with complex interactions that exhibit \emph{resource utilization bottlenecks}, including issues arising from performance regressions, load imbalance, and resource exhaustion, at a variety of scales, from single servers to entire data centers. The sheer size of the system makes it challenging to understand all the components.  In addition, these systems change over time, leading to different bottlenecks presenting themselves at different points in time.

\end{enumerate}

To improve data center utilization, Internet services companies like Amazon, Baidu, Facebook, Google and Microsoft build data centers in different geographic locations and geographically replicate resources (like systems and data). Geographic replication provides flexibility in how requests to a data center can be routed and ensures people in different parts of the Earth can access a nearby data center with low latency. In order to get to one of Facebook's servers, a request takes a journey through several layers of systems. We next describe how a request gets to a Facebook server (\S\ref{subsec:how a request gets to a data center}), what types of workloads the request might be a part of (\S\ref{subsec:data center workloads}), and the design of the server the request ultimately runs on (\S\ref{subsec:Data Center Server Design}).

\subsection{How a Request Gets to a Server}
\label{subsec:how a request gets to a data center}

\begin{figure}[H]
\centering
\includegraphics[width=0.75\columnwidth]{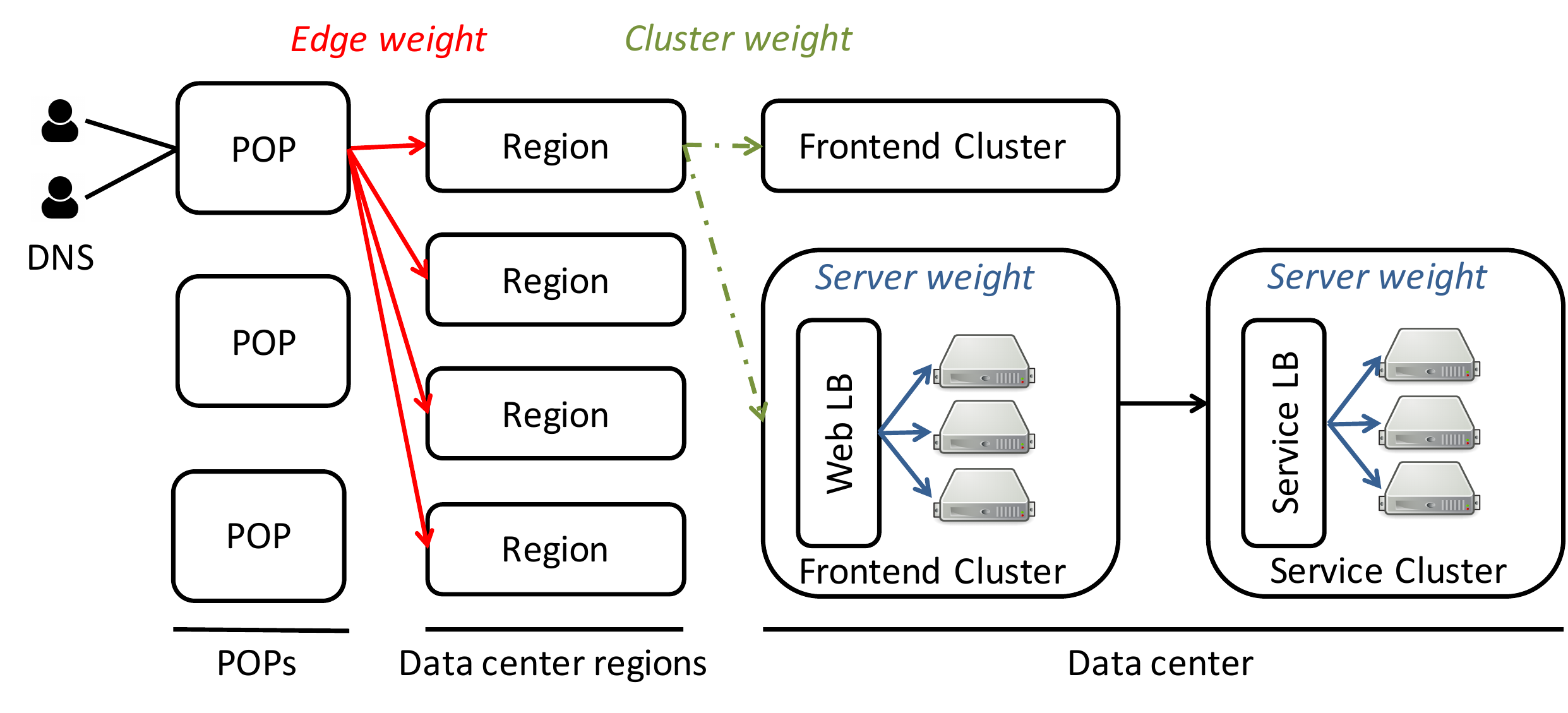}
\caption{An overview of the traffic management components at Facebook. User requests arrive at Edge POPs (points-of-presence). A series of weights defined at the edge, cluster, and server levels are used to route user requests from a POP to a web server in one of Facebook's data centers.}
\label{fig:routing}
\end{figure}

Figure~\ref{fig:routing} provides an overview of how a user request to Facebook is served. The user's request is sent to their \emph{Internet Service Provider (ISP),} which contacts a \emph{Domain Name System (DNS)} resolver to map the \emph{Uniform Resource Link (URL)}, \texttt{facebook.com}, to an \emph{Internet Protocol (IP)} address. This IP address maps to one of tens of edge \emph{Point-Of-Presence (POP)} locations distributed worldwide. A POP consists of a small number of servers on the edge of the network typically located with a local ISP. The user's \emph{Secure Socket Layer (SSL)} session is terminated in a POP at a load balancer that handles packets at the network transport layer, which then forwards the request to a cluster of servers within a data center region. Once a request reaches a cluster, a network application layer load balancer sends the request to to a particular server. The server produces a response that is sent back to the user.

As an example, Facebook groups 1 to 3 data centers in close proximity into a ``region''. Within each data center, Facebook groups machines logical clusters of servers: frontend clusters contain web servers (shown in Figure~\ref{fig:routing}), backend clusters contain storage systems, and ``service'' clusters contain different types of services. We define a ``service'' as a set of systems that provide a particular product either internally within Facebook's infrastructure or externally to users. Each cluster has a few thousand generally heterogeneous machines. Many services span clusters, but web server deployments are confined to a single cluster.

As Figure~\ref{fig:routing} shows, the particular frontend cluster that a request is routed to depends on two factors: (1) the weight used to route traffic from an edge node (also called a POP), to a region (the \emph{edge weight}), and (2) the weight used to route traffic within clusters in a region (the \emph{cluster weight}). To understand why Facebook needs edge weights, consider a request from a user in Hamburg that is terminated at a hypothetical POP in Europe. This POP might prefer forwarding user requests to the Lule{\aa}, Sweden region rather than Forest City, North Carolina, USA to minimize latency, implying that the European POP could assign a higher edge weight to Lule{\aa} than Forest City. A data center might house multiple frontend clusters with machines from different hardware generations. The capability of an Intel\textsuperscript{\textregistered} Skylake cluster will exceed that of an Intel\textsuperscript{\textregistered} Haswell cluster, resulting in differing cluster weights as well as individual servers being assigned different \emph{server weights}.



\subsection{Data Center Workloads}
\label{subsec:data center workloads}

The workloads in Facebook data centers perform a diverse set of operations including web serving, caching~\cite{Nishtala2013}, database management~\cite{datainfra}, video and image processing/storage~\cite{haystack, f4}, and messaging routing/storage.  The resource requirements in Table~\ref{tab:workloads} refer to the relative number of processor cores, memory capacity, and storage capacity for servers for each type of workload.

\begin{table}[H]
  \centering
  \begin{tabular}{|c||c|c|c|} \hline
    \multirow{2}{*}{Workload} & \multicolumn{3}{c|}{Resource requirements} \\ \cline{2-4}
    & Processor & Memory & Storage \\ \hline \hline
    Web & High & Low & Low \\ \hline
    Hadoop~\cite{hadoop} & High & Medium & High \\ \hline
    Ingest~\cite{datainfra} & High & High & Medium \\ \hline
    Database~\cite{datainfra} & Medium & High & High \\ \hline
    Memcache~\cite{Nishtala2013} & Low & High & Low \\ \hline
    Media~\cite{haystack} & Low & Low & High \\ \hline
  \end{tabular}
  \caption{The workloads we examine and their resource requirements.}
  \label{tab:workloads}
\end{table}

Each server runs a single type of workload to eliminate any contention for shared resources among multiple workloads. All the servers configured for a particular workload type have equivalent minimum capabilities and a workload can be run on any of them.

\subsection{Data Center Server Design}
\label{subsec:Data Center Server Design}

\begin{figure}[H]
\centering
\includegraphics[width=1.0\columnwidth]{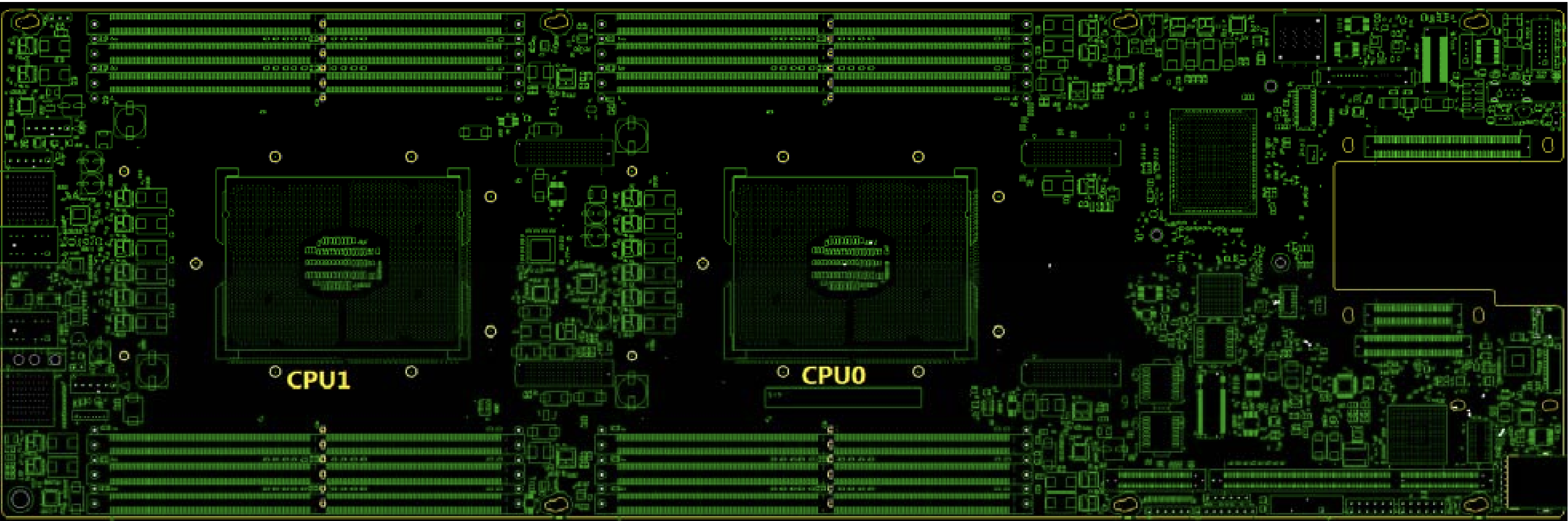}
  \caption{A board layout for a representative Facebook server~\cite{tioga-pass} from the Open Compute Project~\cite{open-compute}.}
\label{fig:tioga-pass}
\end{figure}

Facebook has published the detailed specifications for their base server platform as part of the Open Compute Project~\cite{open-compute}. For example, Figure~\ref{fig:tioga-pass} shows a board layout for a recent generation of server design called ``Tioga Pass''~\cite{tioga-pass}.

We next describe the compute, memory, storage, and network specification for a Tioga Pass server. While we use Tioga Pass as an example, all Facebook servers follow roughly the same organization, with differences in the amount of compute, memory, and storage, depending on their workload (Table~\ref{tab:workloads}).

\subsubsection{Compute}

Tioga Pass measures 6.5 inches by 20 inches and has two \emph{Central Processing Unit (CPU)} sockets. Each CPU socket supports an Intel\textsuperscript{\textregistered} Xeon\textsuperscript{\textregistered} processor~\cite{IntelXeon} (which can have many physical cores per CPU). The CPU sockets are connected via two Intel\textsuperscript{\textregistered} UPI~\cite{IntelUPI} links, which provides a fast socket-to-socket interconnect.

\subsubsection{Memory}

A Tioga Pass server supports six \emph{Dynamic Random Access Memory (DRAM)} channels per CPU socket, for a total of twelve \emph{Dual Inline Memory Modules (DIMMs)}. Communication between the CPU and DIMMs occurs using the \emph{Double Data Rate Fourth Generation (DDR4)} memory protocol~\cite{jedec-ddr4} at up to 2666 MHz frequency. The maximum memory size of 768 GB~\cite{tioga-pass}.

\subsubsection{Storage}

By default, a Tioga Pass server uses a \emph{Hard Disk Drive (HDD)} attached via \emph{Serial AT Attachment (SATA)}~\cite{SATA} for persistent storage. However, a Tioga Pass server can also communicate with \emph{Solid State Drives (SSDs)} attached to the \emph{Peripheral Component Interconnect Express (PCIe)} bus~\cite{PCI}. Only servers that run workloads with large amount of storage \emph{Input/Output (I/O)} are configured to use SSDs.

\subsubsection{Network}

To communicate with other servers, a Tioga Pass server connects to a \emph{Top-Of-Rack (TOR)} switch~\cite{Roy2015} using an expansion card that supports a 40 Gbps \emph{Quad Small Form-factor Pluggable (QSFP)} connection~\cite{mezzanine}. This connection provides network connectivity to other devices in the data center. Routing occurs using the \emph{Border Gateway Protocol (BGP)}.

\section{DRAM Devices}
\label{sec:DRAM Devices}

Computing systems store a variety of data in main memory---program variables, operating system and file system structures, program binaries, and so on. The data stored in main memory is directly accessible via load and store instructions from the programs.  The main memory in modern systems is composed of dynamic random-access memory (DRAM), a technology that, from the programmer's perspective, has the following semantic:  A byte written to an address can be read, repeatedly, until it is overwritten or the machine is turned off.  Indeed, DRAM manufacturers work hard to design reliable devices that obey this semantic, and the International Technology Roadmap for Semiconductors (ITRS) suggests a nominal DRAM cell lifetime of $\mathrm{3\times10^{16}}$ write cycles before failure~\cite{itrs}, or around 47.5 years if accessing data as fast as possible\footnote{The fastest possible write pattern to a DRAM cell is dictated by the latency to change a row, {\em tRC}, in the DDR specifications, with a value of $\sim$\unit[50]{ns} in DDR3~\cite{jedec}, hence, $\mathrm{3\times10^{16}}\times\unit[\mathrm{50}]{ns} \approx \unit[\mathrm{47.5}]{years}$.}.  All correct programs expect to read the same data that was written to a location in DRAM.

\subsection{DRAM Device Architecture}

\begin{figure}[H]
\centering
\includegraphics[width=0.65\columnwidth]{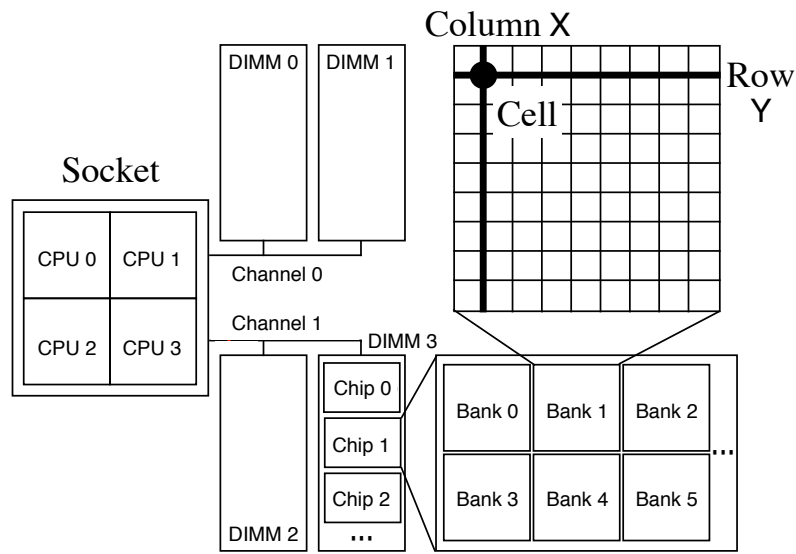}
\caption{Server DRAM architecture.}
\label{fig:memorg}
\end{figure}

Figure~\ref{fig:memorg} shows how memory is organized in a server running in a data center. Modern servers have one or two processor chips that are connected to DRAM via several memory {\it channels} that are operated in parallel.  Attached to each channel are {\em dual in-line memory modules (DIMMs)} \change{which}{that} provide an interface for accessing data stored across multiple DRAM chips.  Processors use the {\em double data rate (DDR)} protocol to communicate with DIMMs.  Chips on a DIMM are logically organized into {\em ranks} and chips within a rank are operated in lockstep.  Chips contain multiple {\em banks} (typically 8 to 16) that are operated in parallel.  Each bank is organized into {\em rows} (typically \unit[16]{K} to \unit[64]{K}) and {\em columns} (each column has a single cell of data in a row, typically \unit[2]{K} to \unit[4]{K} cells total).  At the intersection of a row and a column is a DRAM {\em cell}, which stores a single bit of data. For more detail on DRAM device organization, we refer the reader to~\cite{Mutlu2007, Mutlu2008, salp, Lee2015, tldram, rowhammer, aldram}.

A DRAM cell consists of a capacitor and an access transistor. The access transistor connects the cell to a \emph{word line}, which selects a row of cells, and to a \emph{bit line}, which connects a cell to a \emph{sense amplifier}. The sense amplifier either stores charge in a cell (writes to it) or senses and amplifies the charge stored in a cell (reads from it). For more information on various aspects of DRAM microarchitecture and design, please refer to \cite{Kim2018, Wang2018, Ghose2018, Das2018, Kim2018-2, Seshadri2017, Yu2017, Patel2017, Chang2017, Chang2017-2, Kim2015, Seshadri2016, Luo2018, Lee2016, Lee2016-2, Lee2017, Hassan2017, Khan2016-2, Khan2016-3, Chang2014, Seshadri2013, Kim2010, Chang2016, Chang2016-2, Hassan2016, Seshadri2015, Lee2015, Qureshi2015, salp, tldram, raidr, aldram}.

\subsection{How DRAM Devices Fail}
\label{subsec:How DRAM Devices Fail}

We discuss various device-level failure mechanisms in DRAM including retention failures, disturbance failures, and endurance failures.

\subsubsection{Retention Failures}
\label{subsubsec:Retention Failures}

Over time, charge leaks through a DRAM cell's access transistor, and without intervention, the DRAM cell loses its stored data. To prevent a DRAM cell from losing its data, the charge in its capacitor must be periodically \emph{refreshed}. The refresh interval varies between cells due to manufacturing process variation, with smaller cell sizes making cells more susceptible to failure~\cite{Frank2001, Pellegrini2016, Hamamoto1998, Yamaguchi2000, liu}. While DRAM device manufacturers specify a 64~ms refresh interval, past works have shown the vast majority of cells retain charge for over 256~ms~\cite{Kim2003}. Liu and Khan used this insight to design higher performance and more energy efficient DRAM devices~\cite{raidr, samira, Khan2016-3, Khan2016-2, Khan2017-2}.

To complicate matters, DRAM cells \emph{vary} in how long they store data in a time-dependent manner~\cite{Yaney1987, liu}. These \emph{variable retention time} cells are hard to classify during chip manufacturing because they are affected by the high temperature packaging process~\cite{Seo2002, Kim2004, liu}. Luckily, recent works help devices tolerate and leverage cell failures due to variable retention time~\cite{salp, yixin, Qureshi2015, Patel2017, samira, aldram, Khan2016-3, Khan2016-2, Khan2017-2, Kim2019}. A DRAM cell's retention time also depends on ambient temperature, decreasing retention time exponentially as ambient temperature increases~\cite{Hamamoto1998, liu, Patel2017}.

\subsubsection{Disturbance Failures}
\label{subsubsec:Disturbance Errors}

Even when a DRAM device refreshes its cells frequently enough, external events can \emph{disturb} the contents of DRAM cells. For example, software can disturb a DRAM device's contents.

Fundamentally, software-induced disturbance occurs due to electrical interference between smaller DRAM cells placed closer together~\cite{samsung-intel, Khan2016-2, samira, Kim2005, rowhammer, liu, raidr, Mandelman2002, Mueller2005, Mutlu2017, Mutlu2013, Mutlu2014, Nakagome1988, Patel2017, Redeker2002, Mutlu2013, Mutlu2014, memcon}. Two examples of software-induced disturbance failures include \emph{data pattern disturbance}~\cite{liu} and the \emph{RowHammer}~\cite{rowhammer} phenomenon. An adversarial \emph{data pattern} exploits how DRAM devices store charge to disturb neighboring cells to where an attacker writes the data pattern~\cite{Lee2010, liu}. And regardless of the data pattern, an attacker can disturb nearby rows of cells by writing rapidly to (\emph{hammering}) a victim row~\cite{rowhammer, Kim2016, Gruss2016, Mutlu2017}. Software-induced disturbance threatens process isolation in computer systems and researchers at Google Project Zero used RowHammer to gain kernel privilege escalation on laptops~\cite{Seaborn2015}. Many other recent works developed new RowHammer attacks~\cite{Seaborn2015, Razavi2016, vanderVeen2016, Gruss2016, Kim2016, WikiRowHammer, GoogleGroupRowHammer, TwitterRowHammer, Lanteigne2016, Aweke2016, Seaborn2016, Bosman2016, RowHammerSource, RowHammerTest, Aga2017, Aichinger2014, Bhattacharya2016, Brasser2017, Frigo2018, Gomez2016, Goodin2016, Greenberg2016, Gruss2018, Harris2014, Irazoqui2016, Jang2017, Lipp2018, Qiao2016, Tatar2018, Tatar2018-2, Xiao2016} and defenses~\cite{Aweke2016, Apple2015, HewlettPackard2015, Lenovo2015, Fridley2015, Aweke2016, rowhammer, Lee2018, Seyedzadeh2017, Son2017, Tatar2018, vanderVeen2018}

\subsubsection{Endurance Failures}
\label{subsubsec:Endurance Failures}

DRAM cells can only endure so many writes before they \emph{wear out}. The ITRS, a leading authority on DRAM process technology, suggests a nominal DRAM cell lifetime of $\mathrm{3\times10^{16}}$ write cycles before failure~\cite{itrs}, or around \unit[47.5]{years} if accessing data as fast as possible. As we noted earlier, this is because the fastest possible write cycle to a DRAM cell is dictated by the latency to change a row, \emph{tRC}, in the DDR specifications, with a value of $\sim$\unit[50]{ns} in DDR3~\cite{jedec}, hence, $\mathrm{3\times10^{16}}\times\unit[\mathrm{50}]{ns} \approx \unit[\mathrm{47.5}]{years}$.

Empirical studies on DRAM wearout are scarce, as measuring DRAM wearout in the wild or in a controlled environment is hard. DRAM devices are rarely used in systems for decades and most large scale DRAM studies do not correlate failures to device age~\cite{schroeder-sigmetrics09, schroeder-dsn06, hwang, amd, amd2}. Nevertheless, work by Wang et al.~\cite{Wang2017} as well as our own work~\cite{meza15-2} observe signs of wearout in DRAM devices at a large scale over the course of several years. Endurance failures may be system- and workload-dependent, however, as recent work by Siddiqua et al.~\cite{Siddiqua2017} on a Cielo supercomputer at Los Alamos National Labs, has shown no sign of endurance failures after a five year operational lifetime.

Other prior studies~\cite{const, dram-wearout} have performed small scale laboratory testing to artificially induce wearout in DRAM cells---often going to extreme lengths. As an example, Chia et al.~\cite{dram-wearout} needed to perform experiments with ``low temperature [$-10$\textdegree C], high VDD voltage and stressful data patterns with a low address toggling frequency over a 1000-hour stress time period for a large sample size from three different lots'' before inducing significant endurance issues. These studies concluded that wearout can be induced in DRAM cells in extreme conditions.

\subsubsection{Other Failures}

DRAM devices can also fail to store data correctly despite the devices operating properly. These ``soft'' failures of DRAM occur when charged particles from the environment disturb DRAM cells.

When a charged particle strikes a DRAM cell capacitor, the charged particle transfers part of its charge to the capacitor. A charged particle disturbs the contents of a DRAM cell when its charge inverts the state of the capacitor, causing a capacitor with a low voltage to have a high voltage or a capacitor with a high voltage to have a low voltage. In DRAM's early days, manufacturers noticed impurities in DRAM packaging led to emission of charged \emph{alpha particles} that disturbed the contents of DRAM cells~\cite{intel-dram}. Researchers also observed charged \emph{cosmic rays} had a similar affect to alpha particles and cause more disturbance at higher altitudes due to less protection from the Earth's atmosphere~\cite{amd2, O'Gorman1994, hwang}.

\subsection{How DRAM Errors are Handled}

As prior works have shown, DRAM errors occur relatively commonly due to a variety of stimuli~\cite{intel-dram, schroeder, hwang, amd, amd2, amd3, goodbadugly, dram-wearout, shen1, selse, liu, rowhammer, samira, Mutlu2017, Chang2016, Chang2017, Chang2017-2, Kim2015, Seshadri2016}.  To protect against such errors in servers, additional data is stored in the DIMM (in a separate DRAM chip) to maintain \change{Error Correcting Codes (ECC)}{\textit{error correcting codes (ECC)}} computed over data.  These codes can detect and correct a small number of errors.  For example, {\it single error correction, double error detection (SEC-DED)} is a common ECC strategy that can detect any 2 flipped bits and correct 1 flipped bit per 64 bits by storing an additional 12.5\% of ECC metadata.  An error that can be corrected by ECC is called a correctable error (CE); an error that cannot be corrected by ECC, but which can still be detected by ECC, is called an uncorrectable error (UCE).

The processor's \emph{memory controller} orchestrates access to the DRAM devices and is also responsible for checking the ECC metadata and detecting and correcting errors.  While detecting errors does not add much overhead when performing memory accesses, correcting errors can delay a memory request and disrupt a system.  As an example, on the systems that we examine for DRAM failures in Chapter~\ref{chp:dramfailures}, when an error is corrected, the CPU raises a hardware exception called a \emph{machine check exception (MCE)}, which must be handled by the CPU.

When an MCE occurs, the processor stores information about the memory error in special registers that can be read by the operating system.  This information includes the physical address of the memory access when the error occurred and what type of memory access (e.g., read or write) was being performed when the error occurred.  Note that memory errors do not only occur in DRAM chips: memory errors can occur if the memory controller fails or if logic associated with transmitting data on a memory channel fails. To reduce the impact of DRAM faults, various \emph{Error Correcting Codes (ECC)} for DRAM data~\cite{hamming,ibm-chipkill} have been used to detect---and in some cases correct---memory errors.  However, these techniques require additional DRAM storage overheads and DRAM controller complexity and can not detect or correct all errors.

In addition, {\it error tolerance} techniques, like ECC, do not target the root causes of memory errors.  Permanent single-bit errors can only be tolerated by ECC so long before becoming uncorrectable multi-bit errors, so in practice, at companies like Facebook, these techniques are used only to mask the effects of memory errors long enough for devices to be taken offline and replaced.  At Facebook, memory errors alone contribute to a non-trivial number of failures per month~\cite{meza15-2}, even with server-grade ECC DIMMs, providing a compelling reason for the further study of memory errors and techniques to reduce them.

\subsection{Related Research in DRAM Failures in Modern Data Centers}
\label{subsec:Related Research in DRAM Failures in Modern Data Centers}

Many prior works have examined DRAM failures at a \emph{small scale} or in a \emph{controlled environment.} A selection of such recent work includes ~\cite{Frank2001, Pellegrini2016, Hamamoto1998, Yamaguchi2000, liu, Kim2003, raidr, salp, yixin, Qureshi2015, Patel2017, samira, aldram, Khan2016-3, Khan2016-2, samsung-intel, samira, Kim2005, rowhammer, Mandelman2002, Mueller2005, Mutlu2017, Mutlu2013, Mutlu2014, Nakagome1988, Patel2017, Redeker2002}, which we refer the reader to. As related work to this dissertation, however, we turn to \emph{large scale} studies of DRAM failures from the field.

Schroeder et al.\ performed the first study of memory errors in the field on a majority of Google's servers in 2009~\cite{schroeder}.  The authors' study showed the relatively high rate of memory errors across Google's server population, provided evidence that errors are dominated by hard failures (versus soft failures), and noted that they did not observe any indication that newer generations of DRAM devices have worse error behavior, that CPU and memory utilization are correlated with error rate, and that average server error rate is very high. Their work formed the basis for what was known of DRAM errors in the field. Five years later, we performed a study on all of Facebook's servers~\cite{meza15} and shed new light on Schroeder et al.'s findings, as we will explain in Chapter~\ref{chp:dramfailures}.

Hwang et al.\ analyzed a trace of memory errors from a sample of Google servers and IBM supercomputers, and showed how errors are distributed across various DRAM components~\cite{hwang}. Their work observed a high number of repeat address errors, which led them to simulate the effectiveness of page offlining (proposed in~\cite{solaris-pageoffline}) on the memory error traces. Using page offlining, Hwang et al.\ reduced error rates by 86\% to 94\%. We built on top of Hwang et al.'s work in~\cite{meza15-2} by additionally controlling for the effect of socket and channel failures, as we will explain in Chapter~\ref{chp:dramfailures}. By doing so, we showed that a large number of repeat address failures are due to socket and channel failures. This finding was significant because it meant that page offlining could only help the symptom of socket and channel failures---\emph{not} the cause. In addition, unlike Hwang et al., instead of simulating the effects of page offlining, we evaluated page offlining across a cluster of thousands of Facebook's servers, in \S\ref{sec:reduce}.

Sridharan et al.\ examined memory errors in a supercomputing environment~\cite{amd,amd2,amd3}. Similar to Hwang et al., Sridharan et al.\ found that memory errors are dominated by permanent failures, that DRAM chips are susceptible to large multi-bit failures, that Chipkill~\cite{ibm-chipkill} ECC reduces errors 42$\times$ compared to single-error-correction double-error-detection ECC, that \emph{multi-DIMM errors} occur, and speculated that multi-DIMM errors disrupt access to other DRAM devices that share the same board circuitry. Unable to conclusively identify the source of multi-DIMM errors, Sridharan et al.\ speculated about their origin.

Siddiqua et al.~\cite{selse} suggested a methodology to classify errors that occur \emph{outside of} DRAM chips. Their study examined error data across 30,000 servers in unnamed data centers and taxonomized errors based on whether they were caused by the memory controller, busses, channels, or memory modules, but did \emph{not} examine the background failure rates of components at a finer DRAM chip-level granularity. They found that a small number of faults generate a large number of errors and that faults are predominantly permanent.

Our analysis on all of Facebook's servers~\cite{meza15-2} bridged the gap between Sridharan et al.\ and Siddiqua et al.'s work and painted a complete picture of DRAM failures in the field. We performed the first analysis of DRAM failure trends (on modern DRAM devices using modern data-intensive workloads) that have \emph{not} been identified in prior work (e.g., chip density, transfer width, workload type), presented the first regression-based model for examining the memory failure rate of systems, and performed the first analysis of page offlining in the field.  Prior large scale empirical studies of memory errors analyzed various aspects of memory errors in different systems.
 Like Siddiqua et al.'s work~\cite{selse}, we attributed multi-DIMM errors to socket and channel failures, but we also examined chip-level failures, like Sridharan et al.\ did~\cite{amd,amd2,amd3}. Our work helps to provide a more complete picture of DRAM failures within a server.

Sridharan et al.~\cite{amd2} also found that DRAM vendor and age are correlated with error rate. Around the same time as another study by Sridharan et al.~\cite{goodbadugly}, we, like Sridharan et al., also observed that the average number of errors per server is much larger than the median number of errors per server~\cite{meza15-2}. As we discuss in Chapter~\ref{chp:dramfailures}, we provide the \emph{full distribution} of errors per server and show that the distribution of errors across servers follows a Pareto distribution, with a decreasing hazard rate. This means, roughly, that the more errors a server has had so far, the more errors it is expected to have.

Nightingale et al.\ examined the failure rate of consumer PCs and showed that increased CPU frequency is correlated with increased DRAM error rates~\cite{nightingale}. A pair of works by Li et al.\ analyzed memory errors on 212 Ask.com servers and evaluated their application-level impact~\cite{shen1,shen2}.  They found that most memory errors are permanent, that memory errors affected applications in noticeable ways, and proposed a technique to monitor memory for errors to reduce application impact.

Wang et al.\ reported failure trends across over 290,000 hardware failure reports in state-of-the-art data center hardware and software design~\cite{Wang2017}. A new angle that Wang et al.\ examined DRAM device reliability from is the temporal correlation of memory failures (i.e., certain times of the day and certain days of the week have more failures). Like our study~\cite{meza15-2}, Wang et al.\ corroborated signs of wearout phenomena in older DRAM devices.

Siddiqua et al.\ performed an analysis of memory errors across the entire five-year operational lifetime of a system and reported on its DRAM failures~\cite{Siddiqua2017}. Unlike prior studies, over the course of five years, Siddiqua et al.\ did not observe an increase in DRAM failures, but the authors do not speculate as to why their system was immune to the effects of time. Instead, they found that the primary type of DRAM faults shifts from permanent to transient after around one year of operation time and that the amount of component failures (cells, rows, columns, and so on) does not vary by more than 1.4\% across a device's five-year lifetime. The authors conclude that DRAM devices may be used beyond their planned operational lifetime (five years for the system the authors examined).

\section{SSD Devices}
\label{sec:SSD Devices}

Servers use flash memory-based SSDs as a high-performance alternative to hard disk drives to store persistent data. The flash chips used in SSDs are especially susceptible to endurance failures if used na\:ively, so SSDs make extensive use of error prevention and error correction techniques~\cite{Cai2017-2, Cai2018}. But not all errors can be prevented, and in a data center environment, flash-based SSD failures can lead to downtime and, in the worst case, data loss.

\subsection{SSD Device Architecture}
\label{subsec:SSD Device Architecture}

\begin{figure}[H]
\centering
\onur{\protect\includegraphics[width=0.5\columnwidth]{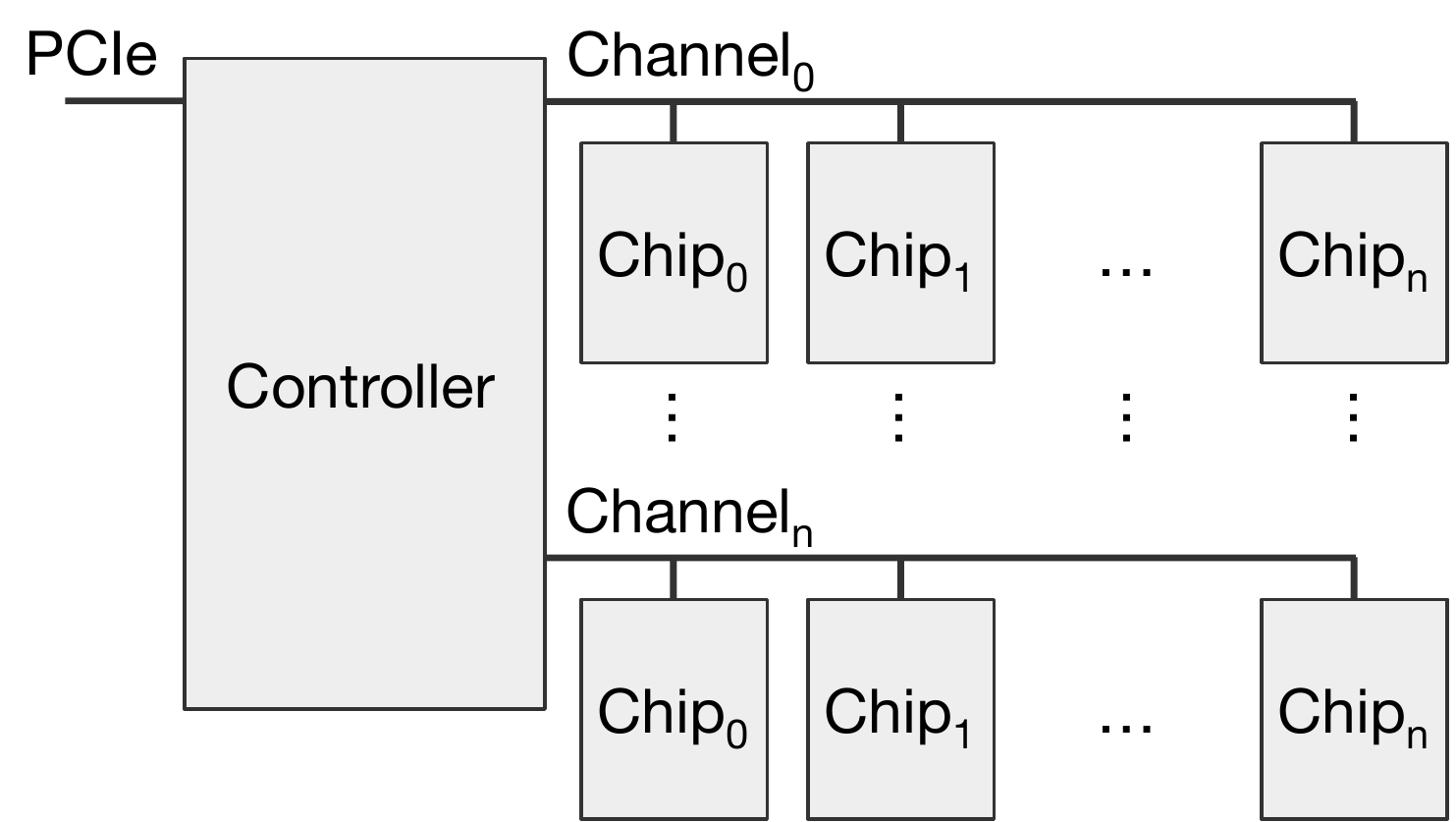}}{\protect\includegraphics[width=0.5\columnwidth]{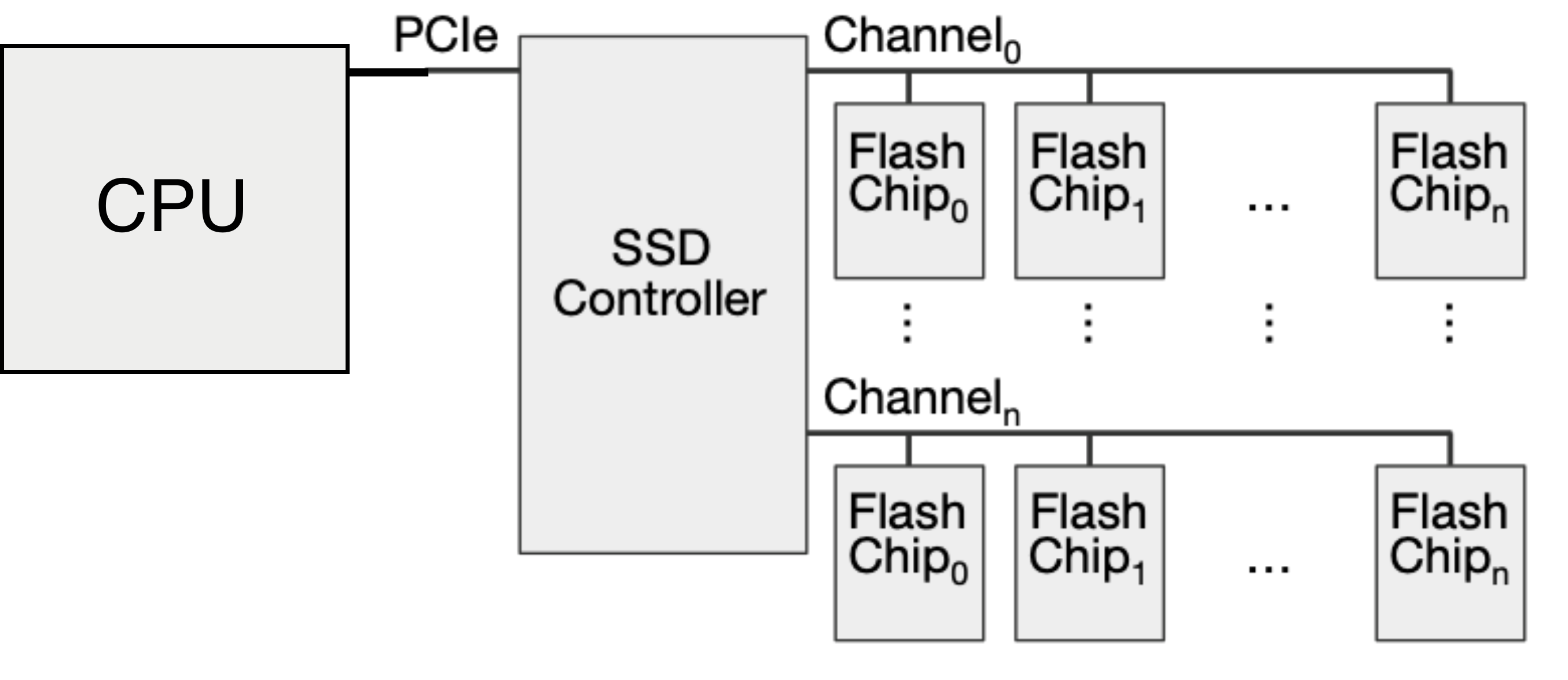}}
\caption{Server \protect\onur{flash device}{SSD} architecture.}
\label{fig:device}
\end{figure}

Figure~\ref{fig:device} provides an illustration of the architecture of a server SSD. Server SSDs differ from consumer SSDs in two ways: (1) by connecting to the server using the PCIe bus to provide high data transfer bandwidth and (2) by providing higher device capacity. The server SSDs we discuss are similar to those available from companies such as Fusion-io~\cite{Fusionio}, Hitachi~\cite{Hitachi}, Intel~\cite{Intel}, Seagate~\cite{Seagate}, Toshiba~\cite{Toshiba}, and Western Digital~\cite{WesternDigital}.

In order to achieve high bandwidth and low access latency, despite potentially long chip read and write latencies, SSDs employ many channels, each of which the SSD controller operates in parallel, with flash chips connected to each channel. Multiple channels provide data in parallel by mapping the data across the channels. This helps to ensure that the SSD can service data at close to the bandwidth of the PCIe connection.

An \emph{SSD controller} coordinates data transfer to and from the server and performs tasks to improve the performance and reliability of flash. In addition to orchestrating flash chip access, the SSD controller manages the reliability of the flash memory chips. Since flash cells wear out after too many writes, SSD controllers perform \emph{wear leveling} to distribute writes (and thus wear) more evenly across the cells in each flash chip. To do this, the SSD controller determines when and where \emph{pages} (around \unit[8]{KB} in size) of flash data should be erased or copied and groups similar pages that need to be erased or copied into \emph{blocks} (a block is around 128 $\times$ \unit[8]{KB} pages). To aid the system software and hide the internal layout of flash data, the flash controller maintains a mapping of logical addresses to physical locations, known as the \emph{flash translation layer (FTL)}~\cite{Cai2017-2, Cai2018, chung-jsa09}. Chung et al.\ provide an overview of the organization and operation of this important part of SSD device operation~\cite{chung-jsa09}. FTL data is stored in a DRAM buffer and is managed by the SSD controller. When blocks of flash cells are deemed unreliable for further use, the SSD controller \emph{discards} them to avoid the risk of encountering an uncorrectable error on them.

As data is written to flash-based SSDs, pages are copied during a process known as \emph{garbage collection} in order to free up blocks with unused data to be erased and to more evenly level wear across the flash chips. Before new data can be written to a page in flash, the entire block needs to be first erased.  Each erase wears out the block as shown in previous works~\cite{mielke-irps08, cai-iccd12, cai-itj13, Cai2017-2, Cai2018, chung-jsa09, cai-date12, cai-hpca15, cai-dsn15, Cai2017, Luo2018, Luo2018-3}. The SSD controller performs \emph{Garbage Collection (GC)} in the background, which compacts data that is in use to ensure that an SSD has enough available space to write new data.

Blocks are discarded by the \onur{flash}{SSD} controller when they are deemed unreliable for further use.  Discarding blocks \onur{directly }{}affects the usable lifetime of \onur{a flash devices}{a flash-based SSD} by reducing the capacity of the \onur{device}{SSD}.  At the same time, discarding blocks has the potential to reduce the amount of errors generated by \onur{a device}{an SSD}, by preventing unreliable cells from being accessed.

Flash-based SSDs also use DRAM to provide buffer space for SSD controller metadata or for data to be written to the flash chips. The SSDs we examine use DRAM buffer space to store metadata related to the FTL~\cite{chung-jsa09} mapping for logical addresses to physical addresses. This allows the SSD controller to locate data on an SSD quickly, reducing the performance impact of address translation.

For more information on SSD architecture, please see~\cite{Cai2017-2, Tavakkol2018, Tavakkol2018-2, Cai2018}.

\subsection{How SSD Devices Fail}
\label{subsec:How SSD Devices Fail}

Within a flash chip, data is stored in the form of charge trapped on a floating-gate transistor. Charge can be incrementally added to the floating-gate transistor until some saturation point. Because charge can be incrementally added to a flash cell, the number of bits stored per cell depends only on how precisely the voltage of the cell can be measured. Some flash devices trade off storing a single bit per cell for faster cell voltage sensing and are called \emph{Single-Level Cell (SLC)} flash devices. Server SSDs typically use flash chips that store more bits per cell but have slower sensing times, called \emph{Multi-Level Cell (MLC)} flash. For more information on flash cells, SSD microarchitecture, and reliability characteristics, we refer the reader to~\cite{ftl, cai-date12, cai-hpca15, cai-dsn15, cai-itj13, cai-iccd13, cai-date13, cai-iccd12, cai-sigmetrics14, Cai2018, Cai2017-2}.

We discuss various device-level failure mechanisms in flash-based SSDs including endurance failures, temperature-dependent failures, and disturbance failures.

\subsubsection{Endurance Failures}

A key trait of flash cells is their low write endurance---flash cells become less reliable each time they are programmed and erased (called a \emph{P/E cycle} for \emph{Program/Erase cycle}). Programming and erasing a flash cell causes the cell to wear out because the high current involved in removing charge from the floating-gate transistor physically degrades the cell over time, causing it to less reliably store its charge. Flash cell erasure is intrinsically tied to flash cell programming because charge can only be added to a flash cell up until it can store no other value at which point it must be erased.

Flash cell endurance failures have been known since the fabrication of some of the earliest devices~\cite{suh-jssc95, jung-isscc96, hur-nvsm04, lee-nvsm06, joo-jjap06, kurata-vlsi06, compagnoni-iedm07, ong-vlsi93, chimenton-ted03, brand-irps93, degraeve-ted04, belgal-irps02, kato-iedm94, yamada-irps00, yamada-vlsi01, lee-irps03, mielke-tdmr04, mielke-rps06, takeuchi-jssc99}. However, the early studies on flash endurance often examined a single type of cell or chip in highly controlled environments. Since these studies, several recent works have quantified the effects of P/E cycles on various error mechanisms in small sets of recent flash chips (e.g.,~\cite{cai-date12, cai-date13, cai-dsn15, cai-hpca15, cai-iccd13, cai-itj13, cai-iccd12, cai-sigmetrics14, grupp-fast12, Cai2018, Cai2017, Cai2017-2, Luo2018, Luo2018-2, Luo2018-3, Luo2015}).

The number of P/E cycles before a cell is unable to retain its contents, or \emph{wears out}, can range from 100,000 (for SLC flash) to 3,000 (for MLC flash). It is therefore important to preserve P/E cycles on flash cells and the FTL plays an important role in doing so. The FTL sits in between the server and the flash chips and helps to manage flash cell wear. It does this in four ways~\cite{chung-jsa09}: (1) by performing \emph{wear-leveling} to ensure no flash cell has much more wear than others, (2) by performing \emph{page offlining} to proactively removing flash pages that have experienced too much wear from the population, (3) refreshing the contents of cells that may have become weak over time, and (4) using error correcting codes to tolerate cell faults.

To detect and protect against endurance errors, the FTL stores and accesses additional ECC metadata. This ECC metadata is also transferred across and computed over the data transmitted in the channel to protect against channel errors. These codes must be sufficiently strong to protect against the errors that may occur on flash chips over time (as discussed in~\cite{Cai2017-2, Cai2018}).

\subsubsection{Temperature-Dependent Failures}
\label{subsubsec:SSD Temperature-Dependent Failures}

Flash cells, like other \emph{Field Effect Transistors (FETs)}, are susceptible to leaking charge and aging more quickly at higher temperatures. Specifically, the physical integrity of flash cells degrades at higher temperatures due to the temperature-activated Arrhenius effect, which ages flash cells at an accelerated rate. Higher temperatures also affect the physical construction of flash devices by shrinking or expanding wires and boards, which can also cause device components to physically degrade. We refer the reader to \cite{Cai2017-2, Cai2018, Luo2018-2} for state-of-the-art studies of flash temperature-dependent failures and techniques in SSDs to fix them. Prior studies have examined the effects of temperature on flash \emph{chips} (e.g.,~\cite{mielke-tdmr04, mielke-rps06, takeuchi-jssc99, cai-date12, cai-iccd12, cai-iccd13, cai-date13, cai-sigmetrics14, cai-hpca15, cai-dsn15, cai-itj13, grupp-fast12, Luo2018, Luo2018-2, Luo2018-3}).

The temperature within the flash devices in modern data centers can range from 30\textdegree~C to 70\textdegree+~C~\cite{meza15}. These temperatures are typically higher than within consumer electronics because modern data centers strive to have efficient data center \emph{Operational Expenditure (OpEx)}~\cite{Barroso2013} and a key component of OpEx is airflow and cooling. Hotter data centers mean less cooling and lower OpEx. This makes the temperature dependence of flash failures an interesting area to explore in modern data centers.

\subsubsection{Disturbance Failures}

A flash cell's value can change because of how \emph{neighboring} cells in the same block are accessed. Cells in the same block are connected by a set of bitlines in series. When reading data from a particular flash cell, the others cells in that cell's bitline must allow its read value to pass through them. To do so, the bitline driver must use a large voltage, called \emph{pass-through voltage,} to ensure the cells in the bitline participating in the read all activate and pass along the read value.

Unfortunately, this large voltage is also enough to cause some charge to leak into the floating gates of the other cells in the bitline~\cite{cai-dsn15}. With enough reads in a short enough period, the charge collected in the neighboring cells can change their stored value. Such \emph{disturbance errors} can be induced in software from adversarial read access patterns~\cite{Sheldon2008, brand-irps93, mielke-irps08, cai-date12, cai-dsn15, Cai2017}. Note that this effect is less common with write access patterns, as the SSD controller typically batches together writes and stores them in a fresh page.

\subsubsection{Other Failures}

In addition to endurance failures, temperature-dependent failures, and disturbance failures, other flash failure modes have been discussed in the literature~\cite{Cai2018, Cai2017-2}. These include program failures and retention failures. Program failures occur when the data read from flash cells contains errors and the errors are used when programming the new data~\cite{Hu2011, Cai2017, Luo2015, Parnell2014}. Programming errors occur in the flash chip, so they cannot be corrected by ECC, which resides in the flash controller. Retention failures occur as charge leaks out of flash cells over time due to flash cell insulation physically degrading over many P/E cycles~\cite{cai-date12, cai-hpca15, cai-iccd12, cai-itj13, mielke-irps08, Tanakamaru2011}. We refer the reader to~\cite{Cai2017, Luo2016} for more information of program failures and to~\cite{Cai2012, cai-date12, cai-hpca15, cai-iccd12, cai-itj13} for more information on retention failures.

\subsection{How SSD Errors are Handled}

To make up for their low write endurance, and ensure reliable operation, the SSD controller uses ECC metadata to detect (and, when possible, correct) errors in a flash chip. SSD flash controllers use a progressive approach to error correction: \emph{small errors} (e.g., several erroneous bits in a KB of data) are quickly corrected using simple logic in the SSD controller while \emph{large errors} (e.g., >10's of erroneous bits in a KB of data) are corrected using more complex controller logic or with assistance from a software driver running on the host.

Thus, large errors that may be uncorrectable from an SSD's perspective, may be correctable from the system's perspective. For example, during the course of a read operation, if the SSD controller is unable to correct the errors in a particular chunk of data, the data and the ECC information is sent to the host machine where a driver uses the host machine's CPU to perform more complex error correction and forwards the result to the host's OS. Errors that are not correctable by the driver result in data loss.

Flash device failure is measured in terms of a device's \emph{Bit Error Rate (BER).} The BER of an SSD is the rate at which errors occur relative to the amount of information that is transmitted from/to the SSD. BER can be used to gauge the reliability of data transmission across a medium.
\begin{equation}
\mathit{BER} = \frac{\mathit{Correctable\ errors} + \mathit{Uncorrectable\ errors}}{\mathit{Bits\ accessed}}
\end{equation}
Because errors can either be \emph{correctable} (with the help of the SSD controller or SSD software driver) or \emph{uncorrectable}, it is helpful to distinguish between the two. \emph{CBER} stands for \emph{Correctable BER} and \emph{UBER} stands for \emph{Uncorrectable BER.}

For flash-based SSDs, UBER is an important reliability metric that is related to the SSD lifetime. SSDs with high UBERs are expected to have more failed cells and encounter more (severe) errors that may potentially go undetected and corrupt data than SSDs with low UBERs.

\subsection{Related Research in SSD Failures in Modern Data Centers}
\label{subsec:Related Research in SSD Failures in Modern Data Centers}

A large body of prior work examined the failure characteristics of flash cells in controlled environments using small numbers (e.g., tens) of raw flash chips (e.g., ~\cite{suh-jssc95, jung-isscc96, hur-nvsm04, lee-nvsm06, joo-jjap06, kurata-vlsi06, compagnoni-iedm07, ong-vlsi93, chimenton-ted03, brand-irps93, degraeve-ted04, belgal-irps02, kato-iedm94, yamada-irps00, yamada-vlsi01, lee-irps03, mielke-tdmr04, mielke-rps06, takeuchi-jssc99, cai-date12, cai-iccd12, cai-iccd13, cai-date13, cai-sigmetrics14, cai-hpca15, cai-dsn15, cai-itj13, grupp-fast12, Cai2017, Cai2017-2, Cai2018, Luo2015, Luo2018, Luo2018-2}). These studies quantified a variety of flash cell failure modes and formed the basis of the community's understanding of flash cell reliability.

However, prior work was limited in its analysis in three ways: (1) the studies were conducted on small numbers of raw flash chips accessed in adversarial manners over short amounts of time, (2) the studies did not examine failures when using real applications running on modern servers and instead used synthetic access patterns or simulated workloads, and (3) the studies did not account for the storage software stack that real applications need to go through to access flash memories. Such conditions assumed in these prior studies are substantially different from those experienced by flash-based SSDs in modern data centers.  Such large-scale systems have five main environmental differences compared to chip-level flash studies:  (1) real applications access flash-based SSDs in different ways over a time span of \emph{years}, (2) applications access SSDs via the storage software stack, which employs various amounts of buffering and hence affects the access pattern seen by the flash chips, (3) flash-based SSDs employ aggressive techniques to reduce device wear and to correct errors~\cite{Cai2018}, (4) factors in platform design, including how many SSDs are present in a node, can affect the access patterns to SSDs, (5) there can be significant variation in reliability due to the existence of a very large number of SSDs and flash chips. All of these real-world conditions present in large-scale systems likely influence the observed reliability characteristics and trends of flash-based SSDs.

We performed an early comprehensive study of flash-based SSD reliability trends across all the flash devices at Facebook~\cite{meza15}. We observed that SSDs go through several lifecycle phases depending on how much data has been written to them, read disturbance errors are uncommon in the SSDs in Facebook's data centers, and higher temperatures lead to higher SSD failure rates, among other findings. We discuss this work in detail in Chapter~\ref{chp:ssdfailures}.

The closest related work prior to our study, by Grupp et al.~\cite{grupp-fast12}, examined the cost, performance, capacity, and reliability trends of flash memory in the context of a prototypical server flash storage device similar in architecture to SSDs deployed in data centers.  Based on their study, the authors projected several challenges for the adoption of flash memory in a server context.  One drawback of the reliability results presented in that study is that the experiments were performed in a controlled environment, under synthetic workloads, while modeling only the latency---but not the function---of the SSD controller on 45 flash chips.

Ouyang et al.~\cite{ouyang-asplos14} performed a study of programmable SSD controllers at a large-scale web services company, Baidu.  While this work examined flash-based SSDs in the field, they did not analyze the reliability characteristics of SSDs and instead focused on bandwidth, capacity, and cost. Wang et al.~\cite{Wang2017} also examined flash cards at a large scale data center (probably at Baidu) and observe infant mortality rates for flash cards during the first year as well as a temporal variation in flash card failures (depending on the time of day or the day of the week).

Narayanan et al.~\cite{Narayanan16} examined the SSD failures over the course of three years at Microsoft. They observed much higher failure rates for SSDs than those quoted by the manufacturer; lower UBER compared to our work~\cite{meza15}, but still much higher than target rates set by the manufacturer; and that device placement, model, and workload were correlated with SSD failure rate.

Schroeder et al.~\cite{Schroeder2016} also performed a study of SSD failures in Google data centers. Among their findings they discovered that uncorrectable errors were the most common observable error on the drives they examined (affecting between 2 to 6 out of 1000 drives), that most drives experience at least one correctable error within a day, that independent of usage a device's age affects reliability, that chips with smaller cell feature size have higher error rates, and that the distribution of bad blocks on drives is bimodal: with drives either having a very small number of bad blocks or having a very large number. Schroeder et al.'s work complements our own~\cite{meza15}, although we reach contrasting findings on two points: infant mortality (we observed signs of early detection and failure of SSDs, which may be due to differences in burn-in procedures for new SSDs at Facebook and Google) and read disturb errors (which we did not find evidence of at Facebook, but which Schroeder et al.\ observed in some errors at Google).

For an overview of recent flash-based SSD studies performed in production data centers, we refer the reader to Schroeder et al.'s survey~\cite{Schroeder2017} that compares and contrasts three contemporary studies of SSD reliability in the field (\cite{meza15, Schroeder2016, Narayanan16}) in an article~\cite{Schroeder2017}.

\section{Network Devices}
\label{sec:Network Devices}

Connecting the many servers in modern data centers and allowing them to communicate with one another is the network. Systems running in modern data centers use \emph{Remote Procedure Calls (RPCs)~\cite{Birrell1984, White1976, Nelson1981, Courier1981, Liskov1979}} (such as Thrift~\cite{thrift}) to handle most communication between software running on different servers. While RPC frameworks handle sending a message between a source server and a destination server, load balancers~\cite{Sommermann2014} running in data centers ensure that requests are distributed evenly across servers. While these systems have timeouts and retries to tolerate transient network outages, failures in the network can be disruptive and cause widespread disruption to software systems running in data centers.

\subsection{Data Center Network Architecture}

\begin{landscape}
\begin{figure*}
  \centering
  \includegraphics[width=\columnwidth]{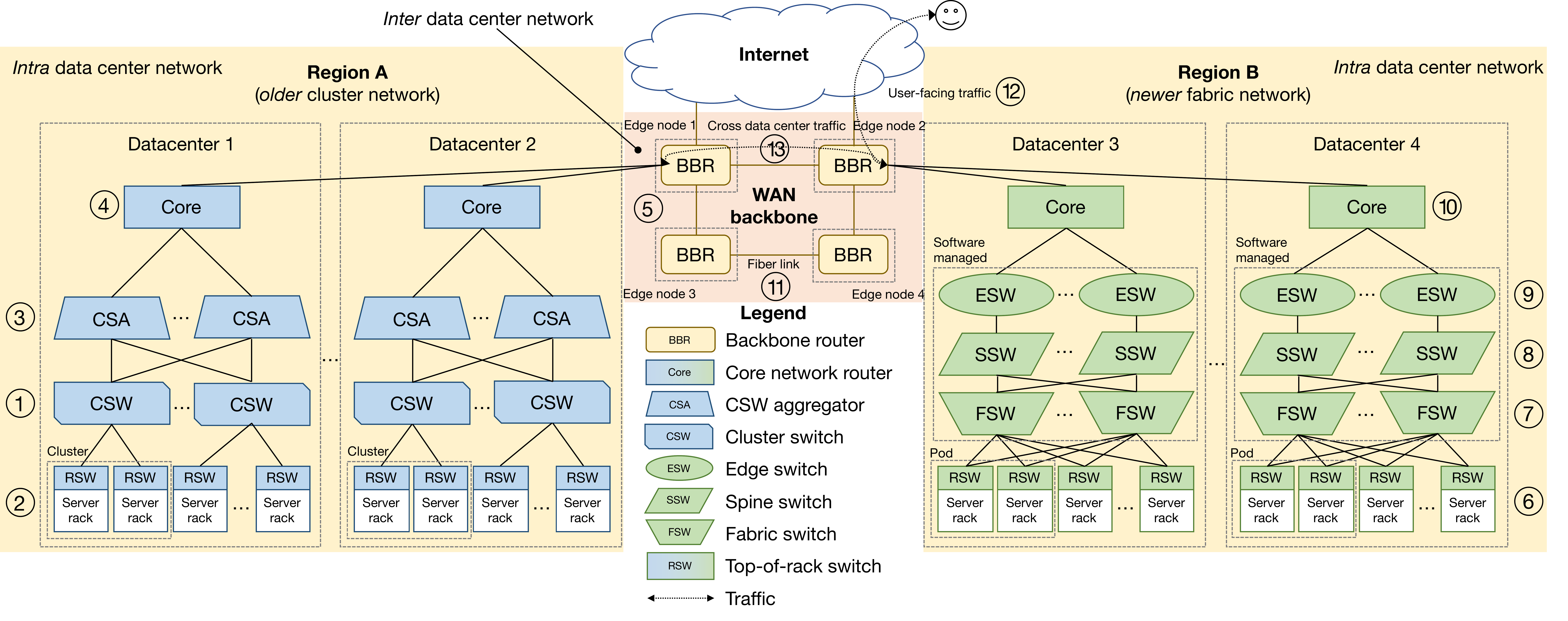}
  \caption{Facebook's network architecture described in \S\ref{subsec:Intra Data Center Networks} and \S\ref{subsec:Inter Data Center Networks}. Data centers use either an older cluster network (Region A) or a newer fabric network (Region B). Cluster networks and fabric networks communicate through the WAN backbone and Internet.}
  \label{fig:network-arch}
\end{figure*}
\end{landscape}

We examine Facebook's network, shown in Figure~\ref{fig:network-arch}~\cite{Farrington2013,Roy2015,Andreyev2014,Jimenez2017}. The network consists of interconnected data center \emph{regions}. Each region contains buildings called \emph{data centers}. This architecture includes both data center and backbone networks. We call the network \emph{within} data centers the \emph{intra} data center network and the backbone network \emph{between} data centers the \emph{inter} data center network.
The diversity of Facebook's network provides an opportunity to compare reliability across different network designs. Though diverse, Facebook's network is by no means unique. Published network architectures from Google and Microsoft use similar design principles and building blocks~\cite{Gill2011, Potharaju2013IMC, Potharaju2013SoCC, Govindan2016, Singh2015, Jain2013}. 

\vspace*{\fill}
\hfill
\begin{center}
\it (This portion of page intentionally left blank.)
\end{center}
\vspace{\fill}

\subsection{Intra Data Center Networks}
\label{subsec:Intra Data Center Networks}

Facebook uses two intra data center network designs: an older \emph{cluster-based} design~\cite{Clos1953, Farrington2013} and a newer \emph{data center fabric} design~\cite{Andreyev2014}. We call these the \emph{cluster network} and the \emph{fabric network}. Unlike the cluster network, the fabric network uses a five-stage Folded Clos~\cite{Fares2008} interconnect design, built from simple commodity hardware, with software-controlled automated repairs. When the older cluster network was designed, most network devices supported little customization (we call these type of devices \emph{third-party devices}), and so a hard-wired topology was used. Later, when Facebook developed its own customizable network devices~\cite{Bachar2015, Bagga2015}, software could be used to dynamically define the network topology.

\subsubsection{Cluster Network Design}

In Facebook's older cluster network (Figure~\ref{fig:network-arch}, Region A), a \emph{cluster} is the basic unit of network deployment. Each cluster comprises four \emph{cluster switches (CSWs, \ding{192})}, each of which aggregates physically contiguous \emph{rack switches (RSW, \ding{193})} via \SI[per-mode=symbol]{10}{\giga\bit\per\second} Ethernet links. In turn, a \emph{cluster switch aggregator (CSA, \ding{194})} aggregates CSWs and keeps inter cluster traffic within the data center. Inter data center traffic flows through {\it core network devices (core devices, \ding{195})}, which aggregate CSAs.

A cluster network has two main limitations:
\begin{enumerate}
  \item \textbf{\emph{Third-party vendor devices}} limit data center scalability. Connecting more devices requires waiting for third-party vendors to produce larger switches. This is a fundamental limiting factor for data center size in cluster networks.
  \item \textbf{\emph{Proprietary software}} is challenging to maintain and customize. Proprietary software on switches makes customization difficult or impossible. Once deployed, proprietary switches must be repaired in-place. When a device becomes unresponsive, a human must power cycle the device. Compared to software, humans perform slow repairs. Slow repairs mean fewer switches to route requests, more traffic on the remaining switches, and more congestion in the network.
\end{enumerate}

Despite its limitations, the cluster networks remain in use in a dwindling fraction of Facebook's data centers. Ultimately, these data centers will join new data centers in using the fabric network design.

\subsubsection{Fabric Network Design}

Facebook's newer network (Figure~\ref{fig:network-arch}, Region B) addresses the cluster network's limitations. A \emph{pod} is the basic unit of network deployment in a fabric network. Unlike the physically contiguous RSWs in a cluster, RSWs in a pod have no physical constraints within a data center. Each RSW (\ding{197}) connects to four \emph{fabric switches (FSWs, \ding{198})}. The 1:4 ratio of RSWs to FSWs maintains the connectivity benefits of the cluster network. \emph{Spine switches (SSWs, \ding{199})} aggregate a dynamic number of FSWs, defined by software. Each SSW connects to a set of \emph{edge switches (ESWs, \ding{200})}. Core devices (\ding{201}) connect ESWs between data centers.

Facebook's fabric networks are managed largely by software and differ from its cluster networks in four ways:
\begin{enumerate}
  \item \textbf{\emph{Simple, custom switches.}} Fabric devices contain simple, commodity chips and eschew proprietary firmware and software.
  \item \textbf{\emph{Fungible resources.}} Fabric devices are {\it not} connected in a strict hierarchy. Control software manages FSWs, SSWs, and ESWs as a fungible pool of resources. Resources dynamically expand or contract based on network bandwidth and reliability needs.
  \item \textbf{\emph{Automated repair mechanisms.}} Failures on data center fabric devices can be repaired automatically by software~\cite{Power2011}. Centralized management software continuously checks for device misbehavior. A skipped heartbeat or an inconsistent network setting raises alarms for management software to handle. Management software triages the problem and attempts to perform automated repairs. Repairs include restarting device interfaces, restarting the device itself, and deleting and restoring a device's persistent storage. If the repair fails, management software opens a support ticket for investigation by a human. 
  \item \textbf{\emph{Stacked devices.}} The same type of fabric device can be stacked in the same rack to create a higher bandwidth virtual device~\cite{Ferreira2018}. Stacking allows fabric networks to have higher port density than the port density of proprietary network devices~\cite{Bachar2014,Simpkins2015,Bachar2015,Bagga2015}.
\end{enumerate}

Both cluster networks and fabric networks use \emph{backbone routers (BBRs)} located in \emph{edge nodes} (\ding{196}) to communicate across the WAN backbone and Internet.

\subsection{Inter Data Center Networks}
\label{subsec:Inter Data Center Networks}
\label{sec:backbone}

Facebook's WAN backbone consists of \emph{edge nodes} connected by \emph{fiber links} (\circled{11} in Figure~\ref{fig:network-arch}). Edge nodes are locations where Facebook deploys hardware to route backbone traffic. Fiber links are optical fibers that connect edge nodes, formed by optical \emph{circuits} made of optical \emph{segments}. An optical segment corresponds to a fiber optic cable and carries multiple channels, where each channel corresponds to a wavelength mapped to a router port.

Fiber link reliability is important to software systems that run in multiple data centers, especially those requiring relatively strong consistency and high availability~\cite{Brewer2017,Bailis2014}. Without careful planning, fiber cuts (e.g., due to natural disasters) can separate entire data centers or regions from the rest of the network. Common results of fiber cuts include lost capacity from edge nodes to regions or lost capacity between regions. In these cases, network operators typically reroute backbone traffic using other links, possibly with increased latency.


On top of the physical fiber-based backbone, multiple WAN backbone networks satisfy the distinct requirements of two types of traffic labeled in Figure~\ref{fig:network-arch}:

\begin{enumerate}

  \item \textbf{\emph{User-facing traffic}} (\circled{12} in Figure~\ref{fig:network-arch}) connects a person using Facebook applications like those hosted at \texttt{facebook.com}, to software systems running in Facebook data centers. To reach a Facebook data center, user-facing traffic goes through the Internet via a {\it peering}~\cite{Yap2017} process. There, {\it Internet Service Providers (ISPs)} exchange traffic among Internet domains. User-facing traffic uses the {\it Domain Name System (DNS)} to connect users to geographically local servers operated by Facebook called {\it edge nodes} (also known as \emph{points of presence})~\cite{Yap2017,Schlinker2017}. From edge nodes, user traffic arrives at Facebook's data center regions through the backbone network.

  \item \textbf{\emph{Cross data center traffic}} (\circled{13} in Figure~\ref{fig:network-arch}) connects a software service in one Facebook data center to a software service in another Facebook data center. The backbone network interconnects both cluster networks and fabric networks. By volume, cross data center traffic consists primarily of \emph{bulk data transfer} streams for replication and consistency. Bulk transfer streams are generated by \emph{backend} services that perform batch processing~\cite{Dean2004,Malewicz2010}, distributed storage~\cite{Beaver2010, f4}, and real-time processing~\cite{Huang2017,Chen2016}.

\end{enumerate}

To serve user-facing traffic, backbone networks support a range of protocols and standards to connect a variety of external networks from different ISPs. Facebook uses a traditional WAN backbone design consisting of backbone routers placed in every edge node (e.g., the BBRs in Edge 1 through 4 in the WAN backbone, \ding{196} in Figure~\ref{fig:network-arch}). In contrast, cross data center traffic is managed by software systems that route traffic between backbone routers (BBRs, \ding{196} in Figure~\ref{fig:network-arch}) built from commodity chips. The design is, in principle, similar to Google's B2 and B4 that are described in~\cite{Jain2013,Jimenez2017,Govindan2016}.

\subsection{How Network Devices Fail}
\label{subsec:How Network Devices Fail}

Network devices fail in a variety of ways ranging from faulty hardware to human operator error. The most disruptive network failures are those that disrupt the software systems running on the network, and we focus on those failures in this dissertation. (For more information on the types of failures that affect individual network devices, we refer the reader to \cite{Zhuo2017,Gill2011,Potharaju2013IMC, Potharaju2013SoCC, Ghobadi2016,Markopoulou2008,Potharaju2013SoCC}.) We call network failures that disrupt software systems {\it network incidents}. Network incidents affect software systems, causing data corruption, connection time outs, and excessive latency, for example. Software systems at Facebook include frontend web servers~\cite{Evans2011}, cache systems~\cite{Bronson2013,Nishtala2013}, storage systems~\cite{Beaver2010, f4}, data processing systems~\cite{Huang2017,Chen2016}, and real-time monitoring systems~\cite{Pelkonen2015,Kaldor2017}.

\label{sec:service-level-events}

Facebook engineers document incidents that affect software systems in reports called \emph{\underline{S}ite \underline{Ev}ents (SEVs)}.\footnote{Pronounced \textipa{[sEv]} in the \emph{International Phonetic Alphabet}~\cite{IPA1999}, rhyming with ``rev.''} SEVs fall into three severity categories ranging from SEV3 (the lowest severity, no external outage) to SEV1 (the highest severity, widespread external outage). Engineers who responded to a SEV, or whose service the SEV affected, write the SEV's \emph{report}. The report contains the incident's root cause, the root cause's effect on software systems, and steps to prevent the incident from happening again~\cite{Maurer2015}. Each SEV goes through a review process to verify the accuracy and completeness of the report. SEV reports help engineers at Facebook prevent similar incidents from happening again.

SEVs come in many shapes and sizes. We summarize three representative example SEVs in increasing site event severity:

\begin{description}

  \item[SEV3] \textbf{\textit{Switch crash from software bug.}} A bug in the switching software triggered an RSW to crash whenever the software disabled a port. The incident occurred on August 17, 2017 at 11:52~am PDT after an engineer updated the software on a RSW and noticed the behavior. The engineer identified the root cause by reproducing the crash and debugging the software: an attempt to allocate a new hardware counter failed, triggering a hardware fault. On August 22, 2017 at 11:51~am PDT the engineer fixed the bug and confirmed the fix in production.

  \item[SEV2] \textbf{\textit{Traffic drop from faulty hardware module.}} A faulty hardware module in a CSA caused traffic to drop on October 25, 2013 between 7:39~am PDT and 7:44~am PDT. After the drop, traffic shifted rapidly to alternate network devices. Web servers and cache servers, unable to handle the influx of load, exhausted their CPUs and failed 2.4\% of requests. Service resumed normally after five minutes when web servers and cache servers recovered. An on-site technician diagnosed the problem, replaced the faulty hardware module, verified the fix, and closed the SEV on October 26, 2013 at 8:22~am PDT.

  \item[SEV1] \textbf{\textit{Data center outage from incorrect load balancing.}} A core device with an incorrectly configured load balancing policy caused a data center network outage on January 25, 2012 at 3:46~am PST. Following a software upgrade, a core device began routing traffic on a single path, overloading the ports associated with the path. The overload at the core device level caused a data center outage. Site reliability engineers detected the incident with alarms. Engineers working on the core device immediately attempted to downgrade the software. Despite the downgrade, core device load remained imbalanced. An engineer resolved the incident by manually resetting the load balancer and configuring a particular load balancer setting. The engineer closed the SEV on January 25, 2012 at 7:47~am PST.

\end{description}

\subsection{How Network Errors are Handled}
\label{sec:incident_def}

Facebook shields software systems from common network failures with \emph{automated repair software}~\cite{Power2011}. Automated repair software prevents common network \emph{failures} from causing network \emph{incidents}. It runs on RSWs, FSWs, and core devices. We list automated repair software data from April 1 to May 1, 2018 in Table~\ref{tab:auto-remediation}. During this time, automated repair software fixed 99.7\% of RSW failures, 99.5\% of FSW failures, and 75\% of core device failures.\footnote{Automated repair software is less effective for core devices because many core devices run third-party vendor software that is \emph{incompatible} with automated repairs.}

\begin{table}[H]
  \vspace{-1ex}
  \centering
  \begin{tabular}{ccc}
    \toprule
    \textbf{Device} & \textbf{Repair Ratio} & \textbf{Avg Priority / Wait / Repair Time} \\
    \midrule
    Core & 75\% & 0 (highest priority) / 4 m / 30.1 s \\
    FSW & 99.5\% & 2.25 / 3 d / 4.45 s \\
    RSW & 99.7\% & 2.22 / 1 d / 2.91 s \\
    \bottomrule
  \end{tabular}
  \caption{The repair ratio (fraction of issues repaired with automated repair versus all issues), average priority (0 = highest, 3 = lowest), average wait time, and average repair time for the network device types that automated repair software supports.}
  \label{tab:auto-remediation}
  \vspace{-4.5ex}
\end{table}

Automated repair software schedules a repair based on its \emph{priority}: low priority repairs likely wait longer than high priority repairs. Engineers assign repairs a priority from 3 (the lowest priority) to 0 (the highest priority). Core device repairs have the highest priority, and wait only minutes on average, because core devices connect data centers. FSW and RSW repairs have lower priorities on average, 2.25 and 2.22, respectively, and wait \emph{days}. Repairs happen relatively fast once they run, taking less than a minute on average. Core device repairs take around \SI{30.1}{\second} on average; FSW and RSW repairs take around \SI{4.45}{\second} and \SI{2.91}{\second} on average, respectively.

If automated repair software cannot fix a device's failure, the software alerts a human technician to investigate the device. Four root causes constitute the top 90.9\% of failures handled by automated repairs at Facebook. (1) 50\% of repairs fix device port ping failures by turning the port off and on again. (2) 32.4\% of repairs fix configuration file backup failures by restarting the configuration service and reestablishing a secure shell connection. (3) 4.5\% of repairs handle fan failures by extracting failure details and alerting a technician to examine the faulty fan. (4) 4.0\% of repairs handle entire device ping failures by collecting device details and assigning a task to a technician.


Network failures in inter data center networks are handled by the third-party vendors that lease fiber and \emph{do not} use automated repair software. These types of failures typically involve teams of operators deployed to the site of failure to perform physical repairs to the failed infrastructure, like a fiber optic cable or an optical amplifier. Depending on the location of the repairs, the time for the operators to get to the failure site and perform fixes can vary widely. For example, fiber vendors can respond to a network cable failure in a big city much faster than a network cable failure in the middle of the Atlantic Ocean.

Network errors that are neither detected nor fixed by automated repair software can result in network incidents, which may lead to SEVs. Engineers perform work to fix the root cause of the SEV. This work sometimes involves the engineer updating configurations, patching software, or repairing devices.

\subsection{Related Research in Network Failures in Modern Data Centers}
\label{subsec:Related Research in Network Failures in Modern Data Centers}


Several large scale data center failure
studies~\cite{Barroso2013,Gunawi2016,Oppenheimer2003,Brewer2017, Wang2017} report that
network incidents are among the major causes of web service outages. However,
none of these studies systematically analyze network incidents at a large
scale, focusing on the availability of an \emph{entire web service}, across both
\emph{inter} and \emph{intra} data center networks, in a long term, longitudinal study.

Other studies examine the failure characteristics of
network links and devices in \emph{different} types of networks than the modern data center networks described in this dissertation,
including traditional data center networks (similar to the cluster-based data center network design)~\cite{Zhuo2017,Gill2011,Potharaju2013IMC,
Potharaju2013SoCC} and optical
backbones~\cite{Ghobadi2016,Markopoulou2008,Potharaju2013SoCC}.

Potharaju and Jain~\cite{Potharaju2013SoCC} and Turner et al.~\cite{Turner2010} study data center network
infrastructure by characterizing device and link failures in intra and inter
data center networks. Their studies characterize the failure impact,
including connectivity losses, high latency, packet drops, and so on.  These studies
significantly boost the understanding of network failure characteristics, and
provide insights for network engineers and operators for improving the fault
tolerance of existing networks and for designing more robust networks. 

Govindan et al.~\cite{Govindan2016} study over 100 failure events in Google WAN
and data center networks, offering insights into why maintaining high levels of
availability is challenging for content providers. Their study, similar
to~\cite{Barroso2013,Gunawi2016,Oppenheimer2003,Brewer2017}, focuses on network
management and the design principles for building robust networks.  Many of the
high-level design principles mentioned in~\cite{Govindan2016}, such as using
multiple layers of fallback (defense in depth), continuous prevention, and
fast recovery, are applicable to large scale software systems to protect
against network incidents.

We perform a large scale study of all the network incidents in Facebook's intra and inter data center networks~\cite{Meza17}, described in detail in Chapter~\ref{chp:networkfailures}. While our work is closely related to Potharaju and Jain~\cite{Potharaju2013SoCC} and Turner et al.~\cite{Turner2010}, it also
\emph{fundamentally} differs from and \emph{complements} prior work in the following three
aspects.  First, our work has a different goal.  Unlike prior studies
that focus on understanding fine-grained per-device, per-link failures and
their impact on the system-level services above the network stack, our work
focuses on how network incidents affect \emph{the availability of an Internet
service}.  Our goal in this work is to reveal and quantify the incidents that \emph{cannot}
be tolerated despite industry best practices, and shed light on how large scale
systems can operate reliably in the presence of these incidents.  Second, prior studies
only examine data center and backbone networks with traditional cluster network designs, whereas our work presents a comparative study of the reliability
characteristics of data center network infrastructure with \emph{both} a
traditional cluster network design and a contemporary fabric network design with
smaller, commodity switches. We achieve this due to the heterogeneity of the data
center network infrastructure of Facebook where networks with different
designs co-exist and co-operate.  Third, we present a long-term (seven years
for intra data center networks and eighteen months for inter data center networks)
longitudinal analysis to reveal the evolution of network reliability
characteristics, while prior studies typically provide only aggregated results,
often over a much shorter period or with orders of magnitude fewer devices~\cite{Turner2010}.


\section{Other Devices}

We did not analyze other devices commonly found in data centers, such as CPUs, GPUs, non--flash-based non-volatile memory, HDDs, and power delivery devices. Some of these devices have been examined at a large scale before. For example, Nightingale et al.~\cite{nightingale} examined CPUs and HDDs across a large number of consumer computers running Microsoft operating systems. More recent studies in High-Performance Computing (HPC) environments examined the failure rates of GPUs~\cite{Tiwari2015, Nie2016}. Researchers from Facebook (including the author of this dissertation) also performed a limited study of the failure modes of breakers in data center power delivery devices~\cite{Wu2016}. It is our hope that future studies shed light on the failure characteristics of emerging devices and periodically update the community's understanding of mainstream devices.

In this chapter, we discussed how modern data centers are organized and provided a comprehensive background on the design and reliability of three devices that are important for the operation of modern data centers: DRAM, flash-based SSDs, and the network. We also provided a comprehensive overview of data center reliability studies each device. We next discuss our work to shed light on new trends from the field, starting with DRAM devices.

\chapter{DRAM Failures}
\label{chp:dramfailures}

Computing systems use dynamic random-access memory (DRAM) as main memory.  As prior works have shown, failures in DRAM devices are an important source of errors in modern servers~\cite{schroeder, hwang, amd, amd2, amd3, meza15-2, Wang2017}.  To reduce the effects of memory errors, error correcting codes (ECC) have been developed to help detect and correct errors when they occur.  In order to develop effective techniques, including new ECC mechanisms, to combat memory errors, it is important to understand the memory reliability trends in modern systems.

In this chapter, we analyze the memory errors in the \emph{entire fleet} of servers at Facebook over the course of fourteen months, representing billions of device days. The systems we examine cover a wide range of devices commonly used in modern servers, with DIMMs manufactured by 4 vendors in capacities ranging from \unit[2]{GB} to \unit[24]{GB} that use the modern DDR3 communication protocol.

We report on several new reliability trends for memory systems that had not been discussed in literature before our work~\cite{meza15-2}.  We show that (1) memory errors follow a power-law, specifically, a Pareto distribution with decreasing hazard rate, with average error rate exceeding median error rate by around {55$\times$}; (2) non-DRAM memory failures from the memory controller and memory channel cause the majority of errors, and the hardware and software overheads to handle such errors cause a kind of denial of service attack in some servers; (3) using our detailed analysis, we provide the first evidence that more recent DRAM cell fabrication technologies (as indicated by chip density) have substantially higher failure rates, increasing 1.8$\times$ over the previous generation; (4) DIMM architecture decisions affect memory reliability:  DIMMs with fewer chips and lower transfer widths have the lowest error rates, likely due to electrical noise reduction; (5) while CPU and memory utilization do not show clear trends with respect to failure rates, \emph{workload type} can influence failure rate by up to {6.5$\times$}, suggesting certain memory access patterns may induce more errors; (6) we develop a model for memory reliability and show how system design choices such as using lower density DIMMs and fewer cores per chip can reduce failure rates of a baseline server by up to 57.7\%; and (7) we perform the first implementation and real-system analysis, on a cluster of thousands of servers, of page offlining at scale, showing that it can reduce memory error rate by 67\%, and identify several real-world impediments to the technique. We also evaluate a new technique to improve DRAM reliability, \emph{physical page offlining}, and discuss the overheads of physical page offlining.

\section{Motivation for Understanding DRAM Failures}

Computing systems store a variety of data in memory---program variables, operating system and file system structures, program binaries, and so on.  The main memory in modern systems is composed of dynamic random-access memory (DRAM), a technology that, from the programmer's perspective, has the following property:  a byte written to an address can be read correctly, repeatedly, until it is overwritten or the machine is turned off.  All correct programs rely on DRAM to operate in this manner and DRAM manufacturers work hard to design reliable devices that obey this property.

Unfortunately, DRAM does not always obey this property.  Various events can change the data stored in DRAM, or even permanently damage DRAM, as we discussed in \S\ref{subsec:How DRAM Devices Fail}. Some documented events include transient charged particle strikes from the decay of radioactive molecules in chip packaging material, charged alpha particles from the atmosphere~\cite{intel-dram}, and wear-out of the various components that make up DRAM chips (e.g.,~\cite{const,dram-wearout}).  Such faults, if left uncorrected, threaten program integrity.  To reduce this problem, various error correcting codes (ECC) for DRAM data~\cite{hamming,ibm-chipkill} have been used to detect and correct memory errors.  However, these techniques require additional DRAM storage overheads~\cite{yixin}, require additional  DRAM controller complexity, and cannot detect or correct all errors.

Much past research analyzed the causes and effects of memory errors in the field (see our comprehensive discussion in \S\ref{subsec:Related Research in DRAM Failures in Modern Data Centers}).  These past works identified a variety of DRAM failure modes and formed the basis of the community's understanding of DRAM reliability.  Our goal is to strengthen the understanding of DRAM failures in the field by comprehensively studying \emph{new trends} in DRAM errors in a large-scale production datacenter environment using \emph{modern} DRAM devices and workloads.  To this end, this chapter presents our analysis of memory errors across Facebook's \emph{entire fleet} of servers over the course of fourteen months and billions of device days.

We provide four main contributions.  We: \emph{(1) analyze new DRAM failure trends in modern devices and workloads that have not been identified in prior work, (2) develop a model for examining the memory failure rates of systems with different characteristics, (3) describe and perform the first analysis of a large-scale implementation of a software technique proposed in prior work to reduce DRAM error rate (page offlining~\cite{solaris-pageoffline}), and (4) examine a new technique to improve DRAM reliability, physical page offlining.}  Specifically, we observe several new reliability trends for memory systems that have not been discussed before in literature and evaluate two techniques to improve DRAM reliability:

\begin{enumerate}

  \item The number of memory errors per machine follows a \emph{power-law} distribution, specifically a Pareto distribution, with decreasing hazard rate.  While prior work reported the \emph{average} memory error rate per machine, we find that the average exceeds the \emph{median} amount by around $55\times$, and thus may not be a reliable number to use in various studies.

  \item Non-DRAM memory failures, such as those in the memory controller and the memory channel, are the source of the \emph{majority} of errors that occur.  Contrary to popular belief at the time when we first published this finding~\cite{meza15-2}, memory errors are \emph{not} always isolated events and can bombard a server (if not handled appropriately), creating a kind of \emph{denial of service attack}.  

  \item DRAM failure rates increase with newer cell fabrication technologies (as indicated by chip density, which is a good indicator of technology node): \unit[4]{Gb} chips have $1.8\times$ higher failure rates than \unit[2]{Gb} chips.  Prior work that examined DRAM \emph{capacity}, which is \emph{not} closely related to fabrication technology, observed inconclusive trends.  Our empirical finding is that the quadratic rate at which DRAM density increases with each generation makes maintaining or reducing DRAM failure rate untenable, as a recent paper by Samsung and Intel~\cite{samsung-intel} also indicated.

  \item DIMM architecture characteristics, such as the number of data chips per DIMM and the transfer width of each chip, affect memory error rate.  The best architecture for device reliability occurs when there are both low chips per DIMM and small transfer width.  This is likely due to reductions in the amount of electrical disturbance within the DIMM.

  \item The \emph{type} of work that a server performs (i.e., its workload), and \emph{not} CPU and memory utilization, affects failure rate.  We find that the DRAM failure rate of different workloads varies by up to $6.5\times$.  This large variation in workloads is potentially due to memory errors induced by certain access patterns, such as accessing the same memory location in rapid succession, as shown in controlled studies in prior work~\cite{rowhammer}.

  \item We develop a model for quantifying DRAM reliability across a wide variety of server configurations and show how it can be used to evaluate the server failure rate trends for different system designs.  We show that using systems with lower density DIMMs or fewer CPUs to access memory can reduce DRAM failure rates by 57.7\% and 34.6\%, respectively.  We make this model publicly available at~\cite{safari-tools}.

  \item We describe our implementation of page offlining~\cite{solaris-pageoffline} at scale and evaluate it on a fraction (12,276) of the servers that we examine.  We show that it can reduce memory error rate by around 67\%.  While prior work reported larger error rate reductions in simulation~\cite{hwang}, we show that real-world factors such as memory controller and memory channel failures and OS-locked pages that cannot be taken offline can limit the effectiveness of this technique.

  \item We examine a new technique to improve DRAM reliability that periodically moves the contents of memory pages in physical memory, \emph{physical page randomization}. We report on the overheads of a proof of concept prototype running in the Linux 3.10.17 kernel and find the average page randomization time to be $374.9\;\mu\mathrm{s}$, and derive a formula for the amount of memory utilization the technique requires to randomize a fixed amount of physical memory over the course of a certain number of days.
\end{enumerate}

\section{Methodology for Understanding DRAM Failures}
\label{sec:analytical}

We examine all of the DRAM devices in Facebook's server fleet, which have operational lifetimes extending across four years and comprise billions of device days of usage.  We analyze data over a fourteen month period.  We examine six different system types with hardware configurations based on the resource requirements of the workloads running on them.  Table~\ref{tab:workloads}, reprinted below, lists the workloads and their resource requirements.

\begin{table}[H]
  \centering
  \begin{tabular}{|c||c|c|c|} \hline
    \multirow{2}{*}{Workload} & \multicolumn{3}{c|}{Resource requirements} \\ \cline{2-4}
    & Processor & Memory & Storage \\ \hline \hline
    Web & High & Low & Low \\ \hline
    Hadoop~\cite{hadoop} & High & Medium & High \\ \hline
    Ingest~\cite{datainfra} & High & High & Medium \\ \hline
    Database~\cite{datainfra} & Medium & High & High \\ \hline
    Memcache~\cite{Nishtala2013} & Low & High & Low \\ \hline
    Media~\cite{haystack} & Low & Low & High \\ \hline
  \end{tabular}
\end{table}

\subsection{The Systems We Examine}

The memory in these systems covers a wide range of devices commonly used in servers.  The DIMMs were manufactured by 4 vendors in capacities ranging from \unit[2]{GB} to \unit[24]{GB} per DIMM.  DDR3 is the protocol used to communicate with the DIMMs.  The DIMM architecture spans devices with 1, 2, and 4 ranks with 8, 16, 32, and 64 chips.  The chip architecture consists of 8 banks with \unit[16]{K}, \unit[32]{K}, and \unit[64]{K} rows and \unit[2]{K} to \unit[4]{K} columns, and has chips that transfer both 4 and 8 bits of data per clock cycle.  We analyze three different chip densities of \unit[1]{Gb}, \unit[2]{Gb}, and \unit[4]{Gb}, which are closely related to DRAM fabrication technology.

The composition of the modules we examine differs from prior studies (e.g., \cite{schroeder,hwang,amd,amd2,amd3,selse,nightingale}) in three ways:  (1) it consists of a current DRAM access protocol (DDR3, as opposed to older generation protocols with less aggressive memory bus clock frequencies, such as DDR and DDR2 in~\cite{schroeder}); (2) it consists of a more diverse range of DRAM device organizations (e.g., DIMMs with a variety of ranks, chips, rows, and columns versus the more homogeneous DIMMs of~\cite{schroeder,hwang,amd,amd2,amd3,selse}); and (3) it contains DIMMs with characteristics that have never been analyzed at a large-scale (such as density, number of chips, transfer width, and workload).

Some of the systems we examined have hardware memory scrubbing\more{}{~\cite{scrubbing}} enabled, which causes the memory controller to traverse memory, detecting (but not correcting) memory errors to determine faulty memory locations.  The system enables the hardware scrubber when the machine enters a low enough idle state, so the scrubbing rate of machines varies depending on their utilization.

\subsection{How We Measure DRAM Failures}

We use the \texttt{mcelog} Linux kernel module to log memory errors in a file.  We do this for every machine in the fleet.  For each correctable memory error, we collect: (1) the time when the error occurred; (2) the physical memory address being accessed when the error occurred; (3) the name of the server the memory error occurred on; (4) the socket, channel, and bank the physical memory address is located on; and (5) the type of memory access performed when the error occurred (e.g., read or write).  Uncorrectable errors halt the execution of the processors on the machines we examine, and cause the system to crash.  We do not have detailed information on uncorrectable errors, so we measure their occurrence by examining a separate log of uncorrectable errors that is kept in non-volatile memory on the servers that we examine.  We use a script to collect and parse log data that is then stored in a Hive~\cite{hive} table.  The script runs every ten minutes.

In addition to information about correctable errors, we also collect information about the systems that have errors (e.g., CPU utilization and system age; see Table~\ref{tab:characteristics} for details).  Collecting system data is done in a separate step and is later combined with error data.

We do not distinguish between transient and permanent faults in our study. This is because permanent faults can not be identified without performing a more intensive analysis of failing DIMMs. We \emph{do} provide circumstantial evidence of failures that exhibit characteristics of permanent failures: those that lead to a repeat errors over a longer period of time.

The scale of the systems we analyze and the amount of data we collect pose make analyzing all of the data challenging. To process billions of device days of information, we use a cluster of machines to perform a parallel aggregation of the data using MapReduce jobs.  The aggregation produces a set of statistics for each of the devices we analyze.  We then process this summary data in R~\cite{r} to collect our results.

\subsection{How We Analyze DRAM Failure Trends}

When we analyze the reliability trends with respect to a system characteristic (e.g., chip density or CPU utilization), we group systems into buckets based on the particular characteristic and plot the failure rate of the systems in each bucket.  When performing bucketing, we round the value of a device's characteristic to the nearest bucket and we eliminate buckets that contain less than 0.1\% of the systems we analyze in order to have a statistically significant sample of systems in our measurements.  We show the 95th percentile confidence interval for our data when relevant.

Due to the size of the fleet, we could not collect detailed information for all the systems \emph{without} errors (we \emph{do} collect detailed information for every system \emph{with} errors).  So, in \S\ref{sec:factors} and \S\ref{sec:model}, instead of examining the \emph{absolute} failure rate among different types of servers, we examine the \emph{relative} failure rate compared to a more manageable size of servers that we call the \emph{control group}.  The servers in the control group are uniformly randomly selected from among all the servers that did not have memory errors, and we collect detailed information on the servers in this group.

Note that such a selection process preserves the distribution of server types in the underlying fleet, and you can think of our analysis in \S\ref{sec:factors} and \S\ref{sec:model} as if it were on a ``scaled down'' version of the fleet.  The size of the control group is equal to the size of the error group, and the sizes of these groups are sufficiently large to be statistically significant.  We bucket servers in each group based on their value for a given characteristic (e.g., age) and plot the fraction of servers \emph{with} errors compared to servers \emph{without} errors in each bucket, which we call the \emph{relative server failure rate}.  With relative server failure rate, we perform \emph{relative comparisons} between failure rates that we compute for different factors, but the absolute value of the metric does not have any substantial meaning.  We find that the \emph{relative} failure rates we examine are in the range $[0, 1]$, and we plot our data within this range.  In \S\ref{sec:factors} and \S\ref{sec:model}, when we refer to \emph{failure rate} we mean \emph{relative server failure rate} and as a reminder that our data should not be confused with absolute failure rates, we label our graphs with \emph{relative server failure rate}.

\subsection{Limitations and Potential Confounding Factors}

While we attempt to control for a variety of factors in our analysis, our study has several limitations and potentially confounding factors that we would like to discuss.

\begin{itemize}
  \item \textbf{\emph{Silent data corruption.}} The servers we analyze do not allow us to determine if bits flipped in a memory location in such a way that the bit flips were undetectable using ECC metadata.
  \item \textbf{\emph{Linear correlations in our model.}} The statistical technique we use to build our model, a logistic regression, only models linear correlations between factors and DRAM errors. If a factor has a non-linear correlation (e.g., a linear increase in the factor, such as chip density, results in a quadratic increase in the error rate), our model will find it to be not statistically significant.
  \item \textbf{\emph{Relative failure rate.}} The output of our model is a failure rate to be used to make \emph{relative} comparisons between server configurations (e.g., the model can predict if a configuration, $A$, has a higher rate of failures than a configuration, $B$, per year). Our model \emph{can not} predict the \emph{absolute} failure rate of a server configuration.

  \item \textbf{\emph{Workloads and system configuration.}} The different workloads that we examine run on different system configurations, as we show in Table~\ref{tab:workloads}. We do not distinguish between whether the effects of the \emph{workload} or the effects of the \emph{system configuration} have a more predominant effect on the trends that we observe.
\end{itemize}

\section{DRAM Failure Trends}
\label{sec:baseline}

We will first focus on the overall error rate and error distribution among the systems that we analyze and then examine correlations between different factors and failure rate.

\subsection{Incidence Error Rate and Error Count}
\label{sec:incidence}

\begin{figure}[H]
  \newcommand{\curfig}{timeline}
  \centering
  \includegraphics[width=0.5\columnwidth]{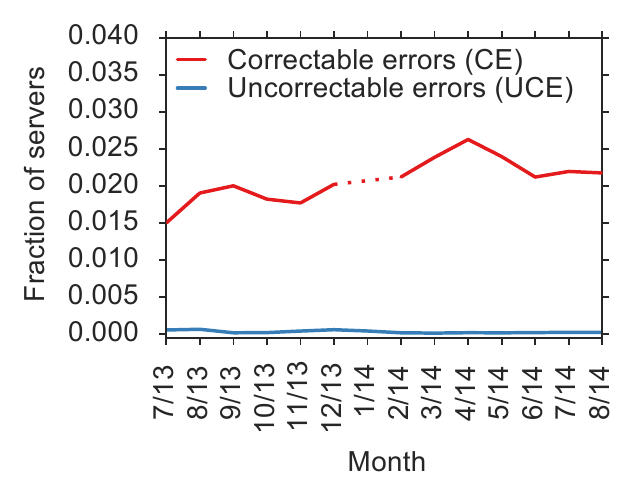}
  \caption{Timeline of correctable and uncorrectable memory errors.}
  \label{fig:timeline}
\end{figure}

Figure~\ref{fig:timeline} shows the monthly incidence error rate for memory over a span of fourteen months.  The \emph{incidence error rate} is the fraction of servers in the fleet that have memory errors compared to the total size of the fleet.\footnote{Correctable error data for January 2014 (1/14) is not available.  Note that if a server has multiple errors in multiple months, it will be represented in multiple data points.}  We observe three trends with respect to incidence error rate.

First, correctable errors occur relatively commonly each month, affecting 2.08\% of servers on average.  Though such errors do not corrupt data, they do reduce machine performance due to the hardware required to reconstruct the correct data.  While a single correctable error may not negatively affect system performance, a large number of correctable errors could lead to performance degradation.  We examine the distribution of the number of correctable errors among machines at the end of this section.

To compare against prior work, we measure the correctable error incidence rate over the course of twelve months (7/13 up to and including 7/14, excluding 1/14) and find that, cumulatively across all months, around 9.62\% of servers experience correctable memory errors.  This is much lower than the yearly correctable error incidence rate reported in work from the field seven years before our study (32.2\% in Table~1 in~\cite{schroeder}) and comparable with the 5.48\% to 9.10\% failure rate reported in more recent work~\cite{amd2} from two years before our study.  Thus, though the overall correctable error incidence rate \emph{decreased} over the better part of a decade of device improvements, our measurements corroborate the trend that \emph{memory errors are still a widespread problem in the field}.

In addition, we find that the correlation between a server having a correctable error in a given month, depending on whether there were correctable errors observed in the previous month is 31.4\% on average.  In comparison, prior work from the field found around a 75\% correlation in correctable errors between two consecutive months~\cite{schroeder}.  Our lower observed amount of correlation is partially due to how the servers we evaluate \emph{handle} memory errors:  repair software flags a server for memory repair if the server has more than 100 correctable errors per week, whereas prior work (e.g.,~\cite{schroeder}) \emph{only} examined the effects of replacing components with \emph{uncorrectable} errors.  Under the more aggressive and proactive repair policy that we examine, we find that on average around 46\% of servers with errors are repaired each month.  As a result, in contrast to prior work, we find that a majority (69.6\%) of the machines that report errors each month are \emph{not} repeat offenders from the previous month.

Second, the rate of uncorrectable errors is much smaller than the rate of correctable errors, with uncorrectable errors affecting 0.03\% of servers each month on average.  Recall that uncorrectable errors cause a server to crash, increasing downtime and potentially causing data loss.  Therefore, it is desirable to decrease the rate of uncorrectable errors as much as possible.

Schroeder et al.\ conjectured that repair policies ``where a DIMM is replaced once it experiences a significant number of correctable errors, rather than waiting for the first uncorrectable error'' could reduce the likelihood of uncorrectable errors~\cite{schroeder}.  To test this hypothesis in the field on our systems that have automated repair software that repairs servers with more than 100 correctable errors, we compare the rate of uncorrectable errors relative to the rate of correctable errors, in order to control for the change in rate of correctable errors between the two studies.  Interestingly, in Schroeder et al.'s study, uncorrectable error rate is only $25.0\times$ smaller than the correctable error rate, while in our study it is $69.3\times$ smaller.  If more aggressive repair policies indeed lead to higher server reliability, then our results suggest that we can lower uncorrectable error rate by up to $2.8\times$ (i.e., $69.3\times$ / $25.0\times$). We achieve this by repairing around 46\% of the machines with errors (those with more than 100 correctable errors).  System designers must decide whether the benefit in reduction of potential data loss is worth the higher repair rate.

Third, the incidence error rate for correctable errors fluctuates little (its standard deviation is $\pm0.297\%$) and is relatively stable over the fourteen months that we examine.  Uncorrectable errors also remain low in comparison to correctable errors (with a standard deviation of $\pm0.018\%$).  We attribute the low standard deviation in error behavior over time to the large population size that we examine.

Figure~\ref{fig:distribution-dram} (left) shows the distribution of correctable errors among servers that had at least one correctable error.  The x axis is the normalized device number, with devices sorted based on the number of errors they had during a month.  The y axis shows the total number of errors a server had during the month in log scale.  Notice that the maximum number of logged errors is in the millions.  We observe that a small number of servers have a large number of errors.  For example, the top 1\% of servers with the most errors have over 97.8\% of all the correctable errors we observe.  We also find that the distribution of number of errors among servers is similar to that of a power-law distribution with exponent $-2.964$.  Prior work observed that some failed devices, such as the memory controller or bus, can account for a large number of errors (e.g., \cite{selse}), though the \emph{full distribution} of errors has \emph{not} been quantified.

\begin{figure}[H]
  \newcommand{\curfig}{distribution-dram}
  \centering
  \includegraphics[width=0.4\columnwidth]{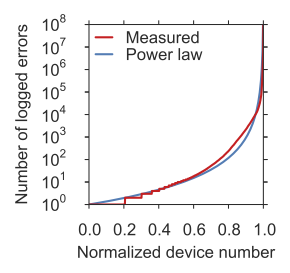}
  \includegraphics[width=0.4\columnwidth]{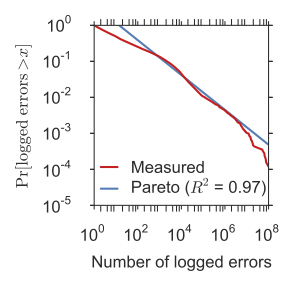}
  \caption{The distribution of memory errors among servers with errors (left) resembles a power-law distribution.  Memory errors also follow a Pareto distribution among servers with errors (right).}
  \label{fig:\curfig}
\end{figure}

  Figure~\ref{fig:distribution-dram} (right) shows the probability density distribution of correctable errors.  The x axis is the number of errors per month and the y axis is the probability of a server having at least that many errors per month.  A Pareto distribution (a special case of the power law) has been fit to the data we measure.  Similarly to past works that found decreasing hazard rates in the behavior of systems (e.g., Unix process lifetimes~\cite{dhr1}, sizes of files transferred through the Web~\cite{dhr2, dhr3}, sizes of files stored in Unix file systems~\cite{dhr4}\also{}{, durations of FTP transfers in the Internet}~\cite{dhr5}, CPU requirements for supercomputing jobs~\cite{dhr6}, and memory access latencies~\cite{atlas}), we find that the distribution of errors across servers follows a Pareto distribution, with a decreasing hazard rate. This means, roughly, that the more errors a server has so far, the more errors we expect it to have in the future.\footnote{Note that one can take advantage of this property to potentially predict which servers may have errors in the future. We leave this as a direction for future research, discussed in Chapter~\ref{sec:Lessons Learned Across Many Billions of Device-Hours}. For more information on the Pareto distribution, decreasing hazard rate, and their properties, we refer the reader to~\cite{atlas,mor}.}

Quantifying the skewed distribution of correctable errors is important as it helps us to diagnose the severity of a memory failure relative to the population.  For comparison, Schroeder et al.\ reported a mean error rate of 22,696 correctable errors per server per year (Table 1 in~\cite{schroeder}), or 1,891 correctable errors per server per month.  Without knowing the underlying distribution, however, it is not clear whether \emph{all} servers in the study had such a large number of errors each month or whether this average is dominated by a small number of outliers (as we observe here).

If we compute the mean error rate as in prior work, we observe 497 correctable errors per server per month.  However, if we examine the error rate for the \emph{majority} of servers (by taking the median errors per server per month), we find that most servers have at most 9 correctable errors per server per month.\footnote{Concurrent work by Sridharan et al.~\cite{goodbadugly} makes a similar observation, though we quantify and provide a model for the \emph{full distribution} of errors per server.}  In this case, using the mean value to estimate the value for the majority overestimates by over $55\times$.  We therefore conclude that, for memory devices, the skew in how errors are distributed among servers, means that we must examine the full distribution of memory errors per server.  If we do this, we see that memory errors follow a power-law distribution, which we can use to accurately assess the severity of machine failures.  Therefore, we hope future studies that use error data from the field take into account the new distribution we observe and openly provide.

In addition, we find that hardware scrubbing detects 13.1\% of the total number of errors.  While we did not monitor how many servers use scrubbing, we observe that 67.6\% of the servers with errors found at least one error with scrubbing.  We do not have details on memory access information, so the interaction between scrubbing and different workloads is not clear, and requires further examination.

\subsection{Component Failure Analysis}
\label{sec:component}

Memory errors occur due to failures in a DRAM device as well as if the memory controller in the processor fails or if logic for transmitting data on a memory channel fails.  While prior works examined DRAM chip-level failures (\cite{hwang, amd, amd2}) and memory controller/channel failures (\cite{selse}) \emph{separately}, no prior work comprehensively examined failures across the \emph{entire memory system}.

We adopt a methodology for classifying component failures similar to prior work (e.g.,~\cite{hwang, amd, amd2, selse}).  We examine all of the correctable errors across the fleet each month.  We begin by determining each correctable error's corresponding processor socket, memory channel, bank, row, column, and byte offset.  Then, we group errors by the component that failed and caused the error to occur.  For grouping errors by components, we use the following criteria:

\begin{description}

\item[Socket]  If there are $>\unit[1]{K}$ errors across $>1$ memory channel on the same processor socket, we classify those errors as being due to a socket failure.  The $>\unit[1]{K}$ error threshold was chosen so as to ensure that the failures we classify are not due to a small number of independent cell failures.  To verify how we classify socket failures, we examine repair logs of the servers we classify with socket failures and find that 50\% of the servers have a large number of errors that require replacing the processor to eliminate the errors and 50\% contain intermittent bursts of errors that cause the server to become unresponsive for long periods of time---both of these are characteristics of socket failures that generate a large number of machine check exceptions, as prior work~\cite{selse} observed.

\item[Channel]  After excluding the above errors, if there were $>\unit[1]{K}$ errors across $>1$ DRAM banks on the same memory channel, we classify those errors as being due to a channel failure.  Similar to sockets, we examine repair logs of the servers we classify with channel failures and find that 60\% of the servers that we classify as being due to channel failures did not have any repair action in the repair software's logs (replacing or reseating the DIMM). This suggests that some channel failures failures are transient, potentially due to temporary misalignment of the transmission signal on the channel.  The other 40\% of servers that we classify as being due to channel failures require DIMM replacement, suggesting permanent failures due to the channel transmission logic (e.g., the I/O circuitry) within the DIMM.

\item[Bank]  After excluding the above errors, we repeat the procedure for banks, classifying a bank failure as $>\unit[1]{K}$ errors across $>1$ row in a bank.  Note that our study examines monthly failure trends, and we assume that multiple row failures in the same bank in the same month may be more indicative of a bank failure than multiple independent row failures in the bank.

\item[Row]  After excluding the above errors, we classify a row failure as $>1$ column in the same row having errors.

\item[Column]  After excluding the above errors, we classify a column failure as $>1$ error in the same column.

\item[Cell]  After excluding the above errors, we classify a cell failure as $>1$ error in the same byte within 60 seconds.  We choose this amount of time because we find that 98.9\% of errors at a particular byte address have another error at the same address within 60 seconds if they ever have an error at the same byte address again in the same day.

\item[Spurious]  After excluding the above errors, we are left with what we call \emph{spurious} errors.  Spurious errors are appear on individual cells that do not share a common component failure and do not repeat in a short amount of time.  Potential causes of spurious errors include alpha particle strikes from the atmosphere or chip packaging~\cite{intel-dram} and cells with weak or variable charge retention times~\cite{liu, samira, samsung-intel}.

\end{description}

\begin{figure}[H]
\centering
\begin{minipage}[t]{0.45\columnwidth}
  \newcommand{\curfig}{faults-error}
  \centering
  \includegraphics[width=0.9\columnwidth]{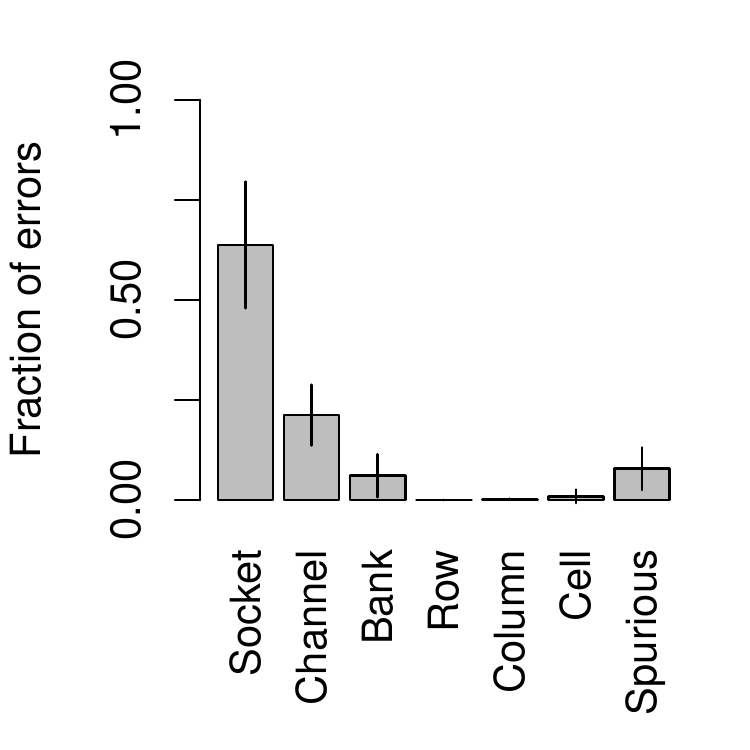}
  \caption{The distribution of errors among different memory components.  Error bars show the standard deviation of total errors from month to month.}
  \label{fig:faults-error}
\end{minipage}
\hspace{0.5em}
\begin{minipage}[t]{0.45\columnwidth}
  \newcommand{\curfig}{faults-servers}
  \centering
  \includegraphics[width=0.9\columnwidth]{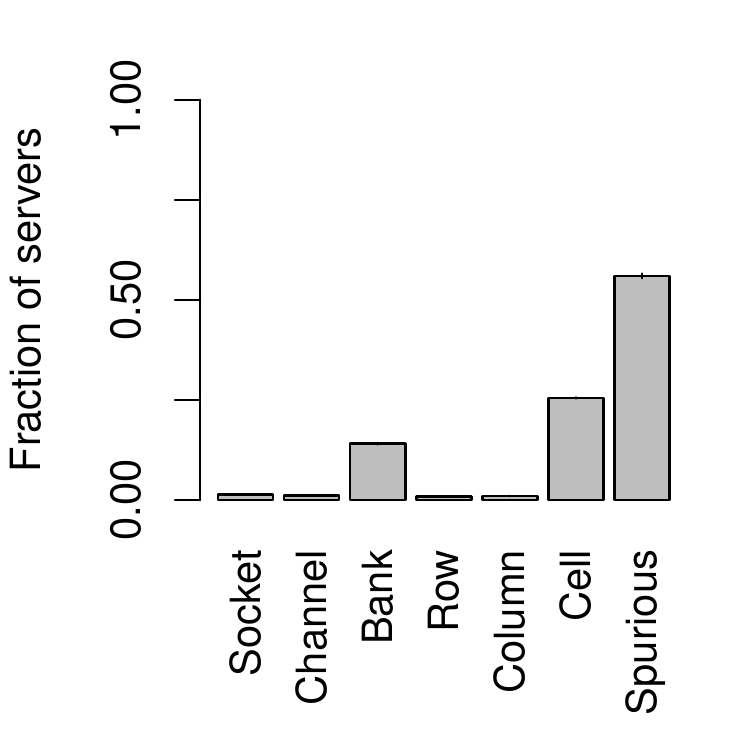}
  \caption{The fraction of servers with memory errors that have each type of memory component failure.}
  \label{fig:faults-servers}
\end{minipage}

\end{figure}

Figure~\ref{fig:faults-error} shows the fraction of errors each month due to different types of failures.  Error bars show the standard deviation between months.

Sockets and channels generate the most errors when they fail, 63.8\% and 21.2\% of all errors each month, respectively.  This is because when these components fail, they affect a large amount of memory.  Compared to a prior work that examined socket (memory controller) and channel failures~\cite{selse} (but did not examine DRAM chip-level failures), we find that our systems have $2.9\times$ more socket errors and $5.3\times$ more channel errors.  This could be due to differences in the server access patterns to memory or how quickly servers crash when experiencing these types of failures.

That sockets and channels cause a large number of errors when they fail helps explain the skew in the distribution of errors among servers (Figure~\ref{fig:distribution-dram}\also{}{, left}).  For example, servers with socket failures had the highest number of errors in the distribution.  This large source of errors, if not accounted for, can confound memory reliability conclusions by artificially inflating the error rates for memory and creating the appearance of more DRAM chip-level failures than in reality.  Besides the work that only measured socket and channel failures, but \emph{not} DRAM chip-level failures (\cite{selse}), we did not find mention of \emph{controlling for socket and channel errors} in prior work that examined errors in the field (e.g.,~\cite{schroeder, hwang, amd, amd2, amd3, shen1, shen2}).

%
%

We observe that DRAM chip-level (banks, rows, columns, cells, and spurious) failures contribute a relatively small number of errors versus sockets and channels:  6.06\%, 0.02\%, 0.20\%, 0.93\%, and 7.80\%, respectively.  This is because when these components fail, they affect only a relatively small amount of memory.  Based on these findings, to help with the diagnosis of memory failures, we recommend that memory error classification \emph{should always} include components such as sockets and channels.

So far, we have examined how component failures relate to the number of \emph{errors}.  We next turn to how component failures \emph{themselves} (the underlying source of errors) are distributed among servers.  Figure~\ref{fig:faults-servers} shows what fraction of servers with correctable errors each month have each type of failure that we examine.  We plot error bars for the standard deviation in fraction of servers that report each type of error between months, though we find that the trends are remarkably stable, and the standard deviation is correspondingly very low (barely visible in Figure~\ref{fig:faults-servers}).


Notice that though socket and channel failures account for a large fraction of errors (Figure~\ref{fig:faults-error}), they occur on only a small fraction of servers with errors each month: 1.34\% and 1.10\%, respectively (Figure~\ref{fig:faults-servers}).  This helps explain why servers that have socket failures often appear unresponsive in the repair logs that we examine.  Socket failures bombard a server with a large flood of MCEs that the operating system must handle, creating a kind of \emph{denial of service attack} on the server.  Systems that have socket failures appear unresponsive for minutes at a time while correcting errors and handling MCEs.  We believe that context switching to the operating system kernel to handle the MCE contributes largely to the unresponsiveness.

Thus, memory errors do \emph{not} always happen in isolation, and correcting errors in hardware and handling MCEs in the system software (as current architectures do) can easily cause a machine to become unresponsive.  We suspect that simple hardware changes such as caching error events and having system software poll the contents of the error cache once in a while, instead of \emph{always} invoking the system software on \emph{each} error detection, could greatly reduce the potential availability impact of socket and channel failures.  In addition, the DDR4 standard~\cite{jedec-ddr4} allows the memory controller to \emph{retry} memory access (by using a cyclic redundancy check on the command/address bits and asserting an ``alert'' signal when the memory controller detects an error) \emph{without interrupting the operating system}, which can help reduce the system-level unavailability resulting from socket and channel failures.

Bank failures occur relatively frequently, on 14.08\% of servers with errors each month.  We observe a larger failure rate for banks than prior work that examined DRAM chip-level failures on Google servers, which found 2.02\% of banks failed over the course of their study (Table 2 in~\cite{hwang}\footnote{Other studies (e.g.,~\cite{amd, amd2}) had similar findings, so, we compare against~\cite{hwang}, which is representative.}).  One reason for this difference could be the different composition of the servers prior work evaluated.  For example, while the prior work examined older DDR and DDR2 DIMMs from over five years ago, we examine newer DIMMs that use the DDR3 protocol.  The relatively large occurrence of bank failures suggests that devices that support single chip failures (e.g., Chipkill~\cite{ibm-chipkill}) can provide additional protection to help ensure that such failures do not lead to uncorrectable errors.

We find that row and column failures are relatively infrequent, occurring in 0.92\% and 0.99\% of servers each month.  Prior work on Google servers found much larger rate of row (7.4\%) and column (14.5\%) failures~\cite{hwang}.  We believe that the much larger estimate in prior work could potentially be due to the confounding effects of socket and channel errors.  If we ignore socket and channel errors, we artificially increase the number of row and column errors (e.g., we may incorrectly the socket and channel errors in Figure~\ref{fig:faults-error} as other types of errors).

%
%

We observe that a relatively large fraction of servers experience cell failures, 25.54\%.  Similar to row and column failures, prior work found a much larger amount of cell failures, 46.1\%.  As with rows an columns, the large amount of cell failures in prior work could be due to the work not accounting for socket and channel failures.  The prevalence of cell failures prompted the prior work to examine the effectiveness of page offlining, where the operating system (OS) removes pages that contain failed cells from the physical address space.  While the prior study evaluated page offlining in simulation using the same memory traces from their evaluation, we evaluate page offlining on a fraction (\change{a cluster of }{}12,276) of the servers we examine in \S\ref{sec:page-offline} and find it to be less effective than reported in prior work (\cite{hwang}).

While prior work, which may not have controlled for socket and channel failures, found repeat cell errors to be the dominant type of failure (e.g.,~\cite{hwang, amd2, amd3}); when controlling for socket and channel failures (by identifying and separately accounting for socket and channel errors), we find \emph{spurious failures} occur the most frequently, across 56.03\% of servers with errors.  Such errors occur due to random DRAM-external events such as alpha particle strikes from the atmosphere or chip packaging~\cite{intel-dram} and DRAM-internal effects such as cells with weak or variable charge retention times~\cite{liu, samira, samsung-intel}.  This is significant because, as we will show in \S\ref{sec:page-offline}, spurious failures limit the effectiveness of page-offlining.  To deal with these type of failures, we require more effective techniques for detecting and reducing the reliability impact of weak cells (\cite{samira, samsung-intel} discuss some options).

\section{The Role of System Factors}
\label{sec:factors}

We next examine how various system factors correlate with the occurrence of failures in the systems we examine.  For this analysis, we examine systems that failed over a span of three months from 7/13 to 9/13.  We focus on understanding DRAM failures and exclude systems with socket and channel failures from our study.  We examine the effects of DRAM density and DIMM capacity, DIMM vendor, DIMM architecture, age, and workload characteristics on failure rate.

\subsection{DIMM Capacity and DRAM Density}
\label{subsec:DIMM Capacity and DRAM Density}

We measure DRAM density in the number of bits per chip. DRAM density relates closely to the DRAM cell technology and manufacturing process technology~\cite{samsung-intel}.  As DRAM cell and fabrication technology improves, chip manufacturers can fabricate devices with higher densities.  The most widely-available chip density at the time of our study in 2014 was \unit[4]{Gb}, with \unit[8]{Gb} chips gaining adoption.

DRAM density is different from DIMM \emph{capacity}.  A DIMM of a certain capacity can be built \emph{in multiple ways} depending on the density and transfer width of its chips.  For example, a \unit[4]{GB} capacity DIMM could have $16\times\unit[2]{Gb}$ chips or $8\times\unit[4]{Gb}$ chips.  Prior work examined DIMM \emph{capacity} when drawing conclusions~\cite{schroeder, nightingale}, and observed trends that were, in the authors' own words, either ``not consistent''~\cite{schroeder} or a ``weak correlation''~\cite{nightingale} with error rate.  This led the prominent Schroeder et al.\ work to conclude that ``unlike commonly feared, we don't observe any indication that newer generations of DIMMs have worse error behavior.''  Our results with DRAM density stand to refute this claim as we explain below.

\begin{figure}[H]
\centering
\begin{minipage}[t]{0.45\columnwidth}
  \newcommand{\curfig}{capacity}
  \centering
  \change{
  \includegraphics[width=0.9\columnwidth]{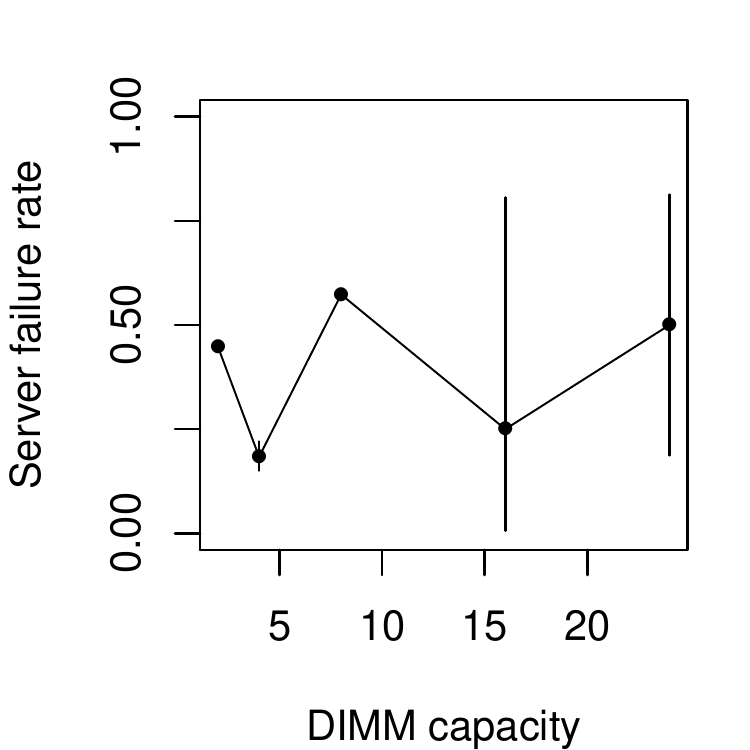}
  }{
  \includegraphics[width=0.9\columnwidth6]{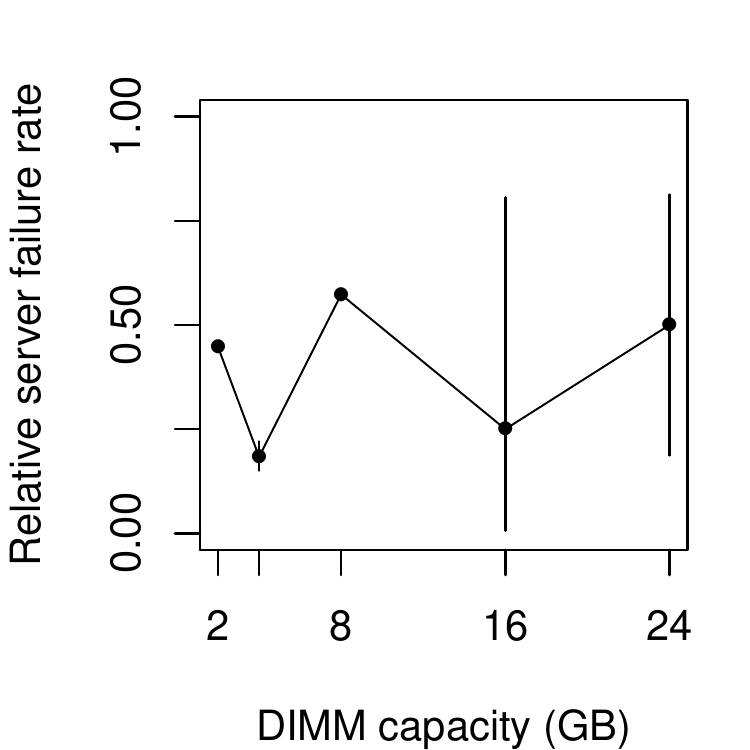}
  }
  \caption{The relative failure rate for servers with different DIMM capacities.  Similar to prior work, we find no consistent reliability trend.}
  \label{fig:capacity}
\end{minipage}
\hspace{0.5em}
\begin{minipage}[t]{0.45\columnwidth}
  \newcommand{\curfig}{density}
  \centering
  \change{
  \includegraphics[width=0.9\columnwidth]{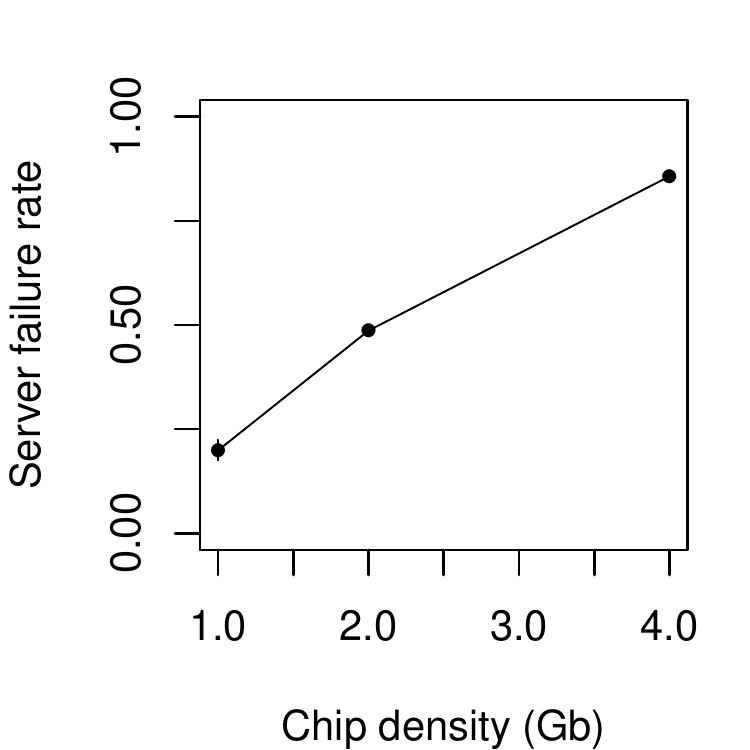}
  }{
  \includegraphics[width=0.9\columnwidth]{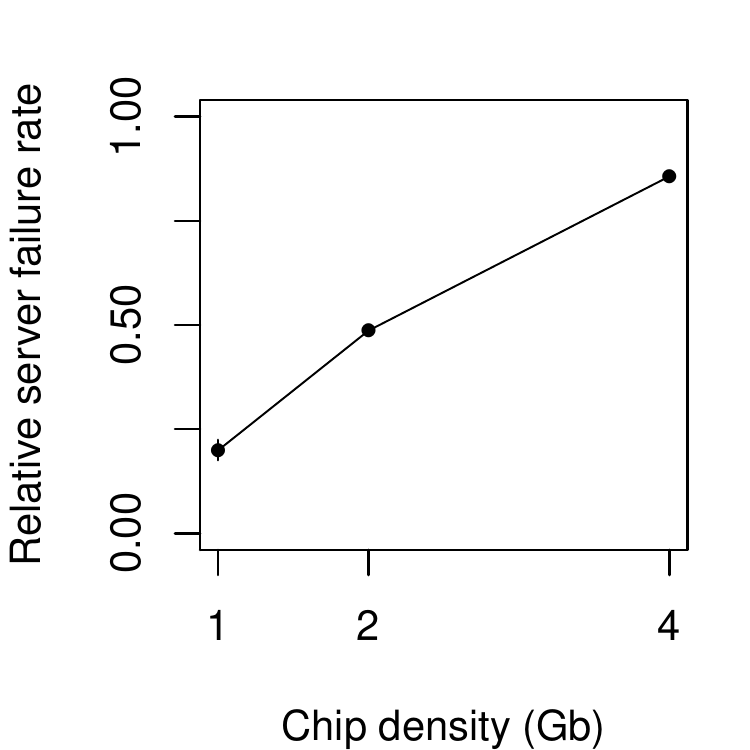}
  }
  \caption{The relative failure rate for servers with different DRAM chip densities.  Newer densities (i.e., newer technology nodes) show a trend of higher failure rates.}
  \label{fig:density}
\end{minipage}
\end{figure}

Similar to these works, we also find that the error trends with respect to \emph{DIMM capacity} are \emph{not consistent}.  Figure~\ref{fig:capacity} shows how the different capacities of DIMMs we examine relate to device failure rate.\footnote{Recall from \S\ref{sec:analytical} that we examine \emph{relative} server failure rates versus a sample control group.  Though relative failure rates happen to be in the range $[0, 1]$, we must not confuse them with absolute failure rates across the fleet.}  The large error bars for \unit[16]{GB} and \unit[24]{GB} DIMMs are due to the relatively small number of DIMMs of those types.  Notice that there is no consistent trend across DIMM capacities.

In contrast to prior works~\cite{schroeder, nightingale}, we \emph{do} observe indication that \emph{newer generations of DRAM chips have worse error behavior} by examining failure rate as a function of DRAM chip density.  The servers we analyze contain three different types of DRAM chip densities:  \unit[1]{Gb}, \unit[2]{Gb}, and \unit[4]{Gb}. Figure~\ref{fig:density} shows \emph{how different DRAM chip densities relate to device failure rate}.  We see that there is a clear trend of increasing failure rate with increasing chip density, with \unit[2]{Gb} devices having $2.4\times$ higher failure rates than \unit[1]{Gb} devices and \unit[4]{Gb} devices having $1.8\times$ higher failure rates than \unit[2]{Gb} devices.  This is troubling because it indicates that business-as-usual practices in DRAM design will likely lead to higher memory failure rates in the future, as both industry~\cite{samsung-intel} and academia~\cite{samira, rowhammer, memcon, kiist} predict in recent works.  To understand the source of this trend, we next examine the failure rate for DRAM cells.

\begin{figure}[H]
  \newcommand{\curfig}{cell}
  \centering
  \change{
  \includegraphics[width=0.4\columnwidth]{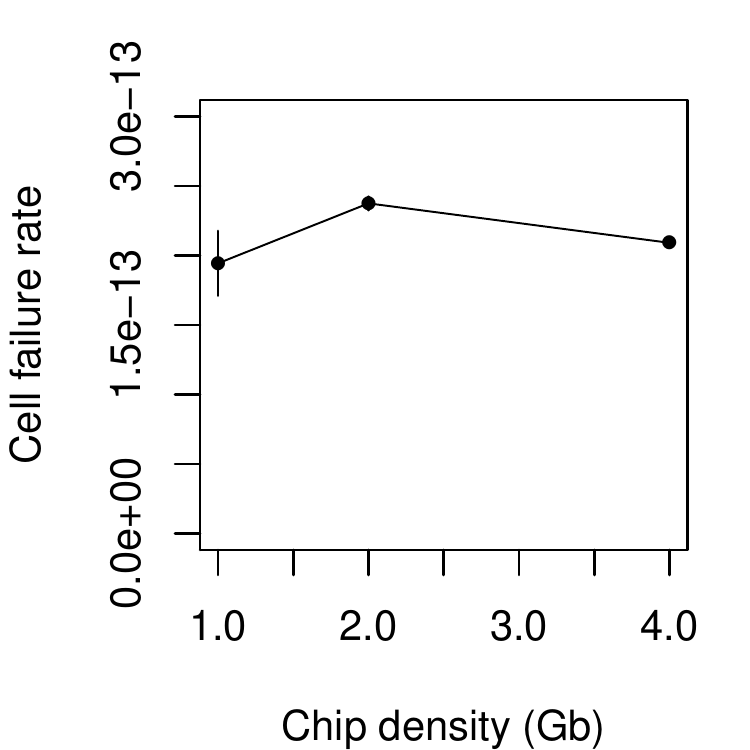}
  }{
  \includegraphics[width=0.4\columnwidth]{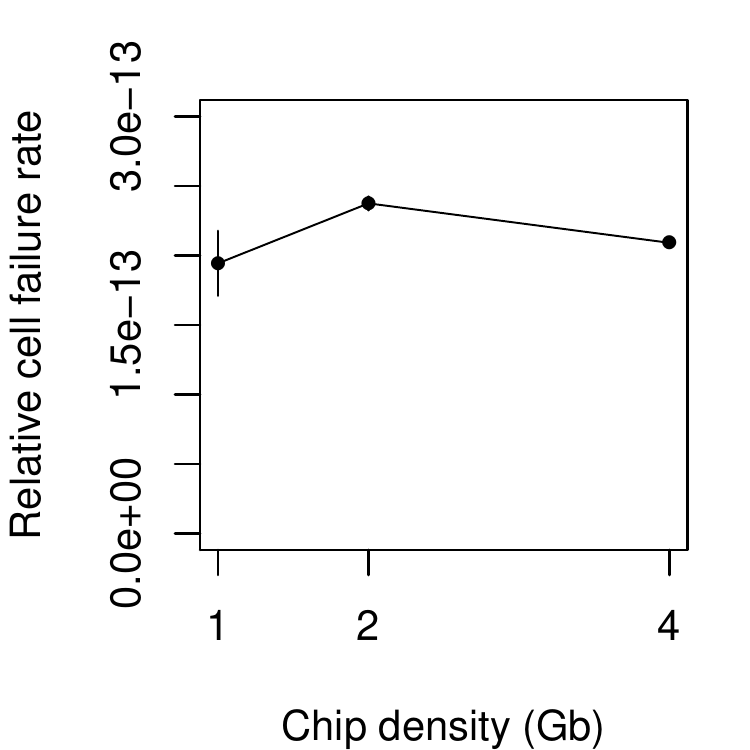}
  }
  \caption{The relative per-cell failure rate at different DRAM technology nodes (chip densities).}
  \label{fig:cell}
\end{figure}

Figure~\ref{fig:cell} shows the cell failure rate what we compute by normalizing the failure rates in Figure~\ref{fig:density} by the number of cells in each chip.  Cell failure rate increases briefly going from \unit[1]{Gb} chips to \unit[2]{Gb} chips but decreases going from \unit[2]{Gb} chips to \unit[4]{Gb} chips.  Our data shows that the reliability of \emph{individual DRAM cells} may be improving recently.  This is likely due to the large amounts of effort that DRAM manufacturers put into designing faster and more reliable DRAM cell architectures.  Our insight is that \emph{the quadratic increase in number of cells per chip outpaces small improvements in DRAM cell reliability}, leading to the trend of net decrease in DRAM reliability as shown by the server failure rate data in Figure~\ref{fig:density}.  Unless DRAM manufacturers achieve more-than--quadratic improvements in DRAM cell reliability in future devices, maintaining or decreasing DRAM server failure rates in the future (while still increasing DRAM chip capacity) will be untenable without stronger hardware and/or software error correction.

\subsection{DIMM Vendor}

DIMM vendors purchase chips from DRAM chip manufacturers and assemble them into DIMMs.  While we have information on DIMM manufacturer, we do not have information on the DRAM chip manufacturers in our systems.

\begin{figure}[H]
  \newcommand{\curfig}{vendor}
  \centering
  \change{
  \includegraphics[width=0.4\columnwidth]{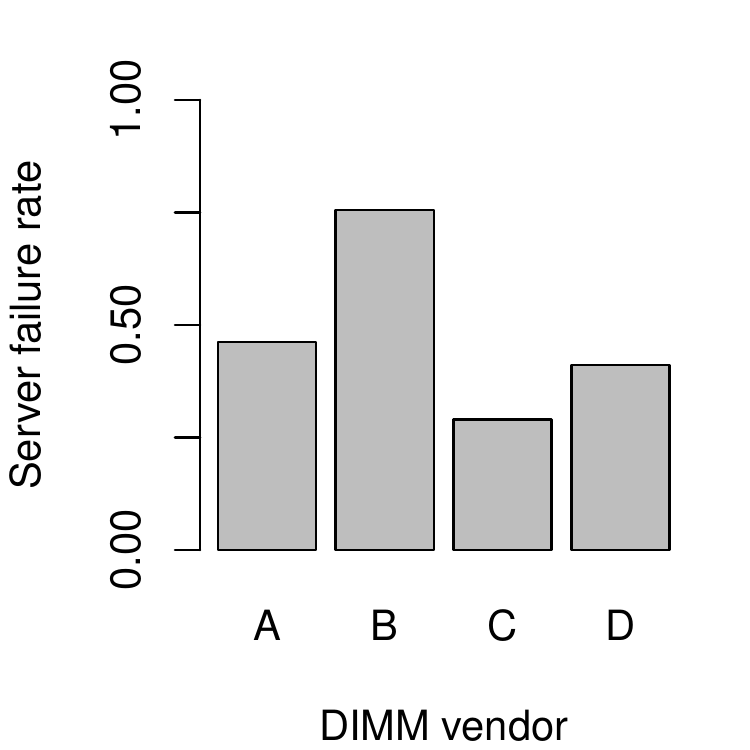}
  }{
  \includegraphics[width=0.4\columnwidth]{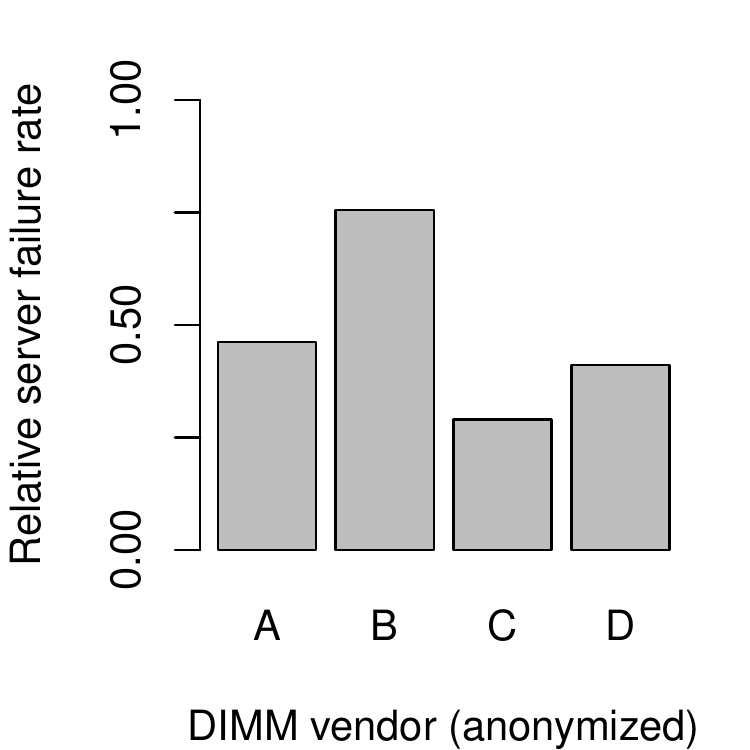}
  }
  \caption{The relative server failure rate for different DRAM vendors varies widely.}
  \label{fig:vendor}
\end{figure}

Figure~\ref{fig:vendor} shows the failure rate for servers with different DIMM vendors.\footnote{We have made the vendors anonymous.}  We observe that failure rate varies by over $2\times$ between vendors (e.g., Vendor B and Vendor C).  The differences between vendors arise if vendors use less reliable chips from a particular foundry or build DIMMs with less reliable organization and manufacturing.  Prior work~\cite{amd2, amd3} also found a large range in the server failure rate among vendors of $3.9\times$.

\subsection{DIMM Architecture}
  \label{subsec:DIMM Architecture}

We next examine how DIMM architecture affects server failure rate.  We examine two aspects of DIMM design that have not been studied in published literature before:  the number of data chips (not including chips for ECC) per DIMM and the transfer width of each chip.

%
%

\begin{figure}[H]
\centering
\begin{minipage}[t]{0.45\columnwidth}
  \newcommand{\curfig}{chips-per-density}
  \centering
  \change{
  \includegraphics[width=0.9\columnwidth]{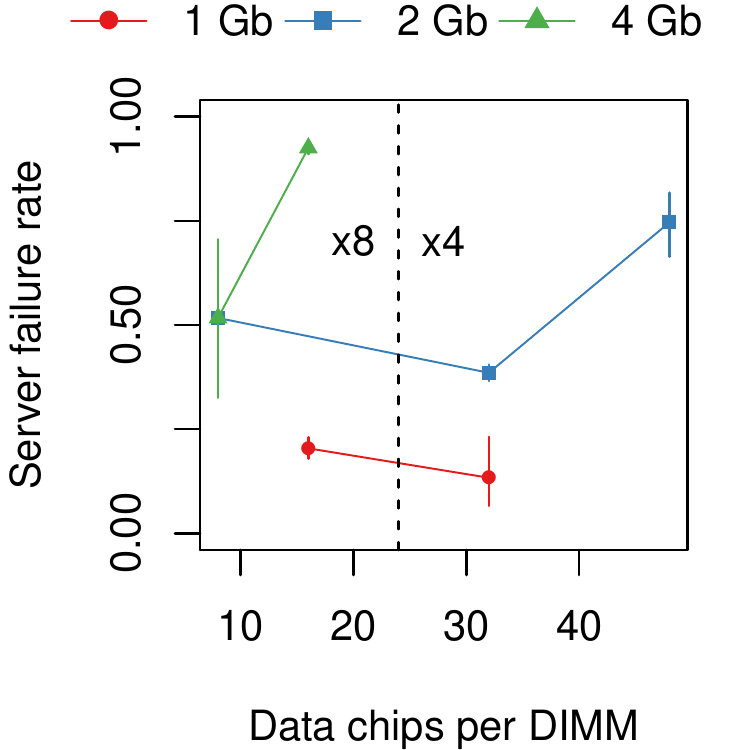}
  }{
  \includegraphics[width=0.9\columnwidth]{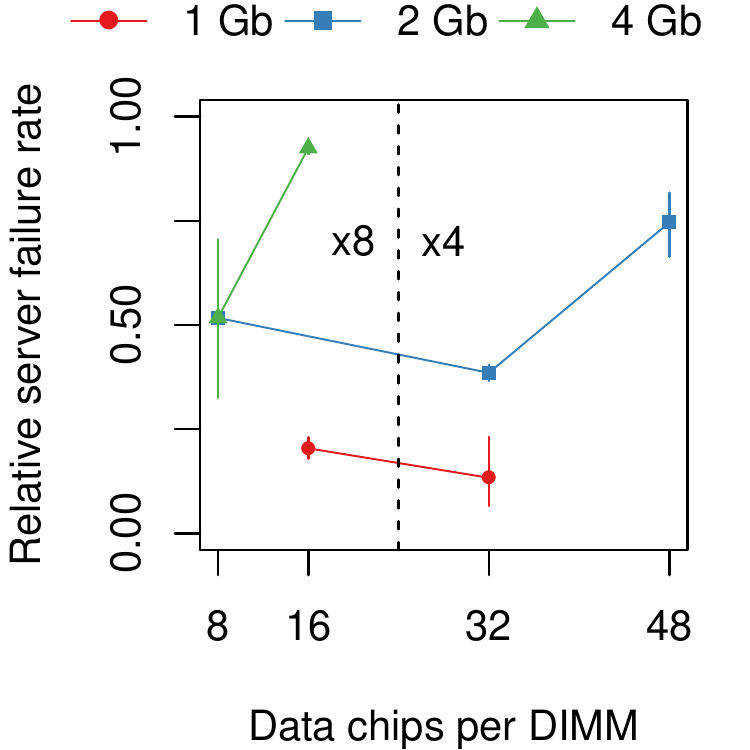}
  }
  \caption{The \protect\also{}{relative }failure rate of servers with DIMMs with different numbers of data chips. We plot each chip density separately.}
  \label{fig:chips-per-density}
\end{minipage}
\hspace{0.5em}
\begin{minipage}[t]{0.45\columnwidth}
  \newcommand{\curfig}{width-per-density}
  \centering
  \change{
  \includegraphics[width=0.9\columnwidth]{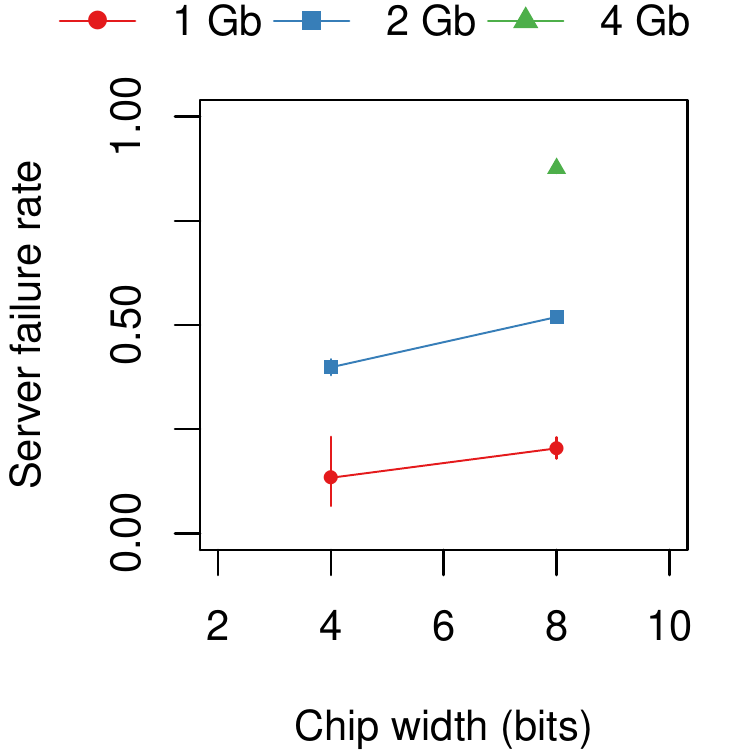}
  }{
  \includegraphics[width=0.9\columnwidth]{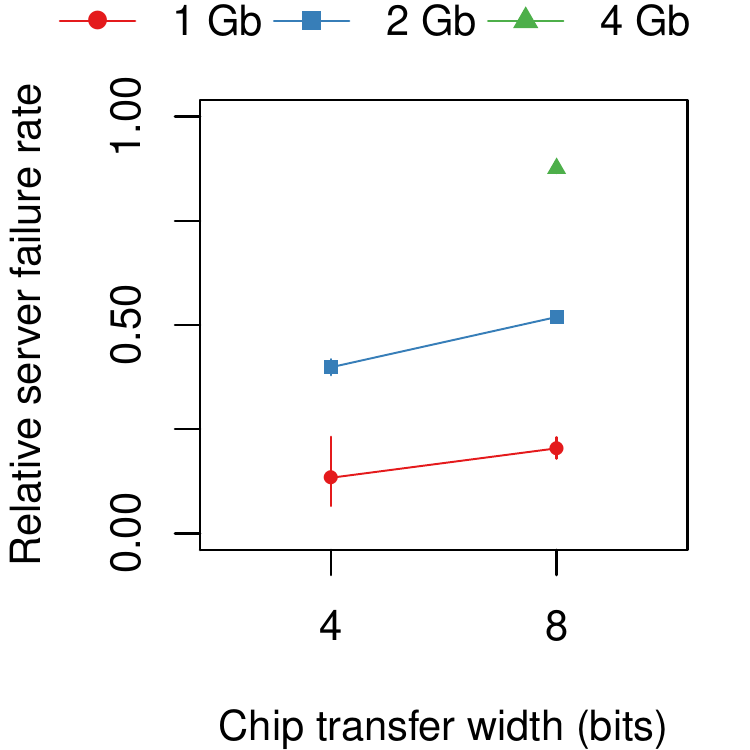}
  }
  \caption{The \protect\also{}{relative }failure rate of servers with DIMMs with different chip transfer widths. We plot each chip density separately.}
  \label{fig:width-per-density}
\end{minipage}
\end{figure}

Figure~\ref{fig:chips-per-density} plots the failure rate for servers with DIMMs with different numbers of data chips for each of the densities that we examine.  The DIMMs that we examine have 8, 16, 32, and 48 chips.  We make two observations from Figure~\ref{fig:chips-per-density}.

First, for a given number of chips per DIMM, servers with higher chip densities generally have higher average failure rates.  This illustrates how chip density is a first-order effect when considering memory failure rate (as we show in Figure~\ref{fig:density}).

Second, we find that server failure rate trends with respect to chips per DIMM are dependent on the \emph{transfer width} of the chips---the number of data bits each chip can transfer in one clock cycle.  In order to transfer data at a similar rate, DIMMs with fewer (8 or 16) chips must compensate by using a larger transfer width of 8 bits per clock cycle (we call these $\times8$ devices) while DIMMs with more chips (32 or 48) can use a smaller transfer width of 4 bits per clock cycle (we call these $\times4$ devices).  We annotate the graph to show which chip counts have transfer widths of $\times4$ bits and $\times8$ bits.

We observe two trends depending on whether chips on a DIMM have the same or different transfer widths.  First, among chips of the \emph{same} transfer width, we find that increasing the number of chips per DIMM increases server failure rate.  For example, for \unit[4]{Gb} devices, increasing the number of chips from 8 to 16 increases failure rate by 40.8\% while for \unit[2]{Gb} devices, increasing the number of chips from 32 to 48 increases failure rate by 36.1\%.  Second, once the number of chips per DIMM increases beyond 16 and chips start using a \emph{different} transfer width of $\times8$, there is a decrease in failure rate.  For example, for \unit[1]{Gb} devices, going from 16 chips with a $\times8$ interface to 32 chips with a $\times4$ interface decreases failure rate by 7.1\%.  For \unit[2]{Gb} devices, going from 8 chips with a $\times8$ interface to 32 chips with a $\times4$ interface decreases failure rate by 13.2\%.

To confirm the transfer width trend, we plot the failure rates dependent on transfer width alone in Figure~\ref{fig:width-per-density}.  We find that, in addition to the first-order effect of chip density increasing failure rate (Effect 1), there is a consistent increase in failure rate going from $\times4$ to $\times8$ devices (Effect 2).


We believe that we can explain both effects partially by considering how number of chips and transfer width contribute to the electrical disturbance within a DIMM that may disrupt the integrity of the signal between components.  For example, a larger transfer width increases internal data transfer current (e.g., I$_{\mathrm{DD4R/W}}$ in Table 19 of~\cite{micron}, which compares the power consumption of $\times4$ and $\times8$ DRAM devices), leading to additional power noise across the device.  Such power noise could induce additional memory errors if, for example, components trap charge.  Interestingly, we find that, for a given chip density, \emph{the best architecture for device reliability occurs when there is, first, low transfer width and, second, low chips per DIMM}.  This is shown by the \unit[2]{Gb} devices with 32 chips with a $\times4$ interface versus the other \unit[2]{Gb} devices in Figure~\ref{fig:chips-per-density}.

\subsection{Workload Characteristics}
\label{sec:workload}

We next examine how workload characteristics such as CPU utilization (the average utilization of the CPUs in a system), memory utilization (the fraction of physical memory pages in use), and workload type affect server failure rate.  Prior work examined CPU utilization and memory utilization and found that they were correlated positively with failure rate~\cite{schroeder}.

\begin{figure}[H]
\centering
\begin{minipage}[t]{0.45\columnwidth}
  \newcommand{\curfig}{cpu-per-density}
  \centering
  \change{
  \includegraphics[width=0.9\columnwidth]{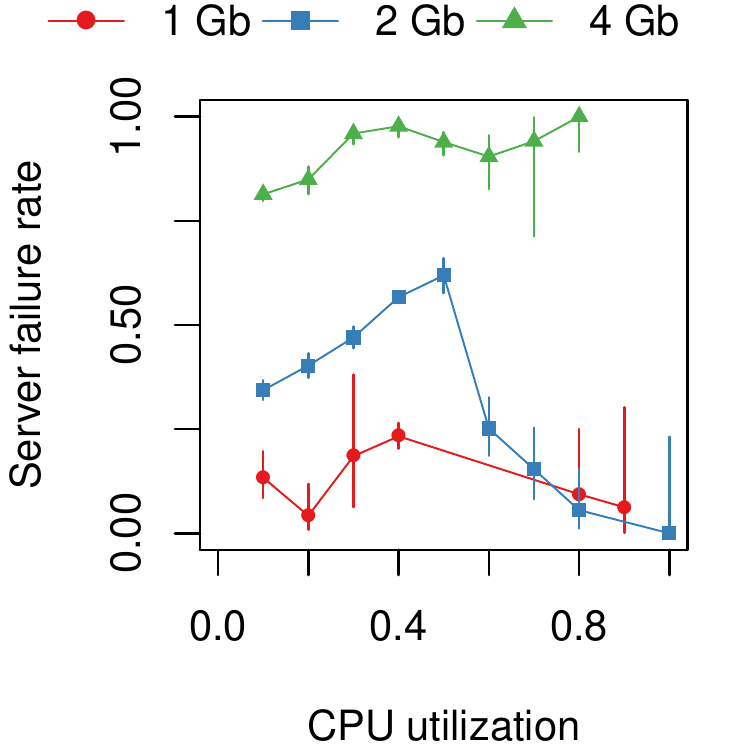}
  }{
  \includegraphics[width=0.9\columnwidth]{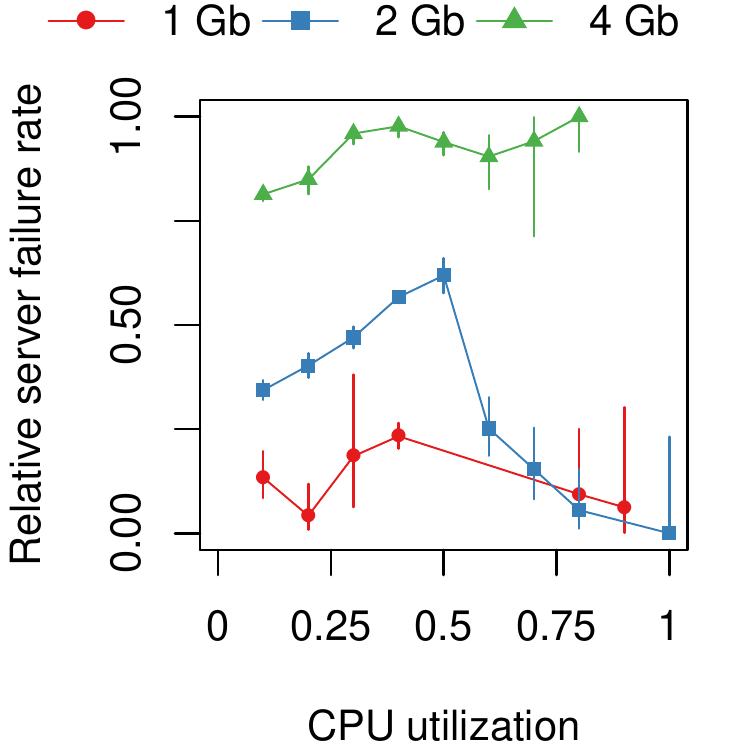}
  }
  \caption{The \protect\also{}{relative }failure rate of servers with different average CPU utilizations.}
  \label{fig:cpu-per-density}
\end{minipage}
\hspace{0.5em}
\begin{minipage}[t]{0.45\columnwidth}
  \newcommand{\curfig}{mem-per-density}
  \centering
  \change{
  \includegraphics[width=0.9\columnwidth]{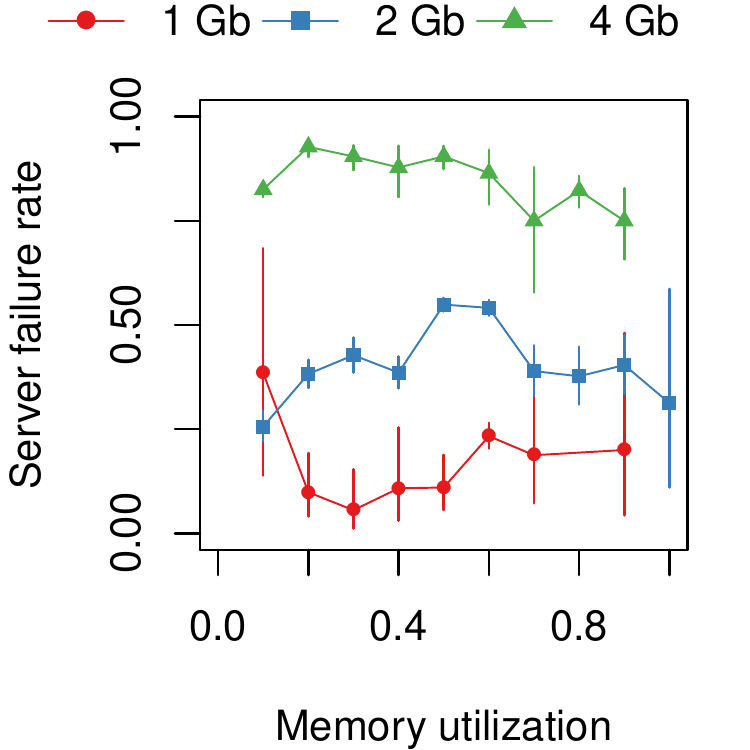}
  }{
  \includegraphics[width=0.9\columnwidth]{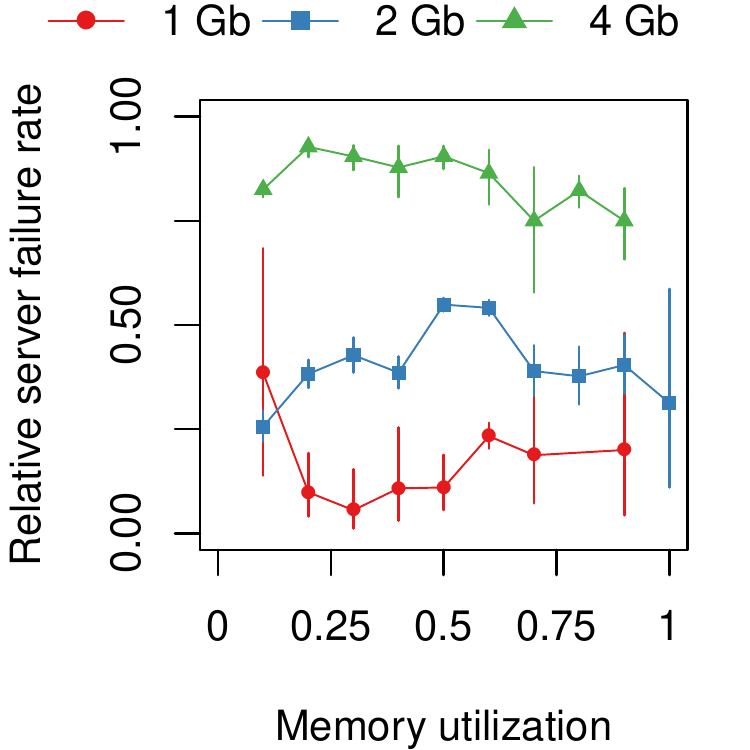}
  }
  \caption{The \protect\also{}{relative }failure rate of servers with different average memory utilizations.}
  \label{fig:mem-per-density}
\end{minipage}
\end{figure}

We measure CPU utilization as the fraction of non-idle cycles versus total cycles across the CPUs in a server.  Due to software load balancing, we find that CPU utilization among cores in a server running the same workload are relatively similar, and so the average utilization across the cores is reasonably representative of each core's individual utilization.  We measure memory utilization as the fraction of pages the OS allocates.  Note that memory utilization does not describe \emph{how} the OS accesses cells in a page.  For this reason, we examine workloads as a proxy for how the OS accesses cells.  We plot how CPU utilization and memory utilization relate to server failure rate in Figures~\ref{fig:cpu-per-density} and \ref{fig:mem-per-density}.

Contrary to what prior work observed, we do \emph{not} find a correlation between either CPU utilization or memory utilization and failure rate.  We observe multiple local maxima for failure rate versus CPU utilization and memory utilization across all the chip densities.  We believe that this is due to the more diverse workloads that we examine (Table~\ref{tab:workloads}) versus prior work~\cite{schroeder, amd, amd2, amd3}, which mainly examined a homogeneous workload.  The implications of this are that memory failure rate may depend more on the \emph{type} of work and not the CPU utilization or memory utilization the work causes.

\begin{figure}[H]
  \newcommand{\curfig}{type}
  \centering
  \includegraphics[width=0.4\columnwidth]{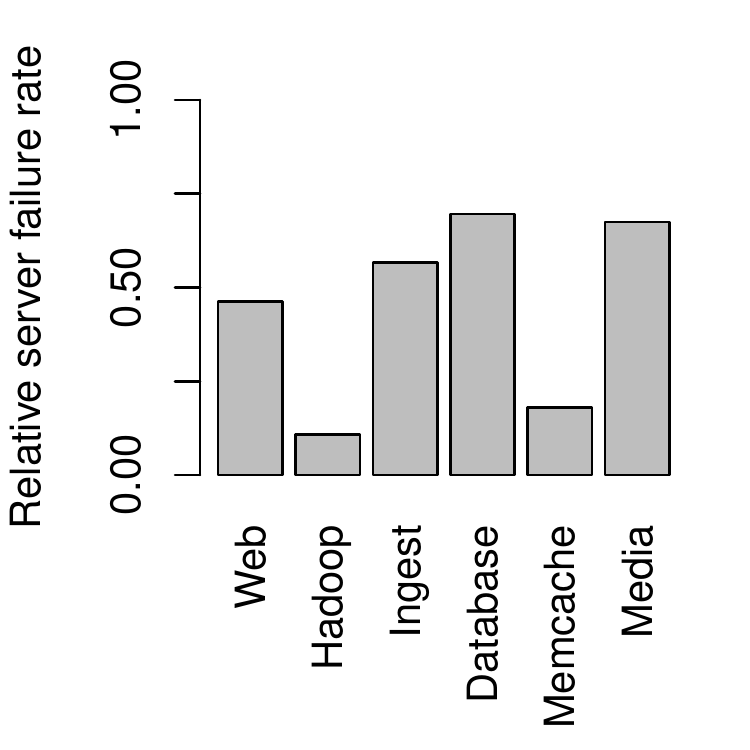}
  \caption{The relative failure rate of servers that run different types of workloads (Table~\ref{tab:workloads}) can vary widely.}
  \label{fig:type}
\end{figure}

To examine how the \emph{type of work} a server performs affects failure rate, we plot the server failure rate for the different workload types at Facebook in Figure~\ref{fig:type}.  We observe that, depending on the workload, failure rate can vary by up to $6.5\times$, as shown by the difference between servers executing a Database-type workload versus those executing a Hadoop-type workload.  While we leave examining in detail how workloads affect memory failure rate to future work, we hypothesize that certain types of workload memory access patterns may increase the likelihood of errors.  For example, prior work has shown that memory errors can be induced in a controlled environment by accessing the same memory row in rapid succession~\cite{rowhammer}.  Such an access pattern involves modifying data and writing it back to memory using the {\tt clflush} and {\tt mfence} instructions (on the Intel\textsuperscript{\textregistered}~$\times$86 instruction set).  We believe it would be interesting to examine what types of workloads exhibit this behavior.

\subsection{Server Age}

\begin{figure*}
\centering
\begin{minipage}[t]{0.4\linewidth}
  \newcommand{\curfig}{age-per-density}
  \centering
  \change{
  \includegraphics[width=0.9\columnwidth]{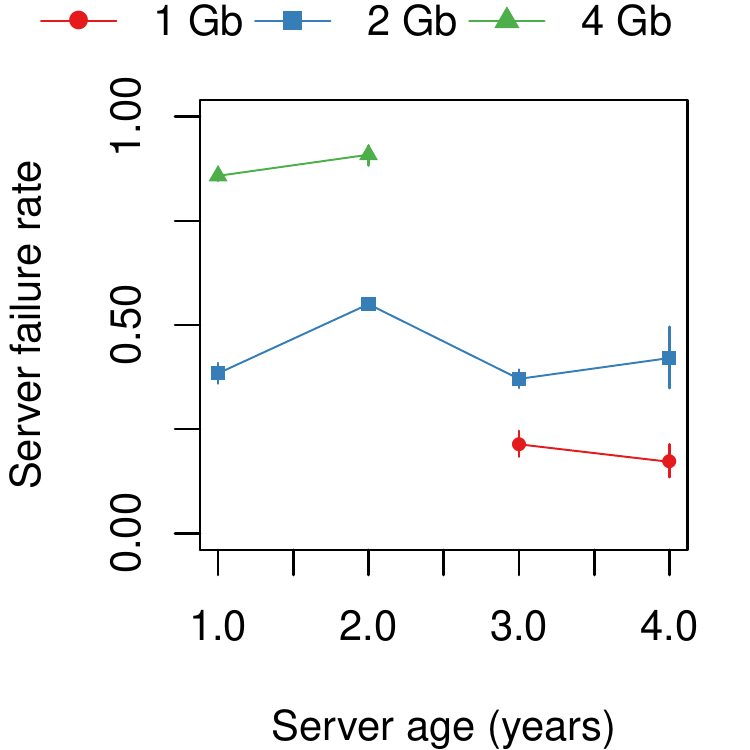}
  }{
  \includegraphics[width=1.0\columnwidth]{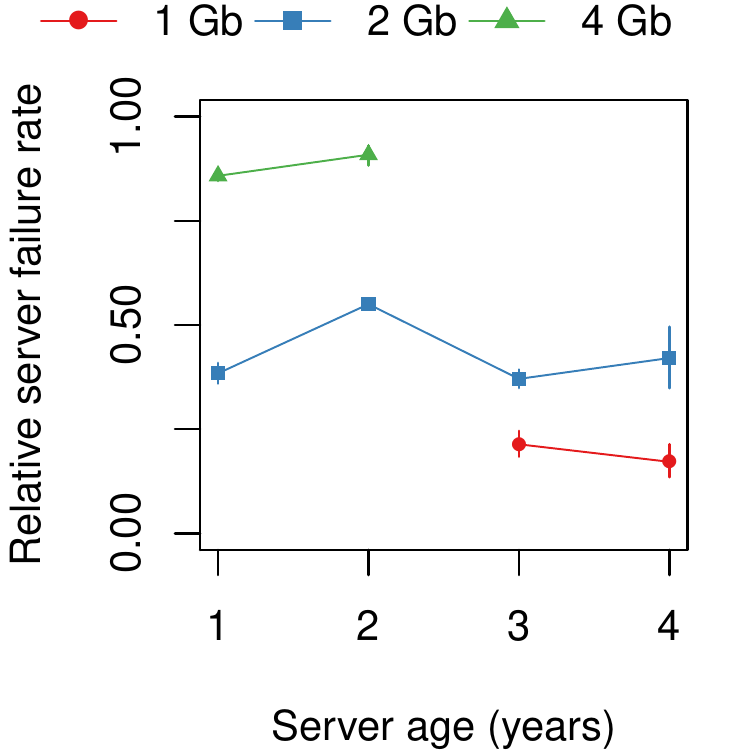}
  }
  \caption{The relative failure rate of servers of different ages.  There is no clear trend when controlling \emph{only} for chip density.}
  \label{fig:age-per-density}
\end{minipage}
\hspace{0.5em}
\begin{minipage}[t]{0.4\linewidth}
  \newcommand{\curfig}{cores-per-density}
  \centering
  \change{
  \includegraphics[width=0.9\columnwidth]{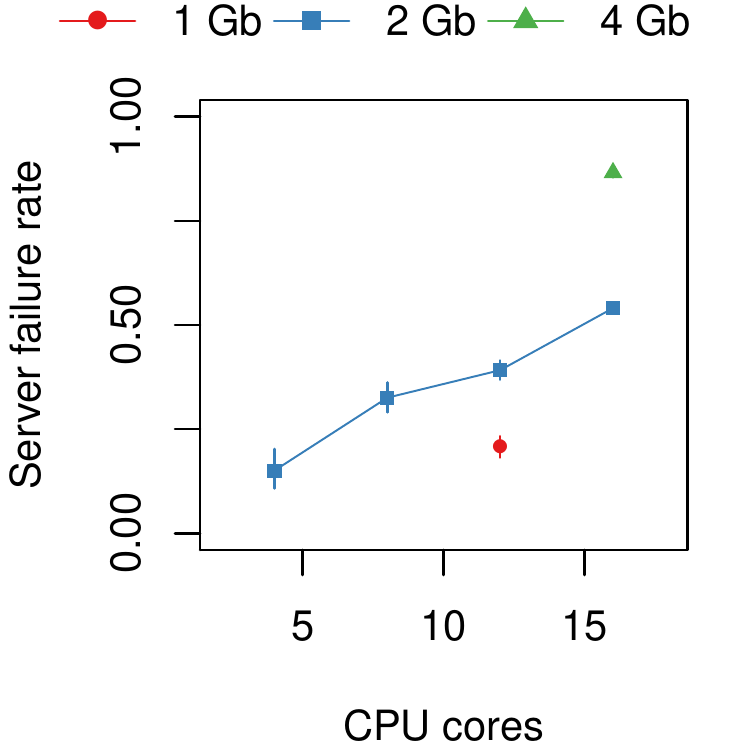}
  }{
  \includegraphics[width=1.0\columnwidth]{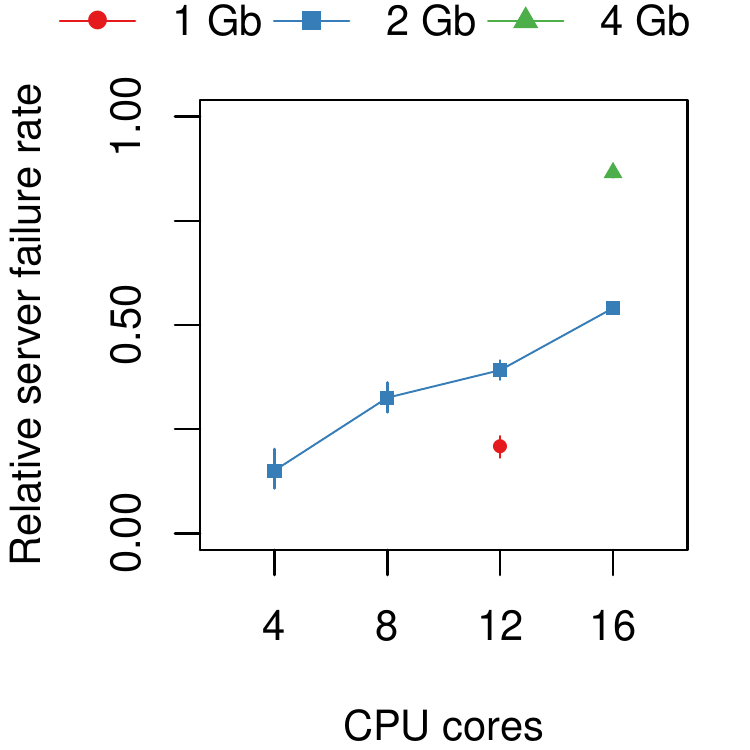}
  }
  \caption{The relative failure rate of servers with different numbers of CPU cores.  Servers with more CPUs have higher failure rates.}
  \label{fig:cores-per-density}
\end{minipage}
\\
\vspace{2.0em}
\begin{minipage}[t]{0.8\linewidth}
  \newcommand{\curfig}{age-density-cores}
  \centering
  \change{
  \includegraphics[width=0.85\columnwidth]{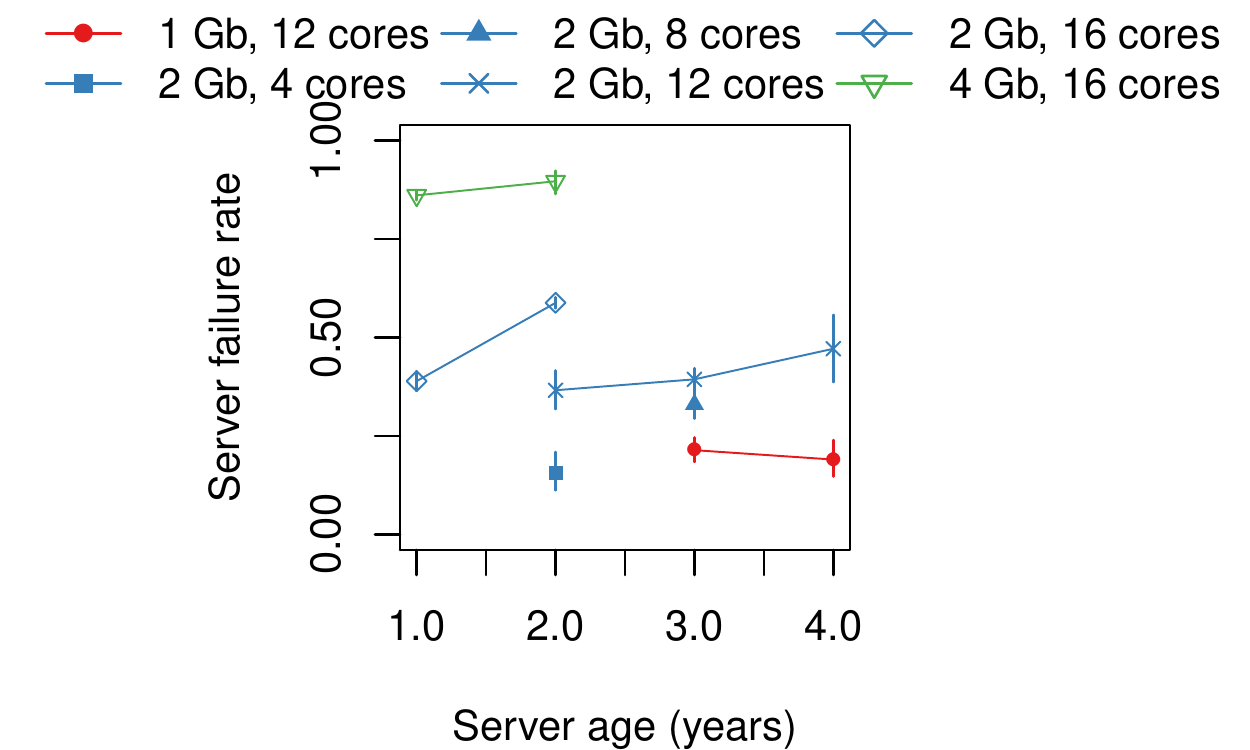}
  }{
  \includegraphics[width=0.85\columnwidth]{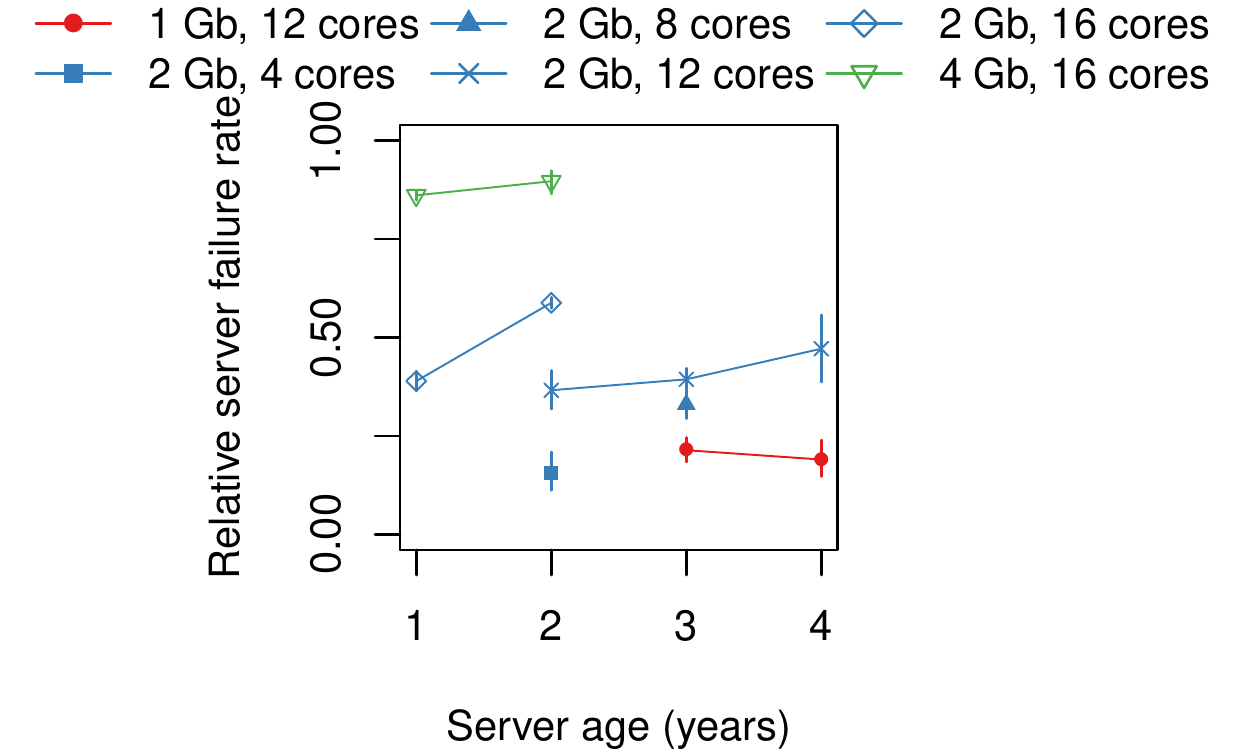}
  }
  \caption{The relative failure rate of servers of different $\langle \textrm{chip\ density}, \textrm{CPU\ count} \rangle$ configurations.  When controlling for density and CPUs together, older devices usually have higher failure rates.}
  \label{fig:\curfig}
\end{minipage}
\end{figure*}

We examine next how age affects server failure rate.  The servers we analyze are between one and four years old, with an average age of between one and two years.  Figure~\ref{fig:age-per-density} shows the monthly failure rate for servers of different ages.  We observe that chip density once again plays a large role in determining server failure rate:  For a given age, servers with \unit[4]{Gb} devices have a 15.3\% higher failure rate on average than \unit[2]{Gb} devices, and servers with \unit[2]{Gb} devices have a 23.9\% higher failure rate on average than \unit[1]{Gb} devices.

%
%

We do not observe any general \emph{age-dependent} trend in server failure rate when controlling for the effects of density alone.  One reason for this is that age correlates with other server characteristics.  For example, we find that in addition to correlating with chip density (correlation coefficient of $-0.69$), age also correlates with the number of CPUs in a system (correlation coefficient of $-0.72$).  Figure~\ref{fig:age-density-cores} shows the trend for age for different combinations of chip density and CPUs (which we will denote as $\langle x,y \rangle$ where $x$ is chip density in Gb and $y$ is number of CPUs).  We make two observations from Figure~\ref{fig:age-density-cores}.


First, we find that among systems of \emph{the same age}, more cores lead to higher failure rates.  For example, consider the $\langle 2, * \rangle$ systems that are two years of age:  going from $4 \rightarrow 12$ cores increases failure rate by 21.0\% and going from $12 \rightarrow 16$ cores increases failure rate by 22.2\%.  Figure~\ref{fig:cores-per-density}, which plots the server failure rate with respect to different numbers of CPUs confirms this trend, with \unit[2]{Gb} systems with 16 cores having a 40.0\% higher failure rate than \unit[2]{Gb} systems with 4 cores.  This could be due to more cores accessing DRAM more intensely and wearing out DRAM cells at a faster rate, a failure mode that was shown in a prior controlled study~\cite{dram-wearout}.


The most related trend observed in prior work was that CPU frequency was shown to be correlated with error rate~\cite{nightingale}.  The trend we observe with respect to CPU count is significant because the number of CPU cores per processor is increasing at a much faster rate than CPU frequency and so our results allow us to predict that future processor generations will likely continue to induce higher rates of errors in DRAM devices.

Second, among systems with \emph{the same number of cores}, older machines generally have higher failure rates than younger machines.  For example, for the $\langle 2, 12 \rangle$ system, average failure rate increases by 2.8\% going from $2 \rightarrow 3$ years of age, and average failure rate increases by 7.8\% going from $3 \rightarrow 4$ years of age.  This is consistent with prior observations from the field that showed that failure rates can increase with age~\cite{schroeder}, though we observe a much clearer trend versus prior work (e.g., Figure 10 in~\cite{schroeder} shows large fluctuations in failure rate over time) because we control for correlated factors such as chip density and CPU count.  Not all systems exhibit this trend, however:  the $\langle 1,12 \rangle$ system shows a small decrease in failure rate going from $3 \rightarrow 4$ years of age, which could be due to second-order effects on failure rate from other factors that may correlate with age, such as transfer width.

We note that other effects besides wearout could explain the age-dependent trends we observe. For example, if servers of different ages have different types of hardware that are more susceptible to faults, or, if the workload running on a server changes over time in a way that affects the server's memory error rate. We hope that future work can help shed light onto age-dependent failure trends in DRAM devices in the field.

\section{Modeling DRAM Failures}

We next develop a model for DRAM failures using the data we collect in study.  We use a statistical regression analysis to determine which server characteristics have a statistically significant effect on failure rate and how much they contribute to failure rate.  We can use the resulting model to examine how relative server failure rate changes for servers with different characteristics, which allows us to reason about the relative reliability of different server configurations.

We use R~\cite{r} for our statistical analysis.  We perform a logistic regression~\cite{logit-1,logit-2} on a binary characteristic that represent whether a server was part of the error group or control group of servers (see \S\ref{sec:analytical} for our error and control group classification/formation).  We include most of the characteristics we analyze in \S\ref{sec:factors} in our regression with the exception of DIMM vendor because we anonymize it and workload type because it is difficult to apply outside the context of Facebook's fleet.\footnote{This results in the model expressing these contributions \emph{indirectly} though other factors, whose values the model computes, in part, by how they correlate with different vendors/workloads.}  One limitation of the logistic regression model is that it is able to identify only \emph{linear relationships} between characteristics and failure rates.  On the other hand, using a logistic regression made analyzing our large data set of errors across many variables tractable.

\subsection{An Open Model for DRAM Failures}
\label{sec:model}

Table~\ref{tab:characteristics}, on the next page, shows the parameters and output of the regression and the resulting model (in the last row).  The first two columns describe the factors we include in the regression.  The third column lists the resulting $p$-value for each factor after performing the logistic regression.  The $p$-value is the likelihood that the model accurately models a characteristic:  lower $p$-values indicate the model more accurately models a characteristic.  The fourth column describes whether the $p$-value is $<0.01$, corresponding to a $<1\%$ chance that the model models the characteristic inaccurately.  The fifth column, $\beta$-coefficient, is the characteristic's contribution to error rate and the last column, standard error, is how much the model differs from the values we measure for a characteristic.

\vspace*{\fill}
\hfill
\begin{center}
\it (This portion of page intentionally left blank.)
\end{center}
\vspace{\fill}

\begin{landscape}

\newcommand{\factorterm}[2]{(\mathit{#1} \cdot \beta_{\mathit{#1}})}
\begin{table*}
  \centering
  \begin{tabular}{lp{0.40\columnwidth}S[table-format=1.3e2,table-comparator=true]cS[table-format=1.3e1]S[table-format=1.3e1]}
    \toprule
    \bf Characteristic & \bf Description & \bf {$\mathit{p}$-value} & \bf Significant? & \bf {$\beta$-coefficient} & \bf {Standard error} \\
    \midrule
    Intercept & A baseline server with \unit[1]{Gb} chips with a $\times$4 interface & <2.000e-16 & Yes & -5.511 & 3.011e-1 \\
    & and 0 for all other factors. & & & & \\
    Capacity & DIMM capacity (GB). & <2.000e-16 & Yes & 9.012e-2 & 2.168e-2 \\
    Density2Gb & 1 if the server has \unit[2]{Gb} density chips; 0 otherwise. & <2.000e-16 & Yes & 1.018 & 1.039e-1 \\
    Density4Gb & 1 if the server has \unit[4]{Gb} density chips; 0 otherwise. & <2.000e-16 & Yes & 2.585 & 1.907e-1 \\
    Chips & Number of chips per DIMM. & <2.000e-16 & Yes & -4.035e-2 & 1.294e-2 \\
    Width8 & 1 if the server has $\times8$ DRAM chips; 0 otherwise. & 0.071 & No & 2.310e-1 & 1.277e-1 \\
    CPU\% & Average CPU utilization (\%). & <2.000e-16 & Yes & 1.731e-2 & 1.633e-3 \\
    Memory\% & Average fraction of allocated physical pages (\%). & 0.962 & No & 5.905e-5 & 1.224e-3 \\
    Age & Server age (years). & <2.000e-16 & Yes & 2.296e-1 & 3.956e-2 \\
    CPUs & Number of physical CPU cores in the server. & <2.000e-16 & Yes & 2.126e-1 & 1.449e-2 \\
    \midrule
    \multirow{2}{*}{\bf Failure model} & \multicolumn{5}{c}{$\ln\left[\mathcal{F} / (1 - \mathcal{F})\right] = \beta_{\mathit{Intercept}} + \factorterm{Capacity}\ + \factorterm{Density2Gb}\ + \factorterm{Density4Gb}\ + \factorterm{Chips}\ $} \\
    & \multicolumn{5}{c}{$+ \factorterm{CPU\%}\ + \factorterm{Age}\ + \factorterm{CPUs}\ $} \\
    \bottomrule
  \end{tabular}
  \caption{The factors in our regression analysis and the resulting error model.  $p$-value is the likelihood that the model inaccurately models a characteristic:  lower $p$-values indicate more accurate modeling.  ``Significant?'' represents whether the $p$-value is $<0.01$, corresponding a $<1\%$ chance that the model inaccurately models the characteristic.  $\beta$-coefficient is the characteristic's contribution to error rate and standard error is how much the model differs from the values we measure for a given characteristic.  The model is publicly available at~\cite{safari-tools}.}
  \label{tab:characteristics}
\end{table*}

\end{landscape}

The \emph{Intercept} is a byproduct of the regression and helps the model better fit the data we measure.  It represents a server with a set of baseline characteristics (those we show in Table~\ref{tab:characteristics}) and 0 for all other factors (0 CPUs, 0 years old, and so on).  The factors $\mathit{Density2Gb}$ and $\mathit{Density4Gb}$ take on the value 0 or 1 depending on whether the server has the characteristic (in which case the value is 1) or does not (0).  Note that the regression analysis computes the $\beta$-coefficients for these variables in such a way that when we add them to the model, they replace the default values we show in $\beta_{\mathit{Intercept}}$ (e.g., though $\beta_{\mathit{Intercept}}$ represents a server with \unit[1]{Gb} chips, when $\mathit{Density2Gb}$ is set to 1, the model computes the failure rate of servers with \unit[2]{Gb} chips).

\begin{figure}[H]
  \centering
\change{}{
  \includegraphics[scale=0.45]{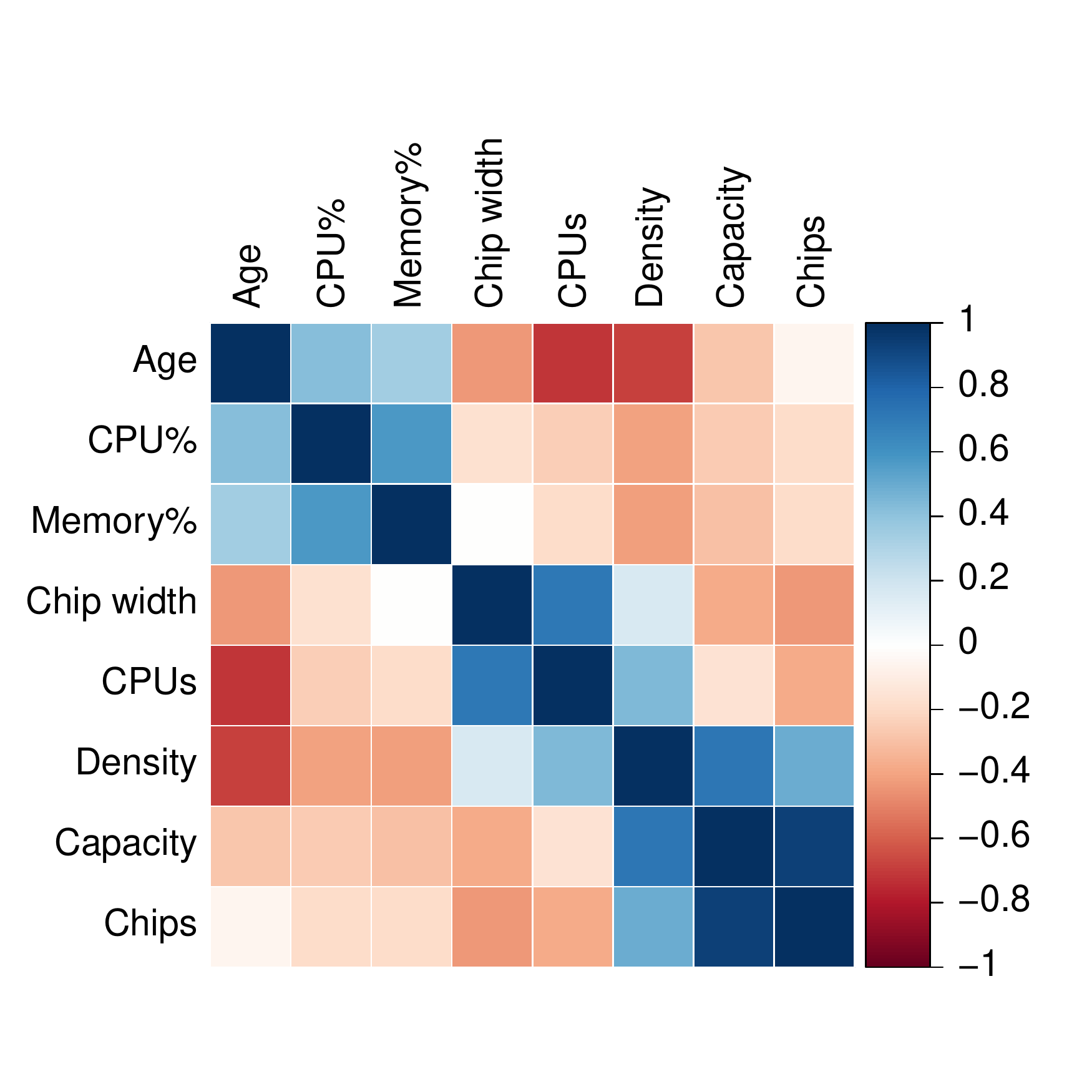}
}
  \caption{The correlation between different measured factors.}
  \label{fig:correlation}
\end{figure}

Note that if we include a characteristic in our model, it does not mean that it will affect failure rates in the real world.  It may mean that it only correlates with other characteristics that \emph{do} affect failure rate.  The opposite is true as well:  A characteristic that we do \emph{not} include in the model may in fact contribute to failure rate but the model does not capture its affects in other characteristics in the model.  For example, Figure~\ref{fig:correlation} shows a heat map representing the correlation between the different factors we measure: darker colors correspond to a stronger correlation while lighter colors correspond to a weaker correlation.  Blue color corresponds to positive correlation and red color corresponds to negative correlation. While some factors that are independent of one another have weak or no correlation (i.e., close to $0$, such as CPUs and Chips), others show strong correlations (i.e., more/less than $\pm0.8$, such as Capacity and Chips).  We discuss these factors and how we attempt to control for the factors correlations in \S\ref{sec:factors}.

\subsection{Case Study: Server Design Reliability Tradeoffs}

Using the equation in Table~\ref{tab:characteristics}, we can solve for $\mathcal{F}$, the rate of memory failure for a server with a given set of characteristics.  For example, Table~\ref{tab:model} compares the failure rates we predict using our model for four different server types:  (1) a \emph{low-end server} with low density DIMMs and few CPUs, (2) a \emph{high-end (HE) server} with high density DIMMs and twice as many CPUs as the low-end server, (3) a high-end server that uses \emph{lower-density DIMMs (HE/$\downarrow$density)}, and (4) a high-end server that uses \emph{half as many CPUs (HE/$\downarrow$CPUs)}.  So that the workload is kept roughly similar across the configurations, we double the CPU utilization for servers with half as many CPUs.

\begin{table}[H]
\centering
\begin{tabular}{p{3cm}rrrr} \hline
  \toprule
\bf Factor     & \bf Low-end          & \bf High-end (HE)    & \bf HE/$\downarrow$density & \bf HE/$\downarrow$CPUs \\
\midrule
Capacity   & \unit[4]{GB}     & \unit[16]{GB}    & \unit[4]{GB}     & \unit[16]{GB}  \\
Density2Gb & 1                & 0                & 1                & 0              \\
Density4Gb & 0                & 1                & 0                & 1              \\
Chips      & 16               & 32               & 16               & 32             \\
CPU\%      & 50\%             & 25\%             & 25\%             & 50\%           \\
Age        & 1                & 1                & 1                & 1              \\
CPUs       & 8                & 16               & 16               & 8              \\
\midrule
\bf Predicted relative & \multirow{2}{*}{{\bf 0.12}} & \multirow{2}{*}{{\bf 0.78}} & \multirow{2}{*}{{\bf 0.33}} & \multirow{2}{*}{{\bf 0.51}} \\
{\bf failure rate} & & & & \\
\bottomrule
\end{tabular}
\caption{The relative failure rates our model predicts for different server types.}
\label{tab:model}
\end{table}

\begin{figure}
  \centering
  \includegraphics[width=0.8\columnwidth]{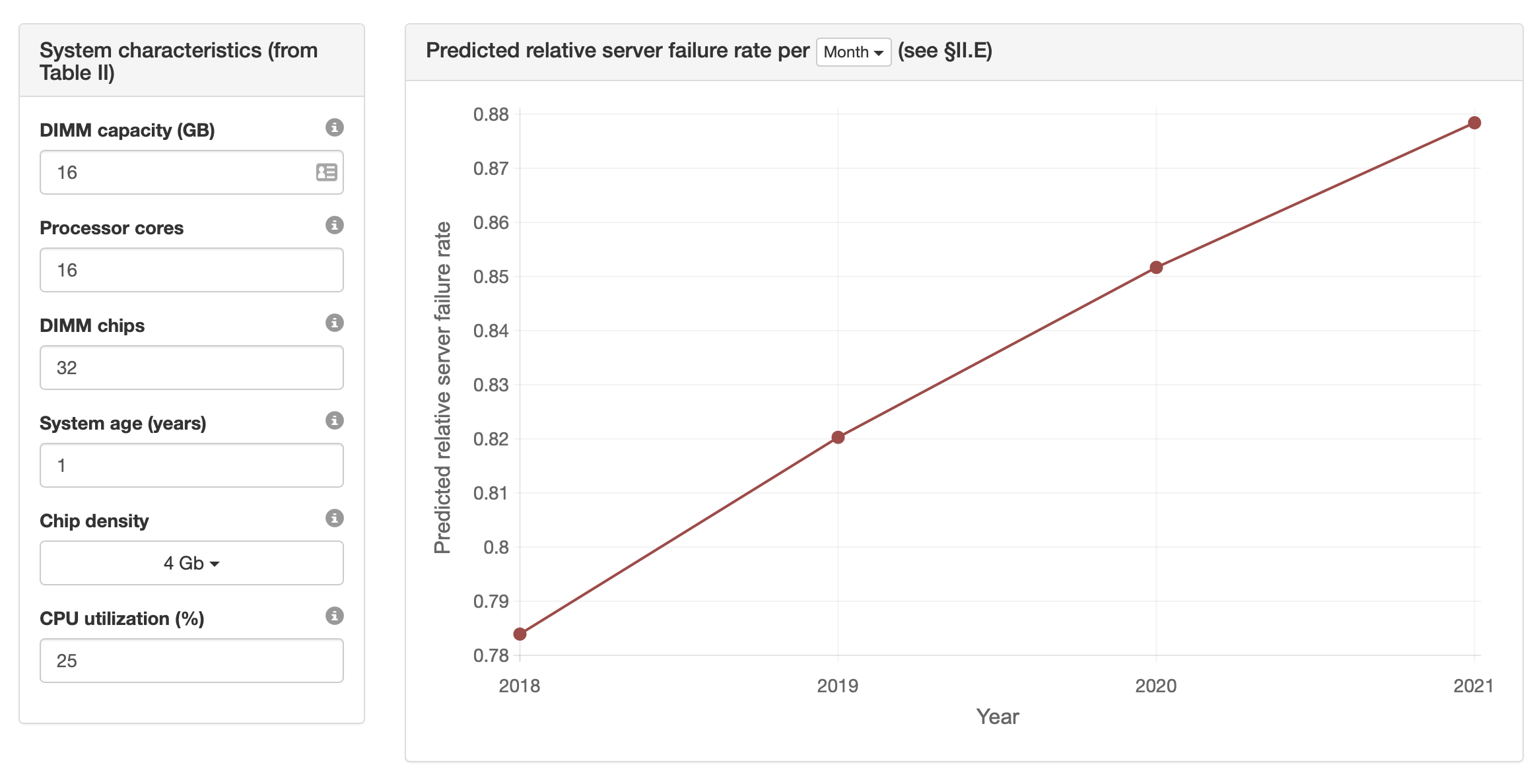}
  \caption{Using the interactive DRAM failure model site~\protect\cite{safari-tools}, you can compute the relative reliability between different server designs.}
  \label{fig:memerr-site}
\end{figure}

We can see that the modeled failure rate of the high-end server is $6.5\times$ that of the low-end server.  This agrees with the trends that we observe in \S\ref{sec:factors}, which show, for example, increasing failure rates with increasing chip density and number of CPUs.  Interestingly, we can use the model to provide insight into the relative change in error rate for \emph{different system design choices}.  For example, we can reduce the failure rate of the high-end server 57.7\% by using lower density DIMMs and by 34.6\% by using half as many cores.  This indicates that designing systems with lower density DIMMs can provide a larger DRAM reliability benefit than designing systems with fewer CPUs.  In this way, the model that we develop allows system architects to easily and quickly explore a large design space for memory reliability.  We hope that, by using the model, system architects can evaluate the reliability trade-offs of their own system configurations in order to achieve better memory reliability. Figure~\ref{fig:memerr-site} shows a screenshot of an interactive version of the model that we make available online at~\cite{safari-tools}.

\section{DRAM Page Offlining at Scale}
\label{sec:page-offline}

We next discuss the results of a study we perform to examine ways to reduce memory errors using page offlining~\cite{solaris-pageoffline, hwang}.  Page offlining removes a physical page of memory that contains a memory error from the set of physical pages of memory that the operating system can allocate.  This reduces the chance of a more severe uncorrectable error occurring on that page versus leaving the faulty page in the physical address space.  While prior work evaluated page offlining using simulations on memory traces~\cite{hwang}, we deploy page offlining on a fraction of the machines we examine (12,276 servers) and observe the results.  We next describe the system design decisions we identify to make page-offlining work well at a large scale, and analyze its effectiveness.

\subsubsection{Design Decisions and Implementation}

The three main design decisions we explore with respect to utilizing page offlining in practice are: (1) {\em when} to take a page offline, (2) {\em for how long} to take a page offline, and (3) {\em how many} pages to take offline (the first and last of which were also identified in~\cite{hwang}).

\emph{(1) When to take a page offline?}  ECC DIMMs provide flexibility for tolerating correctable errors for a certain amount of time.  In some settings, it may make sense to wait until a certain number of memory errors occur on a page in a certain amount of time before taking the page offline.  We examine a conservative approach and take any page that had a memory error offline immediately (the same as the most aggressive policy examined in prior work~\cite{hwang}).  The rationale is that if we leave a page with an error in use, it increases the risk of an uncorrectable error occurring on that page.  Another option is to leave pages with errors in use for longer and, for example, design applications that can tolerate  memory errors.  Such an approach is taken by Flikker~\cite{flikker}, which developed a programming model for reasoning about the reliability of data, and by heterogeneous-reliability memory systems where parts of memory can be less reliable and we can allocate application data that is less vulnerable to errors there~\cite{yixin}.

\emph{(2) For how long to take a page offline?}  One question that arose when designing page offlining at a large scale was how to make an offline page persist across machine reboots (both those we plan for and those we do not) and hardware changes (e.g., disk replacement).  Existing techniques handle neither of these cases.  Allowing an offline page with a permanent error to come back online can defeat the purpose of page offlining by increasing the window of vulnerability for uncorrectable errors.  We examine a policy that takes pages offline \emph{permanently}.  To keep track of offline pages across machine reboots, we store offline pages by host name in a distributed database that the OS queries when the OS kernel loads.  This allows us to take offline bad pages before the kernel allocates them to applications.  We must update entries in this database when we replace DRAM parts in a system.

\emph{(3) How many pages to take offline?}  Taking a page offline reduces the size of physical memory in a system and could increase swapping of pages to storage.  To limit the negative performance impact of this, we place a cap on the number of  physical pages that may be taken offline.  Unlike prior work, as we show in \S\ref{sec:component}, socket and channel failures can potentially cause page offlining to remove large portions of the physical address space, potentially causing large amounts of swapping to storage and degrading performance.  To check how many pages have been taken offline, we routinely inspect logs on each machine.  When the amount of physical memory taken offline is greater than 5\% of a server's physical memory capacity, automated repair software generates a repair ticket for the server.

\subsubsection{Effectiveness}

\begin{figure}[H]
  \centering
  \includegraphics[width=0.5\textwidth]{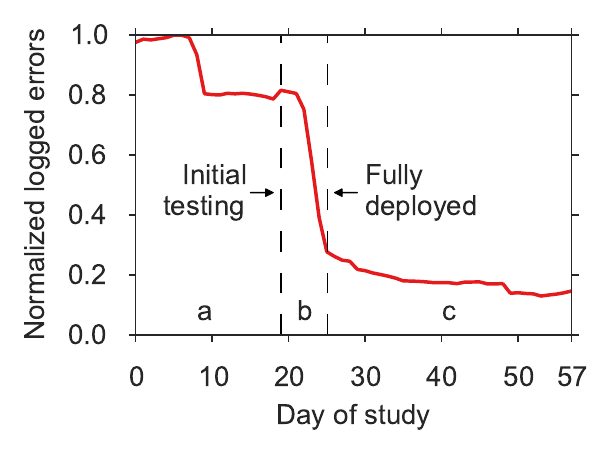}
  \caption{The effect of page offlining on error rate.}
  \label{fig:pageofflining}
\end{figure}

Figure~\ref{fig:pageofflining} shows a timeline of the normalized number of errors in the 12,276 servers that we examine (unlike the rest of this study, we only examine a small number of servers for this technique).  We performed the experiment for 86 days and we measured the number of errors as a moving average over 30 days.  As it was a production environment, we deployed page offlining on all of the machines over the course of several days.  We divide the graph into three regions corresponding to the different phases of our experiment.  Region {\sf a} shows the state of the servers before we deployed page offlining.  Region {\sf b} shows the state of the servers while we deployed page offlining gradually to 100\% of the servers (so that we could detect any malfunctions of the deployment in a small number of machines and not all of them).  Region {\sf c} shows the state of the servers after we fully deployed page offlining.

The initial hump in Region {\sf a} from days 0 to 7 was due to a bank failure on one server that generated a large number of errors.  By day 8 its effects were no longer noticeable in the moving average and we compare the effectiveness of page offlining to the error rate after day 8.

There are three things to note from Figure~\ref{fig:pageofflining}.  First, after deploying page offlining to 100\% of the fleet at day 25, error rate continues to decrease until day 50.  We believe this is because some pages contain errors, but the OS does not access them immediately after we deployed page offlining, instead, the OS accesses them at a later time, triggers an error, and takes the page offline.  In addition, some pages cannot be taken offline immediately due to restrictions in the OS, which we will describe in the next section.  Second, comparing the error rate at day 18 (right before initial testing) to the error rate at day 50 (after deploying page offlining and letting the servers run for a couple of weeks), the error rate decreases by around 67\%.  This is smaller than the 86\% to 94\% error rate reduction reported in Hwang et al.'s study~\cite{hwang}.  One reason for this could be that the prior study may have included socket and channel errors in their simulation---including socket and channel errors would increase the number of errors that page offlining could avoid.  Third, we observe a relatively large rate of error occurrence (e.g., at day 57 the error rate is still around 18\% of the maximum amount), even after page offlining.  This suggests that it is important to design devices and other techniques that help reduce the error rate that aggressive page offlining does not seem to affect.

\subsubsection{Limitations}

While page offlining is relatively effective at reducing DRAM errors, we find that it has two main limitations that were not addressed in prior work.  First, it reduces memory capacity, which requires repairing a machine after a certain fraction of its pages have been taken offline.  Second, it may not always succeed in real systems.  We additionally log the failure rate of page offlining and find that around 6\% of the attempts to offline a page initially fail.  One example we found of why the OS may fail to take a page offline in the Linux kernel is if the OS cannot lock its content for exclusive access.  For example, if the OS is prefetching data for the page into the page cache at the time when the page should be taken offline, locking the page could result in a deadlock, and so the Linux kernel does \emph{not} allow this.  We could, however, increase the rate of success for page offlining by retrying page-offlining at a later time, at the expense of additional complexity to system software.

Despite these limitations, however, we find that page offlining---when we adapt it to function at scale---provides reasonable memory error tolerance benefits, as we demonstrate.  We look forward to future works that analyze the interaction of page offlining with other error correction methods.

\section{Physical Page Randomization}
\label{sec:reduce}


Prior work has shown that DRAM cells can be worn out due to repeated access~\cite{dram-wearout} and that transistors (which comprise every DRAM cell) also wear out~\cite{bti,nbti}.  Our results in \S\ref{sec:model} corroborate these findings in the field:  We observe {\em circumstantial evidence} that device {\em wearout}---the reduction of component reliability over time due to repeated access that, for example, degrades the data retention time of capacitors or degrades the switching speed of transistors---relates to DRAM error rates.  This is shown by the statistically-significant correlation between device age and error rate, and also between the statistically-significant correlation between number of cores (the entities that access memory) and error rate, while controlling for a variety of factors (such as memory size, density, and those in Table~\ref{tab:characteristics}).

We therefore hypothesize that {\em DRAM device wearout} is a fundamental trend that underlies many of the faults seen across the machines we examine.  We call these errors ``wearout-like'' errors (since we cannot be absolutely sure the cause is wearout without component-level device analysis).  We next examine ways of reducing wearout-like errors in DRAM.

\subsection{Wear Reduction Techniques in Other Devices}

Techniques to reduce or tolerate cell wear have been examined for flash and emerging non-volatile memory devices.  In general, these techniques fall into three categories depending on if they reduce wear by (1) throttling writes to a device (e.g.,~\cite{throttle}), by (2) converting data to a representation that does not cause as much wear (e.g.,~\cite{filpnwrite}), or by (3) wear-leveling writes to a device (e.g.,~\cite{sandisk-wl,flash-wl1, kaist-wl, start-gap, mattan, free-p, paygo}).  The feasibility of these techniques depends on device characteristics and system software requirements.  For example, throttling (technique 1) directly impacts performance, making it less desirable for reducing wear in DRAM, while data conversion (technique 2) does not apply to DRAM devices.  We therefore explore a form of wear-leveling (technique 3) for use in DRAM.


Techniques to reduce or tolerate cell wear per unit time have been examined extensively for storage technologies, most notably non-volatile storage and memory devices~\cite{sandisk-wl,flash-wl1, kaist-wl, start-gap, mattan, free-p, paygo}.  In general, can classify these techniques into three categories based on whether they reduce wear by (1) throttling writes to a device, by (2) wear-leveling writes to a device, or by (3) converting data to a representation that does not cause as much wear.  The feasibility of these techniques depends on device characteristics and system software requirements.  We briefly discuss these three approaches.

  \begin{enumerate}
    \item \textbf{\emph{Throttling-based approaches.}}
One way to reduce cell wear per unit time and guarantee a minimum bound on device lifetime is to limit the rate at which wear occurs on the cells~\cite{throttle}.  This is typically done by reducing the rate at which an application can issue requests to a device, thereby reducing performance.  This characteristic makes them less desirable for reducing wear in DRAM due to their direct impact on program performance.

\item \textbf{\emph{Data translation approaches.}}
Some memory devices wear out cells differently depending on the symbol that is being read or written.  For example, in flash, changing a cell from 0 to 1 may only require applying a small amount of charge to the cell, but changing a cell from 1 to 0 may require resetting the contents of a cell, a more charge-intensive operation that exhibits higher amounts of wear~\cite{cai-date12}.  In such devices, it can be favorable to change the representation of the data the device stores to reduce wear (in our example, this could involve writing either the original data or its inverse, depending on which version requires fewer 1 to 0 transitions)~\cite{kumar, moshovos}.  Unfortunately, data translation is only effective for devices that exhibit asymmetric write characteristics, which DRAM does not.

\item \textbf{\emph{Wear-leveling--based approaches.}}
Because non-volatile storage and memory devices have a noticeably short lifetime if the device does not manage wearout, many wear-leveling techniques have been proposed for flash~\cite{sandisk-wl,flash-wl1, kaist-wl} and non-volatile memory~\cite{start-gap, mattan, free-p, paygo}.  The key observation behind this idea is that not all physical regions are worn out at the same rate.  Therefore, techniques that redistribute wear evenly across all of physical memory can improve device lifetime.  If we can employ such techniques with minimal interference and cost in DRAM, they may help reduce the wearout problem.
  \end{enumerate}

\subsection{Challenges and Key Observations}

Two challenges with applying prior wear-leveling approaches in DRAM are (1) the cost of adding the required custom hardware to DRAM, and (2) the metadata overhead and execution overhead of employing these techniques in software.  To help solve these challenges, we leverage two key observations to design a technique to reduce wearout-like errors in DRAM: (1) server systems do not necessarily use all of their physical memory all the time, and in these systems, data remains resident in memory for long amounts of time; and (2) unlike non-volatile devices, the slower wearout process of DRAM means that we can apply techniques at a coarser time granularity.


Based on these observations, we propose {\it physical page randomization}.  The key idea is to  occasionally migrate physical pages to different {\it random} locations, spreading wear across the physical address space.  The slower wearout process of DRAM allows us to perform this leveling at a relatively coarse granularity, for example, throughout a week, when the machine is idle, or only for the most frequently-accessed data.  We could implement our technique in either hardware or software, but for our evaluation, we focus on our initial software implementation.

We identify two design decisions for physical page randomization: (1) coarse versus fine granularity and (2) static versus dynamic movement.  In terms of granularity, we can migrate data at a granularity from the program's working set of data (coarse) to a single page (fine).
We opt for migrating at a fine granularity to reduce the potential for performance degradation from migrating many pages at once and to provide more flexibility in scheduling randomizations.
In terms of movement, we could statically remap data when we load into physical memory from storage (static), or we could proactively remap throughout a device's operation at a certain frequency (dynamic).
A downside of static movement is that, in servers, the pages an OS allocates typically remain in the same location in memory until the machine reboots, leaving long-running systems prone to wearout-like errors.  We therefore adopt dynamic physical layout randomization, and analyze its performance impact.

\subsection{Proof of Concept Prototype}

We implement dynamic physical page randomization in the Linux 3.10.17 kernel.  At a high level, the basic building block of our implementation performs the following algorithm:
\begin{algorithm}
\KwIn{The address of a physical page to randomize.}
Lock the page.

\Indp

Flush any pending updates to the page.

Randomly select a new free page to allocate.

Migrate the contents of the old page to the new page.

Update the page table mappings and remove any stale TLB entries.

\Indm

Unlock the page.

\end{algorithm}
\vspace{-0.25em}

Our current implementation traverses memory at a certain interval and randomly remaps pages in the physical address space.  Given that there is room to remap pages in the physical address space, our technique will level accesses across memory in the steady state.  It is also possible to apply this same technique to only the most frequently accessed pages by tracking or sampling page access information.

\subsection{Overhead}

We ran dynamic physical page randomization on a workstation running Debian 6.0.7 with the kernel with our modifications.  The system had 4$\,$GB of DRAM and a Core 2 Duo processor running at 3$\,$GHz.  Our primary interest is to understand the OS overhead of a single page randomization, as additional techniques built on top of this primitive would incur the same basic latency.

\begin{figure}
  \centering
  \includegraphics[width=0.4\columnwidth]{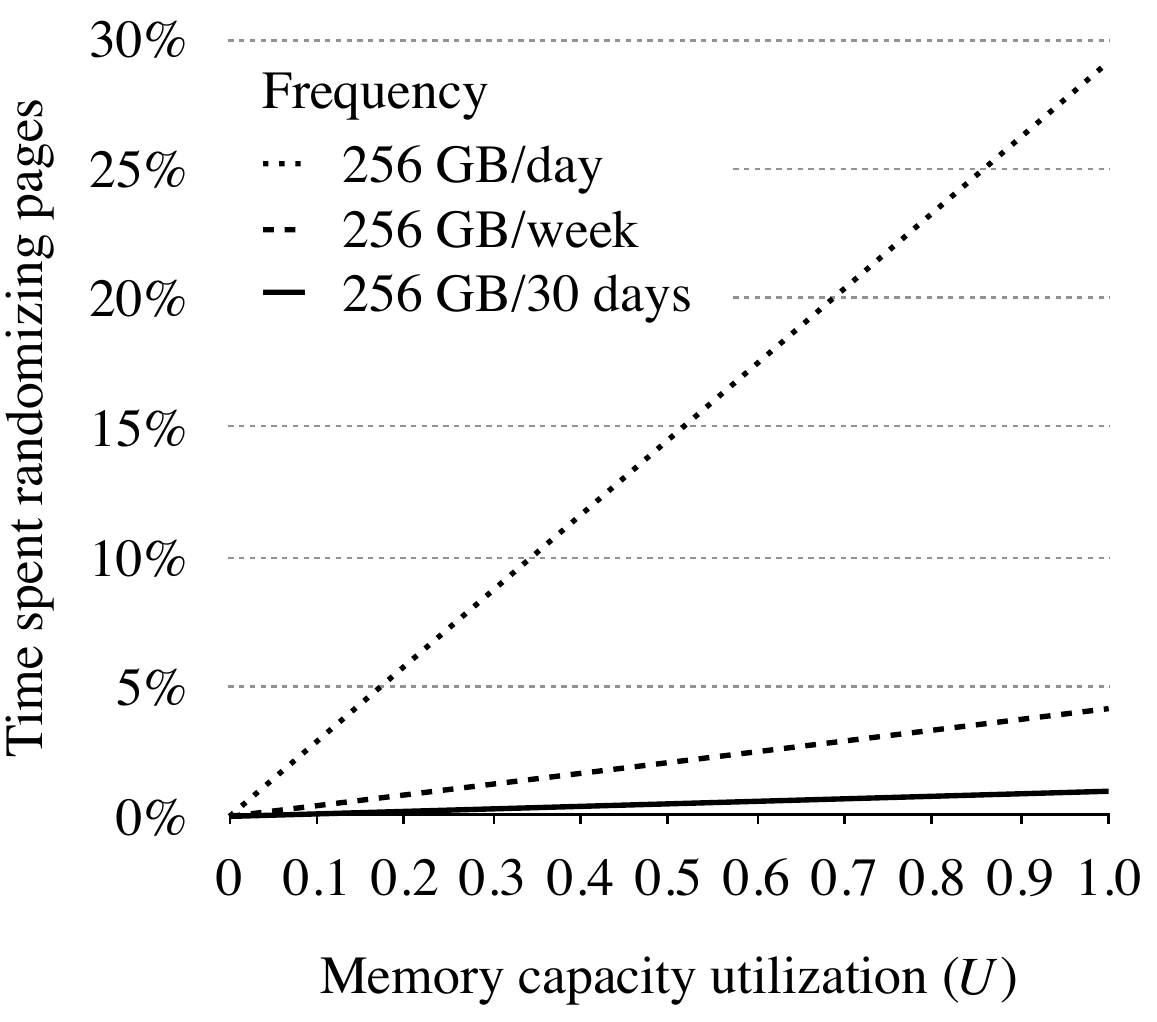}
  \caption{Dynamic physical page randomization overhead.}
  \label{fig:randomize}
\end{figure}

We measure several hundred randomizations of physical pages and found the average time to randomize a page was 374.9$\;\mu$s, with a standard deviation of 193.3$\;\mu$s.  A server with $256\,$GB of memory must remap
\begin{equation}
(256\,\mathrm{GB} \times U) / 4\,\mathrm{KB} / D\,\mathrm{days} \approx 777\times U / D \mathrm{pages/s}
\end{equation}

\noindent
in order to remap the fraction of physical memory the OS uses, $U$, in $D$ days.  Figure~\ref{fig:randomize} shows the worst-case performance overhead of dynamic physical page randomization (assuming no parallelism) for several different randomization frequencies.


From Figure~\ref{fig:randomize} we can see that we can employ physical page randomization with low performance impact if we randomize pages at a frequency of 1 week, for example.  Furthermore, we can tune this overhead depending on how much wear we want to tolerate in the memory device.
We plan to further study techniques to enable more efficient dynamic physical page randomization, but we believe that others can employ it with relatively low overhead to more evenly spread wear across DRAM devices.


\section{Summary}

In this chapter, we examine the memory errors across all of Facebook's servers over fourteen months. We analyze a variety of factors and how they affect server failure rate and observe several new reliability trends for memory systems that have not been discussed before in literature.  We identify several important trends:

\begin{description}

\item[Lesson D.1] Memory errors follow a power-law distribution, specifically, a Pareto distribution with decreasing hazard rate, with average error rate exceeding median error rate by around $55\times$. [\S\ref{sec:incidence}]

\item[Lesson D.2] Non-DRAM memory failures from the memory controller and memory channel contribute the majority of errors and create a kind of \emph{denial of service attack} in servers. [\S\ref{sec:component}]

\item[Lesson D.3] More recent DRAM cell fabrication technologies (as indicated by chip density) show higher failure rates (prior work that measured DRAM \emph{capacity}, which is not closely related to fabrication technology, observed inconclusive trends). [\S\ref{subsec:DIMM Capacity and DRAM Density}]

\item[Lesson D.4] DIMM architecture decisions affect memory reliability: DIMMs with fewer chips and lower transfer widths have the lowest error rates, likely due to electrical noise reduction. [\S\ref{subsec:DIMM Architecture}]

\item[Lesson D.5] While CPU and memory utilization do not show clear trends with respect to failure rates, workload type can influence server failure rate by up to $6.5\times$. [\S\ref{sec:workload}]

\item[Lesson D.6] We show how to develop a model for memory failures and show how system design choices such as using lower density DIMMs and fewer processors can reduce failure rates of baseline servers by up to 57.7\%. [\S\ref{sec:model}]
  
\item[Lesson D.7] We perform the first analysis of page offlining in a real-world environment, showing that error rate can be reduced by around 67\% identifying and fixing several real-world challenges to the technique. [\S\ref{sec:page-offline}]

\item[Lesson D.8] We evaluate the efficacy of a new technique to reduce DRAM faults, \emph{physical page randomization}, and examine its potential for improving reliability and its overheads. [\S\ref{sec:reduce}]
\end{description}

We hope that the data and analyses we present in this chapter can aid in (1) clearing up potential inaccuracies and limitations in past studies' conclusions, (2) understanding the effects of different factors on memory reliability, (3) the design of more reliable DIMM and memory system architectures, and (4) improving evaluation methodologies for future memory reliability studies.

While the servers in modern data centers use DRAM to store \emph{volatile} data, servers use SSDs to store \emph{persistent} data. The next chapter explores the reliability of flash-based SSDs in modern data centers.

\chapter{SSD Failures}
\label{chp:ssdfailures}

Servers use flash-memory--based SSDs as a high-performance alternative to hard disk drives to store persistent data. Unfortunately, recent increases in flash density have also brought about decreases in chip-level reliability. In a data center environment, flash-based SSD failures can lead to downtime and, in the worst case, data loss. As a result, it is important to understand flash memory reliability characteristics over flash lifetime in a realistic production data center environment running modern applications and system software.

In this chapter, we analyze data we collect from across a majority of flash-based solid state drives at Facebook data centers over nearly four years and many millions of operational hours in order to understand failure properties and trends of flash-based SSDs.  Our study considers a variety of SSD characteristics, including: the amount of data written to and read from flash chips; how the flash controller maps data within the SSD address space; the amount of data the flash controller copies, erases, and discards; and flash board temperature and bus power.

Our field analysis shows how flash memory errors manifest when running modern workloads on modern SSDs~\cite{meza15}, we make several major observations: (1) \emph{SSD} failure rates do \emph{not} increase monotonically with \emph{flash chip} wear but instead go through several phases, (2) the effects of read disturbance errors are \emph{not} prevalent in the field, (3) sparse logical data layout across an SSD's physical address space (e.g., non-contiguous data), as we measure by the amount of metadata the flash controller stores to track logical address translations in an SSD-internal DRAM buffer, can greatly affect SSD failure rate, (4) higher temperatures lead to higher failure rates, but techniques that throttle SSD operation appear to greatly \emph{reduce} the negative reliability impact of higher temperatures, and (5) data written by the operating system to flash-based SSDs does \emph{not} always accurately indicate the amount of wear on flash cells due to optimizations in the SSD controller and buffering in the system software.


\section{Motivation for Understanding SSD Failures}

Servers use flash memory for persistent storage due to the low access latency of flash chips versus hard disk drives.  Historically, flash capacity lags behind hard disk drive capacity, limiting the use of flash memory.  In the past decade, however, advances in NAND flash memory technology increased flash capacity by more than $1000\times$.  This rapid increase in flash capacity has brought both an increase in flash memory use and a decrease in flash memory reliability.  For example, the number of times that an SSD can reliably program and erase a cell before wearing out and failing decreased from 10,000 times for \unit[50]{nm} cells to only 2,000 times for \unit[20]{nm} cells~\cite{maislos-fms11}.  We expect this trend to continue for newer generations of flash memory.  Therefore, if we want to improve the operational lifetime and reliability of flash memory-based devices, we must first fully understand their failure characteristics.

In the past, a large body of prior work examined the failure characteristics of flash cells in controlled environments using small numbers (e.g., tens) of raw flash chips (we cover this work in detail in \S\ref{subsec:Related Research in SSD Failures in Modern Data Centers}).  These works quantified a variety of flash cell failure modes and formed the basis of the community's understanding of flash cell reliability.  Yet prior work was limited in its analysis since these studies:  (1) were conducted on small numbers of raw flash chips accessed in adversarial manners over short amounts of time, (2) did not examine failures when using real applications running on modern servers and instead used synthetic access patterns, and (3) did not account for the storage software stack that real applications go through to access flash memories.

Such conditions assumed in these prior studies are substantially different from the conditions flash-based SSDs experience in large-scale installations in the field.  In such large-scale systems: (1) real applications access flash-based SSDs in different ways over a time span of \emph{years}, (2) applications access SSDs via the storage software stack, which employs various amounts of buffering and hence affects the access pattern seen by the flash chips, (3) flash-based SSDs employ aggressive techniques to reduce device wear and to correct errors, (4) factors in platform design, including how many SSDs are present in a node, affect the access patterns to SSDs, (5) there can be significant variation in reliability due to the existence of a very large number of SSDs and flash chips. All of these real-world conditions present in large-scale systems likely influence the reliability characteristics and trends of flash-based SSDs.

Our goal is to understand the nature of flash-based solid state drive (SSD) failures in the field.  To this end, we provide the \emph{first comprehensive study of flash memory reliability trends in a large-scale production data center environment}.  We base our study on data we collect from a majority of flash-based SSDs in Facebook's server fleet, with operational lifetimes extending over nearly four years and comprising many millions of device-days of usage.  We analyze SSDs of different capacities and data transfer technologies, with a focus on understanding how various \emph{internal factors} (i.e., those that relate to how the device operates) and \emph{external factors} (i.e., those that relate to the environment the SSD operates in) affect flash-based SSD reliability.

Our main contribution is a rigorous characterization of the reliability trends of flash-based SSDs in the field.  We observe several reliability trends for flash-based SSDs that have not been discussed before in prior works:

\begin{enumerate}

	\item Flash-based SSDs do \emph{not} fail at a monotonically increasing rate with wear.  They instead go through \emph{several} distinct reliability periods corresponding to how failures emerge and how the flash controller detects failures.  Unlike the failure trends for \emph{individual flash} \emph{chips}~\cite{cai-date12, cai-itj13, grupp-fast12}, across a large number of \emph{flash-based SSDs} we observe \emph{early detection}, \emph{early failure}, \emph{usable life}, and \emph{wearout} periods, whose failure rates can vary by up to 81.7\%.

	\item Read disturbance errors (i.e., when a read operation~\cite{cai-iccd12, cai-dsn15} causes errors in neighboring pages) are \emph{not} prevalent in the field. SSDs that have read the most data do not show a statistically significant increase in failure rates.

\item Sparse logical data layout across an SSD's physical address space (e.g., non-contiguous data), as we measure by the amount of SSD-internal DRAM buffer utilization for flash translation layer metadata, greatly affects device failure rate.  In addition, dense logical data layout (e.g., contiguous data) with adversarial access patterns (e.g., small, sparse writes) also negatively affect SSD reliability.

\item Higher temperatures lead to higher failure rates, but techniques that modern SSDs use that throttle SSD operation (and, consequently, the amount of data written to flash chips) appear to greatly reduce the reliability impact of higher temperatures by reducing access rates to the raw flash chips.

\item The amount of data written by the operating system to \emph{an SSD} is not the same as the amount of data that is eventually written to flash \emph{cells}.  This is due to system-level buffering and techniques that the storage software stack employs to reduce wear.  It is important that system-level flash reliability studies account for these effects.

\end{enumerate}



\begin{table*}
\small
\centering
\begin{tabular}{|c||c|c|c|c|c|c|c|}
\hline
\multirow{2}{*}{Platform} & \multirow{2}{*}{SSDs} & \multirow{2}{*}{PCIe} & \multicolumn{5}{c|}{Per SSD} \\ \cline{4-8}
& & & Capacity & Age (years) & Data written & Data read & UBER \\ \hline \hline
A & 1 & \multirow{2}{*}{v1, $\times$4} & \multirow{2}{*}{\unit[720]{GB}} & \multirow{2}{1.25cm}{2.4 $\pm$ 1.0} & \unit[27.2]{TB} & \unit[23.8]{TB} & \num{5.2e-10} \\ \cline{1-2} \cline{6-8}
B & 2 & & & & \unit[48.5]{TB} & \unit[45.1]{TB} & \num{2.6e-9} \\ \hline
C & 1 & \multirow{4}{*}{v2, $\times$4} & \multirow{2}{*}{\unit[1.2]{TB}} & \multirow{2}{1.25cm}{1.6 $\pm$ 0.9} & \unit[37.8]{TB} & \unit[43.4]{TB} & \num{1.5e-10} \\ \cline{1-2} \cline{6-8}
D & 2 & & & & \unit[18.9]{TB} & \unit[30.6]{TB} & \num{5.7e-11} \\ \cline{1-2} \cline{4-8}
E & 1 & & \multirow{2}{*}{\unit[3.2]{TB}} & \multirow{2}{1.25cm}{0.5 $\pm$ 0.5} & \unit[23.9]{TB} & \unit[51.1]{TB} & \num{5.1e-11} \\ \cline{1-2} \cline{6-8}
F & 2 & & & & \unit[14.8]{TB} & \unit[18.2]{TB} & \num{1.8e-10} \\ \hline
\end{tabular}
\caption{The platforms we examine in our study.  We show PCIe technology with the notation v$X$, $\times$$Y$ where $X$ = version and $Y$ = number of lanes.  We collect data over the entire age of the SSDs.  Data reads and writes are to the physical storage over an SSD's lifetime.  UBER = uncorrectable bit error rate.}
\label{tab:systems}
\end{table*}



\section{Methodology for Understanding SSD Failures}
\label{sec:errors}

We examine the majority of flash-based SSDs in Facebook's server fleet, which have operational lifetimes extending over nearly four years and comprising many millions of SSD-days of usage.  We collect data over the lifetime of the SSDs.  We found it useful to separate the SSDs depending on the type of \emph{platform} an SSD is in.  We define a platform as a combination of the SSD capacity, the PCIe technology, and the number of SSDs in the system.  Table~\ref{tab:systems} shows the platforms we examine in our study.

\subsection{The Systems We Examine}

We examine a range of high-capacity \emph{planar} \emph{Multi-Level Cell (MLC)} flash-based SSDs with capacities of \unit[720]{GB}, \unit[1.2]{TB}, and \unit[3.2]{TB}.  The technologies we examine span two generations of PCIe, versions 1 and 2.  Some of the systems in the fleet use one SSD while others use two.  Platforms A and B contain SSDs with around two or more years of operation and represent 16.6\% of the SSDs we examine; Platforms C and D contain SSDs with around one to two years of operation and represent 50.6\% of the SSDs we examine; and Platforms E and F contain SSDs with around half a year of operation and represent around 22.8\% of the SSDs we examine.

\subsection{How We Measure SSD Failures}

The flash devices in Facebook's fleet contain registers that keep track of SSD operation statistics (e.g., number of bytes read, number of bytes written, number of errors that could not be corrected by the device). These registers are similar to, but distinct from, the standard \emph{Self-Monitoring, Analysis and Reporting Technology (SMART)} data in some SSDs to monitor their reliability characteristics~\cite{smart}.  We can query the values of these registers using the host machine.  We use a script to retrieve the raw values from the SSD and parse them into a form that we store in a Hive~\cite{thusoo-icde10} table.  This process is done in real time every hour.


Our SSDs allow us to collect information only on large errors that are uncorrectable by the SSD but correctable by the host. For this reason, our results are in terms of such SSD-uncorrectable but host-correctable errors, and when we refer to \emph{uncorrectable errors} we mean these type of errors.  We refer to the occurrence of such uncorrectable errors in an SSD as an \emph{SSD failure}. Note that when the system requests data with host-correctable errors, the SSD sends the data to a driver running on the host to try and correct the data. This operation takes additional latency to perform compared to the host directly reading the data from the SSD.

The scale of the systems we analyze and the amount of data we collect poses challenges for analysis. To process the many millions of SSD-days of information, we use a cluster of machines to perform a parallel aggregation of the data in Hive using MapReduce jobs in order to get a set of lifetime statistics for each of the SSDs we analyze.  We then process this summary data in R~\cite{r} to collect our results.

\subsection{How We Analyze SSD Failure Trends}

Our infrastructure allows us to examine a \emph{snapshot} of SSD data at a point in time (i.e., our infrastructure \emph{does not} store \emph{timeseries} information for the many SSDs in Facebook's fleet).  This limits our analysis to the behavior of the fleet of SSDs at a point in time (for our study, we focus on a snapshot of SSD data taken during November 2014).  Fortunately, the number of SSDs we examine and the diversity of their characteristics allows us to examine how reliability trends change with various characteristics.  When we analyze an SSD characteristic (e.g., the amount of data written to an SSD), we group SSDs into buckets by their value for that characteristic in the snapshot and plot the failure rate for SSDs in each bucket.

For example, if we place an SSD $s$ in a bucket for \unit[$N$]{TB} of data written, we do {\em not} also place $s$ in the bucket for \unit[$(N - k)$]{TB} of data written (even though at some point in its life it did only have \unit[$(N - k)$]{TB} of data written).  When performing bucketing, we round the value of an SSD's characteristic to the nearest bucket and we eliminate buckets that contain less than $0.1$\% of the SSDs we analyze to have a statistically significant sample of SSDs for our measurements.  In order to express the confidence of our observations across buckets that contain different numbers of servers, we show the 95th percentile confidence interval for all of our data (using a binomial distribution when considering failure rates).  We measure SSD failure rate for each bucket in terms of the fraction of SSDs that have had an uncorrectable error versus the total number of SSDs in that bucket.


\subsection{Limitations and Potential Confounding Factors}

\emph{Workload changes due to SSD failure.}  In order to make systems more resilient to SSD failures, Facebook uses data replication at the software level in some workloads.  Such replication stores multiple copies of data (e.g., three copies) across different servers.  In the event that data on one SSD becomes unavailable (e.g., due to an SSD failure), a copy of data can still be readily accessible.

In \S\ref{sec:corr}, we examine the correlation between SSD failures \emph{within} the same server.  Note that, with replication, an SSD failure will not shift the workload to the other SSD in the same server, as replicas of data are spread out among different servers.  Replication will, however, increase the utilization of other SSDs that contain copies of data. As such, correlation between SSD failures in \emph{different} servers may exist, though we do not examine such occurrences in this work.

In addition, we did not examine how the performance the rate of uncorrectable errors on an SSDs affects the performance of the server it connects to and whether this made servers with SSDs reporting more errors less effective at executing their workload.

\emph{SSD access patterns.}  Our data collection infrastructure does not allow us to collect access traces to the SSDs that we examine.  This makes understanding the underlying causes of error behavior challenging.  While we examine the aggregate behavior of SSDs (e.g., the read and write characteristics of SSDs in \S\ref{sec:written} and \S\ref{sec:read}) in order to gain insight into potential underlying access pattern trends, we recommend that future work analyze SSD access patterns to fully understand the underlying hardware-level causes of the reliability trends we identify.

\emph{Error correction scheme.}  The SSDs that we examine use one viable strategy for error correction.  The flash controller forwards data uncorrectable by the SSD to the host for repair.  We note that such an error correction scheme is \emph{not} on all SSDs (and may not even be possible for some SSDs, such as those that use the NVMe interface~\cite{nvme}).  While our results are still representative of the types of uncorrectable errors that such SSDs will face, we believe that that future studies will benefit from examining the effects of different error correction schemes in SSDs.

\emph{SSD data loss and repair.}  Recall from \S\ref{sec:errors} that when an SSD cannot correct an error, the SSD forwards the error for repair to the host that the SSD connects to.  If the host cannot repair an error, data loss ensues.  We do not examine the rate of data loss (nor did we examine the rate of SSD replacement).  However, we believe that such an analysis (potentially with various error correction schemes, as we mention above) would be useful to perform in future work.

As we also mention in \S\ref{sec:errors}, the amount of errors that the SSDs we examine can tolerate without the assistance of the host is on the order of bits per KB of data (similar to data reported in~\cite{cooke}).  However, we do not have visibility into the error correction schemes the host uses for larger amounts of errors (>10's of bits per KB of data).

\emph{Transient and permanent failures.} We do not distinguish between transient and permanent failures in our study. However, we note that the types of \emph{large errors} that we examine, that \emph{cannot} be corrected using simple ECC mechanisms, provide circumstantial evidence of failure modes that may appear more frequently with permanent faults.

\emph{Firmware bugs.} We are not able to determine whether any of the errors we observe are due to bugs in the firmware of the SSDs that we examine. Prior work~\cite{Bairavasundaram2008} has found such bugs in hard disk drives lead to silent data corruption and we do not have a way to measure silent data corruption in the systems that we examine. Such errors would \emph{not} be included in our analysis.

\section{SSD Failure Trends}
\label{sec:baseline}

We next focus on the overall error rate and error distribution among the SSDs in the platforms we analyze and then examine correlations between different failure events.

\subsection{Bit Error Rate}

As we discuss, the bit error rate (BER) of an SSD is the rate at which errors occur relative to the amount of information that the system software transmits from/to the SSD. We can use BER to gauge the reliability of data transmission across a medium.  We measure the rate of uncorrectable bit errors (UBER) from the SSD as:
\begin{equation}
\mathit{UBER} = \frac{\mathit{Uncorrectable\ errors}}{\mathit{Bits\ accessed}}
\end{equation}

For flash-based SSDs, UBER is an important reliability metric that relates to the SSD lifetime.  We expect SSDs with high UBERs to have more cell failures and encounter more (severe) errors that the SSD cannot detect (leading to corrupt data) than SSDs with low UBERs.  As we discuss in \S\ref{subsec:Related Research in SSD Failures in Modern Data Centers}, recent work by Grupp et al.\ examined the BER of \emph{raw} MLC flash chips (which did not perform error correction) in a controlled environment~\cite{grupp-fast12}.  They found the raw BER to vary between \num{1e-1} for the least reliable flash chips down to \num{1e-8} for the most reliable, with most chips having a BER in the \num{1e-6} to \num{1e-8} range. Their study did \emph{not} analyze the effects of the use of chips \emph{in SSDs under real workloads}.

Table~\ref{tab:systems} shows the UBER of the platforms that we examine.  We find that for flash-based SSDs in the servers we examine, the UBER ranges from \num{2.6e-9} to \num{5.1e-11}.  While we expect that the UBER of the SSDs that we measure (which correct small errors, perform wear leveling, and are not at the end of their rated life) should be less than the raw chip BER in Grupp et al.'s study (which did not correct any errors, exercised flash chips until the end of their rated life, and accessed flash chips in an adversarial manner), we find that in some cases the BERs were within around one order of magnitude of each other.  For example, the BER of Platform B, \num{2.6e-9}, comes close to the lower end of the raw BER range reported in prior work, \num{1e-8}.

Thus, we observe that in flash-based SSDs that employ error correction for small errors and wear leveling, the UBER ranges from around 1/10 to 1/1000$\times$ the raw BER of similar flash chips examined in prior work~\cite{grupp-fast12}.  This is likely due to the fact that our flash-based SSDs correct small errors, perform wear leveling, and are not at the end of their rated life.  As a result, the error rate we see is smaller than the previous study observed.

As shown by the SSD UBER, the effects of uncorrectable errors are noticeable across the SSDs that we examine.  We next turn to understanding the distribution of errors among a population of SSDs and how failures occur within SSDs.

\subsection{Failure Rate and Error Count}

\begin{figure}
\centering
\includegraphics[width=0.4\textwidth]{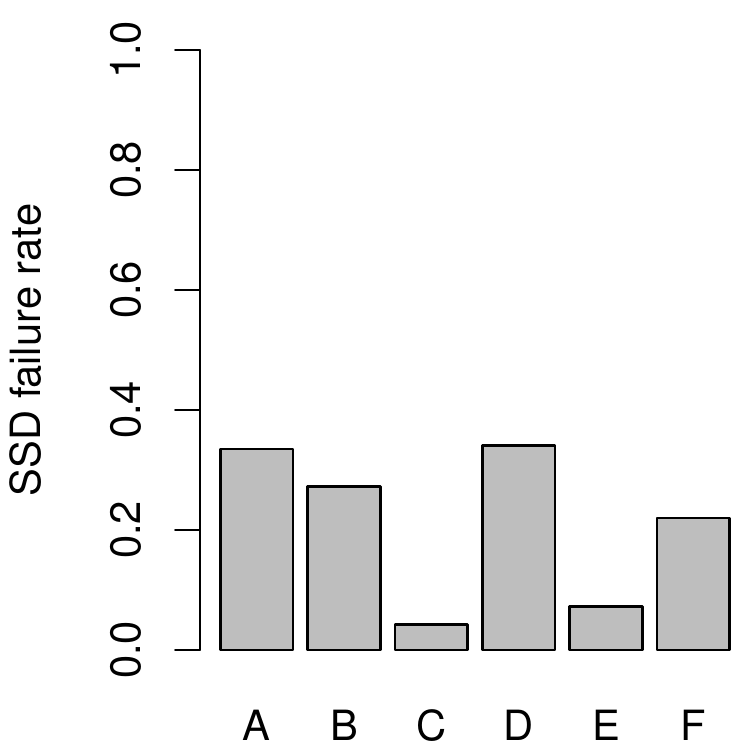}
\hspace{0.5cm}
\includegraphics[width=0.4\textwidth]{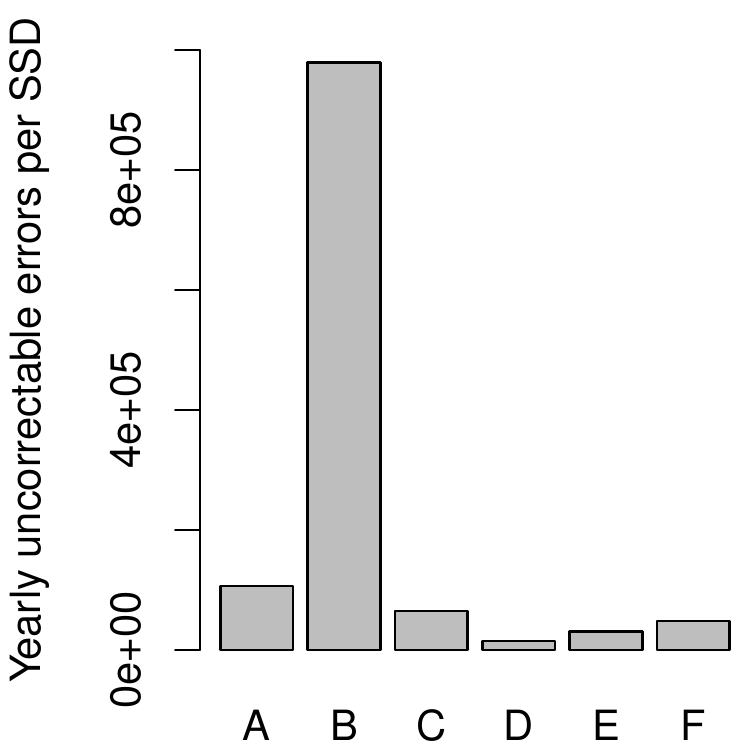}
\caption{The failure rate (left) and average yearly rate of uncorrectable errors (right) among SSDs within each platform.  Note the y axis magnitude differences between the left and right graphs.}
\label{fig:rate}
\end{figure}

Figure~\ref{fig:rate} (left) shows the SSD incidence failure rate within each platform---the fraction of SSDs in each platform that have at least one uncorrectable error.  We find that SSD failures are relatively common events with between 4.2 and 34.1\% of the SSDs in the platforms we examine reporting uncorrectable errors.  Interestingly, the failure rate is much lower for Platforms C and E despite their comparable amounts of data written and read (cf.\ Table~\ref{tab:systems}).  This suggests that there are differences in the failure process among platforms.  We analyze which factors play a role in determining the occurrence of uncorrectable errors in \S\ref{sec:internal} and \S\ref{sec:external}.

Figure~\ref{fig:rate} (right) shows the average yearly uncorrectable error rate among SSDs within the different platforms---the sum of errors that occur on all servers within a platform over 12 months ending in November 2014, which we divide by the number of servers in the platform.  The yearly rates of uncorrectable errors on the SSDs we examine range from 15,128 for Platform D to 978,806 for Platform B.  The older Platforms A and B have a higher error rate than the younger Platforms C through F, which suggests that the incidence of uncorrectable errors increases when we use SSDs more.  We will examine this relationship further in \S\ref{sec:internal}.

\begin{figure}
\centering
\includegraphics[width=0.5\columnwidth]{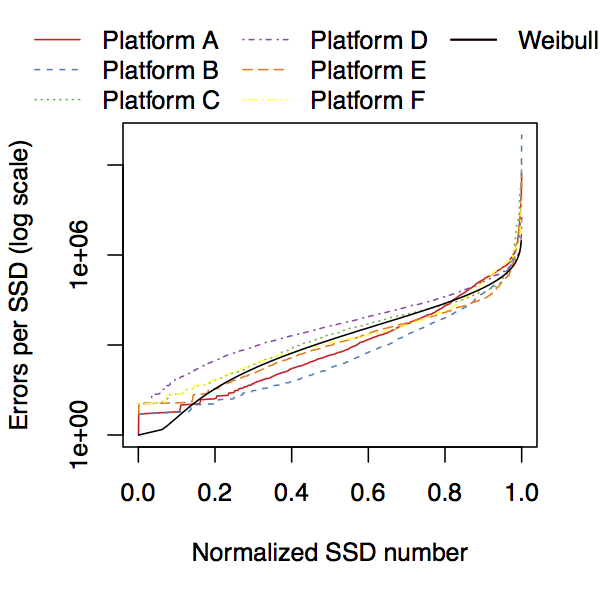}
\caption{The distribution of uncorrectable error count across SSDs.  The total number of errors per SSD skews toward a few SSDs accounting for a majority of the errors.  The solid dark line plots a Weibull distribution that resembles the error trends.}
\label{fig:distribution}
\end{figure}

Platform B has a higher average yearly rate of uncorrectable errors (978,806) versus other platforms (the second highest, Platform A, is 106,678).  This is due to a small number of SSDs having a much higher number of errors in that platform:  Figure~\ref{fig:distribution} quantifies the distribution of errors among SSDs in each platform.  The x axis is the normalized SSD number within the platform, and we order SSDs by the SSD's total number of errors.  The y axis plots the number of errors for a given SSD in log scale.  For every platform, we observe that the top 10\% of SSDs with the most errors have over 80\% of all uncorrectable errors seen for that platform.  For Platforms B, C, E, and F, the distribution has a higher skew, with 10\% of SSDs with errors making up over 95\% of all uncorrectable errors.  We also find that the distribution of number of errors among SSDs in a platform is similar to that of a Weibull distribution with a shape parameter of 0.3 and scale parameter of \num{5e3}.  The solid black line on Figure~\ref{fig:distribution} plots this distribution.


An explanation for the relatively large differences in errors per machine could be that error events are correlated.  Examining the data shows that this \emph{is} the case:  during a recent two weeks, 99.8\% of the SSDs that had an error during the first week also had an error during the second week.  We therefore conclude that an SSD that has had an error in the past is highly likely to continue to have errors in the future.  With this in mind, we next turn to understanding the correlation between errors occurring on both devices in the two-SSD systems we examine.

\subsection{Correlations Between Different SSDs}
\label{sec:corr}

Given that some of the platforms we examine have two flash-based SSDs, we would like to understanding if the likelihood of one SSD failing affects the other SSD failing.  To examine this, we compute the conditional probability of both SSDs failing given that one SSD fails.  We compute the conditional probability by dividing the number of systems in which both SSDs fail over their lifetime by the number of systems in which at least one SSD fails over its lifetime.

We denote the set of servers with two devices where the device in the lower-index PCIe slot as $S_{\mathit{lower}}$ and the set of servers with two devices where the device in the higher-index PCIe slot as $S_{\mathit{higher}}$, and we compute the conditional probability as
\begin{equation}
\Pr[\mathit{both\ devices\ fail} \mid \mathit{one\ device\ fails}] = \frac{|S_{\mathit{lower}} \cap S_{\mathit{higher}}|}{|S_{\mathit{lower}} \cup S_{\mathit{higher}}|}.
\end{equation}

We find that the conditional probability of both SSDs failing given one SSD fails is 42.2\% for Platform B, 59.9\% for Platform D, and 39.8\% for Platform F.  For comparison, if there were no correlation between SSD failures in the same machine and failures were uniformly distributed among SSDs, we would expect the conditional probability of both devices failing given that one SSD fails to be similar to the uncorrectable error incidence rate.  For example, if one SSD in a server in Platform B fails, and that failure has no influence on the other SSD in the same server failing, the probability of the other SSD failing should be the same as the first SSD failing, 27.3\%.
Instead, we find that one SSD failing in a machine \emph{does} increase the probability of the other SSD failing (by 14.9\% for Platform B, 25.8\% for Platform D, and 17.8\% for Platform F).  This suggests that an SSD platform's operational conditions contribute to the SSD failure trends we observe.

\begin{figure}
\centering
\includegraphics[width=0.4\columnwidth]{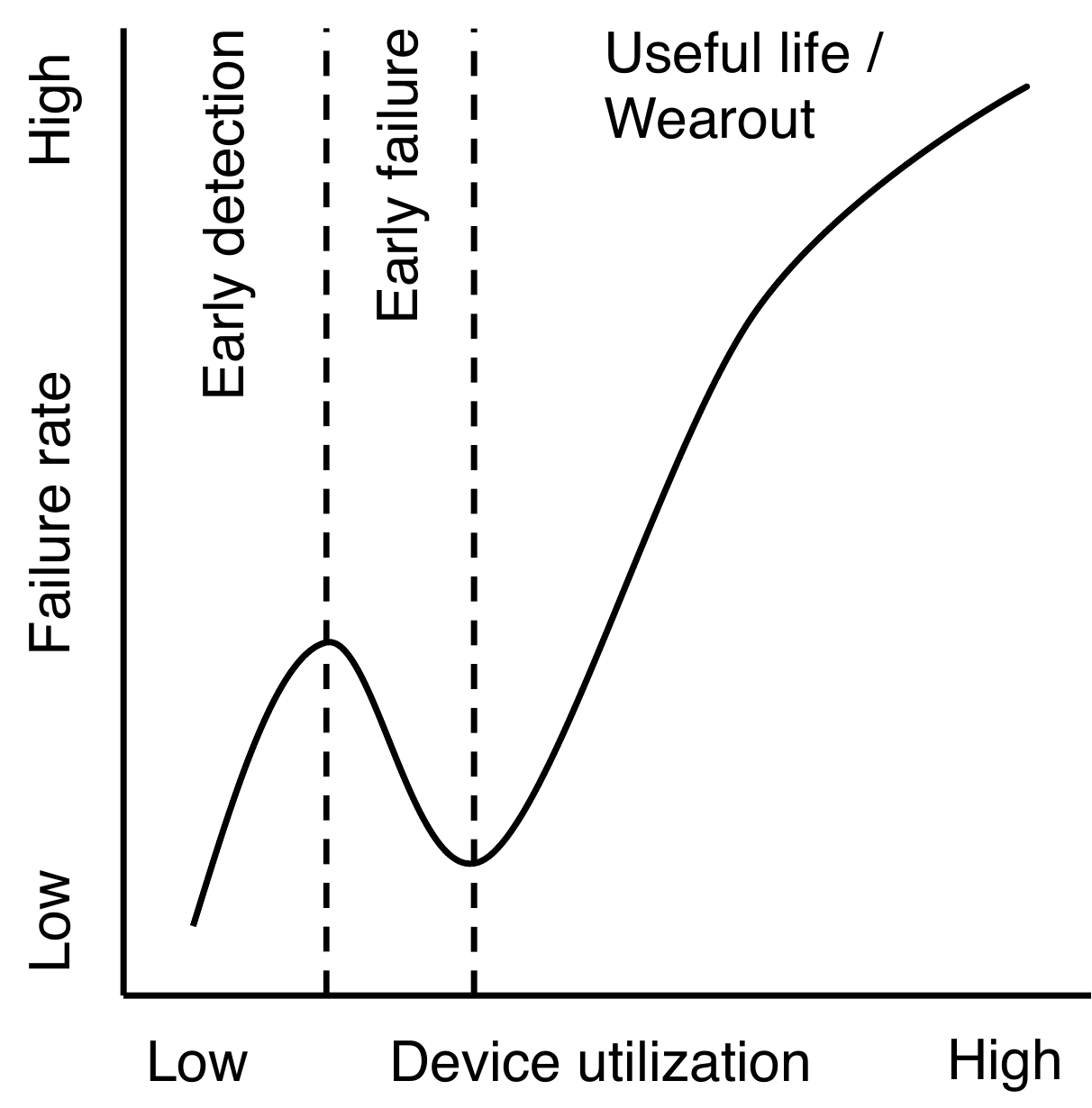}
\caption{SSDs fail at different rates during several distinct periods throughout their lifetime (which we measure by usage): early detection, early failure, useful life, and wearout.}
\label{fig:lifecycle}
\end{figure}

\begin{figure*}
\centering
\includegraphics[width=0.4\textwidth]{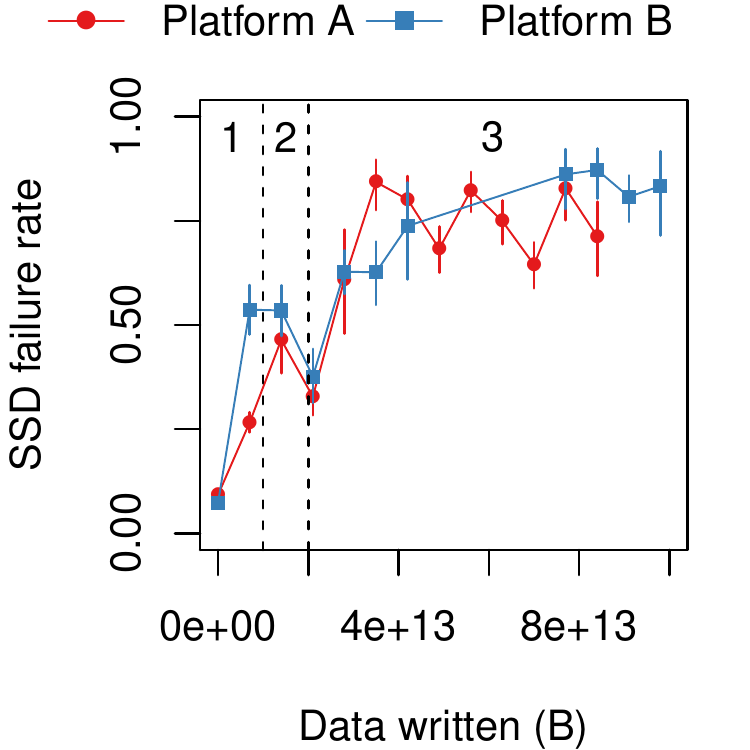}
\includegraphics[width=0.4\textwidth]{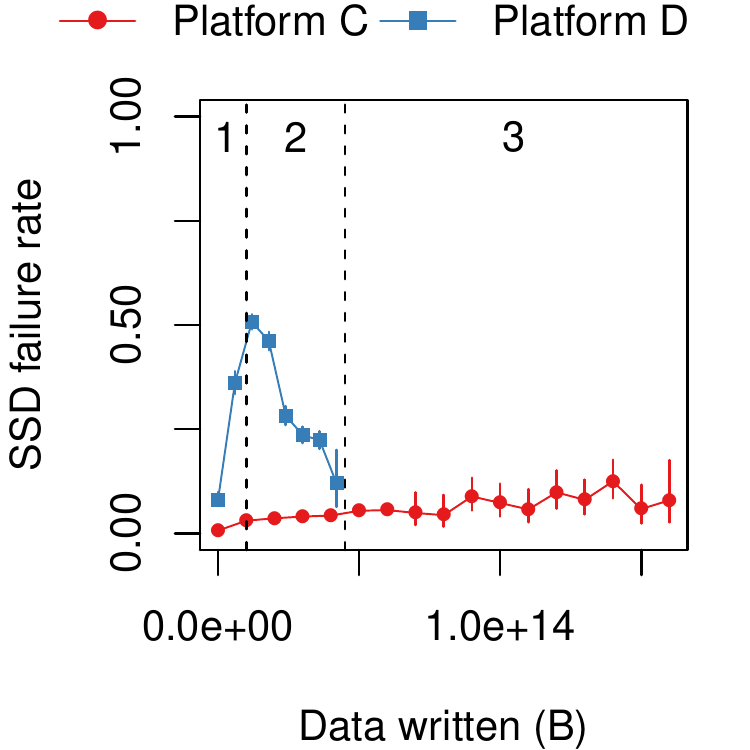}
\\
\vspace{0.5cm}
\includegraphics[width=0.4\textwidth]{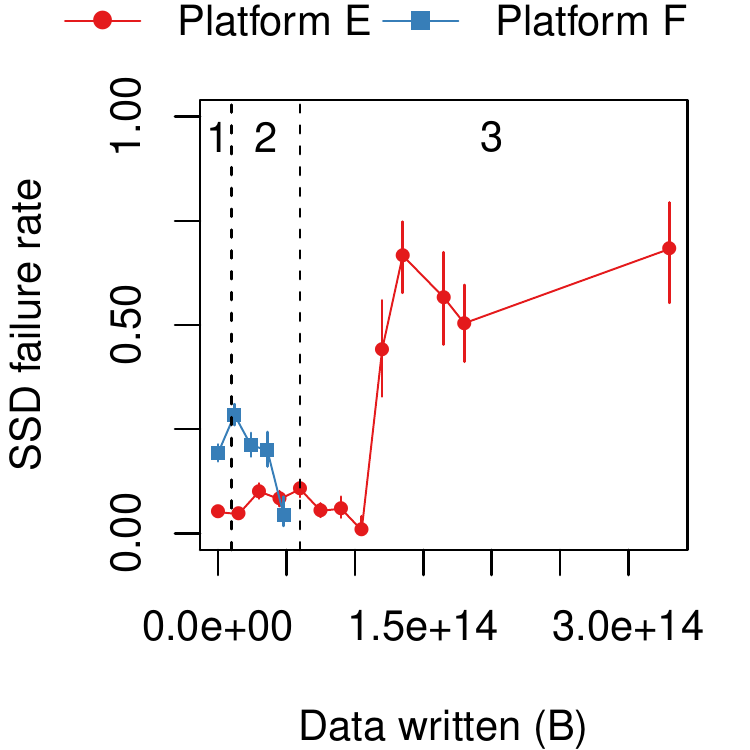}
\caption{SSD failure rate versus the amount of data written to flash cells.  SSDs go through several distinct phases throughout their life:  increasing failure rates during early detection of less reliable cells ({\sf 1}), decreasing failure rates during early cell failure and subsequent removal ({\sf 2}), and eventually increasing error rates during cell wearout ({\sf 3}).}
\label{fig:written}
\end{figure*}

These baseline statistics we examine raise questions about SSD failure in the field.  For example, \emph{``How do workload-dependent characteristics (such as the amount of data written or read) affect failure rates?},  \emph{``What role does the SSD controller play in SSD failure?}, and \emph{How do external factors (such as temperature) affect SSD failure trends?}  We seek to answer these questions by examining how a variety of \emph{internal} and \emph{external} characteristics affect SSD failure trends, in \S\ref{sec:internal} and \S\ref{sec:external}.

\label{sec:internal}
We examine next how internal factors of the flash chips and SSD controller affect uncorrectable errors over an SSD's lifetime.  We examine the effects of writing data to flash cells, reading data from flash cells, copying and erasing data, discarding unusable blocks, and internal DRAM buffer utilization.

\subsection{Data Written to Flash Cells}
\label{sec:written}

Recall from \S\ref{subsec:SSD Device Architecture}, that due to the physical properties that govern them, flash cells have been shown to become less reliable the more times their contents are programmed and erased, i.e., P/E cycles the flash cell has endures (we refer the reader to~\cite{cai-date12} which provides a good summary of flash cell operation and characteristics).  Several recent works have quantified the effects of P/E cycles on various error mechanisms in small sets of recent flash chips (e.g.,~\cite{cai-date12, cai-date13, cai-dsn15, cai-hpca15, cai-iccd13, cai-itj13, cai-iccd12, cai-sigmetrics14, grupp-fast12}).  The higher capacity MLC chips commonly in flash-based SSDs may exacerbate the effects of P/E cycles on flash reliability, so it is important for us to understand how P/E cycles affect flash reliability.

In order to examine the effect of P/E cycles on flash reliability, we consider the amount of data \emph{written directly to} \emph{flash cells} over each SSD's lifetime.  Our framework allows us to measure this value, which more accurately portrays SSD utilization versus software-level writes.  Note that software-level writes (i.e., write requests sent to the storage device) are not directly written to the flash cells due to layers of buffering in the storage software stack.

Prior reliability studies on hard disk drives in the field (e.g.,~\cite{schroeder-fast07}) observed a trend with respect to writes called the ``bathtub curve''.  The bathtub curve gets its name from its shape with respect to device failure rate over time: devices initially experience a high failure rate during an early failure period, then devices experience low rate of failure during the useful life period, and ultimately a high failure rate once again during their wearout period.  In the SSDs we examine, we notice an additional period before the early failure period that we call the \emph{early detection period}.  During the early detection period, which occurs when SSDs are young (in terms of how much use they have), the SSD controller identifies unreliable cells, which results in an initially high failure rate among devices.  Figure~\ref{fig:lifecycle} pictorially illustrates the lifecycle failure pattern that we observe, which we quantify in Figure~\ref{fig:written}.

Figure~\ref{fig:written} plots how the failure rate of SSDs varies with the amount of data written to the flash cells.  We group the platforms by the individual capacity of their SSDs.  Notice that across most platforms, the failure rate is low when little data is written to flash cells, then increases (corresponding to the early detection period, which we label by the region {\sf 1} in the figures), then the failure rate decreases (corresponding to the early failure period, which we label by the region {\sf 2} in the figures).  Finally, the error rate generally increases for the remainder of the SSD's lifetime (corresponding to the useful life and wearout periods, which we label by the region {\sf 3} in the figures).  An obvious outlier for this trend is Platform C---in \S\ref{sec:external}, we observe that some \emph{external} characteristics of this platform leads to its atypical failure rates.

Note that different platforms are in different stages in their lifecycle depending on the amount of data written to flash cells.  For example, SSDs in Platforms D and F, which have the least amount of data written to flash cells on average, are mainly in the early detection or early failure periods.  On the other hand, SSDs in Platforms A and B, which have more data written to flash cells, span all stages of the lifecycle failure pattern (which we show in Figure~\ref{fig:lifecycle}).  In the case of SSDs in Platform A, we observe up to an 81.7\% difference between the failure rates of SSDs in the early detection and wearout periods of the lifecycle.

\begin{figure*}
\centering
\includegraphics[width=0.4\textwidth]{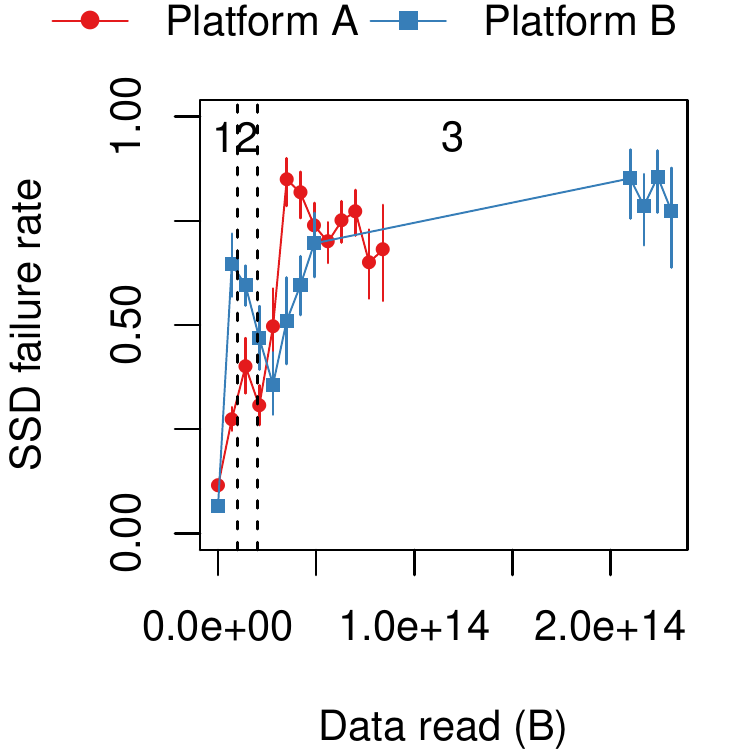}
\includegraphics[width=0.4\textwidth]{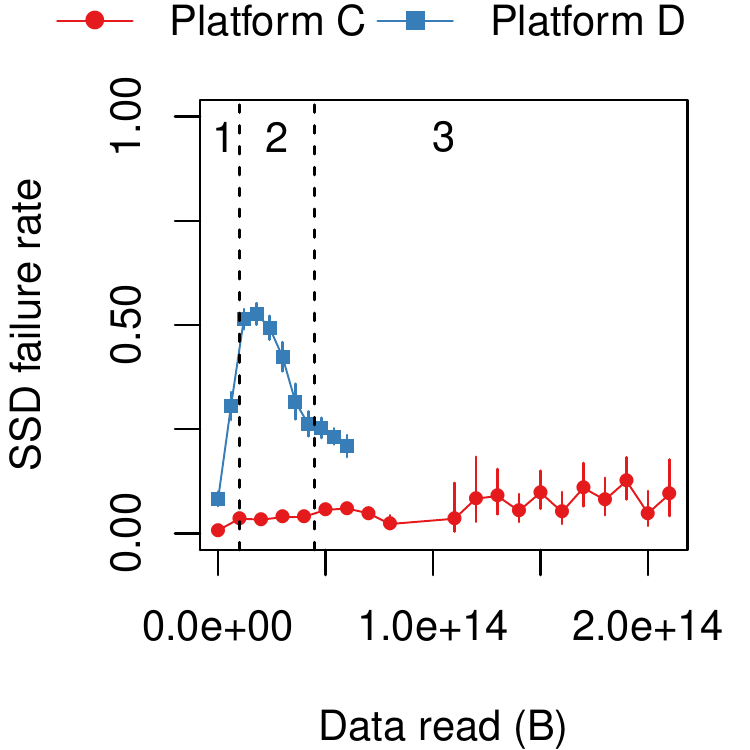}
\\
\vspace{0.5cm}
\includegraphics[width=0.4\textwidth]{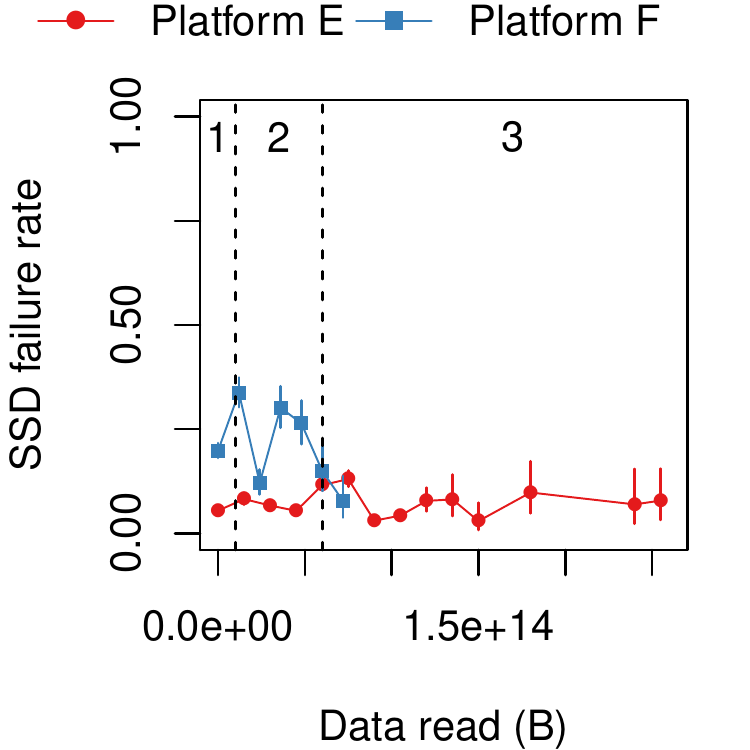}
\caption{SSD failure rate versus the amount of data read from flash cells.  SSDs in Platform E, that have over twice as many reads from flash cells as writes to flash cells, do not show failures dependent on the amount of data read from flash cells.}
\label{fig:read}
\end{figure*}

\begin{figure*}
\centering
\includegraphics[width=6in, trim={0 0 4.95in 0}, clip]{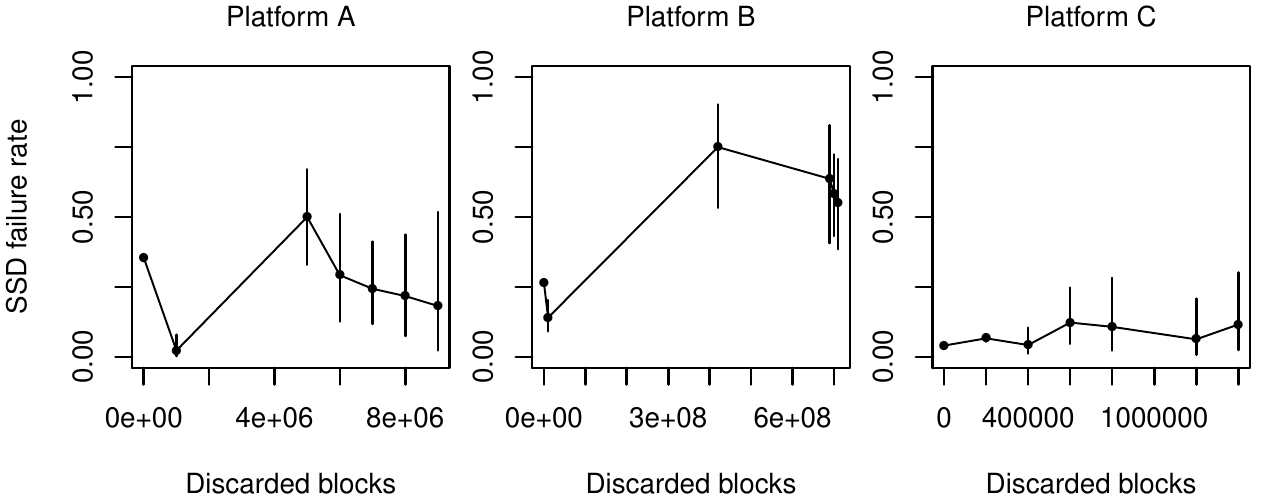}
\vspace{0.5cm}

\hspace{1cm}\includegraphics[width=6in, trim={5.05in 0 0 0}, clip]{fig/pdf/discarded}
\caption{SSD failure rate versus the number of discarded blocks.  We find that (1) failure rate is relatively high among SSDs that have discarded few blocks (far left of plots), (2) some SSDs seem to effectively mask failures by discarding blocks (initial decrease in failure rates), and (3) discarding a large amount of blocks indicates higher failure rates (toward the right of plots).}
\label{fig:discarded}
\end{figure*}


As we explain and depict in Figure~\ref{fig:lifecycle}, the lifecycle failure rates we observe with the amount of data written to flash cells does \emph{not} follow the conventional bathtub curve.  In particular, the new \emph{early detection period} we observe across the large number of devices leads us to investigate why this ``early detection period'' behavior exists.  Recall that the early detection period refers to failure rate increasing early in lifetime (i.e., when a small amount of data is written to the SSD).  After the early detection period, failure rate starts decreasing.

We hypothesize that this non-monotonic error rate behavior during and after the early detection period is due to a \emph{two-pool} model of flash blocks: one pool of blocks, that we call the \emph{weaker pool}, consists of cells whose error rate increases much faster than the other pool of blocks, that we call the \emph{stronger pool}.  The weaker pool quickly generates uncorrectable errors (leading to increasing failure rates we observe in the early detection period as these blocks keep failing).  The cells comprising this pool ultimately fail and are taken out of use early by the SSD controller.  As the SSD exhausts blocks in the weaker pool, the overall error rate starts decreasing (as we observe \emph{after} the end of what we call the \emph{early detection period}) and it continues to decrease until the more durable blocks in the stronger pool start to wear out due to typical use.


We notice a general increase in the duration of the lifecycle periods (in terms of data written) for SSDs with larger capacities.  For example, while the early detection period ends after around \unit[3]{TB} of data written for \unit[720]{GB} and \unit{1.2}{TB} SSDs, it is roughly \unit[10]{TB} for \unit[3.2]{TB} SSDs.  Similarly, the early failure period ends after around \unit[15]{TB} of data written for \unit[720]{GB} SSDs, \unit[28]{TB} for \unit[1.2]{TB} SSDs, and \unit[75]{TB} for \unit[3.2]{TB} SSDs.  This is likely due to higher capacity SSDs being able to reduce wear across a larger number of flash cells.


\subsubsection{Lifecycle Opportunities for Future Study}

While we establish circumstantial evidence of a lifecycle trend resembling the bathtub curve for flash-based SSDs, we also note that there could be other interpretations of the lifecycle data that we collect that could be useful:

\begin{itemize}
	\item \emph{\textbf{A completely new lifecycle process.}} While we use the bathtub curve as an analogy to explain the trends we observe for flash-based SSDs in the field, we may, in fact, observe a \emph{new} lifecycle process. In this lifecycle, flash-based SSDs may spend the majority of their operational lifetime in the wearout period and spend little time in the usable period. It may also imply that wearout does not happen as a spike at the end of an SSD's lifetime (as with the mechanical failure of physical components in hard disk drives, like the mechanical head), but instead becomes a normal mode of operation.
	\item \emph{\textbf{Workload-dependent effects on aging.}} Most of the systems we examine only have written data amounting to 10's of times the capacity of the SSD they use. Part of the reason for this is that the SSDs we examine make up much larger distributed systems which may use more SSDs than absolutely necessary to provide higher throughput or lower latency. Using more SSDs may also reduce the overall amount of data written to the SSDs and reduce the effects of wearout on the systems we analyze. We believe that field studies of devices that have written 100's or more times the capacity of the SSDs used will provide additional insights into SSD lifecycle trends.
\end{itemize}

We await evidence from future studies to shed light on the underlying aging process that flash-based SSDs follow in the field.

\subsection{Data Read from Flash Cells}
\label{sec:read}

Similar to writes, our framework allows us to measure the amount of data directly read from flash cells over each SSD's lifetime. We would like to understand how prevalent this effect is across the SSDs we examine.

Figure~\ref{fig:read} plots how failure rate of SSDs varies with the amount of data read from flash cells.  For most platforms (i.e., A, B, C, and F), the failure rate trends we see in Figure~\ref{fig:read} are very similar to those we observe in Figure~\ref{fig:written}.  We find this similarity when SSDs in a platform have written more data to flash cells than they have read from the flash cells. (Platforms A, B, C, and F show this behavior.)

In Platform D, where more data is read from flash cells than written to flash cells (\unit[30.6]{TB} versus \unit[18.9]{TB} on average), we notice error trends during the early detection and early failure periods.  Unfortunately, even though devices in Platform D are prime candidates for observing the occurrence of read disturbance errors (because more data is read from flash cells than written to flash cells), the effects of early detection and early failure appear to dominate the types of errors we observe on this platform.

In Platform E, however, the SSDs read twice as much data as the SSDs write (\unit[51.1]{TB} versus \unit[23.9]{TB} on average).  In addition, unlike Platform D, devices in Platform E display a diverse amount of utilization. Under these conditions, we \emph{do not} observe a statistically significant difference in the failure rate between SSDs that read the most amount of data versus those that read the least amount of data.  This suggests that the effect of reads causing errors in the SSDs we examine is not predominant versus other effects such as writes causing errors.  This corroborates prior flash cell analysis that showed that most errors occur due to retention effects and not read disturbance~\cite{mielke-irps08, cai-iccd12, cai-itj13}.



\subsection{Block Erases}
\label{subsec:Block Erases}

Recall that before new data can be written to a page in flash, the flash controller needs to erase an entire block.  (A block is around 128 $\times$ \unit[8]{KB} pages.)  Each erase wears out the block as shown in previous works~\cite{mielke-irps08, cai-iccd12, cai-itj13}. Our infrastructure tracks the average number of blocks that the SSD controller erases when it performs \emph{garbage collection}.  Recall from \S\ref{subsec:SSD Device Architecture} that garbage collection occurs in the background and compacts data that is in use to ensure that an SSD has enough available space to write new data.

While we do not have statistics on how frequently the SSD performs garbage collection, we examine the number of erases \emph{per garbage collection period} metric as an indicator of the average amount of erasing that occurs within each SSD.


We examine the relationship between the number of erases the SSD perform versus the failure rate (which we do not plot).  Across the SSDs we examine, we find that the trends in terms of erases correlate with the trends due to data written.  This behavior reflects the typical operation of garbage collection in SSD controllers:  As more data is written, the SSD must level wear across the flash cells within an SSD, requiring the SSD to erase more blocks and copy more pages to different locations (an effect we analyze next).

\begin{figure*}
\centering
\includegraphics[width=0.4\textwidth]{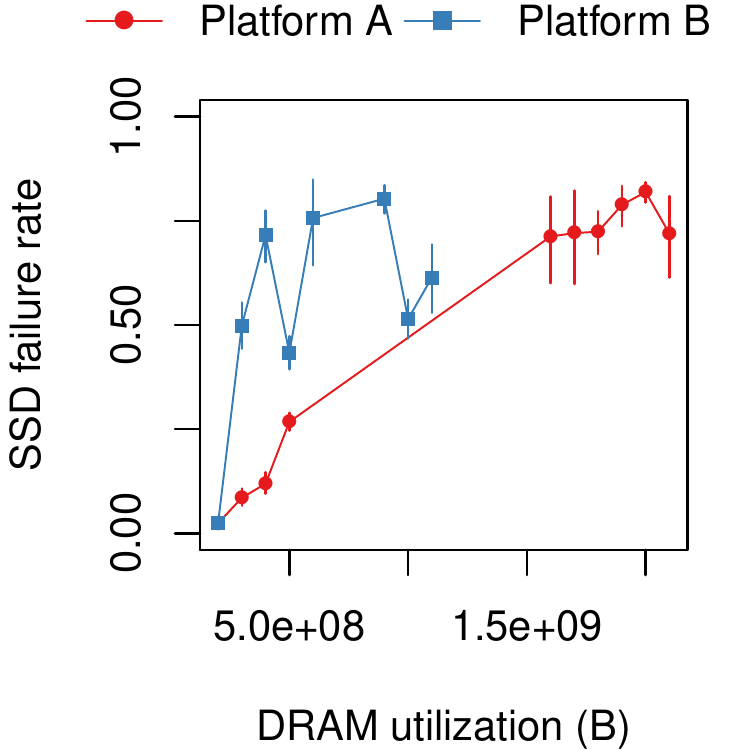}
\includegraphics[width=0.4\textwidth]{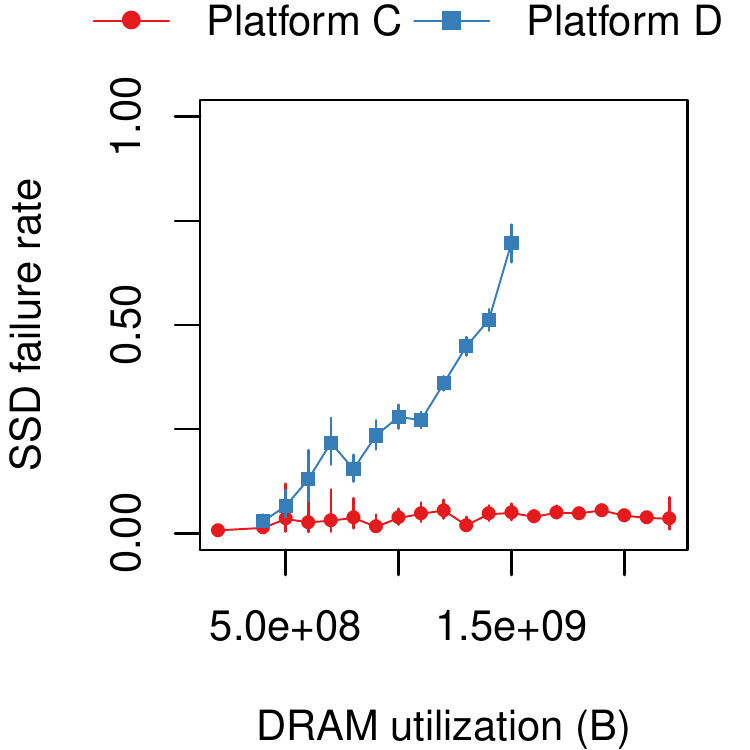}
\\
\vspace{0.5cm}
\includegraphics[width=0.4\textwidth]{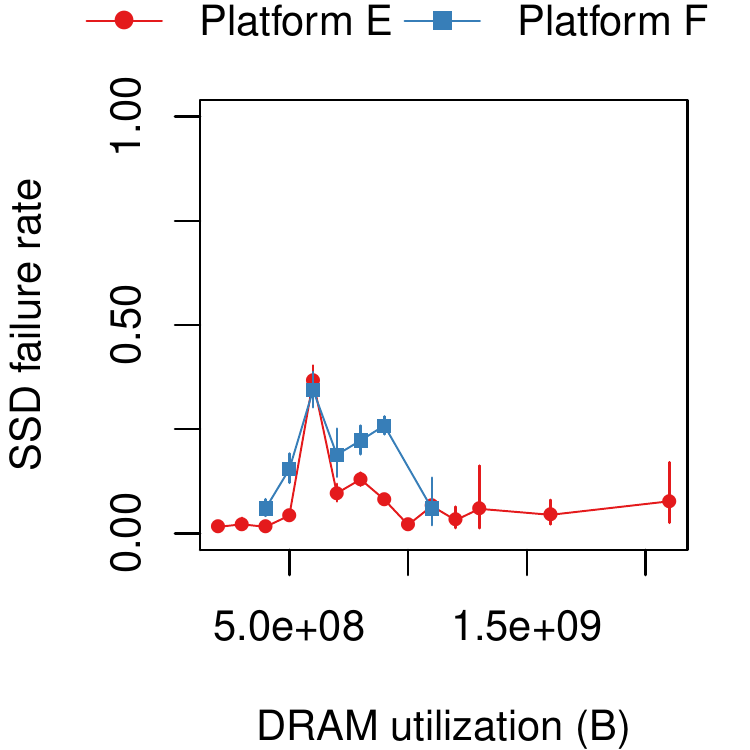}
\caption{SSD failure rate versus DRAM buffer utilization.  Sparse data mappings (e.g., non-contiguous data, that has a high DRAM utilization to store flash translation layer metadata) affect SSD reliability the most (Platforms A, B, and D).  Additionally, some dense data mappings (e.g., contiguous data in Platforms E and F) also negatively affect SSD reliability, likely due to the effect of small, sparse writes.}
\label{fig:ram}
\end{figure*}

\begin{figure*}
\centering
\includegraphics[width=6in, trim={0 0 4.95in 0}, clip]{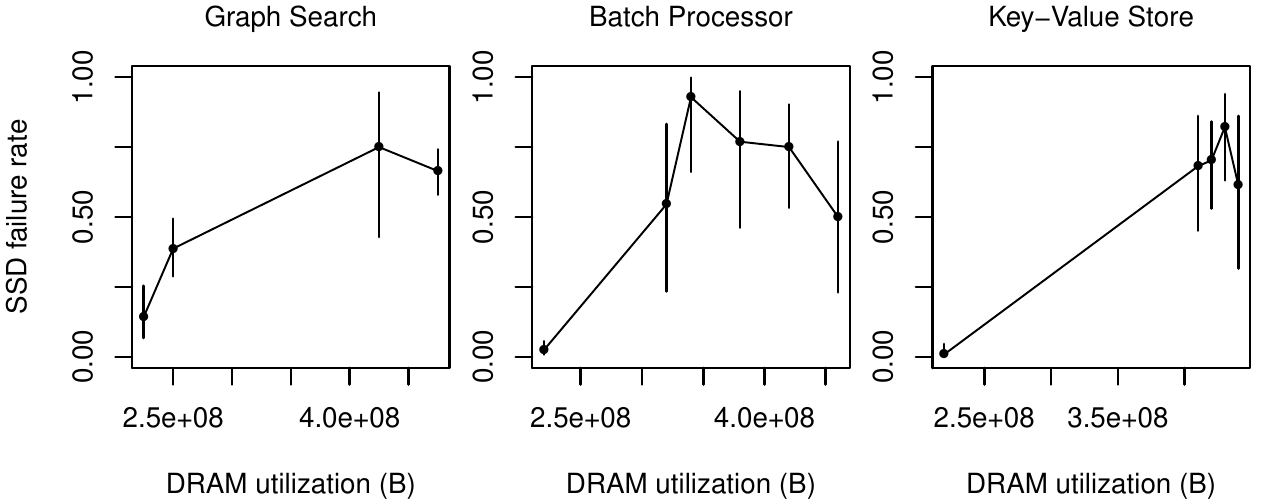}
\vspace{0.5cm}

	\hspace{1cm}\includegraphics[width=6in, trim={5.05in 0 0 0}, clip]{fig/pdf/ram-app_B}
\caption{SSD failure rate versus DRAM utilization across six applications that run on Platform B.  We observe similar DRAM buffer effects even among SSDs running the same application.}
\label{fig:ram-app_B}
\end{figure*}

\subsection{Page Copies}

Recall from \S\ref{subsec:SSD Device Architecture} that as data is written to flash-based SSDs, the SSD copies pages (around \unit[8]{KB} in size) during garbage collection in order to erase unused blocks and more evenly level wear across the flash chips.  Our infrastructure measures the number of pages the SSD copies across an SSD's lifetime.


We examine the relationship between the number of pages an SSD copies versus the SSD failure rate (which we do not plot).  Since the SSD uses the page copying process to free up space to write data and also to balance the wear due to writes, the amount of data written to an SSD dictates its operation.  Accordingly, we observe similar SSD failure rate trends with the number of pages the SSD copies as we observe with the amount of data written to flash cells.


\subsection{Discarded Blocks}

Recall that the SSD controller discards a block when the SSD deems the block unreliable for use.  Discarding blocks affects the usable lifetime of a flash-based SSD by reducing the amount of over-provisioned capacity of the SSD.  At the same time, discarding blocks has the potential to reduce the amount of SSD errors, by preventing the SD from accessing unreliable cells.  Understanding the reliability effects of discarding blocks is important, as this process is the main defense that SSD controllers have against cells that fail and a major impediment for SSD lifetime.

Figure~\ref{fig:discarded} plots the SSD failure rate versus the number of blocks an SSD discards over its lifetime.  For SSDs in most platforms, we observe an initially decreasing trend in SSD failure rate with respect to the number of blocks the SSD discards and then an increasing and finally decreasing failure rate trend.

We attribute the initially high SSD failure rate when the SSD discards few blocks to the two-pool model of flash block failure, that we discuss in \S\ref{sec:written}.  In this case, the weaker pool of flash blocks (corresponding to weak blocks that fail early) cause errors before the SSD controller can discard them, causing an initially high SSD failure rate.  On the other hand, the stronger pool of flash blocks (which fail due to gradual wearout over time) contribute to the subsequent decrease in SSD failure rate.  Some SSDs that discard a large number of blocks have high error rates (e.g., Platforms A, B, and E), indicating that discarding a large number of blocks is a good indicator of the likelihood of SSD failure.

We further examine the devices with the largest number of block discards and found that the number of blocks that the SSD discards does \emph{not} correlate with the amount of data written to or read from flash cells (in contrast with \S\ref{sec:written} and~\ref{sec:read}).  In other words, we observe SSDs of both low \emph{and} high utilization across their lifetime (which we measure by flash cell reads and writes) that discard large amounts of blocks and have high failure rates.  This suggests that the relationship between block discards and failure rate is to some extent intrinsic to the SSD.  Thus, some SSDs, despite discarding many blocks, continue to encounter uncorrectable errors at a much higher rate than their peers.

\begin{figure*}
\centering
\includegraphics[width=0.4\textwidth]{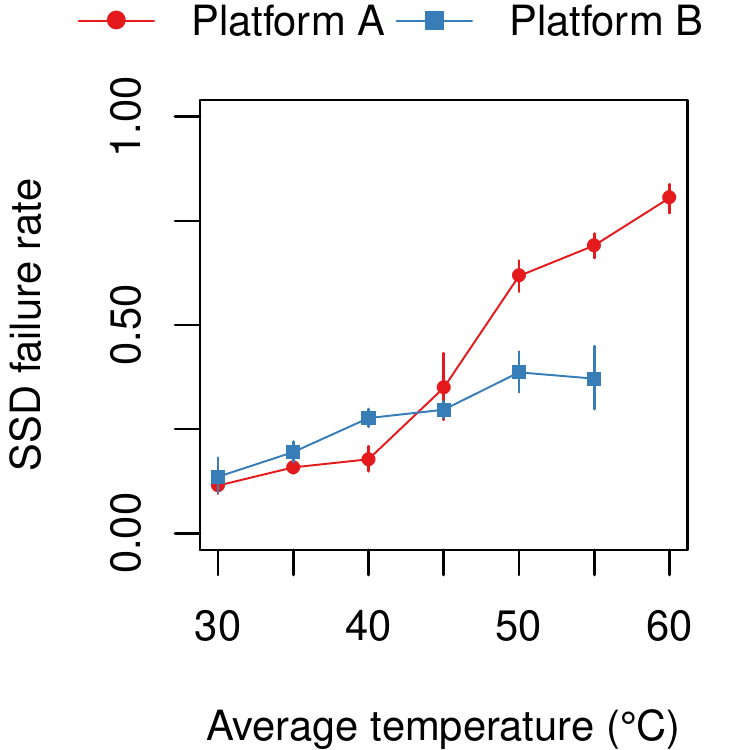}
\includegraphics[width=0.4\textwidth]{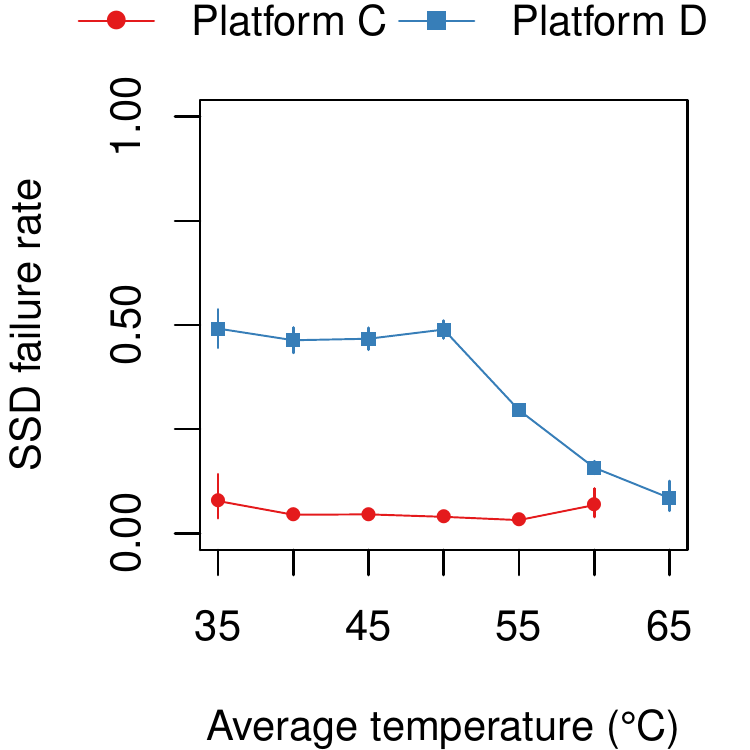}
\\
\vspace{0.5cm}
\includegraphics[width=0.4\textwidth]{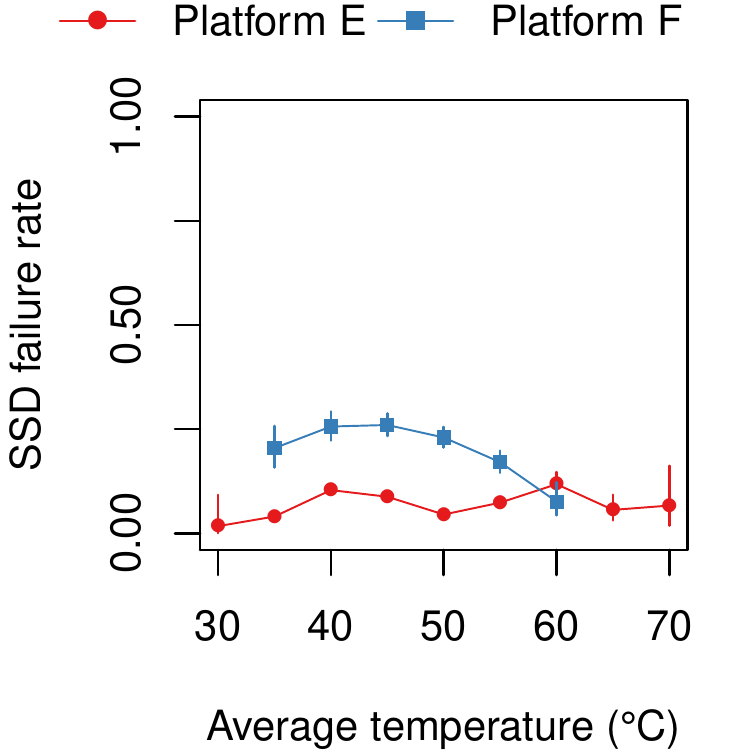}
\caption{SSD failure rate versus temperature.  Operational temperatures of 30 to 40\degree C generally show increasing failure rates.  Failure trends at and above 45\degree C follow three distinct failure rate patterns: increasing, not sensitive, and decreasing.}
\label{fig:avg-temp}
\end{figure*}

\subsection{DRAM Buffer Usage}
\label{subsec:DRAM Buffer Usage}

Recall that flash-based SSDs use DRAM to provide buffer space for SSD controller metadata or for data to write to the flash chips.  The SSDs we examine use DRAM buffer space to store metadata for the flash translation layer mapping for logical addresses to physical addresses.  This allows the SSD controller to locate data on an SSD quickly, reducing the performance impact of address translation.

The SSDs we examine utilize the DRAM buffer \emph{less} when data is \emph{densely allocated} (e.g., contiguous data) and utilize the DRAM buffer \emph{more} when data is \emph{sparsely allocated} (e.g., non-contiguous data).  As an illustrative example, using an SSD to read from a \emph{large} file would lead to a contiguous allocation of data on the SSD and a \emph{small} DRAM buffer utilization.  On the other hand, using an SSD to read from and write to many \emph{small} files would lead to many non-contiguous allocations of data on the SSD and a \emph{large} DRAM buffer utilization.

Examining DRAM buffer utilization can therefore provide an indication of how system data allocation behavior affects flash reliability.  To capture the DRAM buffer usage of SSDs, we examine the average amount of DRAM buffer space the SSD uses over two recent weeks of SSD operation, and we sample the data at one hour intervals.

Figure~\ref{fig:ram} plots the failure rate for SSDs that use different amounts of DRAM buffer space on average.  For some platforms (A, B, and D), we observe that as the SSD uses more DRAM buffer space, SSD failure rate increases.  Such a trend indicates that some systems that allocate data more sparsely have higher failure rates.  Such behavior is potentially due to the fact that sparse data allocation can correspond to access patterns that write small amounts of non-contiguous data, causing the SSD controller to more frequently erase and copy data versus writing contiguous data.


While Platform C does not appear to be sensitive to DRAM buffer utilization, Platforms E and F demonstrate a trend of higher error rates at lower DRAM buffer utilizations.  We attribute a separate (but similarly adversarial) pattern of write behavior to these SSDs.  In these platforms, we believe that the SSDs allocate large, contiguous regions of data (resulting in low DRAM buffer utilization) but write to them sparsely but intensely (leading to cell wearout).  This could occur, for example, when frequently updating small fields in a large, contiguously-allocated data structure.

Interestingly, we observe similar behavior at the application level (Figure~\ref{fig:ram-app_B}).  We examine six of the largest distributions of applications on Platform B:  Graph Search is a distributed graph search service; Batch Processor executes long-running asynchronous jobs; Key--Value Store stores persistent mappings of keys to values; Load Balancer is a programmable traffic routing framework; Distributed Key--Value Store is like a Key--Value Store with stronger reliability, availability, and consistency guarantees; and Flash Cache is a cache for large working set, low access frequency data~\cite{gartrell-13}.   Applications across SSDs have increasing failure rates with increasing DRAM buffer usage (sparse data mappings) and in some cases have increases in failure rates at lower DRAM utilizations (dense data mappings, e.g., Batch Processor and Flash Cache).

We conclude that small, sparse writes affect SSD failure rates the most for sparse data mappings but are also noticeable for dense data mappings.  Given this behavior, we believe that there is the potential for developing more effective write coalescing techniques to handle the adversarial access pattern of small, sparse writes.

\begin{figure*}
\centering
\includegraphics[width=6in, trim={0 0 4.95in 0}, clip]{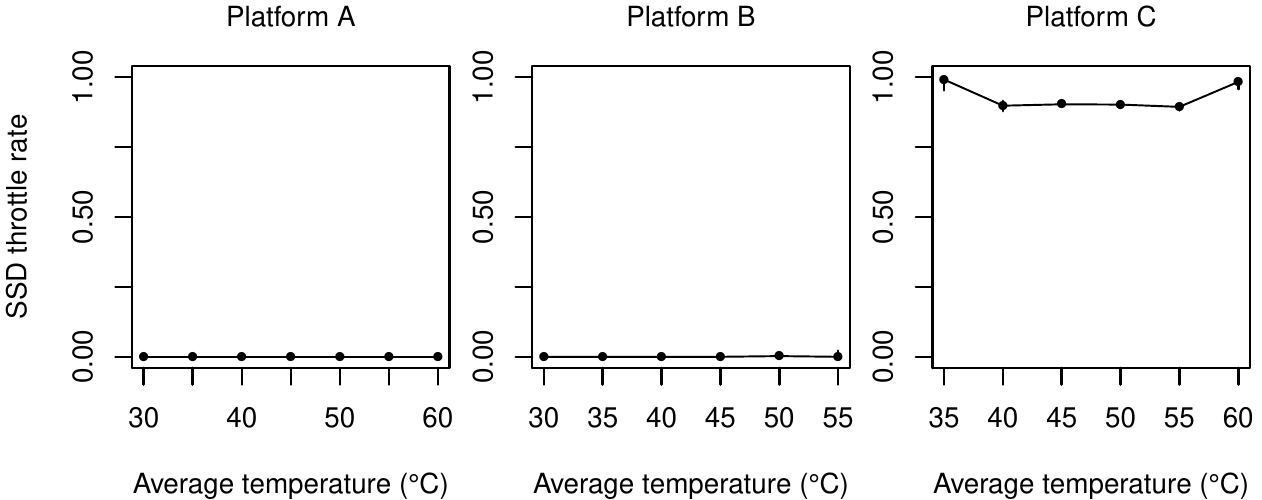}
\vspace{0.5cm}

	\hspace{1cm}\includegraphics[width=6in, trim={5.05in 0 0 0}, clip]{fig/pdf/avg-temp-throttles}
\caption{Fraction of throttled SSDs versus SSD temperature.  While SSDs in some platforms are never throttled (A and B), others are throttled more aggressively (C and E).}
\label{fig:avg-temp-throttles}
\end{figure*}

\begin{figure*}
\centering
\includegraphics[width=0.4\textwidth]{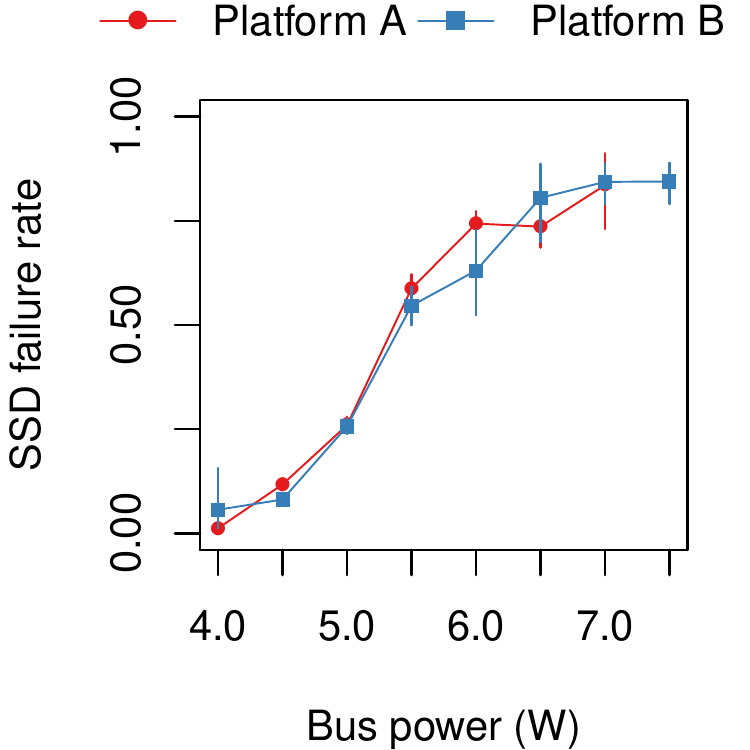}
\includegraphics[width=0.4\textwidth]{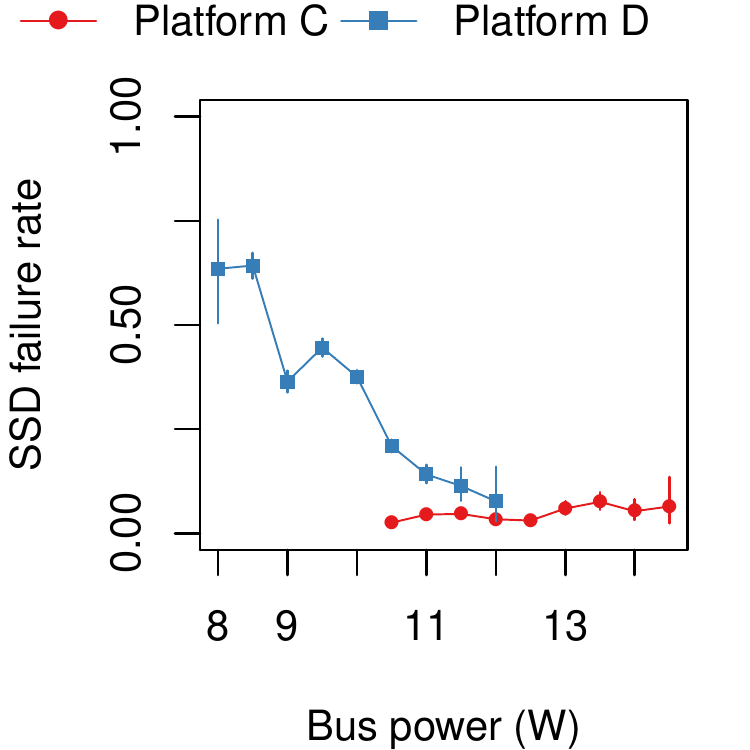}
\\
\vspace{0.5cm}
\includegraphics[width=0.4\textwidth]{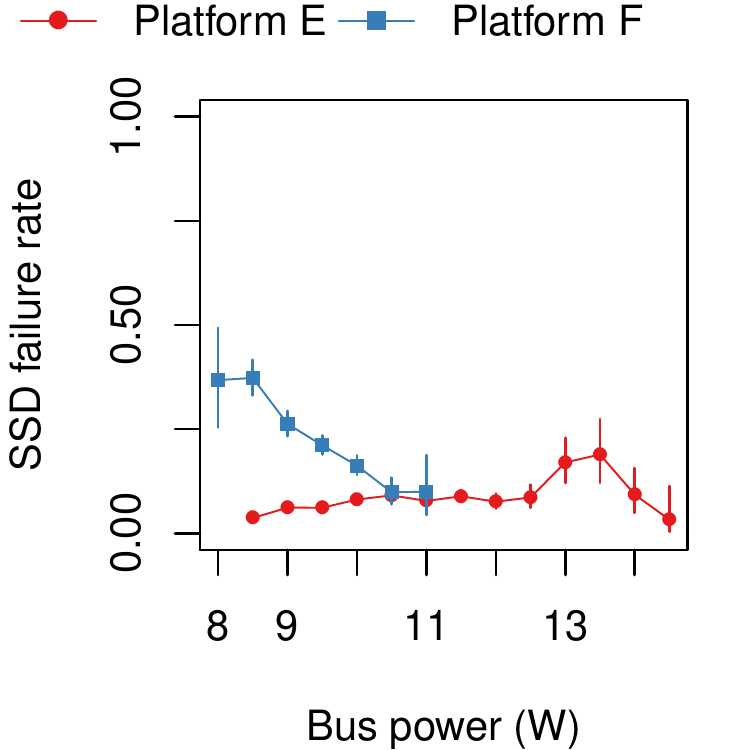}
\caption{SSD failure rate versus bus power consumption.  PCIe v2 SSDs (Platforms C through F) consume more bus power than PCIe v1 SSDs (Platforms A and B) due to their higher frequency.  Bus power consumption range (1) spans up to around 2$\times$ for SSDs within a platform (e.g., Platform B) and (2) is typically different from the nominal PCIe bus power of \unit[10]{W}.}
\label{fig:bus-power}
\end{figure*}

\subsection{Temperature}
\label{sec:external}

We next examine how external factors influence the errors we observe over an SSD's lifetime.  We examine the effects of temperature, PCIe bus power, and system-level writes the OS reports.

Recall from \S\ref{subsubsec:SSD Temperature-Dependent Failures}, that higher temperature negatively affects the operation of flash-based SSDs.  In flash \emph{cells}, higher temperatures have been shown to cause cells to age more quickly due to the temperature-activated Arrhenius effect~\cite{xu-apl03} (for more information, see the references in \S\ref{subsec:How SSD Devices Fail}).  Temperature-dependent effects are especially important to understand for flash-based SSDs in order to make adequate data center provisioning and cooling decisions.  To examine the effects of temperature, we use temperature measurements from temperature sensors embedded on the SSD cards, which provide a more accurate portrayal of the temperature of flash cells than temperature sensors at the server or rack level.

Figure~\ref{fig:avg-temp} plots the failure rate for SSDs that have various average operating temperatures.  We find that at operating temperatures of 30 to 40\degree C, SSDs across server platforms see a similar or slight increase in failure rates as temperature increases.



Outside of this range (at temperatures of 45\degree C and up), we find that SSDs fall into one of three categories with respect to their reliability trends with temperature:  (1) temperature-sensitive with increasing failure rate (Platforms A and B), (2) less temperature-sensitive (Platforms C and E), and (3) temperature-sensitive with decreasing failure rate (Platforms D and F).  There are two factors that may confound the trends we observe with respect to SSD temperature.

One potentially confounding factor when analyzing the effects of temperature is the operation of the SSD controller in response to changes in temperature.  The SSD controllers in some of the SSDs we examine attempt to ensure that SSDs do not exceed certain temperature thresholds (starting around 80\degree C).  Similar to the techniques a processor employs to reduce the amount of processor activity in order to keep the processor within a certain range of temperatures, our SSDs attempt to change their behavior (e.g., reduce the frequency of SSD access or, in the most extreme case, shut down the SSD) in order not to exceed temperature thresholds.

The thermal characteristics of servers in each platform also confound analysis.  Two SSDs in a machine (in Platforms B, D, and F) versus one SSD in a machine both increases the thermal capacity of the machine (causing its SSDs to reach higher temperatures more quickly and increase the work to cool the SSDs) and reduces airflow to the components, prolonging the effects of high temperatures when they occur.

One hypothesis is that temperature-sensitive SSDs with increasing error rates, such as Platforms A and B, may not employ as aggressive temperature reduction techniques as other platforms.  While we cannot directly measure the actions the SSD controllers take in response to temperature events, we examine an event that can correlate with temperature reduction: whether or not an SSD throttles its operation in order to reduce its power consumption.  Performing a large number of writes to an SSD consumes power and increases the temperature of an SSD.  Figure~\ref{fig:avg-temp-throttles} plots, for each temperature, the fraction of machines that have ever been throttled.  Examining this figure confirms that Platforms A and B, where no machines or few machines have been throttled, exhibit behavior that is typical for SSDs without much preventative action against temperature increases.  In these platforms, as temperature increases, failure rate of SSDs increases.

In contrast to Platforms A and B, Platforms C and E, which are less temperature-sensitive, throttle their SSDs more aggressively across a range of temperatures.  From Figure~\ref{fig:avg-temp} we can see that throttled SSDs have lower failure rates (in Platforms C and E) versus SSDs that are throttled less or not throttled at all (Platforms A and B).  We attribute the relatively low failure rate for Platforms C and E we observe in our measurements to the very aggressive throttling that occurs for SSDs in Platforms C and E.  Such throttling could potentially reduce performance, though we are not able to examine its impact.

SSDs in Platforms D and F employ a relatively low amount of throttling (Figure~\ref{fig:avg-temp-throttles}), but exhibit the counter-intuitive trend of decreasing failure rate with higher temperature.  Recall from \S\ref{sec:written} that these SSDs are predominantly in their early detection and early failure periods and so the failure rates for most SSDs in these platforms are relatively high versus their peers in Platforms C and E.  It is likely that a combination of power throttling and some other form of temperature-dependent throttling the SSD controller uses that we are not able to measure is responsible for reducing the failure rate among the SSDs in Platforms D and F as temperature increases.


\begin{figure*}
\centering
\includegraphics[width=0.4\textwidth]{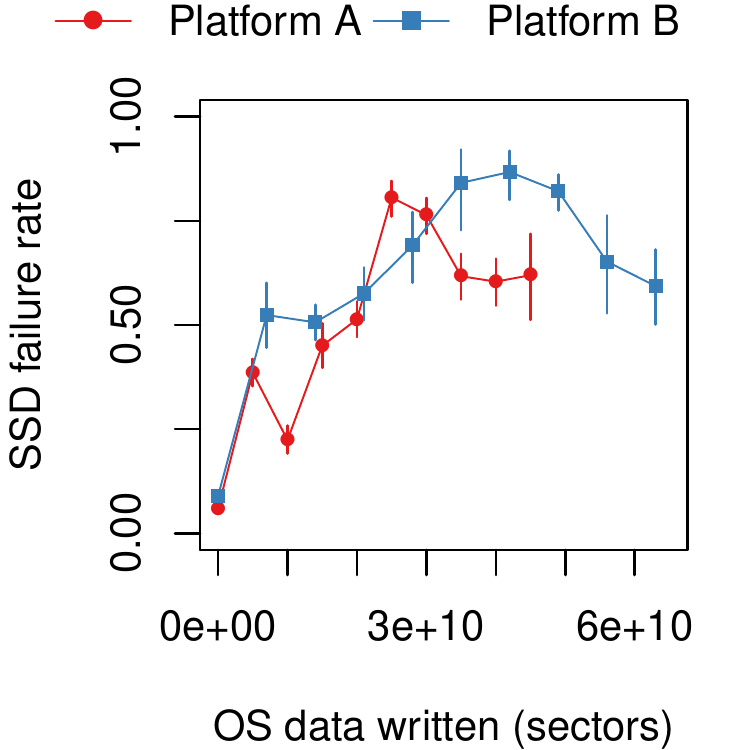}
\includegraphics[width=0.4\textwidth]{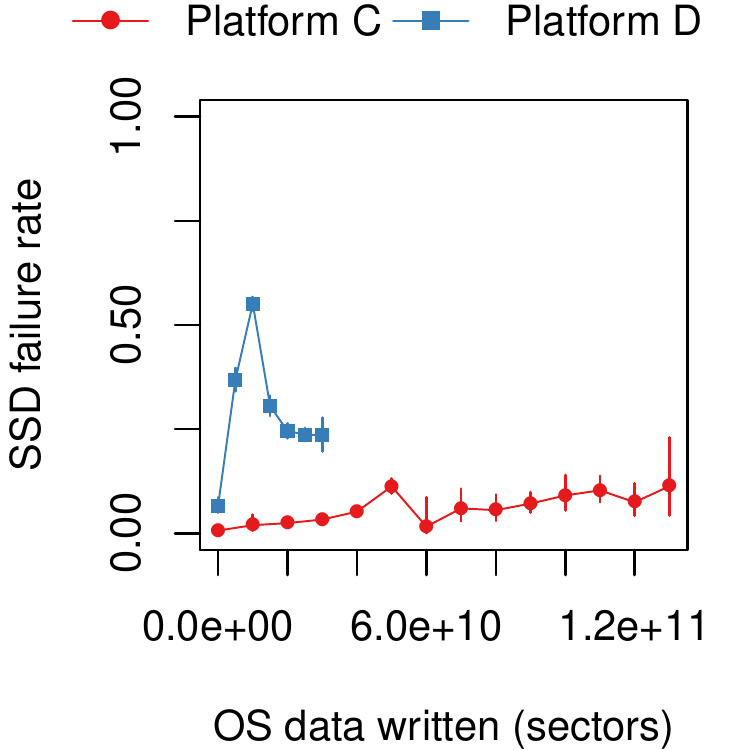}
\\
\vspace{0.5cm}
\includegraphics[width=0.4\textwidth]{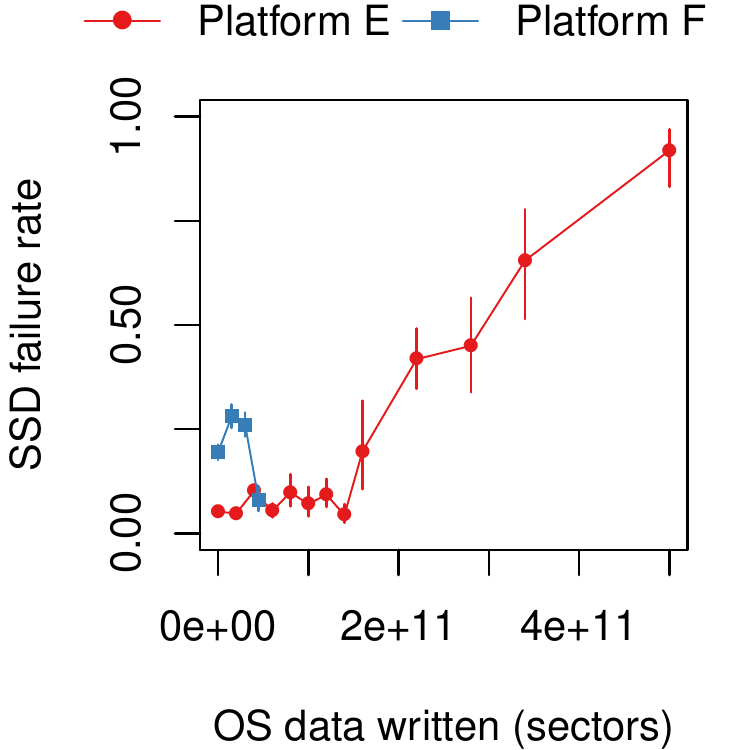}
\caption{SSD failure rate versus OS data written by system software.  Interestingly, the amount of data written by the OS is not always an accurate indication of the amount of SSD wear, as seen in Platforms A and B, where more data written by the OS can correspond to lower failure rates.}
\label{fig:os-writes}
\end{figure*}

\begin{figure*}
\centering
\includegraphics[width=6in, trim={0 0 4.95in 0}, clip]{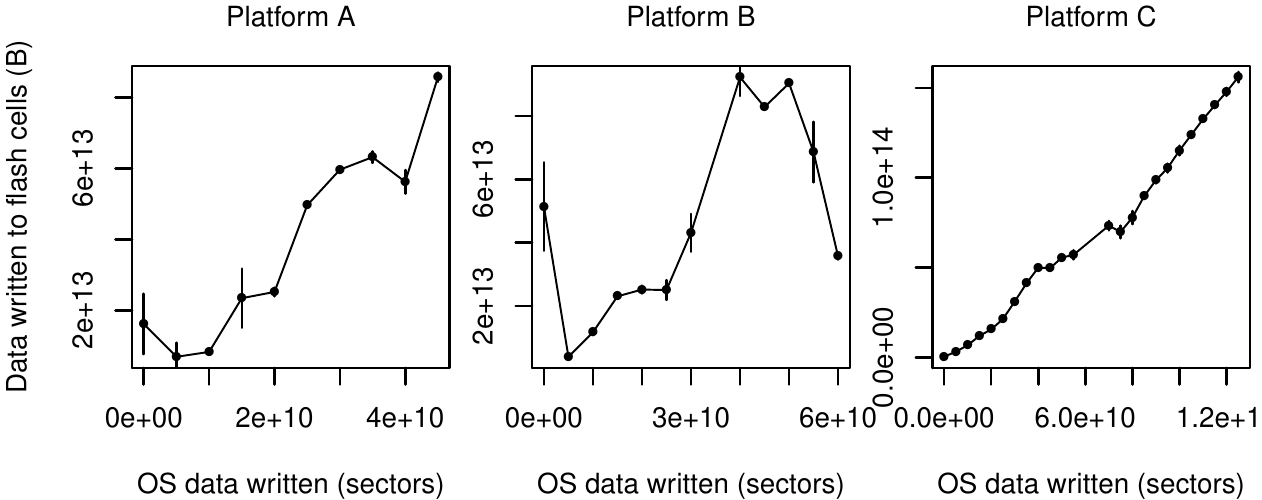}
\vspace{0.5cm}

	\hspace{1cm}\includegraphics[width=6in, trim={5.05in 0 0 0}, clip]{fig/pdf/os-writes-written}
\caption{Data written to flash \emph{cells} versus data the OS reports as having been written.  Due to buffering present in the system software (e.g., the page cache), the amount of data written at the OS does not always correspond to the amount of data written to the SSD (e.g., in Platforms A, B, and F).}
\label{fig:os-writes-written}
\end{figure*}


\subsection{Bus Power Consumption}

According to the PCIe standard, the nominal bus power consumption for the PCIe $\times$4 SSDs that we analyze is \unit[10]{W} (regardless of PCIe version 1 or 2).  Our infrastructure allows us to measure the average amount of power SSDs consume on the PCIe bus.  As power consumption in servers can lead to higher electricity use for operation and cooling in data centers, we seek to understanding the role that SSD power draw plays with respect to errors.

Figure~\ref{fig:bus-power} plots the failure rate for SSDs that operate at different average amounts of bus power consumption.  Recall that Platforms A and B use PCIe v1 and that Platforms C through F use PCIe v2.  We make three observations about the operation of these SSDs.  First, PCIe v2 SSDs (Platforms C through F) support twice the bandwidth of PCIe v1 SSDs (Platforms A and B) by operating at twice the frequency, leading to around twice the amount of power consumption between the two sets of SSDs: SSDs supporting PCIe v1 operate in the range of 4 to \unit[7.5]{W} and SSDs supporting PCIe v2 operate in the range of 8 to \unit[14.5]{W}.  Second, we find that the bus power that SSDs consume can vary over a range of around 2$\times$ between the SSDs that consume the least bus power and those that consume the most within a platform.  Third, we find that PCIe v2 SSDs in systems that use two SSDs (Platforms D and F) tend to consume lower bus power versus PCIe v2 SSDs in systems that use one SSD (Platforms C and E).  This may be due to SSD accesses spread across two SSDs (note, however, that the \emph{total} bus power consumption of a two-SSD system is larger than a single-SSD system).

With regard to failure rates, we observe that the trends for bus power consumption correlates with those of average temperature.  We believe that the higher bus power consumption is due to more data transfer and more data transfer requires more power to read or write from the SSD, which in turn increases the SSD temperature.  One interesting result of this correlation is that techniques that wish to reduce SSD temperature in the absence of precise temperature sensors may be able to use bus power consumption as a proxy.

\subsection{Data Written by the System Software}
\label{subsec:Data Written by the System Software}

While we have shown that the amount of data that is written to flash cells has a direct effect on flash-based SSD lifetime, it is less clear what effect system-level writes (e.g., those the system software initiates on behalf of user programs or the file system) have on SSD error characteristics.  The reduction of writes the system software performs has been used as a figure of merit in research techniques that evaluate flash reliability in lieu of modeling the wear reduction techniques present in SSDs.  We seek to examine whether or not different amounts of writes the system software performs lead to different error characteristics in SSDs.  Note that not all data written by system software may get written to flash chips due to system-level buffering.

To do so, we measure the number of sectors (\unit[512]{B} in size) the OS modifies on behalf of the software on our machines over their lifetime, which we refer to as data written by the system software.  Figure~\ref{fig:os-writes} plots the failure rate for SSDs whose OS has written different numbers of sectors over their lifetime.

In general, we observe indications that the total amount of data written by the system software correlates with higher failure rates (and follow similar trends as the SSD lifecycle from Figure~\ref{fig:lifecycle}), with one important caveat:  some systems where a large amount of data has been written by the system software, have lower failure rates.  This can be seen, for example, in Platforms A, B, and F:  while there is a general trend toward higher SSD failure rates with larger amounts of data written by the system software for other platforms, the SSDs in Platforms A, B, and F can have lower failure rates with larger amounts of data written by the system software.  We believe that this is due to the fact that systems that write more data may be able to benefit more from system-level buffering (e.g., the OS page cache) along with techniques the SSD controller uses to reduce the amount of wear on flash cells (e.g., data encoding schemes that help reduce the number of program or erase operations).

To confirm this trend, Figure~\ref{fig:os-writes-written} plots the amount of data \emph{actually written to the flash hardware} versus the amount of data the system software reports as having written.  SSDs in Platforms B and F clearly show that writing more data at the system software-level does not always imply physically writing more data to SSDs.  In fact, writing more data at the system software level may actually offer the system software more opportunities to \emph{coalesce} data in system-level buffers before writing it to the flash cells, resulting in smaller numbers of actual writes to the flash chips.  This observation suggests that work to improve flash reliability \emph{must} consider the effects of buffering across the system when evaluating the efficacy of different techniques.

%
%
%
%
%
%
%
%
%
%
%
%
%
%
%
%

\section{Summary}

In this chapter, we analyze flash-based SSD reliability across a majority of the SSDs at Facebook. We examine how a variety of internal and external characteristics affect the trends for uncorrectable errors and identifies several important lessons:

\begin{description}

\item[Lesson S.1] We observe that SSDs go through several distinct periods---early detection, early failure, usable life, and wearout---with respect to the factors that relate to the amount of data written to flash chips. Due to pools of flash blocks with different reliability characteristics, failure rate in a population does not monotonically increase with respect to amount of data written.  This is unlike the failure rate trends seen in raw flash chips.
	
\item[Lesson S.2] We should design techniques to help reduce or tolerate errors throughout SSD operation, not only toward the end of their lives. For example, additional error correction at the beginning of an SSD's life could help reduce the failure rates we see during the \emph{early detection} period. [\S\ref{sec:corr}]

\item[Lesson S.3]  We find that the effect of read disturbance errors is not a predominant source of errors in the SSDs that we examine. While prior work has shown that such errors can occur under certain access patterns in controlled environments~\cite{brand-irps93, mielke-irps08, cai-date12, cai-dsn15}, we do {\em not} observe this effect across the SSDs we examine.  This corroborates prior work which showed that the effect of write errors in flash cells dominate error rate compared to read disturbance~\cite{mielke-irps08, cai-date12}.  It may be beneficial to perform a more detailed study of the effect of these types of errors in flash-based SSDs that servers use. [\S\ref{sec:read}]

\item[Lesson S.4] Sparse logical data layout across an SSD's physical address space (e.g., non-contiguous data) greatly affects SSD failure rates; dense logical data layout (e.g., contiguous data) can also negatively impact reliability under certain conditions, likely due to adversarial access patterns.
	
\item[Lesson S.5] Further research into flash write coalescing policies with information from the system level may help improve SSD reliability.  For example, information about write access patterns from the operating system could potentially inform SSD controllers of non-contiguous data that the system software accesses most frequently, which may be one type of data that adversely affects SSD reliability and is a candidate for storing in a separate write buffer. [\S\ref{subsec:DRAM Buffer Usage}]

\item[Lesson S.6] Higher temperatures lead to higher failure rates, but do so most noticeably for SSDs that do not employ throttling techniques.  In general, we find techniques like \emph{throttling}, which likely correlate with techniques to reduce SSD temperature, to be effective at reducing the failure rate of SSDs.  We also find that SSD temperature correlates with the power the system uses to transmit data across the PCIe bus, which we can use as a proxy for temperature in the absence of SSD temperature sensors. [\S\ref{sec:external}]

\item[Lesson S.7] The amount of data written by the system software can overstate the amount of data written to flash cells due to system-level buffering and wear reduction techniques. Simply reducing the rate of software-level writes without considering the qualities of the write access pattern to system software is not sufficient for assessing SSD reliability.  Studies seeking to model the effects of reducing software-level writes on flash reliability should also consider how other aspects of SSD operation, such as system-level buffering and SSD controller wear leveling, affect the actual data written to SSDs. [\S\ref{subsec:Data Written by the System Software}]

\end{description}

We hope that our new observations, with real workloads and real systems from the field can aid in (1) understanding the effects of different factors, including system software, applications, and SSD controllers on flash memory reliability, (2) the design of more reliable flash architectures and system designs, and (3) improving the evaluation methodologies for future flash memory reliability studies.

While servers use flash-based SSDs to store persistent data running in modern data centers, servers must also send and receive data from each other using the networks within and between data centers. The next chapter examines the reliability of the networks that connect modern data centers.

\chapter{Network Failures}
\label{chp:networkfailures}

The ability to tolerate, remediate, and recover from network incidents (due to device failures and fiber cuts, for example) is critical for building and operating highly-available web services. Achieving fault tolerance and failure preparedness requires system architects, software developers, and site operators to have a deep understanding of network reliability at scale, along with its implications on the software systems that run in data centers. Unfortunately, little has been reported on the reliability characteristics of large scale data center network infrastructure, let alone its impact on the availability of services powered by software running on that network infrastructure.

In this chapter, we discuss an expanded version of a large scale, longitudinal study we performed~\cite{Meza17} of data center network reliability using operational data we collect from the production network infrastructure at Facebook. Our study covers reliability characteristics of both intra and inter data center networks. For intra data center networks (\S\ref{sec:intra-dc}), we study seven years of operation data comprising thousands of network incidents across two different data center network designs, a cluster network design and a state-of-the-art fabric network design. For inter data center networks (\S\ref{sec:inter}), we study eighteen months of recent repair tickets from the field to understand reliability of Wide Area Network (WAN) backbones. In contrast to prior work, we study the effects of network reliability on software systems, and how these reliability characteristics evolve over time (\S\ref{subsec:Related Research in Network Failures in Modern Data Centers} provides an overview of where our work stands in relation to other large scale network studies). We discuss the implications of network reliability on the design, implementation, and operation of large scale data center systems and how it affects highly-available web services.

\section{Motivation for Understanding Network Failures}

The reliability of data center network infrastructure is critically important for building and operating highly available and scalable web services~\cite{Bailis2014,Brewer2017}. Despite an abundance of device- and link-level monitoring, the effects of network infrastructure reliability on the software systems that run on them is not well understood.  The fundamental problem lies in \emph{the difficulty of correlating device- and link-level failures with software system impact.} First, many network failures do \emph{not} cause software system issues due to network infrastructure redundancy (including device, path, and protocol redundancy). Second, large scale network infrastructure uses automated repair mechanisms that take action to resolve failures when they occur.

To understand the behavior of network failures, we must be able to answer questions such as: \emph{``How long do network failures affect software when they occur?''}, \emph{``What are the root causes of the network failures that affect software?''}, and \emph{``How do network failures manifest themselves in software systems?''}.  Unfortunately, past efforts to understand network incidents in large scale network infrastructure are limited to informal surveys and a small number of public postmortem reports~\cite{Bailis2014,Gunawi2016}, which could be biased toward certain types of failures and not comprehensive.  As noted in~\cite{Bailis2014}, due to scant evidence and even less data, it is hard to discuss the reliability of software systems in the face of network incidents because \emph{``much of what we believe about the failure modes is founded on guesswork and rumor.''} We perform a comprehensive overview of prior work in \S\ref{subsec:Related Research in Network Failures in Modern Data Centers}.

Even for large web and cloud service providers, understanding the reliability of network infrastructure is challenging, given the complex, dynamic, and heterogeneous nature of large scale networks.  With complex and constantly evolving network designs built from a wide variety of devices, it is hard to reason about the end-to-end reliability of network infrastructure under different failure modes, let alone how the network affects the software that uses it.  As far as we know, from what limited information has been publicly discussed, this is a common challenge across the industry.

Facebook attempts to address this challenge by seeking to understand the reliability of its data center network infrastructure.  At Facebook, software system events that affect reliability (known as \emph{SEVs}, which we discuss in detail in \S\ref{sec:service-level-events}) are rigorously documented and reviewed to uncover their root causes, duration, software system impact, as well as mitigation and recovery procedures~\cite{Maurer2015}.  These postmortem reports form an invaluable source of information for analyzing and understanding network reliability from the perspective of large scale web services.

\textbf{\emph{Our goal}} is to shine light on the network reliability incidents, both \emph{within} and \emph{between} data centers, which affect the software systems that power large scale web services. We hope that our work helps researchers and practitioners anticipate and prepare for network incidents, and inspires new network reliability solutions.

\section{Methodology for Understanding Network Failures}

We next describe how we measure and analyze the reliability of intra and inter data center networks. In Chapter~\ref{chp:backgroundrelated}, we detail the scope of our study (\S\ref{sec:incident_def}) and our network incident dataset (\S\ref{sec:service-level-events}). We now discuss our analytical methodology (\S\ref{sec:analytical_method}), and limitations and conflating factors in our study (\S\ref{sec:limitations-and-conflating-factors}).

\subsection{How We Measure and Analyze Network Failures}
\label{sec:analytical_method}

We use two sets of data for our study that we introduce in \S\ref{sec:service-level-events}. For \emph{intra} data center reliability, we examine seven years of service-level event data we collect from a database of SEV reports. For \emph{inter} data center reliability we examine eighteen months of data we collect from vendors on fiber repairs the vendors performed between October 2016 and April 2018. We describe the analysis for each data source below.

\textbf{\emph{Intra data center networks.}} For intra data center reliability, we study the network incidents in three aspects:

\begin{enumerate}
  \item \textbf{\emph{Root cause.}} We use the root causes chosen by the engineers who authors the
corresponding SEV reports. The root cause category
(we list root causes in Table~\ref{tab:root-cause}) is a mandatory field in our
    SEV authoring workflow, although the root cause may be \emph{undetermined}.

  \item \textbf{\emph{Device type.}} To classify a network incident by the implicated device's type, we rely on the naming convention Facebook enforces where each network device has a name with a unique, machine-understandable string prefixed with the device type.  For example, every rack switch has a name prefixed with ``{\tt rsw}''.  By parsing the prefix of the name of the offending device, we are able to classify SEVs using device type.

  \item \textbf{\emph{Network design.}} We also classify network incidents using network architecture.
Recall from Figure~\ref{fig:network-arch} that CSA and CSW devices
belong to cluster networks,
while ESW, SSW, and FSW devices are a part of fabric networks.
\end{enumerate}

The SEV report dataset we analyze comprises thousands of SEVs and resides in a MySQL database. Network SEV reports contain details on the network incident: the network device implicated in the incident, the duration of the incident (which we measure from when the root cause manifests until when engineers fix the root cause), and the incident's effect on software systems (for example, load increase from lost capacity, message retries from corrupt packets, downtime from connectivity partitions, and higher latency from links congestion). We use SQL queries to analyze the SEV report dataset for our study.

\textbf{\emph{Inter data center networks.}}
For inter data center reliability, we study the reliability of
edge nodes and fiber links based on repair tickets from
fiber vendors whose links form Facebook's backbone networks that
connect the data centers. Facebook has monitoring systems that
check the health of every fiber link, as unavailability of the links could
significantly affect the traffic or partition a data center from
the rest of Facebook's infrastructure.

\label{sec:meth:inter-dc}

When a vendor starts repairing a link (after something severs the link) or
performing maintenance on a fiber link, the vendor notifies Facebook via email.
The email has structure, including the
logical IDs of the fiber link, the physical location of the affected fiber
circuits,
the starting time of the repair or maintenance,
and an estimate of the duration of the maintenance.
Similarly, when the vendor completes the repair or maintenance of a
fiber link, the vendor sends a confirmation email.
Facebook automatically parses the emails and stores the emails in a database for later analysis.
We examine eighteen
months of repair data in this database from October 2016 to April
2018. From this data, we measure fiber link mean time between failures (MTBF) and mean time
to repair (MTTR).

\subsection{Limitations and Potential Confounding Factors}
\label{sec:limitations-and-conflating-factors}

We found it challenging to control for all variables in a longitudinal study of failures at a company of Facebook's scale. So, our study has limitations and conflating factors, some of which we briefly discuss below. Throughout our analysis, we state when a factor that is \emph{not} under our control may affect our conclusions.

\begin{itemize}
  \item \textbf{\emph{Absolute versus relative number of failures.}} We cannot report the absolute number of failures. Instead, we report failure rates using a fixed baseline when the trend of the absolute failures aids our discussion.
  \item \textbf{\emph{Logged versus unlogged failures.}} Our intra data center network study relies on SEVs reported by employees. While Facebook fosters a culture of opening SEVs for all incidents affecting production, we cannot guarantee our incident dataset is exhaustive.
  \item \textbf{\emph{Technology changes over time.}} Switch hardware consists of a variety of devices sourced and assembled from different vendors. We do not account for these factors in our study. Instead, we analyze trends by switch type when a switch's architecture significantly deviates from others.
  \item \textbf{\emph{Switch maturity.}} Switch architectures vary in their lifecycle, from newly-introduced switches to switches ready for retirement. We do not differentiate the effect a switch's maturity has in Facebook's fleet in our analyses.
  \item \textbf{\emph{More engineers making changes.}} As Facebook has grown, so has the number of engineers performing network operations. While all network software and configuration changes go through code review to reduce the chances of network incidents, more engineers can potentially lead to more opportunities for failure.
\end{itemize}

\section{Intra Data Center Reliability}
\label{sec:intra-dc}

In this section, we study the reliability of data center networks. We analyze network incidents within Facebook data centers over the course of seven years, from 2011 to 2018, comprising thousands of real world events. A network incident occurs when the root cause of a SEV relates to a network device. We analyze root causes (\S\ref{sec:root-causes}), incident rate and distribution (\S\ref{sec:incident_rate}), incident severity (\S\ref{sec:incident_impact}), network design (\S\ref{sec:incidents-by-topology}), and device reliability (\S\ref{sec:mtbi}).

\subsection{Root Causes}
\label{sec:root-causes}


Table~\ref{tab:root-cause} lists network incident root causes.\footnote{We use Govindan et al.~\cite{Govindan2016}'s definition of \emph{root cause}: \emph{``A failure event's root-cause is one that, if it had not occurred, the failure event would not have manifested.''}} If a SEV has \emph{multiple} root causes, we count the SEV toward multiple categories. Human classification of root causes implies SEVs can be misclassified~\cite{Potharaju2013,Medem2009}. While the rest of our analysis does \emph{not} depend on the accuracy of root cause classification, we find it instructive to examine the types of root causes that occur in Facebook's networks.

\begin{table*}
  \centering
  \begin{tabular}{lcp{10cm}}
    \toprule

    \textbf{Category}
    &
    \textbf{Fraction}
    &
    \textbf{Description}
    \\

    \midrule

    Maintenance
    &
    17\%
    &
    Routine maintenance (for example, upgrading the software and firmware of network devices).
    \\

    Hardware
    &
    13\%
    &
    Failing devices (for example, faulty memory modules, processors, and ports).
    \\

    Misconfiguration
    &
    13\%
    &
    Incorrect or unintended configurations (for example, routing rules blocking production traffic).
    \\

    Bug
    &
    12\%
    &
    Logical errors in network device software or firmware.
    \\

    Accidents
    &
    11\%
    &
    Unintended actions (for example, disconnecting or power cycling the wrong network device).
    \\

    Capacity planning
    &
    5\%
    &
    High load due to insufficient capacity planning.
    \\

    \midrule

    Undetermined
    &
    29\%
    &
    Inconclusive root cause.
    \\

    \bottomrule
  \end{tabular}
  \caption{Common root causes of intra data center network incidents at Facebook from 2011 to 2018.}
  \label{tab:root-cause}
\end{table*}

We find the root cause of 29\% of network incidents is \emph{undetermined}. We observe these SEVs correspond typically to transient and isolated incidents where engineers only reported on the incident's symptoms. Wu et al.\ note a similar fraction of unknown issues (23\%, \cite{Wu2012}, Table 1), while Turner et al.\ report a smaller fraction (5\%, \cite{Turner2010}, Table 5).

\emph{Maintenance failures contribute the most documented root causes (17\%).} 
This suggests that in the network infrastructure of a large web service
provider like Facebook,
despite the best efforts to automate and standardize the maintenance
procedures,
maintenance failures still occur and lead to disruptions.
Therefore, it is important to build mechanisms for quickly and reliably
routing around faulty devices or devices under maintenance.

\emph{Hardware failures represent 13\% of the root causes, while human-induced
misconfiguration and software bugs occur at nearly double the rate
(25\%) of those caused by hardware failures}. Turner et al. and Wu et al. observe hardware incident rates similar to the incident rates we observe (18\% in Table 1 in \cite{Wu2012} and 20\% in Table 5 in \cite{Turner2010}), suggesting that hardware incidents remain a fundamental root cause. Misconfiguration causes as many incidents as faulty
hardware.  This
corroborates the findings of prior works that report misconfiguration as a large source of
network failures in data
centers~\cite{Potharaju2013IMC,Govindan2016,Benson2009,Benson2011,
Mahajan2002,Liu2017,Wu2012}, and shows the importance of emulation, verification, and
automated repair techniques to reduce the number of incidents~\cite{Liu2017,
Gember-Jacobson2017,Beckett2017,Beckett22017,Beckett2016,Fayaz2016,Fogel2015,
Khurshid2013}.

We observe a similar rate of misconfiguration incidents (13\%) as Turner et al.\ (9\% in Table 5 in \cite{Turner2010}), and a lower rate of misconfiguration incidents than Wu et al. (38\% in Table 1 in \cite{Wu2012}). We suspect network operators play a large role in determining how misconfiguration causes network incidents. At Facebook, for example, all configuration changes require code review and typically get tested on a small number of devices before being deployed to the fleet. These practices may contribute to the lower misconfiguration incident rate we observe compared to Wu et al..

A potpourri of accidents and capacity planning issues makes up the last 16\% of
incidents (cf.~Table~\ref{tab:root-cause}).
This is a testament to the many sources of
entropy in large scale production data center networks. Designing network
devices to tolerate all of these issues is prohibitively difficult
(if not impossible) in practice.  Therefore, one reliability engineering principle
is to \emph{prepare for the unexpected} in large scale data center networks.

\begin{figure}[H]
  \centering
  \includegraphics[width=0.8\columnwidth]{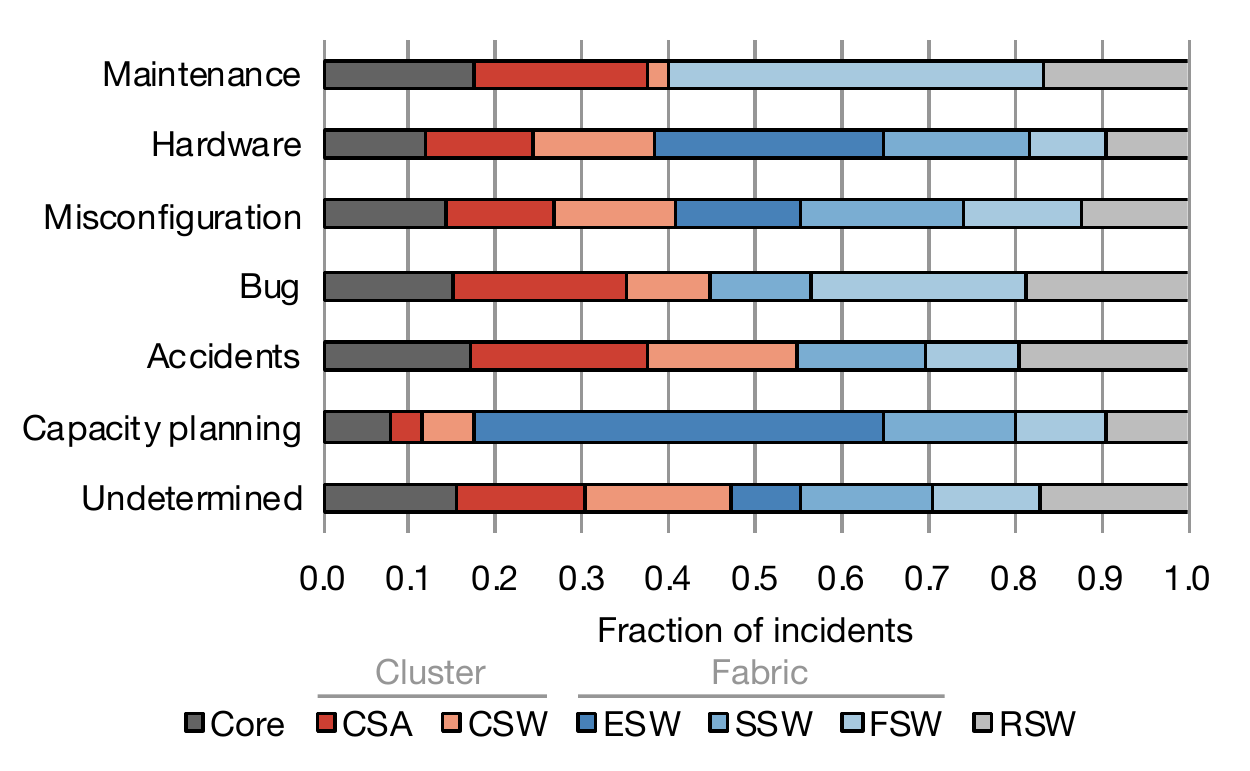}
  \caption{Breakdown of each root cause across the device types it affects.}
  \label{fig:root-cause-device}
\end{figure}

Figure~\ref{fig:root-cause-device} breaks down each root cause across the types of network devices it affects.  Note that the major root cause
categories, including \emph{undetermined, maintenance, hardware, misconfiguration,
bugs, accidents, and capacity planning} affect most network device types.
Some root cause categories are represented unequally among
devices. For example, capacity planning issues tend to affect more ESWs and maintenance issues tend to affect more FSWs. ESWs do \emph{not} have SEVs due to bugs or maintenance, not because ESWs are immune to bugs and maintenance issues, but because the population size is small and such incidents have not yet been observed.

We conclude that maintenance failures contribute the most documented network incidents (17\%) and human-induced failures (misconfiguration and bugs) occur twice as much as hardware-induced failures.

\subsection{Incident Rate and Distribution}
\label{sec:incident_rate}


\textbf{\emph{Incident rate.}} The reliability of
each interconnected network device determines the overall reliability of data center networks.  To measure the frequency of incidents
as they relate to each device type, we define \emph{incident rate} of a device type as
$r = \frac{i}{n}$, where $i$ denotes the number of incidents this
type of network device causes and $n$ is the number of active devices in the network of that type (the \emph{population}).  Note
that the incident rate could be larger than $1.0$, meaning that each device of
that type causes more than one network incident, on average.

Figure~\ref{fig:incident-to-population} shows the incident rate of each type of network
device in Facebook's data centers over the seven-year span of our study.
From Figure~\ref{fig:incident-to-population}, we make four observations:

\begin{figure}[H]
  \centering
  \includegraphics[width=0.8\columnwidth]{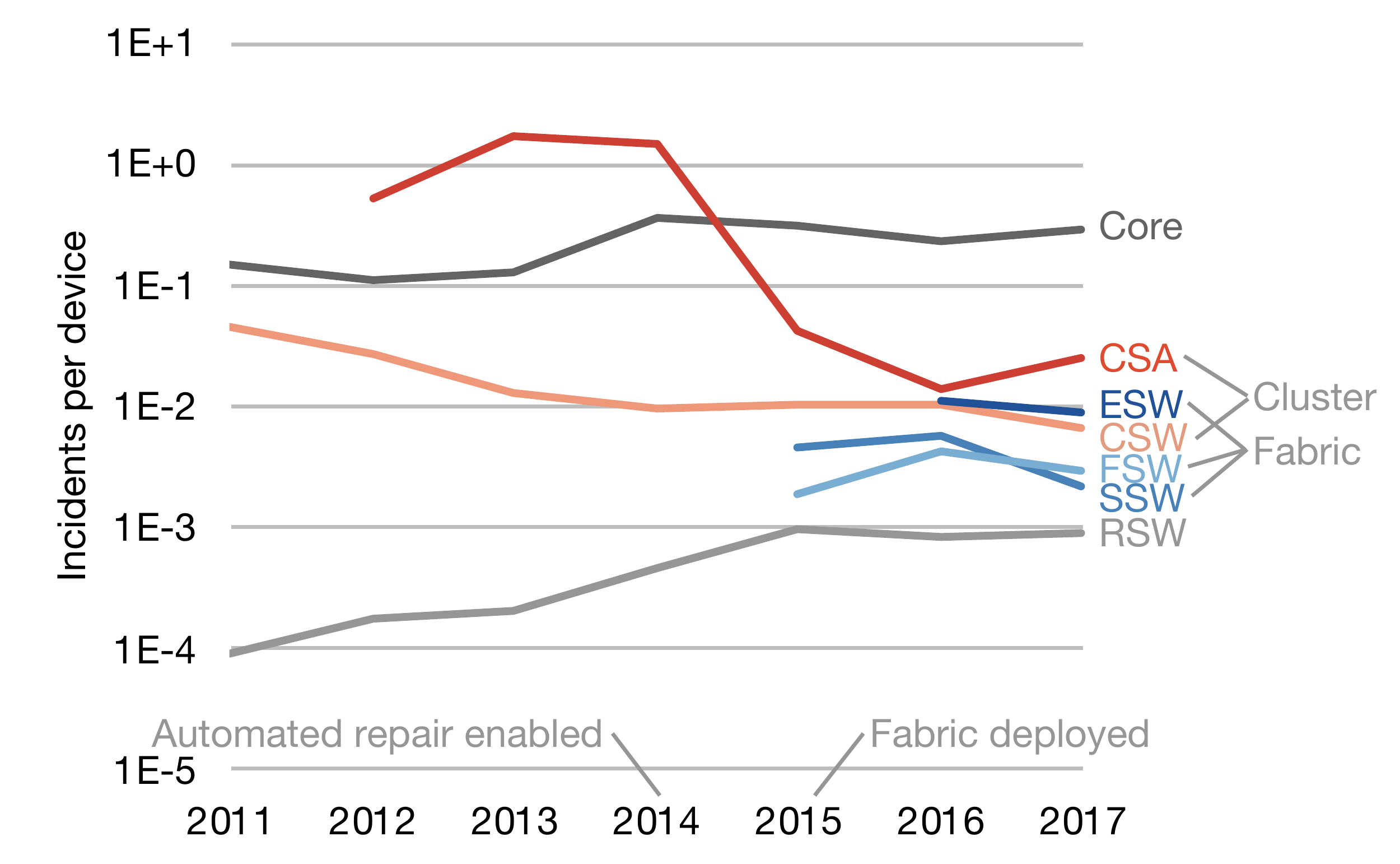}
  \caption{Yearly incident rate of each device type.
           Note that the y axis is in logarithmic scale and some devices
           have an incident rate of 0, which occurs if they did not exist in
           the fleet in a year.}
  \label{fig:incident-to-population}
\end{figure}

\begin{enumerate}
  \item \emph{Network devices with higher bisection bandwidth
(e.g., core devices and CSAs in Figure~\ref{fig:network-arch})
generally have higher incident rates, in comparison to the devices with lower bisection bandwidth (e.g., RSWs).}
Intuitively, devices with higher bisection bandwidth tend to affect a larger
number of connected devices and thus correlate with
more widespread impact when they fail.
The annual incidence rate for
ESWs, SSWs, FSWs, RSWs, and CSWs in 2017 is less than 1\%.

  \item \emph{Fabric network devices (ESWs,
SSWs, and FSWs) have lower incident rates versus
cluster network devices (CSAs and CSWs)}.
There are two differences between fabric network devices and cluster network devices: (a) fabric network devices are built from commodity chips~\cite{Bachar2015,Bachar2014},
while companies purchase cluster network devices from third-party vendors and (b)
 fabric networks use automated repair
software to handle common sources of failures~\cite{Power2011}.

  \item The fact that fabric network devices are less frequently associated with failures
    verifies that a \emph{fabric network design, that uses automated
    failover and repair, is more resilient to device failures.}  Specifically, we
can see a large rate of CSA-related incidents during 2013 and 2014, where the
number of incidents exceeds the number of CSAs (with the incident rate as
high as 1.7 and 1.5, respectively). Such high incidence rates were part of the motivation
to transition from the cluster network to fabric network.

  \item The CSA-related incident rate decreased in 2015, while the core device-related
incident rate has generally increased from pre-2015 levels. We can attribute this trend to two causes: (1) the decreasing size of the CSA population, and (2) new repair practices that Facebook adopted
around the time. For example, prior to 2014, engineers performed network device repairs were often
\emph{without} draining the traffic on their links. This meant that in the
worst case, when things went wrong, maintenance could affect a large volume of
traffic. Draining devices prior to maintenance provides a simple but effective
way to limit the likelihood of a repair affecting production traffic.

  \item RSW incident rate is increasing over time. We analyze this trend when we discuss Figure~\ref{fig:by-device}.
\end{enumerate}

These reliability characteristics influence Facebook's fault tolerant  
data center network design. For example, Facebook provisions eight core devices in
each data center, which allows Facebook data centers to tolerate one unavailable core device (e.g., if engineers must remove a device from operation for maintenance)
without any impact on the data center network.
Note that nearly all of the core devices and CSAs
are third-party vendor devices.
In principle, if we do \emph{not} have direct control of the proprietary software
on these devices,
the network design and implementation must take this lack of control into consideration.\footnote{
Facebook has been manufacturing custom RSWs and modular switches
since 2013. Please refer to the details in~\cite{Bachar2014,Bagga2015,Simpkins2015,Bachar2015}.} For example, it may be more challenging to diagnose, debug, and repair devices that rely on firmware whose source code is unavailable. In these cases, it may make sense to increase device redundancy in case vendors must remove devices for repair.

\textbf{\emph{Incident distribution.}} Figure~\ref{fig:by-device} shows the distribution of incidents 
each type of network device causes on a yearly basis. From Figure~\ref{fig:by-device}, we make two observations:

\begin{figure}[H]
  \centering
  \includegraphics[width=0.8\columnwidth]{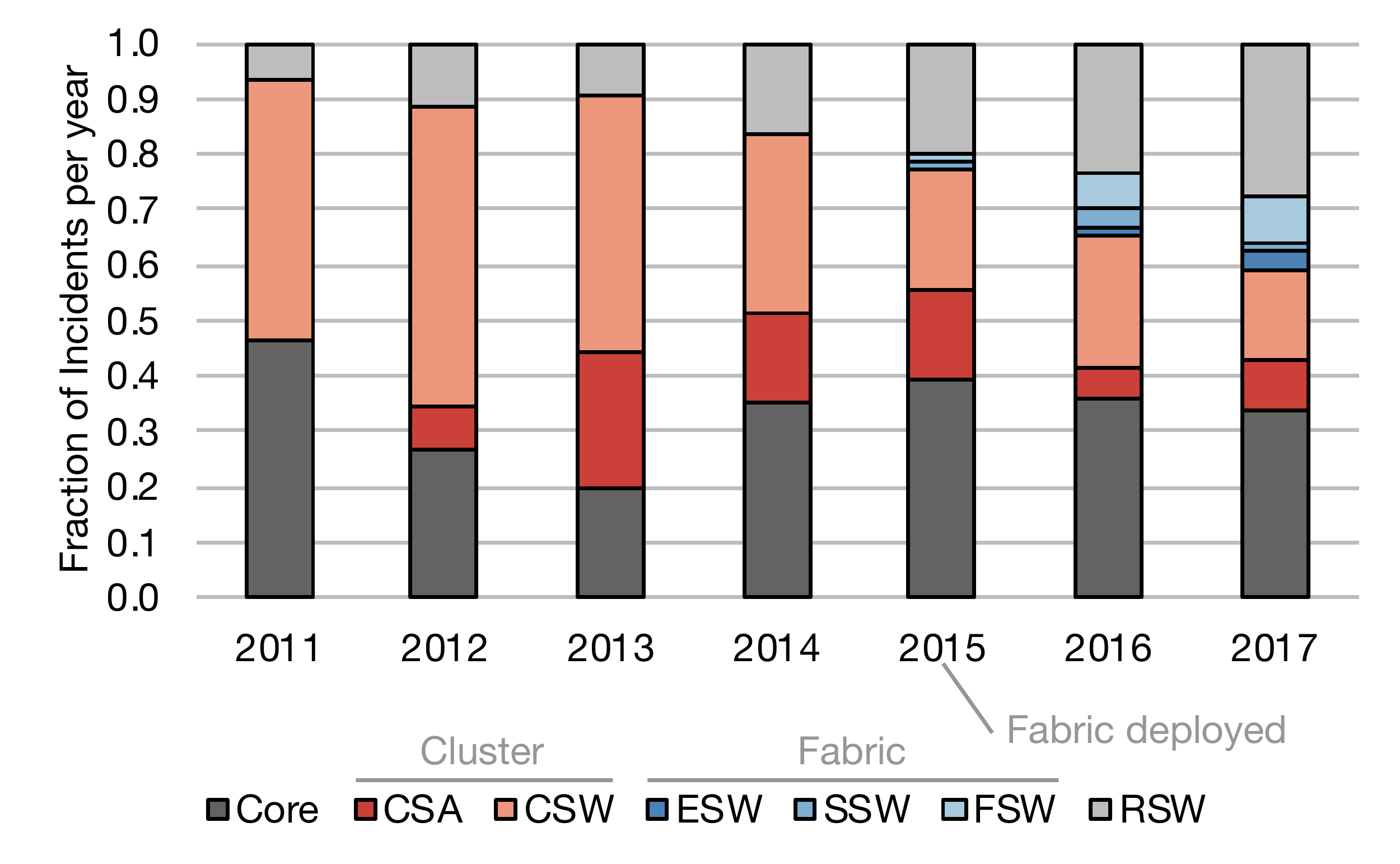}
  \caption{Fraction of network incidents per year broken down by device type.}
  \label{fig:by-device}
\end{figure}

\begin{enumerate}
  \item RSW-related incidents have been steadily increasing over time (a finding that corroborates that of Potharaju et al.~\cite{Potharaju2013SoCC, Potharaju2013IMC}).
This is partially driven by an increase in the size of the rack population over time. In addition, this is also a result of
Facebook's data center network design, where Facebook uses one RSW as the
Top-Of-Rack (TOR) switch. Other companies, such as some cloud service
providers and enterprises,
use two TORs with each server connects to both TORs for
redundancy.
At Facebook, we find that replicating and distributing server resources leads to low RSW incident rates and is more efficient than using redundant RSWs in every rack.

  \item Devices in fabric networks do not demonstrate a
large increase in incidents over time.  This again suggests that fabric-based
data center designs with automated failover provide good fault tolerance. We analyze this trend further in Section~\ref{sec:incidents-by-topology}.
\end{enumerate}

We conclude that (1) higher bandwidth devices have a higher likelihood of causing network incidents, (2) network devices built from commodity chips have much lower incident rates versus devices from third-party vendors due to automated failover and software repairs, (3) better repair practices lead to lower incident rates, and (4) RSW incidents are increasing over time, but they are still relatively low.

\subsection{Incident Severity}
\label{sec:incident_impact}


Not all incidents are created equal. Facebook classifies incidents into three severity levels from SEV3 (lowest severity) to SEV1 (highest severity). A SEV level reflects the high watermark for an incident. Engineers never downgrade a SEV's level reflect progress in resolving the SEV. Table~\ref{tab:sevs} provides examples of incidents for each SEV level.

\begin{table}[H]
  \centering
  \begin{tabular}{lp{6.85cm}}
    \toprule
    \textbf{Level} & \textbf{Incident Examples} \\
    \midrule
    SEV3 & Redundant or contained system failures, system impairments that do not affect or only minimally affect customer experience, internal tool failures. \\
    SEV2 & Service outages that affect a particular Facebook feature, regional network impairment, critical internal tool outages that put the site at risk. \\
    SEV1 & Entire Facebook product or service outage, data center outage, major portions of the site are unavailable, outages that affect multiple products or services. \\
    \bottomrule
  \end{tabular}
  \caption{SEV levels and incident examples.}
  \label{tab:sevs}
\end{table}

Figure~\ref{fig:sev-breakdown} shows how each type of network SEV in 2017
is distributed among network devices. We make two observations from
Figure~\ref{fig:sev-breakdown} that complement our raw incident rate findings
from \S\ref{sec:incident_rate}:

\begin{figure}[H]
  \centering
  \includegraphics[width=0.8\columnwidth]{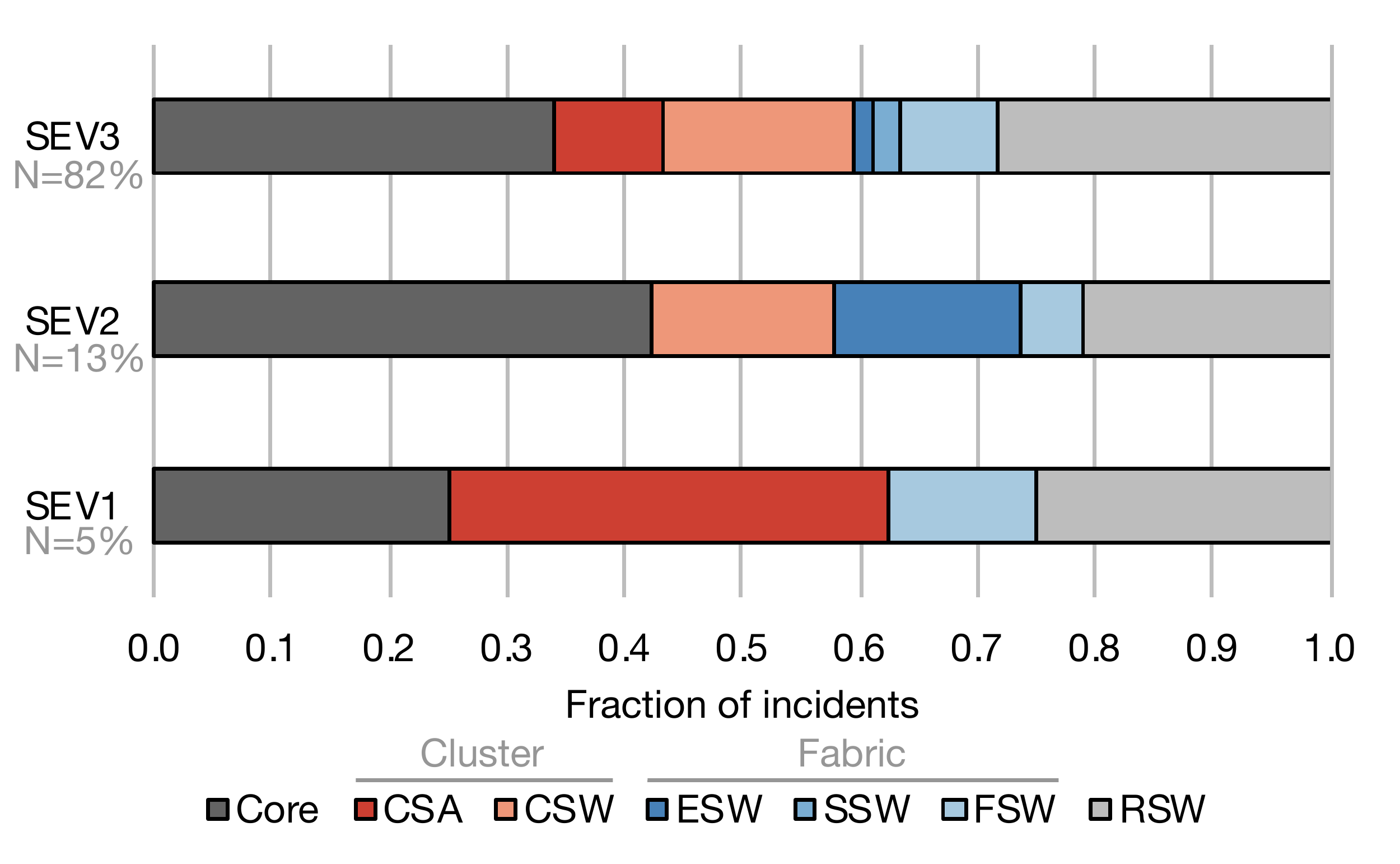}
  \caption{Breakdown of each SEV type across different network devices in 2017.}
  \label{fig:sev-breakdown}
\end{figure}

\begin{enumerate}
 \item While core devices have the highest number of SEVs, the severity of core device SEVs is
   typically low, with around 81\% of SEVs at level 3, 15\% at level 2,
    and 4\% at level 1. RSWs have nearly as many incidents as core 
    devices, with severity distributed in roughly the same proportion (85\%,
    10\%, and 5\% for SEV levels 3, 2, and 1, respectively).
 \item Compared to cluster network devices (CSAs and
   CSWs), fabric network devices typically have lower severity, with 66\%
    fewer SEV1s, 33\% \emph{more} SEV2s (though the overall rate is still
    relatively low), and 52\% fewer SEV3s. The lower severity is due to the automatic failover and repair support in fabric network devices.
\end{enumerate}

\begin{figure}[H]
  \centering
  \includegraphics[width=0.8\columnwidth]{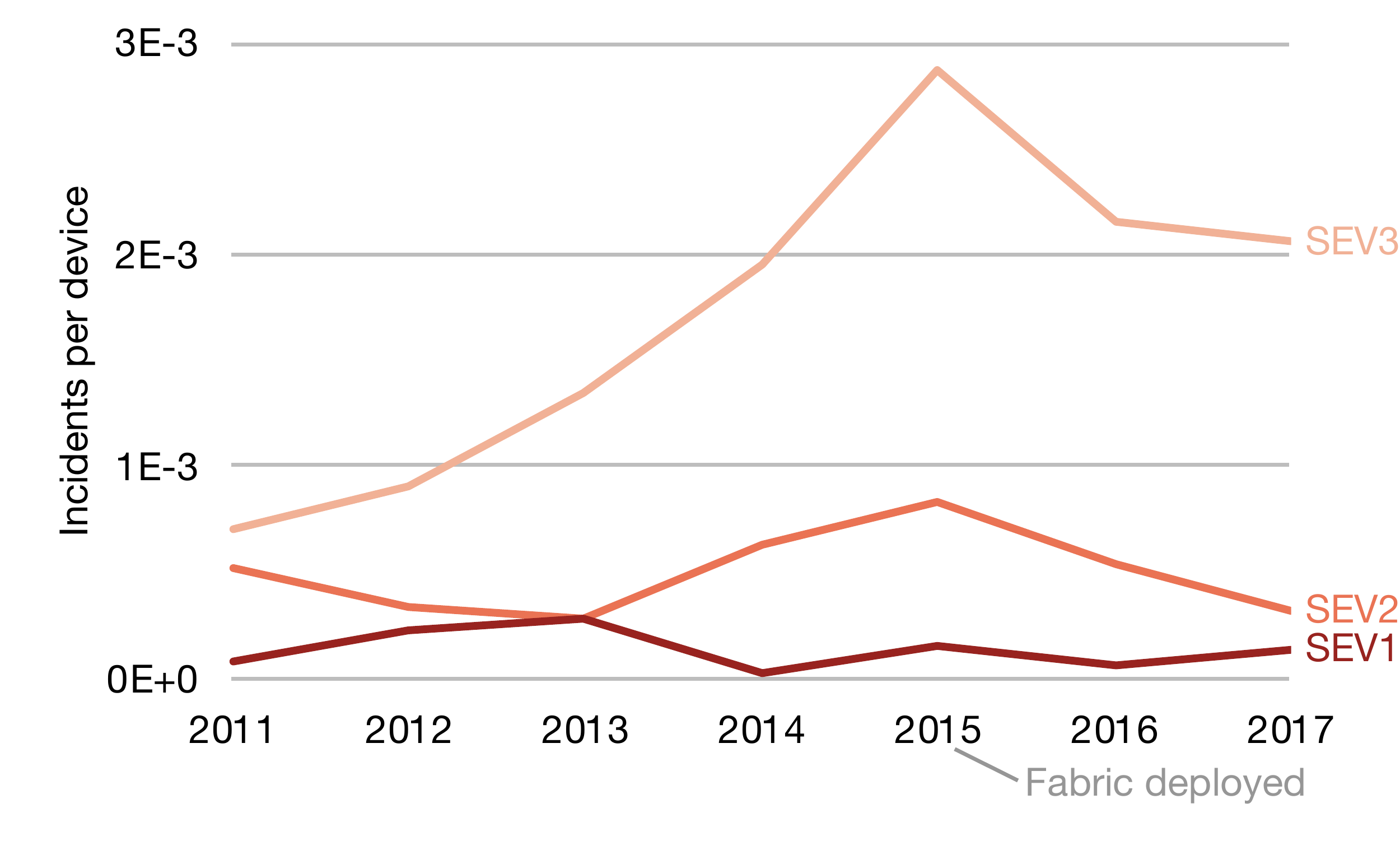}
  \caption{The number of network SEVs over time normalized to the
  number of deployed network devices. Note that the y axis is in logarithmic scale.}
  \label{fig:sev-timeseries}
\end{figure}

Figure~\ref{fig:sev-timeseries} shows how the rate of each SEV level changes over the years, normalized to the total number of devices in the
population during that year. While we cannot disclose the absolute size of the population, we note that it is multiple orders of magnitude larger than similar studies, such as Turner et al.~\cite{Turner2010}. The main conclusion we draw from
Figure~\ref{fig:sev-timeseries} is that the overall rate of SEVs per device
had an inflection point in 2015, corresponding to the deployment of fabric
networks. This was a significant turnaround, as, prior to 2015,
the rate of SEV3s grew at a nearly exponential rate.

\subsection{Network Design}
\label{sec:incidents-by-topology}

\begin{figure}[H]
  \centering
  \includegraphics[width=0.8\columnwidth]{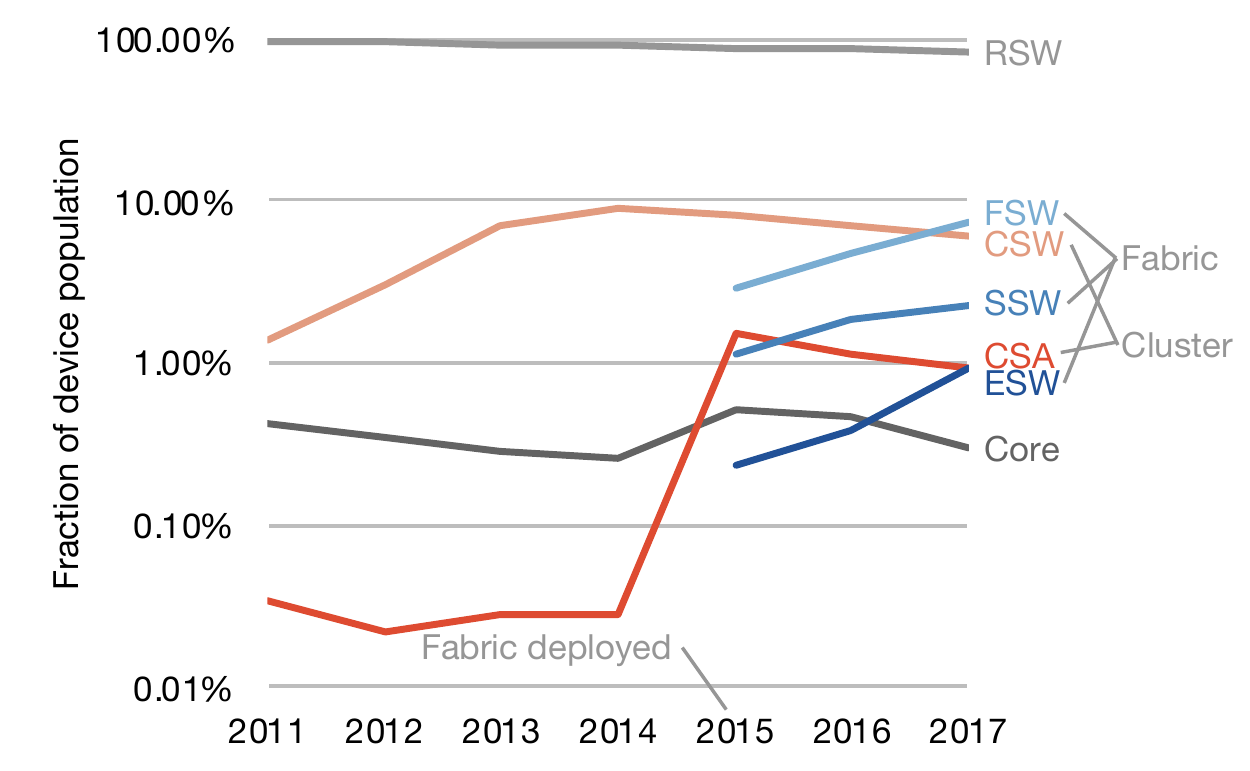}
  \caption{Population breakdown by network device type over the seven-year span of our study. Note that the y axis is in logarithmic scale.}
  \label{fig:population-timeseries}
\end{figure}

We start by describing the composition of Facebook's fleet of network devices. We plot the population breakdown of devices deployed in Facebook's data centers from 2011 to 2017 in Figure~\ref{fig:population-timeseries}. Aside from
showing the proliferation of RSWs in the fleet, Figure~\ref{fig:population-timeseries} shows that an
inflection point occurs in 2015, when the populations of CSWs and CSAs begin to
decrease and the populations of FSWs, SSWs, and ESWs begin to increase. This is due to the adoption of fabric networks across more Facebook data centers. In 2017, fabric network device deployment surpassed cluster network device deployment, with 1.5 fabric network devices for every 1 cluster network device in Facebook data centers.

Data center network design plays an important role in network reliability.
Figure~\ref{fig:by-topology-relative} shows how the fraction of network incidents
from the \emph{older} cluster network design and the \emph{newer} fabric network design changes over time.
Cluster network devices are CSAs
and CSWs; fabric network devices are ESWs, 
SSWs, and FSWs.
We calculate the fraction by summing the network incidents across \emph{all} of the device types
in each network design and dividing it by a common baseline, the number of incidents in 2017. Focusing on 2015, for example, the year fabric networks started being deployed, cluster networks caused nearly the same number of incidents as \emph{all} network incidents in 2017.
 From Figure~\ref{fig:by-topology-relative}, we make two observations:

\begin{figure}[H]
  \centering
  \includegraphics[width=0.8\columnwidth]{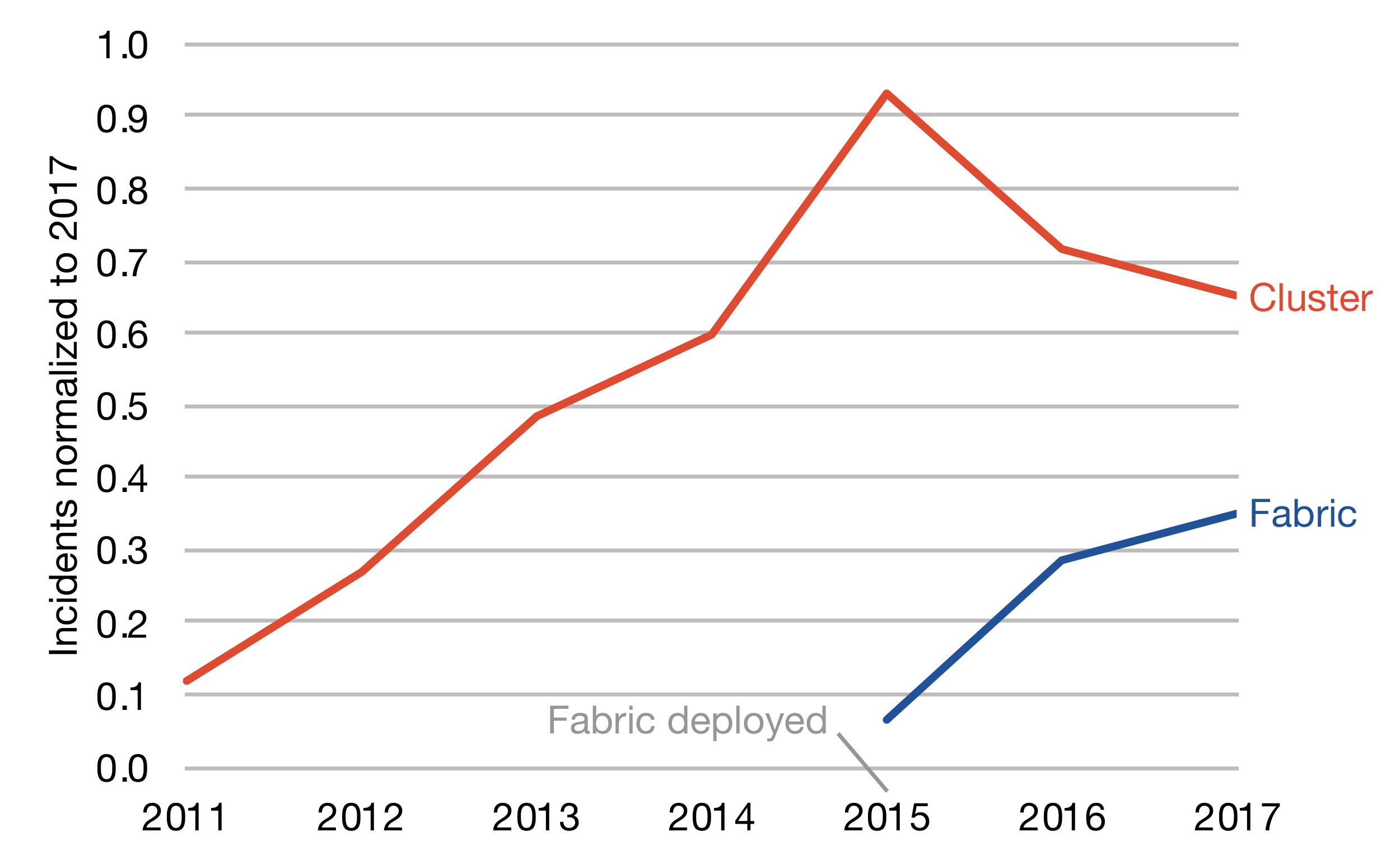}
  \caption{Number of incidents for each network design normalized to a fixed baseline, the total number of SEVs in 2017.}
\label{fig:by-topology-relative} \end{figure}

\begin{enumerate}
  \item \emph{Cluster network incidents increased steadily}
over time until around 2015, when it became challenging to make
additional reliability improvements to the cluster network design.

  \item In 2017, the number of incidents for cluster network devices was $1.87\times$ that of fabric network devices, despite fabric networks having 50\% more devices. Thus, normalized by number of devices, \emph{cluster network devices have $1.87 \times 1.5 = 2.8\times$ as many incidents as fabric network devices}. This is because the software-managed fault tolerance and
automated repair provided by fabric networks can \emph{mask} some
failures that would cause incidents in cluster networks.
\end{enumerate}

We conclude cluster networks have around $2\times$ the number of network incidents as fabric networks. We find that fabric networks are more reliable due to their simpler, commodity-chip
based switches and automated repair software that dynamically adapts
to tolerate device failures.

\subsection{Device Reliability}
\label{sec:mtbi}


We analyze the reliability of Facebook data center network devices. We use the incident start time and incident resolution time from SEVs to measure \emph{mean time between incidents (MTBI)} and \emph{75th percentile (p75) incident resolution time (p75IRT)}. p75IRT deserves additional explanation. Engineers at Facebook document \emph{resolution time}, not \emph{repair time}, in a SEV. Resolution time exceeds repair time and includes time engineers spend on developing and releasing fixes. To prevent occasional months-long incident resolution times from dominating the mean, we examine the 75th percentile incident resolution time.

\textbf{\emph{MTBI.}} We measure the average time between the start of two consecutive incidents for MTBI. Figure~\ref{fig:mtti} plots MTBI for each switch type by year. We draw two conclusions from the data:

\begin{figure}[H]
  \centering
  \includegraphics[width=0.8\columnwidth]{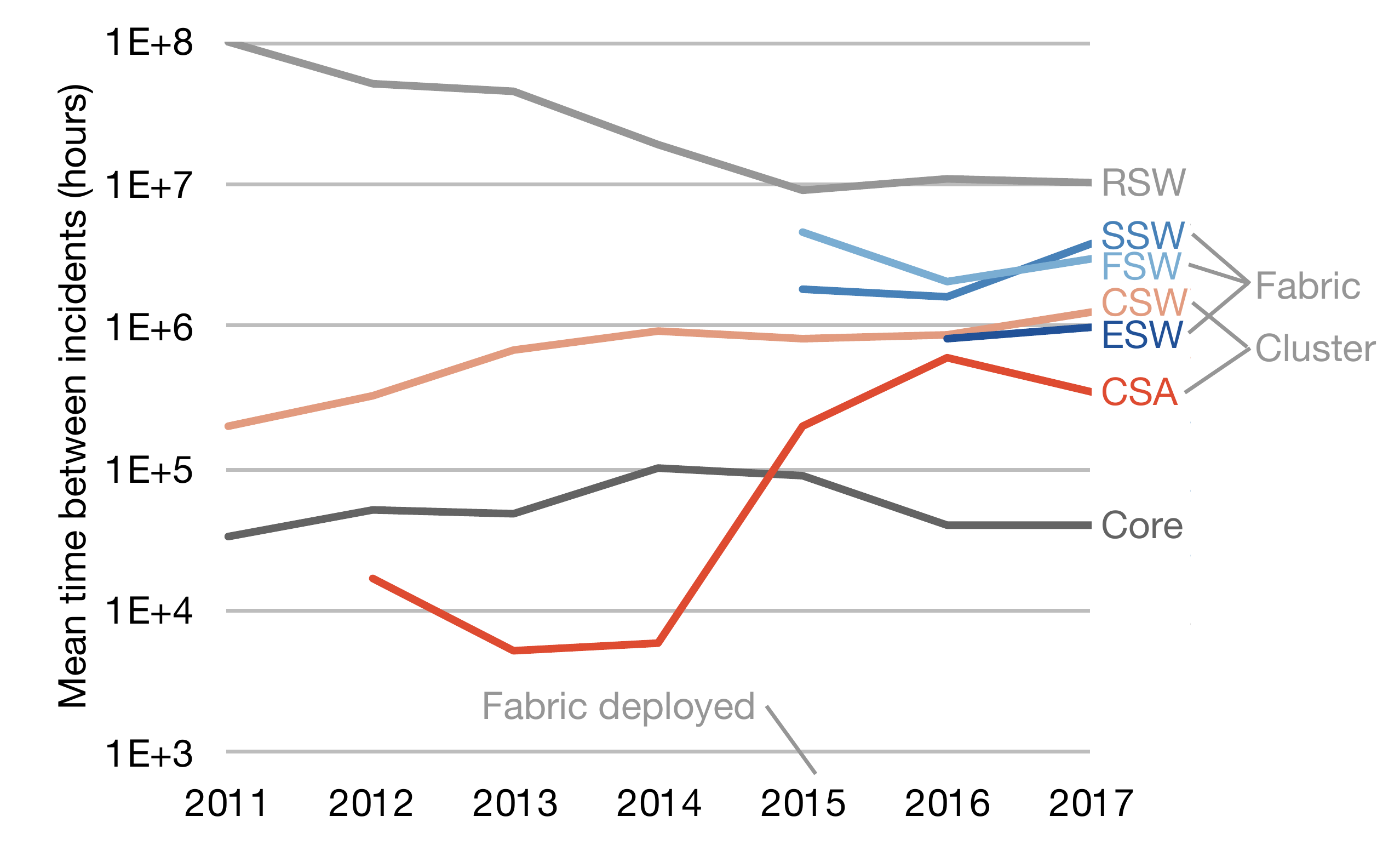}
  \caption{Mean time between incidents in hours for different network device types.
  Note that the y axis is in logarithmic scale.}
  \label{fig:mtti}
\end{figure}

First, we find that, from 2011 to 2017, MTBI did not change by more than 10$\times$ across each switch type, except CSAs. In 2015, in response to frequent CSA maintenance incidents, engineers strengthened CSA operational procedure guidelines, adding checks to ensure that operators drained CSAs before performing maintenance, for example. These operational improvements increased CSA MTBI by two orders of magnitude between 2014 and 2016.

Second, we find that, in 2017, \emph{MTBI varies by three orders of magnitude across switch types:} from \num{39495} device-hours for core devices to \num{9958828} device-hours for RSWs. If we compare MTBI to switch type population size in 2017 (shown in Figure~\ref{fig:population-timeseries}), we find that devices with larger population sizes tend to have larger MTBIs. This is because engineers at Facebook focus on deploying techniques like automated repair mechanisms to devices with large population sizes.

\begin{figure}[H]
  \centering
  \includegraphics[width=0.8\columnwidth]{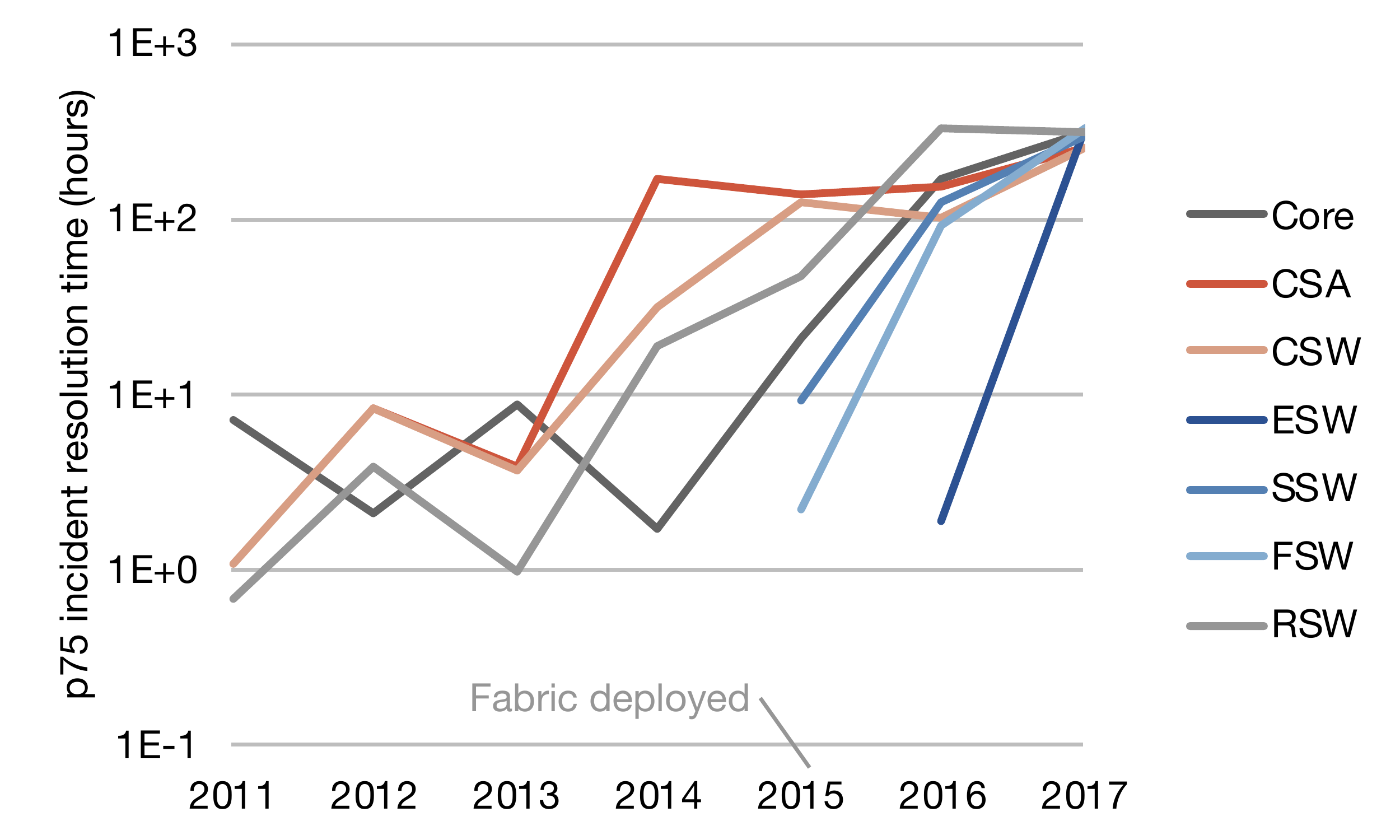}
  \caption{75th percentile incident resolution time in hours for different network device types. Note that the y axis is in logarithmic scale.}
  \label{fig:p75irt}
\end{figure}

\textbf{\emph{p75IRT.}} We measure the average time between the start and the resolution of incidents for p75IRT. Figure~\ref{fig:p75irt} plots p75IRT for each device type by year.

\begin{figure}[H]
  \centering
  \includegraphics[width=0.8\columnwidth]{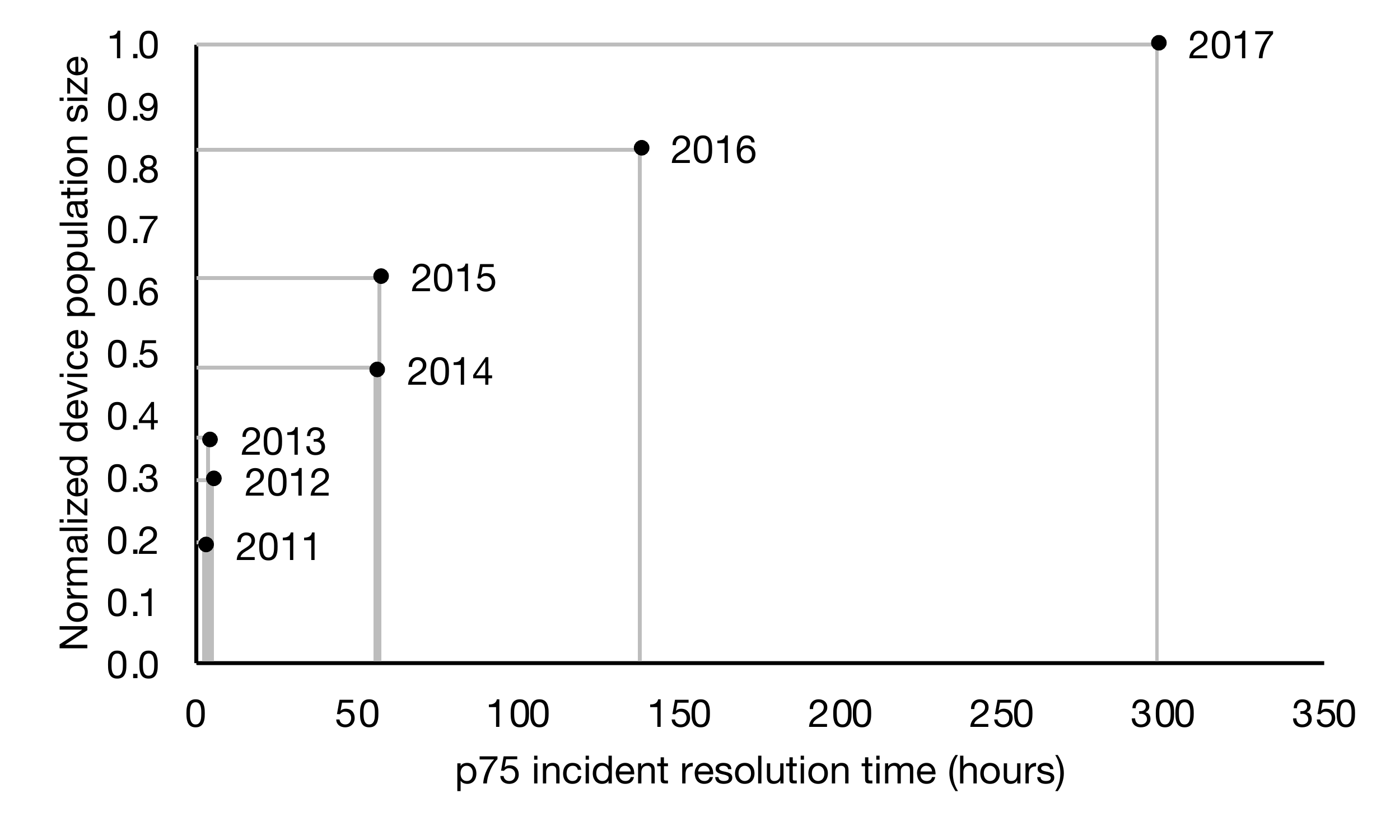}
  \caption{Average p75IRT per year compared to the population size of network devices in Facebook's data centers during that year.}
  \label{fig:switches-vs-p75irt}
\end{figure}

We find that, from 2011 to 2017, p75IRT \emph{increases similarly across device types}. The increase happens without significant changes to individual device design, operation, and management. To explain the overall increase in p75IRT, we plot p75IRT versus the normalized number of devices at Facebook in Figure~\ref{fig:switches-vs-p75irt}.

We observe a positive correlation between p75IRT and number of devices. At Facebook, we find that \emph{larger networks increase the time humans take to resolve network incidents.} We attribute part of the increases in resolution time to more standardized processes for releasing fixes to production infrastructure. Today, device configuration and software changes go through more thorough review processes, testing, and deployment than in the past.

We conclude that in terms of device reliability, incident rates vary by 3 orders of magnitude across device types in Facebook's data centers and incidents that happen in larger networks tend to have longer incident resolution times.

\section{Inter Data Center Reliability}
\label{sec:inter}

In this section, we study the reliability of backbone networks. We analyze network failures \emph{between} Facebook's data centers over the course of eighteen months, from October 2016 to April 2018, comprising tens of thousands of real world events, comparable in size to Turner et al.~\cite{Turner2010} and over three times as long of a timescale as Wu et al.~\cite{Wu2012}. We analyze two types of backbone network failures:

\begin{itemize}
  \item \textbf{\emph{Link failures,}} where an individual bundle of optical fiber linking two edge nodes (Figure~\ref{fig:network-arch}, \ding{196}) fails.
  \item \textbf{\emph{Edge node failures,}} where multiple link failures cause an edge node to fail. An edge node connects to the backbone and Internet using at least three links. When all of an edge node's links fail, the edge node fails.
\end{itemize}

Our backbone network dataset does not contain root causes. We measure \emph{mean time between failures (MTBF)} and \emph{mean time to recovery (MTTR)} for edge nodes and links. We analyze edge node reliability (\S\ref{sec:edge-reliability}), link reliability by fiber vendor (\S\ref{sec:reliability_fiber_vendor}), and edge node reliability by geography (\S\ref{sec:reliability_geo_location}).

\subsection{Edge Node Reliability}
\label{sec:edge-reliability}
\label{sec:mean-time-between-failures}
\label{sec:mean-time-to-recovery}


We first analyze the MTBF and MTTR of the edge nodes in Facebook's backbone network. An edge node fails when a combination of planned fiber maintenance or unplanned fiber cuts sever its backbone and Internet connectivity. An edge node recovers when repairs restore its backbone and Internet connectivity.

\begin{figure}[H]
  \centering
  \includegraphics[width=0.8\columnwidth]{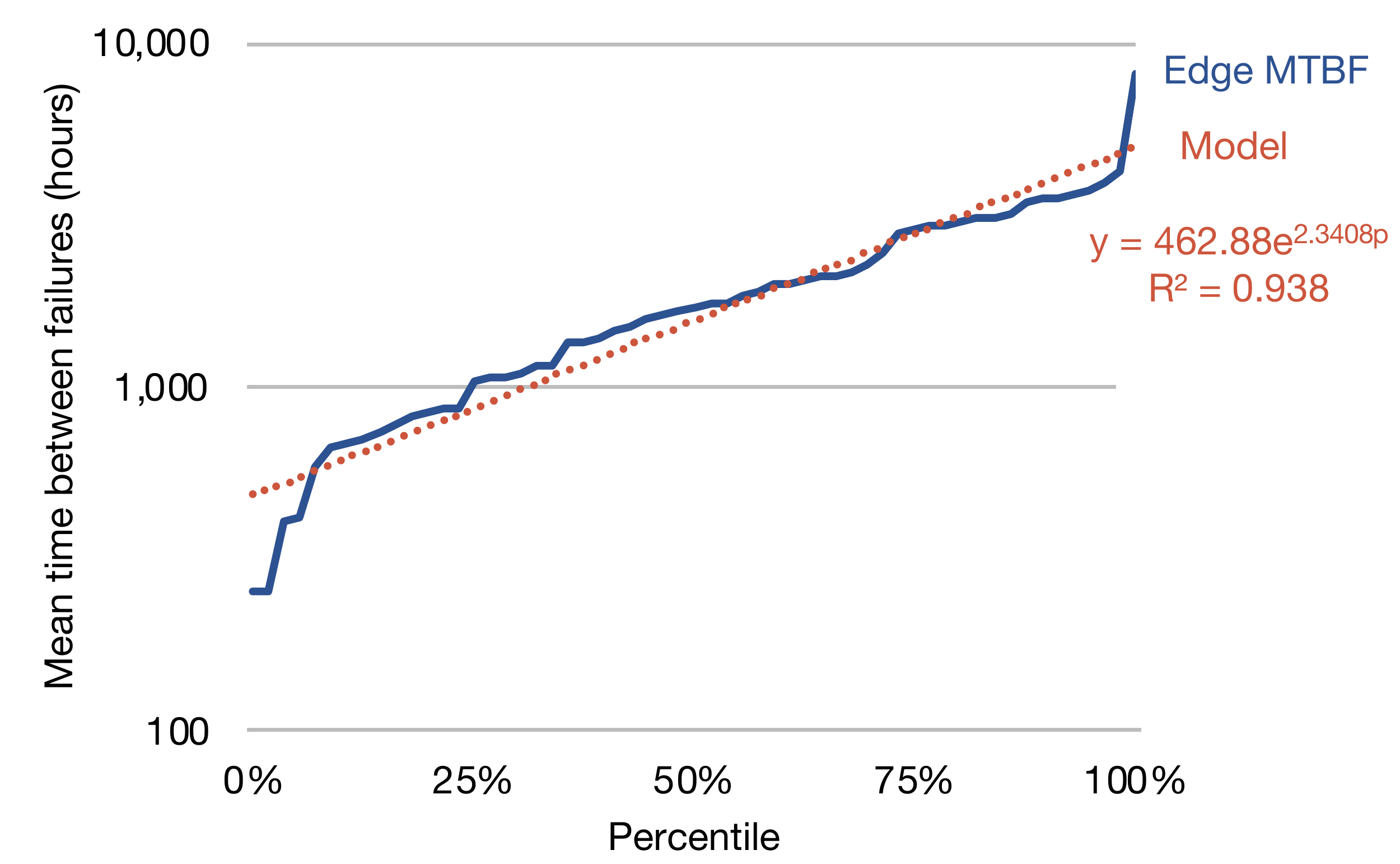}
  \caption{MTBF as a function of percentage of edge nodes connecting Facebook data centers with that MTBF or lower.}
  \label{fig:bb-mtbf}
\end{figure}

\textbf{\emph{MTBF.}} The solid line in Figure~\ref{fig:bb-mtbf} plots edge node MTBF in hours as a function of the percentage of edge nodes with that MTBF or lower. Most edge nodes fail infrequently because fiber vendors strive to maintain reliable links. 50\% of edge nodes fail less than once every \SI{1710}{\hour}, or \SI{2.3}{\months}. And 90\% of edge nodes fail less than once every \SI{3521}{\hour}, or \SI{4.8}{\months}.

Edge nodes exhibit high variance in MTBF due to their diverse fiber vendor makeup and geographic locations (observations we explore in \S\ref{sec:reliability_fiber_vendor} and \S\ref{sec:reliability_geo_location}). The standard deviation of edge node MTBF is \SI{1320}{\hour}, with the least reliable edge node failing, on average, once every \SI{253}{\hour} and the most reliable edge node failing, on average, once every \SI{8025}{\hour}.

We model $\mathit{MTBF}_\mathit{edge}(p)$ as an exponential function of the percentage of edge nodes, $0 \le p \le 1$, with that MTBF or lower. We built the models in this section by fitting an exponential function using the least squares method~\cite{Legendre1805}. At Facebook, we use these models in capacity planning to calculate \emph{conditional risk}, the probability of an edge node or link being unavailable or overloaded. We plan edge node and link capacity to ensure conditional risk is below 0.0001. We find that $\mathit{MTBF}_\mathit{edge}(p) = 462.88e^{2.3408p}$ (the dotted line in Figure~\ref{fig:bb-mtbf}) with $R^2 \approx 0.94$.

\begin{figure}[H]
  \centering
  \includegraphics[width=0.8\columnwidth]{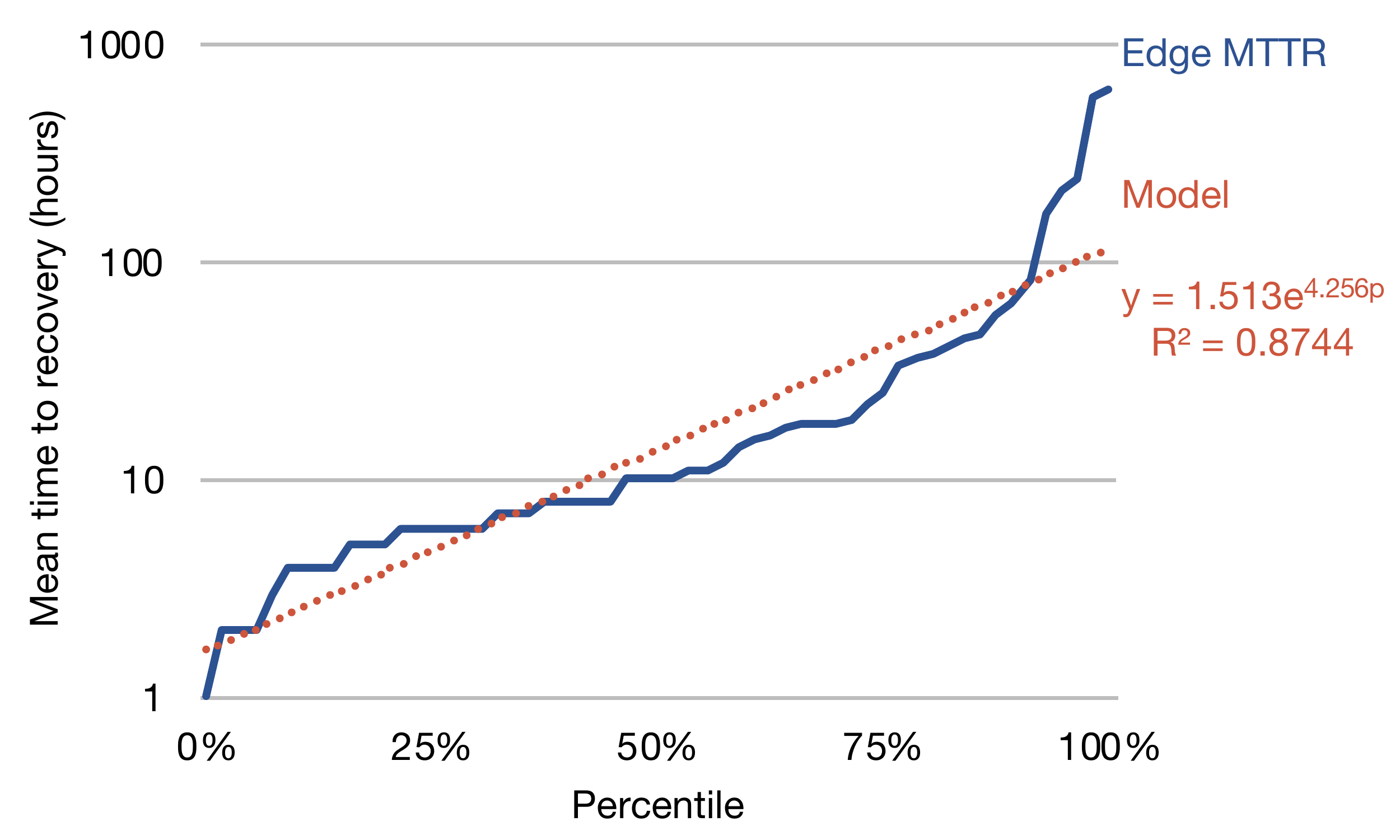}
  \caption{MTTR as a function of percentage of edge nodes connecting Facebook data centers with that MTTR or lower.}
  \label{fig:bb-mttr}
\end{figure}

\textbf{\emph{MTTR.}} The solid line in Figure~\ref{fig:bb-mttr} plots edge node MTTR in hours as a function of the percentage of edge nodes with that MTTR or lower. Edge node recovery occurs much faster than the time between failures because edge nodes contain multiple links (at least three) and fiber vendors work to repair link failures rapidly. 50\% of edge nodes recover within \SI{10}{\hour} of a failure; 90\% within \SI{71}{\hour}.

Edge nodes exhibit high variance in MTTR because some edge nodes are easier to repair than others. Imagine the differences between an edge node on a remote island versus an edge node in a big city. Weather conditions, physical terrain, and travel time affect the time it takes a fiber vendor to repair an edge node's links. The standard deviation of edge node MTTR is \SI{112}{\hour}, with the slowest edge node to recover taking \SI{608}{\hour} and the fastest edge node to recover taking \SI{1}{\hour}.

We model $\mathit{MTTR}_\mathit{edge}(p)$ as an exponential function of the percentage of edge nodes, $0 \le p \le 1$ with that MTTR or lower. We find that $\mathit{MTTR}_\mathit{edge}(p) = 1.513e^{4.256p}$ (the dotted line in Figure~\ref{fig:bb-mttr}) with $R^2 \approx 0.87$.

The high variances in edge node MTBF and MTTR motivate us to study the reliability characteristics of the links connecting edge nodes in \S\ref{sec:reliability_fiber_vendor} and the geographic location of edge nodes in \S\ref{sec:reliability_geo_location}.

\subsection{Link Reliability by Fiber Vendor}
\label{sec:reliability_fiber_vendor}


We analyze the MTBF and MTTR for fiber vendors using when the links they operate fail or recover. For brevity, we shorten \emph{``the MTBF/MTTR of the fiber a vendor operates''} to \emph{``fiber vendor MTBF/MTTR.''}

\begin{figure}[H]
  \centering
  \includegraphics[width=0.8\columnwidth]{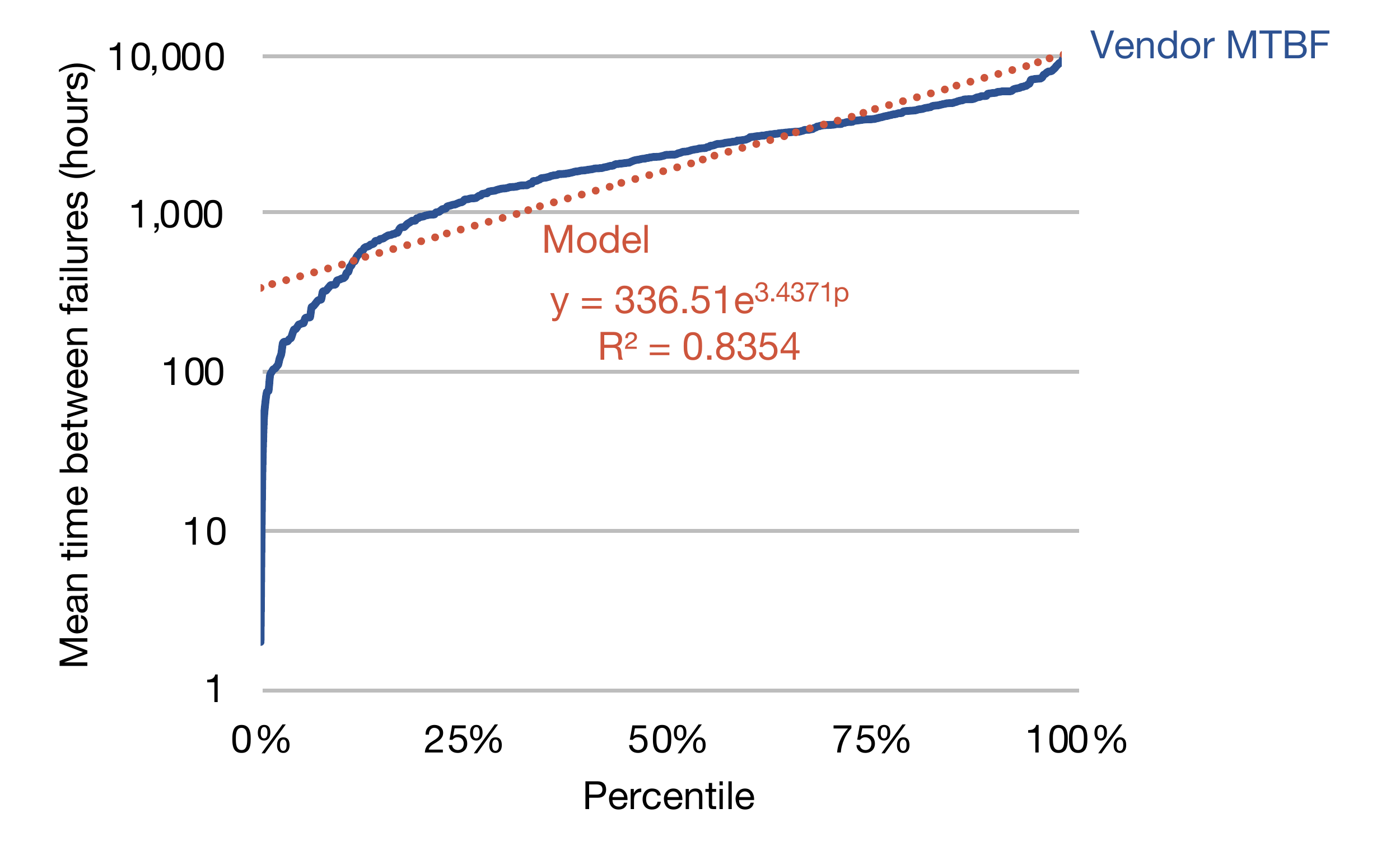}
  \caption{MTBF as a function of percentage of fiber vendors with that MTBF or lower.}
  \label{fig:bb-vendor-mtbf}
\end{figure}

\textbf{\emph{MTBF.}} The solid line in Figure~\ref{fig:bb-vendor-mtbf} plots the fiber vendor MTBF in hours as a function of the percentage of fiber vendors with that MTBF or lower. For most vendors, link failure happens only occasionally due to regular maintenance and monitoring. 50\% of vendors have links that fail less than once every \SI{2326}{\hour}, or \SI{3.2}{\months}. And 90\% of vendors have links that fail less than once every \SI{5709}{\hour}, or \SI{7.8}{\months}.

Fiber vendor MTBF varies by orders of magnitude. The standard deviation of fiber vendor MTBF is \SI{2207}{\hour}, with the least reliable vendor's links failing, on average, once every \SI{2}{\hour} and the most reliable vendor's links failing, on average, once every \SI{11721}{\hour}. Anecdotally, we observe that fiber markets with high competition lead to more incentive for fiber vendors to increase reliability. For example, the most reliable vendor operates in a big city in the USA.

We model $\mathit{MTBF}_\mathit{vendor}(p)$ as an exponential function of the percentage of vendors, $0 \le p \le 1$ with that MTBF or lower. We find that $\mathit{MTBF}_\mathit{vendor}(p) = 336.51e^{3.4371p}$ (the dotted line in Figure~\ref{fig:bb-vendor-mtbf}) with $R^2 \approx 0.84$.

\begin{figure}[H]
  \centering
  \includegraphics[width=0.8\columnwidth]{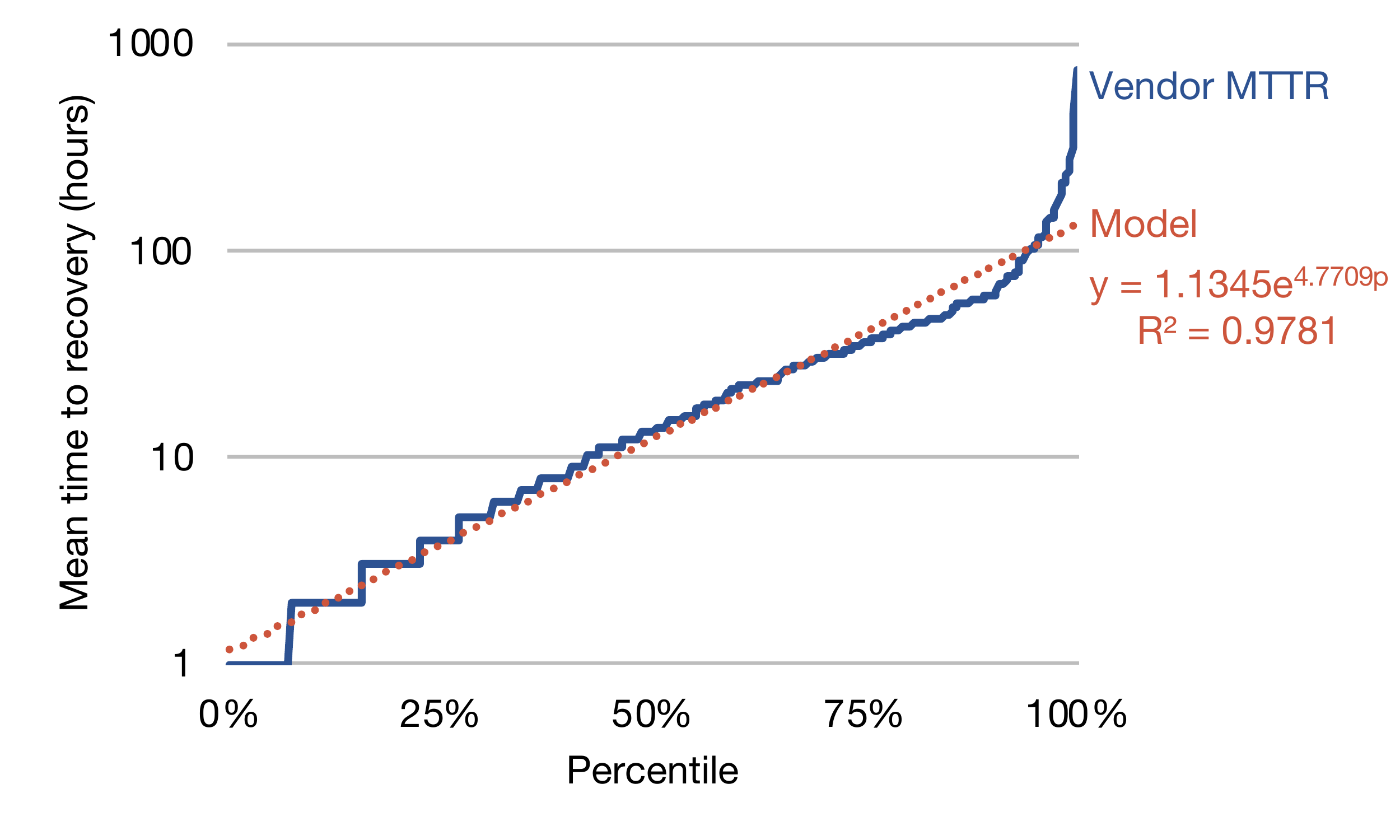}
  \caption{MTTR as a function of percentage of fiber vendors with that MTTR or lower.}
  \label{fig:bb-vendor-mttr}
\end{figure}

\textbf{\emph{MTTR.}} The solid line in Figure~\ref{fig:bb-vendor-mttr} plots fiber vendor MTTR as a function of the percentage of fiber vendors with that MTTR or lower. Most vendors repair links promptly. 50\% of vendors repair links within \SI{13}{\hour} of a failure; 90\% within \SI{60}{\hour}.

Fiber vendors exhibit high variance in MTTR because some fiber vendors operate in areas where they can more easily repair links (an observation we analyze in \S\ref{sec:reliability_geo_location}). The standard deviation of fiber vendor MTTR is \SI{56}{\hour}, with the slowest vendor taking on average \SI{744}{\hour} to repair their links and the fastest vendor taking on average \SI{1}{\hour} to repair their links.

We model $\mathit{MTTR}_\mathit{vendor}(p)$ as an exponential function of the percentage of vendors, $0 \le p \le 1$ with that MTTR or lower. We find that $\mathit{MTTR}_\mathit{vendor}(p) = 1.1345e^{4.7709p}$ (the dotted line in Figure~\ref{fig:bb-vendor-mttr}) with $R^2 \approx 0.98$.

We conclude that {\it not all fiber vendors operate equally}. We summarize the reliability models we develop in Table~\ref{tab:reliability-models}. We next explore how the geographic location of edge nodes affects their reliability.

  \begin{table}[H]
    \centering
    \begin{tabular}{lcr}
      \toprule
      \textbf{Reliability Model} & \textbf{Exponential Function} & $R^2$ \\
      \midrule
  Edge node MTBF    & $462.88e^{2.3408p}$ & 0.94 \\
  Edge node MTTR & $1.513e^{4.256p}$   & 0.87 \\
  Vendor MTBF & $336.51e^{3.4371p}$ & 0.84 \\
  Vendor MTTR & $1.1345e^{4.7709p}$ & 0.98 \\
      \bottomrule
    \end{tabular}
    \caption{Each reliability model is an exponential function expressing the MTBF or MTTR for a given percentile, $0 \leq p \leq 1$, of edge nodes (or vendors).}
    \label{tab:reliability-models}
  \end{table}

\subsection{Edge Node Reliability by Geography}
\label{sec:reliability_geo_location}


We analyze the reliability of edge nodes by their geographic location (the continent they reside in). Table~\ref{tab:geographic-location-reliability} shows the distribution of edge nodes in Facebook's network across continents. Most edge nodes reside in North America, with edge nodes in Europe following closely in size. The continents with the fewest edge nodes are Africa and Australia.

\begin{table}[H]
  \centering
  \begin{tabular}{lrrr}
    \toprule
    \textbf{Continent} & \textbf{Distribution} & \textbf{MTBF (h)} & \textbf{MTTR (h)} \\
    \midrule
    North America & 37\% & 1848 & 17 \\
    Europe        & 33\% & 2029 & 19 \\
    Asia          & 14\% & 2352 & 11 \\
    South America & 10\% & 1579 & 9  \\
    Africa        & 4\%  & 5400 & 22 \\
    Australia     & 2\%  & 1642 & 2  \\
    \bottomrule
  \end{tabular}
  \caption{The distribution and reliability of edge nodes in Facebook's network across continents.}
  \label{tab:geographic-location-reliability}
\end{table}

\textbf{\emph{MTBF.}} We show the average MTBF for the edge nodes in each continent in Table~\ref{tab:geographic-location-reliability}. Edge nodes in Africa are outliers, with an average MTBF of \SI{5400}{\hour}, or \SI{7.4}{\months}. Edge node reliability in Africa is important because edge nodes in Africa are few and connect Europe to Asia. Edge nodes in North America, South America, Europe, Asia, and Australia have average MTBFs ranging from \SI{1579}{\hour} (\SI{2.2}{\months}, for South America) to \SI{2352}{\hour} (\SI{3.2}{\months}, for Asia). The standard deviation of edge node MTBF across continents is \SI{1333}{\hour}, or \SI{1.8}{\months}.

\textbf{\emph{MTTR.}} We show the average MTTR for the edge nodes in each continent in Table~\ref{tab:geographic-location-reliability}. Across continents, edge nodes recover within \SI{1}{\day} on average. Edge nodes in Africa, despite their long uptime, take the longest time, on average, to recover (\SI{22}{\hour}), in part due to their submarine links. Edge nodes in Australia take the shortest time, on average, to recover (\SI{2}{\hour}), due to their locations in big cities. We observe a \SI{7}{\hour} standard deviation in edge node MTTR across continents.

We conclude that edge node failure rate varies by \emph{months} depending on the continent that edge nodes reside in, and edge nodes typically recover in \SI{1}{\day} across continents.

\section{Summary}

In this chapter, we present a large scale, longitudinal study of data center network reliability using operational data we collect from the production network infrastructure at Facebook. Our study spans thousands of intra data center network incidents across seven years, and eighteen months of inter data center network incidents. We show how the reliability characteristics of different network designs and different network device types manifest as network incidents and affect the software systems that use the network. Our key findings include:

\begin{description}
  \item[Lesson N.1] We observe that most failures that software cannot repair involve maintenance, faulty hardware, and misconfiguration. We also find $2\times$ more human errors than hardware errors as devices and routing configurations become more complex and challenging to maintain. [\S\ref{sec:root-causes}]
  \item[Lesson N.2] Network devices with higher bandwidth have a higher likelihood of affecting software systems. Network devices built from commodity chips have much lower incident rates compared to devices from third-party vendors due to the devices' integration with automated failover and remediation software. Rack switch incidents are increasing over time and are currently around 28\% of all network incidents. [\S\ref{sec:incident_rate}]
  \item[Lesson N.3] Although high bandwidth \emph{core} network devices have the most incidents, the incidents they have are low severity. Fabric network devices cause incidents of lower severity than cluster network devices. [\S\ref{sec:incident_impact}]
  \item[Lesson N.4] Cluster network incidents increased steadily over time until the adoption of fabric networks, with cluster networks currently having $2.8\times$ the incidents compared to fabric networks. [\S\ref{sec:incidents-by-topology}]
  \item[Lesson N.5] While high reliability is essential for widely-deployed devices, such as rack switches, incident rates vary by three orders of magnitude across device types. Larger networks tend to have longer incident remediation times. [\S\ref{sec:mtbi}]
  \item[Lesson N.6] We develop models for the reliability of Facebook's WAN, which consists of a diverse set of edges and links that form a backbone. We find that time to failure and time to repair closely follow \emph{exponential} functions. We provide models for these phenomena so that future studies can build on our models and use them to understand the nature of backbone failures. [\S\ref{sec:mean-time-between-failures}--\S\ref{sec:reliability_fiber_vendor}]
  \item[Lesson N.7] Backbone edge nodes that convey traffic between data centers fail on the order of months and recover on the order of hours. However, there is high variance in edge node failure rate and recovery rate.  Path diversity in the backbone topology ensures that large scale networks can tolerate failures with long repair times. [\S\ref{sec:mean-time-between-failures}]
  \item[Lesson N.8] Links that backbone vendors supply typically fail on the order of months, with links in big cities failing less frequently. Both failure rate and recovery rate for links span multiple orders of magnitude among vendors. [\S\ref{sec:reliability_fiber_vendor}]
  \item[Lesson N.9] Edge failure rate varies by months across continents in the world. Edges recover within \SI{1}{\day} on average on all continents. [\S\ref{sec:reliability_geo_location}]
\end{description}

As software systems grow in complexity, interconnectedness, and geographic distribution, unwanted behavior from network infrastructure has the potential to become a key limiting factor in the ability to reliably operate distributed software systems at a large scale.  It is our hope that the research community can build upon our comprehensive study to better characterize, understand, and improve the reliability of large scale data center networks and systems.

We examine DRAM devices in Chapter~\ref{chp:dramfailures}, flash-based SSD devices in Chapter~\ref{chp:ssdfailures}, and data center networks in this chapter, the next chapter summarizes the lessons we learn across our studies.

\chapter{Lessons Learned}
\label{sec:Lessons Learned Across Many Billions of Device-Hours}

We next summarize the key findings of our study, for DRAM, SSD, and network devices. These lessons form the basis for an understanding of the way the devices in modern data centers behave and how their behavior affects the software systems that run in the data centers.

\section{Lessons Learned for DRAM Devices}


In Chapter~\ref{chp:dramfailures}, we perform a comprehensive analysis of the memory errors across all of Facebook's servers over fourteen months.  We analyze a variety of factors and how they affect server failure rate and observe several new reliability trends for memory systems that have not been discussed before in literature.  We identify several important trends:

\begin{description}

\item[Lesson D.1] Memory errors follow a power-law distribution, specifically, a Pareto distribution with decreasing hazard rate, with average error rate exceeding median error rate by around $55\times$. [\S\ref{sec:incidence}]

\item[Lesson D.2] Non-DRAM memory failures from the memory controller and memory channel contribute the majority of errors and create a kind of \emph{denial of service attack} in servers. [\S\ref{sec:component}]

\item[Lesson D.3] More recent DRAM cell fabrication technologies (as indicated by chip density) show higher failure rates (prior work that measured DRAM \emph{capacity}, which is not closely related to fabrication technology, observed inconclusive trends). [\S\ref{subsec:DIMM Capacity and DRAM Density}]

\item[Lesson D.4] DIMM architecture decisions affect memory reliability: DIMMs with fewer chips and lower transfer widths have the lowest error rates, likely due to their lower electrical noise. [\S\ref{subsec:DIMM Architecture}]

\item[Lesson D.5] While CPU and memory utilization do not show clear trends with respect to failure rates, workload type can influence server failure rate by up to $6.5\times$. [\S\ref{sec:workload}]

\item[Lesson D.6] We show how to develop a model for memory failures and show how system design choices such as using lower density DIMMs and fewer processors can reduce failure rates of baseline servers by up to 57.7\%. [\S\ref{sec:model}]
  
\item[Lesson D.7] We perform the first analysis of page offlining in a real-world environment, showing that error rate can be reduced by around 67\% identifying and fixing several real-world challenges to the technique. [\S\ref{sec:page-offline}]

\item[Lesson D.8] We evaluate the efficacy of a new technique to reduce DRAM faults, \emph{physical page randomization}, and examine its potential for improving reliability and its overheads. [\S\ref{sec:reduce}]

\end{description}


We hope that the data and analyses presented in our work can aid in (1) clearing up potential inaccuracies and limitations in past studies' conclusions, (2) understanding the effects of different factors on memory reliability, (3) designing more reliable DIMM and memory system architectures, and (4) improving evaluation methodologies for future memory reliability studies.


\section{Lessons Learned for SSD Devices}

In Chapter~\ref{chp:ssdfailures}, we perform an extensive analysis of the effects of various factors on flash-based SSD reliability across a majority of the SSDs employed at Facebook, running production data center workloads.  We analyze a variety of internal and external characteristics of SSDs and examine how these characteristics affect the trends for large errors that the SSD alone cannot correct. We briefly summarize the key observations from our study and discuss their implications for SSD and system design:

\begin{description}

\item[Lesson S.1] We observe that SSDs go through several distinct periods---early detection, early failure, usable life, and wearout---with respect to the factors related to the amount of data written to flash chips. Due to pools of flash blocks with different reliability characteristics, failure rate in a population does not monotonically increase with respect to amount of data written.  This is unlike the failure rate trends seen in raw flash chips. [\S\ref{sec:corr}]
  
\item[Lesson S.2] We must design techniques to help reduce or tolerate errors \emph{throughout} SSD operation, not only toward the end of life of the SSD. For example, additional error correction at the beginning of an SSD's life could help reduce the failure rates we see during the \emph{early detection} period. [\S\ref{sec:corr}]

\item[Lesson S.3]  We find that the effect of read disturbance errors is not a predominant source of errors in the SSDs that we examine. While prior work has shown that such errors can occur under certain access patterns in controlled environments~\cite{brand-irps93, mielke-irps08, cai-date12, cai-dsn15}, we do {\em not} observe this effect across the SSDs we examine.  This corroborates prior work which showed that the effect of write errors in flash cells dominate error rate compared to read disturbance~\cite{mielke-irps08, cai-date12}.  It may be beneficial to perform a more detailed study of the effect of disturbance errors in flash-based SSDs used in servers. [\S\ref{sec:read}]

\item[Lesson S.4] Sparse logical data layout across an SSD's physical address space (e.g., non-contiguous data) greatly affects SSD failure rates; dense logical data layout (e.g., contiguous data) can also negatively impact reliability under certain conditions, likely due to adversarial access patterns. [\S\ref{subsec:DRAM Buffer Usage}]
  
\item[Lesson S.5] Further research into flash write coalescing policies with information from the system level may help improve SSD reliability.  For example, information about write access patterns from the operating system could potentially inform SSD controllers of non-contiguous data that is accessed most frequently, which may be one type of data that adversely affects SSD reliability and is a candidate for storing in a separate write buffer. [\S\ref{subsec:DRAM Buffer Usage}]

\item[Lesson S.6] Higher temperatures lead to increased failure rates, but do so most noticeably for SSDs that do not employ throttling techniques.  In general, we find techniques like \emph{throttling}, which are likely correlated with techniques to reduce SSD temperature, to be effective at reducing the failure rate of SSDs.  We also find that SSD temperature is correlated with the power used to transmit data across the PCIe bus that connects the SSD to the server's central processors. Power can thus potentially be used as a proxy for temperature in the absence of SSD temperature sensors. [\S\ref{sec:external}]

\item[Lesson S.7] We find that the amount of data written by the system software overstates the amount of data written to flash cells. This is because the operating system and SSD controller buffer certain data, so not every write in the system software translates to a write to SSD cells. Simply reducing the rate of software-level writes without considering the qualities of the write access pattern to system software is not sufficient for improving SSD reliability.  Studies seeking to model the effects of reducing software-level writes on flash reliability should also consider how other aspects of SSD operation, such as system-level buffering and SSD controller wear leveling, affect the actual data written to SSDs. [\S\ref{subsec:Data Written by the System Software}]

\end{description}

We hope that our new observations, with real workloads and real systems from the field can aid in (1) understanding the effects of different factors, including system software, applications, and SSD controllers on flash memory reliability; (2) designing more reliable flash architectures and system designs; and (3) improving the evaluation methodologies for future flash memory reliability studies.


\section{Lessons Learned for Network Devices}

In Chapter~\ref{chp:networkfailures}, we present a large scale, longitudinal study of data center network reliability based on operational data collected from the production network infrastructure at Facebook. Our study spans thousands of intra data center network incidents across seven years, and eighteen months of inter data center network incidents. We show how the reliability characteristics of different network designs and different network device types manifest as network incidents and affect the software systems that use the network. Our key findings include:

\begin{description}
  \item[Lesson N.1] We observe that most failures that software cannot repair involve maintenance, faulty hardware, and misconfiguration. We also find $2\times$ more human errors than hardware errors as devices and routing configurations become more complex and challenging to maintain. [\S\ref{sec:root-causes}]
  \item[Lesson N.2] Network devices with higher bandwidth have a higher likelihood of affecting software systems. Network devices built from commodity chips have much lower incident rates compared to devices from third-party vendors due to the devices' integration with automated failover and remediation software. Software incidents due to rack switches are increasing over time and are currently around 28\% of all network incidents. [\S\ref{sec:incident_rate}]
  \item[Lesson N.3] Although high bandwidth \emph{core} network devices have the most incidents, the incidents they have are low severity. Fabric network devices cause incidents of lower severity than cluster network devices. [\S\ref{sec:incident_impact}]
  \item[Lesson N.4] Cluster network incidents increased steadily over time until the adoption of fabric networks, with cluster networks currently having $2.8\times$ the incidents compared to fabric networks. [\S\ref{sec:incidents-by-topology}]
  \item[Lesson N.5] While high reliability is essential for widely-deployed devices, such as rack switches, incident rates vary by three orders of magnitude across device types. Larger networks tend to have longer incident remediation times. [\S\ref{sec:mtbi}]
  \item[Lesson N.6] We develop models for the reliability of Facebook's WAN, which consists of a diverse set of edges and links that form a backbone. We find that time to failure and time to repair closely follow \emph{exponential} functions. We provide models for these phenomena so that future studies can build on our models and use them to understand the nature of backbone failures. [\S\ref{sec:mean-time-between-failures}--\S\ref{sec:reliability_fiber_vendor}]
  \item[Lesson N.7] Backbone edge nodes that convey traffic between data centers fail on the order of months and recover on the order of hours. However, there is high variance in edge node failure rate and recovery rate.  Path diversity in the backbone topology ensures that large scale networks can tolerate failures with long repair times. [\S\ref{sec:mean-time-between-failures}]
  \item[Lesson N.8] Links that backbone vendors supply typically fail on the order of months, with links in big cities failing less frequently. Both failure rate and recovery rate for links span multiple orders of magnitude among vendors. [\S\ref{sec:reliability_fiber_vendor}]
  \item[Lesson N.9] Edge failure rate varies by months across continents in the world. Edges recover within \SI{1}{\day} on average on all continents. [\S\ref{sec:reliability_geo_location}]
\end{description}

As software systems grow in complexity, interconnectedness, and geographic distribution, unwanted behavior from network infrastructure has the potential to become a key limiting factor in the ability to reliably operate distributed software systems at a large scale.  It is our hope that the research community can build upon our comprehensive study to better characterize, understand, and improve the reliability of large scale data center networks and systems.

\section{Lessons Learned From Performing These Studies}

We performed three large scale studies of device reliability over the course of several years. A key lesson that we learned was that, even in large scale data centers, device failures may be infrequent enough to require years of data collection to identify failure trends, especially those associated with wear and aging. Sound statistical tools helped greatly in understanding how various factors affected device reliability and whether those effects were significant or just noise. Based on our experience, we recommend that future studies report similar sound statistical information when analyzing device failure trends.

\subsection{What We Would Change in These Studies}

If we were to perform these studies again, we would change several aspects of the studies:

\begin{itemize}
  \item \textbf{\emph{We would standardize the factors we measured across devices.}} While some of factors we examine are common across multiple devices (for example, we examine error rate across DRAM devices and SSD devices), most of the factors are disjoint across our studies. This makes it challenging to answer questions like, \emph{``Does a 10\degree C increase in temperature affect DRAM, SSD, or network devices the most?''}
  \item \textbf{\emph{We would standardize the modeling across devices.}} While we use a logistic regression~\cite{logit-1, logit-2} to build a model for DRAM reliability across a variety of factors, for network devices, we use the least squares method~\cite{Legendre1805} for modeling the reliability of network backbone edge nodes and links. We would like to have performed a similar logistic regression for SSD devices and network devices as it is a robust statistical tool for understanding how a large number of factors affect device reliability.
  \item \textbf{\emph{We would examine ways to prevent errors in SSD devices and network devices.}} While we analyzed two techniques for improving the reliability of DRAM devices (page offlining in~\S\ref{sec:page-offline} and physical page randomization in~\S\ref{sec:reduce}), we did not perform similar studies for SSD devices and network devices. We believe that this is a promising area for future research and we describe our recommendations for future studies in \S\ref{sec:Future Research Directions}.
\end{itemize}

\subsection{Limitations of These Studies}

We find it challenging to fully characterize the nature of failures across large populations of devices. To make our analysis tractable, we analyzed a subset of faults and characteristics of devices that we could collect data for. Because of this, our studies have several limitations:

\begin{itemize}

  \item \textbf{\emph{They only examine the data centers of one company.}} While we attempt to focus on the fundamental causes of device failure in our studies, our results also show that a device's \emph{workload} plays a role in its reliability (e.g.,~\S\ref{sec:workload} and \S\ref{subsec:Data Written by the System Software}). We hope that our work inspires others to perform a comparative study of the way that software affects device reliability across the data centers of different companies.

  \item \textbf{\emph{They do not consider the combined effects of failures in DRAM, SSD, and network devices.}} We examine only one device at a time and we do not report how one type of device's faults may influence others. We also do not examine \emph{the combined effect} of multiple types of device faults on software-level reliability. Nor do we examine widespread correlated failures (e.g., ``bad batches'' of devices). We recommend that future work in large scale device failure examine the relationship between \emph{different types} of device failures and how they are distributed across \emph{populations of devices.}

  \item \textbf{\emph{They do not consider silent data corruption.}} The servers we analyze do not allow us to determine if errors were due to bit flips that were undetectable using ECC metadata. We recommend that future large scale device failure studies use techniques like checkpointing~\cite{checkpointing} to examine the effects of silent data corruption.

\end{itemize}

\chapter{Conclusions and Future Research Directions}
\label{chp:conclusions}

In this dissertation, we introduce why device failures in modern data centers are a problem and discuss how we can use large scale device reliability studies to better tolerate device failures. Specifically, our thesis in this dissertation is, \emph{if we measure the device failures in modern data centers, then we can learn the reasons why devices fail, develop models to predict device failures, and learn from failure trends to make recommendations to enable systems to tolerate device failures}.

Our key finding in this study is that \emph{the problem of understanding why data center devices fail \textbf{can} be solved by using the scale of modern data centers to observe failures and by building robust statistical models to understand the implications of the failure trends.}

At the same time, we do not believe that it is tenable to perform such studies manually every several years for a select few types of devices. This is because the rapid pace of device development and deployment makes manual point-in-time studies an inefficient and incomplete way to predict future reliability trends. We conclude by proposing directions for future research that aim to make the process of learning about and improving device reliability self-sustaining using automation and inference.

\section{Future Research Directions}
\label{sec:Future Research Directions}


Our studies shed light on the causes and effects of memory, storage, and network devices in modern data centers.
Looking back over our work, we see a common theme emerge: if we can measure and model how devices fail in modern data centers, we can better tolerate those failures. One downside of this approach is that it requires humans to (1) manually collect and analyze device reliability data, (2) build models from the data to understand device failure trends, and (3) analyze those trends to understand the biggest reliability bottlenecks. We wonder, \emph{``What if systems could be designed to (1) introspect on how their actions affect the reliability of the devices they run on and (2) make decisions on how they operate to improve data center reliability?''} We call the practice of designing systems to be aware of reliability, \emph{introspective reliability system design}, and describe next some open questions and future directions to enable introspective system design.

\subsection{Motivation for Introspective Reliability System Design}

Hardware support for fault detection and performance analysis is providing systems with the opportunity to \emph{introspectively observe and take action toward preventing faults and errors}.  In this context, introspection refers to a system's ability to inspect its various hardware counters that are available in processors and peripherals to monitor the utilization and performance characteristics of itself.  If we can identify how such characteristics (as well as factors related to machine configuration) correlate with hardware failures, we can autonomously control and tune techniques to prevent faults in memory, storage, network, and other data center devices.

As one example, based on the trend we observed of increasing memory error rate with increasing memory utilization in Facebook's fleet (Chapter~\ref{chp:dramfailures}), we proposed a technique called Physical Page Randomization (\S\ref{sec:reduce}) to dynamically randomize physical page mappings, to spread memory accesses more evenly across memory, which can help prevent memory faults from occurring.  While the technique we proposed simply scans through and randomizes memory periodically, the technique could be adapted to benefit from system introspection by identifying and focusing its efforts on only the regions of memory that are the most highly utilized and thus most prone to memory faults.

However, improving data center reliability with such techniques poses several challenges in how fault vectors can be identified and analyzed, how introspection can be done efficiently and accurately using hardware and software monitoring, and how techniques can be designed to help prevent faults with low performance overhead and cost. To enable introspective reliability system designs, we must (1) systematically analyze the trends that cause hardware faults and (2) develop novel fault prevention techniques to solve these key challenges with introspective system monitoring.

The central vision of our proposed future work (based on the understandings we develop in this dissertation) is to develop introspective hardware/software fault monitoring and prevention techniques targeting a wide range of applications in servers, client systems, and mobile systems.  To realize this vision, we have identified three main thrusts of future work.

The first thrust, {\em field study-based statistical fault vector correlation and identification}, explores the correlations and causations of memory, storage, network, and other device type faults in the field using a statistical approach (similar to the studies we have performed in this dissertation for memory, storage, and network devices).  The second thrust, {\em hardware/software cooperative techniques for proactive fault prevention}, develops techniques to help prevent device faults based on the insights gained from the field study-based statistical fault vector correlation and identification.  The final thrust, {\em introspective hardware/software fault monitoring and reduction}, explores ways to use hardware and software monitoring techniques to track the fault vulnerability of systems and determine when to apply proactive fault prevention techniques.  The following sections provide a brief overview of each of the three research thrusts along with their key research challenges and questions. We hope future research tackles these challenges and questions.

\subsection{Field Study-Based Statistical Fault Vector Correlation and Identification}

The focus of this future research direction is on {\em collecting memory, storage, network, and other device type failure data in the field} and developing {\em statistical techniques and models for analyzing component failure trends}.  We have begun this work for memory, storage, and network devices in this dissertation and hope that the work continues for other devices, like emerging \emph{Non-Volatile Memory (NVM)}~\cite{Lee2009, Lee2010-2, Lee2010-3, Qureshi2009}, \emph{Graphics Processing Unit (GPU)}, \emph{Field-Programmable Gate Array (FPGA)}, special-purpose accelerator, disaggregated storage/memory, and \emph{Application-Specific Integrated Circuit (ASIC)} devices, to name a few. The three primary outcomes of this work should be (1) new statistical approaches that can be used to efficiently identify trends in large amount of failure data from the field, (2) the identification of new factors that correlate and cause different types of device failure, and (3) predictive models for analyzing trends in component failure.


There are at least four key research questions to answer in the context of field study-based statistical fault vector correlation and identification:

\begin{enumerate}
  \item \emph{What are field failure characteristics like for device types that have not been examined before, such as NVM, GPU, FPGA, special-purpose accelerator, disaggregated storage/memory, and ASIC devices?}
  \item \emph{What efficient statistical techniques exist to help analyze and correlate the large amounts of failure data that may be collected in the field?}
  \item \emph{What trends exist between failure rate and system characteristics like server age, CPU/memory/network/storage utilization, CPU/memory/network/storage size, ambient/device temperature, data center airflow/layout, and so on?}
  \item \emph{How do trends for data center devices compare to mobile devices, personal devices, and embedded devices?}
\end{enumerate}

\subsection{Hardware/Software Cooperative Techniques for Proactive Fault Prevention}

This future research direction leverages the insights from the data collected and analyzed in the first thrust to {\em design new hardware/software cooperative techniques to proactively provide fault prevention}.  The primary outcomes of this research should be a set of techniques that help prevent the occurrence of device faults.  As an example, in our analysis of memory errors at Facebook, we showed that one class of faults seemed to be caused due to wearout in memory devices over time (\S\ref{sec:reduce}, specifically, we found statistically-significant trends that showed that older devices and devices with more cores had higher error rates).  Based on this analysis, we proposed and implemented a technique in the Linux kernel that periodically shuffles the mappings of memory pages to more evenly spread utilization across main memory.  We also showed that such a technique could be run in the background of a system with low overhead.  While our prior work only examined one type of failure mode (wearout) in DRAM, we hope the community can extend the approach taken to examine other techniques that target different devices and different failure modes.


There are at least four key research questions to answer with regard to enabling proactive fault prevention:

\begin{enumerate}
  \item \emph{Given that we can identify correlations between system characteristics and their affect on device failure rates, how can these be tied back to hardware-level causes?}
  \item \emph{How can techniques be designed to help prevent such faults with high accuracy and low overhead?}
  \item \emph{What are the trade-offs---in terms of performance, energy-efficiency, and/or cost---involved in hardware and software cooperation toward preventing faults?}
  \item \emph{How can techniques be efficiently implemented using existing hardware and software, and what new system and device architectures can enable more efficient fault prevention?}
\end{enumerate}

\newpage

\subsection{Introspective Hardware/Software Fault Monitoring and Reduction}

This future research direction builds upon the previous two to {\em design and implement systems that are able to dynamically monitor and modify their behavior to respond to potential reliability threats using hardware/software cooperative techniques}.   As one example of this approach, consider the memory wearout trend we discussed in \S\ref{sec:workload} (higher error rates with different workload characteristics) and the prevention technique we proposed (dynamically randomizing memory layout to more evenly spread utilization):  If a machine could {\em introspectively monitor its behavior}, it could adapt its behavior based on a set of fault prevention ``rules'' (which we will need to develop) to reduce its fault risk.  In the case of DRAM, a system might occasionally monitor the way its workload accesses different physical memory locations (or the hardware could communicate this information to the system) and invoke memory layout randomization when it senses it would be most beneficial for potentially reducing faults.  This has the benefit of catering fault prevention policies to the needs of individual machines, while still ensuring that all machines remain protected.  Using this approach, we hope that future research can systematically investigate policies to prevent various types of hardware faults across memory and storage devices.


There are at least four key research questions that should be answered in relation to enabling introspective fault monitoring and reduction:

\begin{enumerate}
  \item \emph{Which sources of information about machine operation and behavior should be collected?}
  \item \emph{What hardware or software modifications could enable more accurate observation, communication, and collection of system behavior?}
  \item \emph{What types of monitoring and policy enforcement techniques are effective at helping prevent faults and how can these be designed with low performance, power, and/or cost overhead?}
  \item \emph{Can systems be designed to autonomously learn about and react to new potential fault vectors using a combination of data collection and online/offline analysis?}
\end{enumerate}

    
This future research direction has the potential to enable much more robust and reliable systems, and can help ensure that aggressive performance and power optimizations that may be employed on these systems (for example, workload consolidation~\cite{dean-cacm13} or the temperature-dependent effects of low-power idle states~\cite{Meisner2009}) do not disproportionately expose machines to higher fault rates.  The cross-cutting nature of this research will influence reliability analysis, application and system software, computer architecture, and device architecture, both for the devices within data centers as well as outside of them.

\section{Key Conclusion}

In this dissertation, we set out to validate the thesis, \emph{if we measure the device failures in modern data centers, then we can learn the reasons why devices fail, develop models to predict device failures, and learn from failure trends to make recommendations to enable workloads to tolerate device failures.} We perform a set of three large scale studies of DRAM, SSD, and network device reliability in a modern data center to analyze and model how DRAM, SSD, and network devices fail. Our key conclusion in this dissertation is that \textbf{\emph{we can gain a deep understanding of why devices fail---and how to predict their failure---using measurement and modeling.}} We hope that the analysis, techniques, and models we present in this dissertation will enable the community to better measure, understand, and prepare for the hardware reliability challenges we face in the future.

\appendix

\chapter{Other Works of the Author}
\label{chp:Other Works of the Author}

Before my work on data center reliability, I researched several areas of computer systems and architecture in collaboration with research scientists from Hewlett-Packard Labs and graduate students at CMU. I would like to acknowledge this work and those who contributed to it.

When I was an undergraduate student at the University of California at Los Angeles, I was fortunate to perform several internships under the mentorship of Rich Friedrich, Parthasarathy Ranganathan, Mehul Shah, and Jichuan Chang at Hewlett-Packard Labs. This work focused primarily on energy efficiency benchmarking~\cite{Riviore2007, Harizopoulos2009, Meza2009} and introducing the computer systems community to the concept of the lifecycle environmental impact of data centers~\cite{Meza2010, Chang2010, Chang2012}.

At Carnegie Mellon University, I collaborated with HanBin Yoon, Rachata Ausavarungnirun, Rachael Harding, Yixin Luo, Samira Khan, and Lavanya Subramanian. Our work focused on solving challenges in the use of non-volatile memory devices as the main memory in computing systems. We proposed efficient and scalable ways of using DRAM as a cache to NVM~\cite{Meza2012}, showed how to use row buffer locality to reduce device access latency and wear~\cite{Yoon2012}, proposed efficient device architectures for multi-level cell phase-change memory (PCM)~\cite{Yoon2014}, and examined interfaces to enable efficient persistent main memory systems~\cite{Meza2013}. We examined how to make the main memory system more resilient to bit flips~\cite{yixin} and summarized some of the challenges and opportunities for main memory design~\cite{kiist}. During this time I also collaborated with Jing Li at IBM to examine the effects of smaller row buffers in PCM~\cite{Meza2012-2, Meza2012-poster}.

While working at Facebook, I had the opportunity to help publish the designs of some of Facebook's large scale systems along with Qiang Wu, Qingyuan Deng, Lakshmi Ganesh, Chang-Hong Raymond Hsu, Yun Jin, Sanjeev Kumar, Bin Li, Kaushik Veeraraghavan, Yee Jiun Song, David Chou, Wonho Kim, Sonia Margulis, Scott Michelson, Rajesh Nishtala, Daniel Obenshain, Dmitri Perelman, Tuomas Pelkonen, Scott Franklin, Justin Teller, Paul Cavallero, Qi Huang, Tianyin Xu, Alex Gyori, Shruti Padmanabha, and Ashish Shah. We presented the design of Facebook's systems to manage data center power~\cite{Wu2016}, store operational time series data~\cite{Pelkonen2015}, production load testing~\cite{Veeraraghavan2016}, and disaster recovery~\cite{Veeraraghavan2018}.

\backmatter

\bibliography{ref/ref.bib}

\end{document}